\documentclass[a4paper,useAMS,usenatbib]{mn2e}

\usepackage{graphicx}
\usepackage[T1]{fontenc}
\usepackage{amsmath}
\usepackage{aas_macros}
\usepackage{times}
\usepackage{amssymb}
\usepackage{color}
\usepackage[total={17.8cm,24.0cm},centering]{geometry}

\newcommand{\kms}{km\,s$^{-1}$}
\newcommand{\dgr}{$^\circ$}

\newcommand{\reff}{r$_\mathrm{e}$}
\newcommand{\Qb}{Q$_{\rm_b}$}
\newcommand{\Qkin}{Q$_{\rm kin}$}

\title[The BaLROG project - I. Bar influence on kinematics]
{The BaLROG project - I. Quantifying the influence of bars on the kinematics of nearby galaxies}

\author[Seidel et al.]{M. K. Seidel$^{1,2}$\thanks{E-mail: mseidel@iac.es},
J. Falc\'{o}n-Barroso$^{1,2}$,
I. Mart\'inez-Valpuesta$^{1,2}$,
S. D\'iaz-Garc\'ia$^{3}$ 
\newauthor
E. Laurikainen$^{3,4}$,
H. Salo$^{3}$,
J. H. Knapen$^{1,2}$  \\
$^{1}$Instituto de Astrof\'isica de Canarias, E-38200 La Laguna, Tenerife, Spain\\
$^{2}$Departamento de Astrof\'isica, Universidad de La Laguna, E-38205 La Laguna, Tenerife, Spain\\
$^{3}$Astronomy and Space Physics, P.O. Box 3000, FI-90014 University of Oulu, Finland\\
$^{4}$Finnish Center for Astronomy with ESO (FINCA), University of Turku, Finland}

\begin{document}

\date{ }

\pagerange{\pageref{firstpage}--\pageref{lastpage}} \pubyear{2015}

\maketitle
\label{firstpage} 
\begin{abstract}
We present the \textit{BaLROG} (Bars in Low Redshift Optical Galaxies) sample of 16 morphologically
 distinct barred spirals to characterise observationally the influence of bars on nearby galaxies. Each 
 galaxy is a mosaic of several pointings observed with the IFU spectrograph SAURON leading to 
 a tenfold sharper spatial resolution ($\sim$100\,pc) compared to ongoing IFU surveys. In this paper we focus on
 the kinematic properties. We calculate the bar strength \Qb\ from classical torque analysis using
3.6 $\mu$m $Spitzer$ (S$^{4}$G) images, but also develop a new method based solely on the
    kinematics. A correlation between the two measurements is found and backed up by N-body
     simulations, verifying the measurement of \Qb. We find that bar strengths from ionised 
     gas kinematics are $\sim$2.5 larger than those measured from stellar kinematics and that stronger bars have enhanced influence on inner kinematic features.       
    We detect that stellar angular momentum "dips" at 0.2$\pm$0.1 bar lengths and half of our sample exhibits an anti-correlation of  h$_3$  - stellar velocity (v/$\sigma$) in these central parts. An increased flattening of the stellar $\sigma$ gradient with increasing bar strength supports the notion of bar-induced orbit mixing.
      These measurements set important 
      constraints on the spatial scales, namely an increasing influence in the central regions 
(0.1-0.5 bar lengths), revealed by kinematic signatures due to bar-driven secular evolution in present day galaxies.
\end{abstract}


\section{Introduction}
\label{sec:intro}

The elongated shapes of bars in nearby galaxies were already identified by
Hubble in the mid-30's \citep{1936Hubble}. By the 90s, reported bar fractions reached
values of $\sim$30\% in local galaxies, each in SBs and SABs \citep{1991RC3}. With the arrival of extensive imaging
surveys, several works reported that at least 2/3 of the local disc galaxies
exhibit bars \citep[e.g.,][]{2000Knapen,2000Eskridge, 2008Sheth}, with an increasing fraction
found in the infrared. Our own Galaxy was also found to host a bar
\citep[e.g.][]{1991Blitz}, confirmed by large spectroscopic surveys
detecting prominent cylindrical rotation (e.g. BRAVA \citealt{2009Howard}, ARGOS
\citealt{2013Ness}). Bars are ubiquitous in the local Universe
\citep[e.g.][]{2000Eskridge, 2000Knapen, 2002Whyte, 2007Marinova, 2007MenendezD,
2008Barazza, 2009Aguerri, 2010MendezA, 2011Masters, 2014MendezA, 2014Cisternas}
and have even been found at higher redshifts ($z$) \citep[e.g.,][]{1996Abraham,
2004Elmegreen, 2004Jogee, 2014Simmons}, but their high-$z$ fraction seems to be smaller than
in the local Universe \citep[e.g.,][]{2008Sheth, 2010NairA}. But how do bars form
and how do they influence the evolution of a galaxy?

In the early Universe, galaxies were closer together and their evolution
was dominated by interactions. Nowadays, due to the accelerating expansion of
the Universe, distances between galaxies are increasing.  Therefore, the
internal evolution experienced by the galaxy in isolation -- usually termed
\textit{secular evolution} -- is increasingly important. Theoretical studies propose that
bars have a great influence on this internal evolution (see, e.g., \citealt{korm_rev} for a review). N-body
simulations have helped to understand barred galaxies, their orbital
structures and their influence on different galaxy properties
\citep[e.g.][]{1981CombesSanders,1992Atha,1992Atha2, 2000DebattistaS, 2003Atha, 2005Bureau, 2006MartinezV,
2010Minchev, 2012Wang, 2013Atha}. Because of their significant departure from
axisymmetry and the associated torques, bars are likely to play a key role in
disc galaxy evolution via a number of processes: (1) redistribution of the
dissipative component (as witnessed by e.g., central mass concentration,
\citealt{1999Sakamoto, 1995Knapen}, or flattening of chemical abundance gradients,
\citealt{1994MartinR}); (2) triggered star formation, associated with bar-driven
crowding and shocks (e.g. spiral arms, rings) or central gas accumulation
(nuclear starbursts and AGN, e.g. \citealt{2011CoelhoG}) ; (3) redistribution of
the stellar component \citep[e.g.,][]{2001Gadotti}; (4) heating of the stellar
component and build-up of bulge structures \citep[e.g.,][]{1999Bureau, 2004Chung,
2003Fathi}. 

Many observational studies have investigated barred galaxies in profound detail
as well as statistically and provided insights on their photometry,
morphology, kinematics and stellar populations \citep[e.g.,][]{2005Sheth,
2006GadottideSouza, 2007Laurikainen, 2010Buta, 2011Laurikainen,
Perez2011,2011MNRAS.415..709S, 2014SB2, 2014BarreraB, 2014Kim}. Because of the
relative ease with which gas can be observed, the gas dynamics of barred
galaxies are fairly well known, strongly constraining the models available (hydro, SPH, sticky
particles). In contrast, the stellar component has not been studied extensively
yet, so that the dynamical theories (potential and orbital structure) remain
poorly tested. In particular, our study aims to perform a thorough and direct test of the
increasing effect on the stellar kinematics with increasing bar strength. The
consistency of bar-driven bulge formation scenarios (through gaseous inflow,
recurring bar formation/destruction, recurrent buckling events or other similar
mechanisms) also still needs to be understood in more detail. 

One of the most important parameters in the characterisation of bars (together
with properties such as bar length or pattern speed) is its "strength". Since
our aim is to quantify the influence of the bar on the host galaxy, we
start our study with a detailed determination of this parameter. Analysing our
sample in this way, we hope to find indications of common features that will be
more pronounced in more strongly barred systems. Numerous attempts have been made to
define a bar strength parameter in the past, such as the bar
axis ratio \citep[e.g.,][]{1995Martin, 1997MartinetF}. \citet{1998Aguerri} related the amplitude of $m=2$ and $m=0$
components as a measure of bar strength. 
In 1981, \citet{1981CombesSanders} suggested a measure of bar strength, based on the maximum of the bar induced tangential force, normalised to the axisymmetric radial force field, denoted as \Qb, and widely used in observational studies \citep[e.g.,][]{2001ButaB, 2002Laurikainen2, 2004Block, 2005Buta, 2010Salo}. To calculate \Qb\ in this study, we use the polar method \citep{1999Salo,2002Laurikainen2}.

Numerical simulations \citep[e.g.,][]{2003Atha, 2005Atha, 2005Bournaud2,
2012KimWT, 2012KimWT2, 2014Lokas} often use comparable measurements for bar
strength and find correlations of increased bar strength with increased angular momentum exchange, higher masses and velocity dispersions, but also investigate the influence of magnetic fields and dust
lanes \citep[e.g.,][]{2006Peeples}. In this paper we introduce a new method
of measuring the bar strength using stellar velocity maps, based on the
recent kinematic decomposition developed by \citet{2012Maciejewski}. This will
provide an independent view and check of the assumptions made when computing \Qb\
using only the photometry.

In order to quantify the influence of bars on the properties of their host
galaxies, we present in this paper integral-field spectroscopic observations of
a sample of 16 nearby barred galaxies. The panoramic field of view (FoV), which
we combined to large mosaics for each galaxy, allows us to investigate these
galaxies in unique detail and probe the influence of bars from the inner regions
up until co-rotation. This is the first of a series of papers trying to unravel
the influence of bars in detail combining this high spatial resolution dataset
with recent simulations and novel analysis techniques. In this paper we will focus on the
kinematics and establish a yardstick to measure the strength of the bars in
order to test their influence on other parameters. In a forthcoming paper we
will continue with a stellar population analysis (BaLROG II, Seidel et al. in
prep.).

The paper is organised as follows. In section~\ref{sec:data} we explain the
motivation of our sample, the observations and data reduction.
Section~\ref{sec:methods} summarises the methods employed, particularly
explaining our calculation of the kinematic torque. We present our results in
Section~\ref{sec:results} and discuss their implications in
Section~\ref{sec:discuss}. Our main conclusions are summarised in
Section~\ref{sec:summ}.  Throughout the paper the cosmological parameters used
are H$_0$\,=\,67\,\kms\,s$^{-1}$\,Mpc$^{-1}$, $\Omega_\Lambda$\,=\,0.7 and
$\Omega_m$\,=\,0.3 \citep{2014Planck} .

\begin{figure*}
\includegraphics[width=0.88\linewidth]{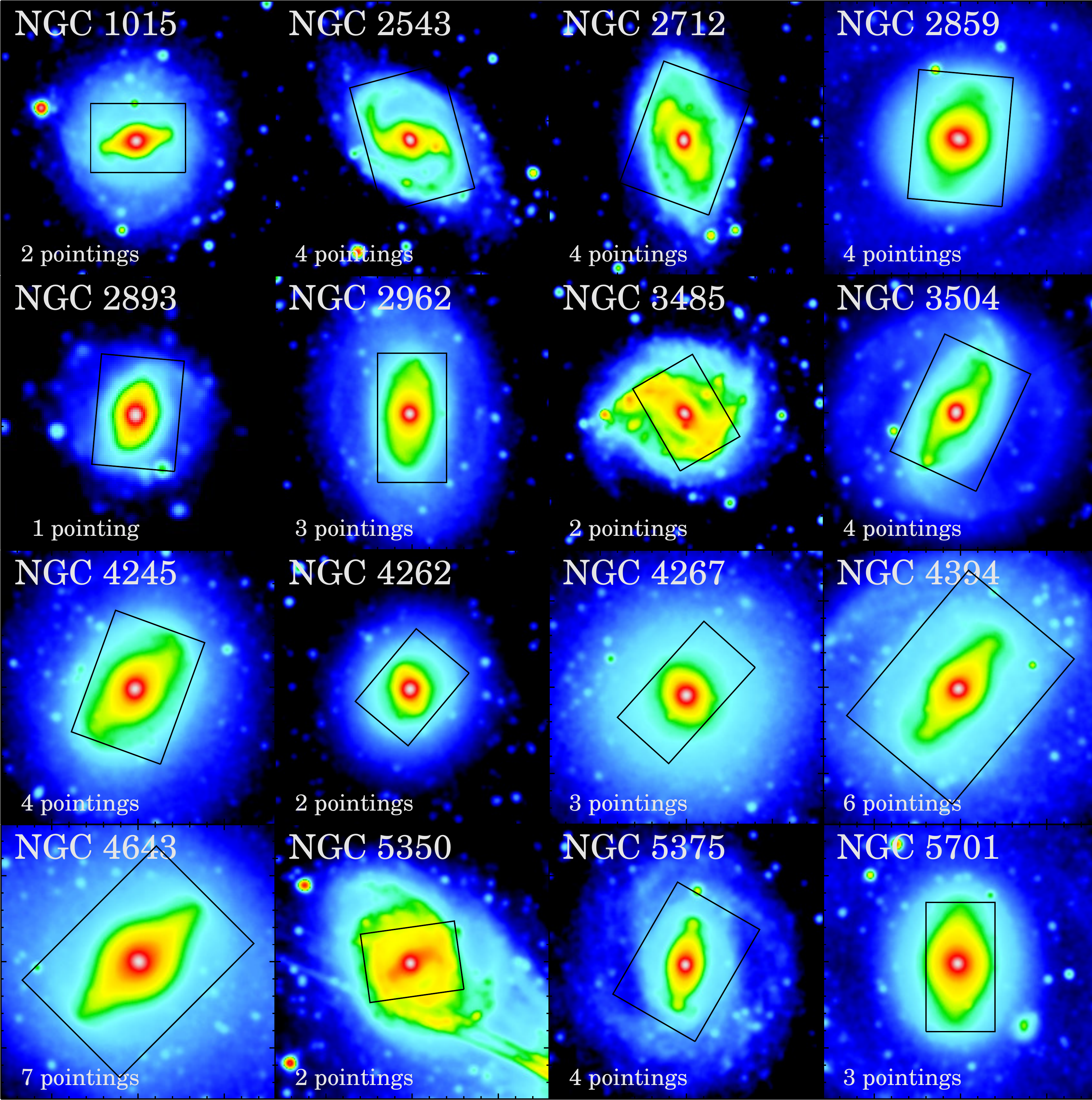}
 \vspace{.5cm}
\caption{$Spitzer$ images drawn from the S$^4$G survey of nearby galaxies
\citep{2010Sheth} of our sample observed with SAURON. The number of IFU
pointings is indicated in the left lower part for each galaxy. The final SAURON
mosaic, composed of individual pointings of 30$\times$40 arcsec FoV, is shown on
top of the images. North is up and East is towards the left in all cases. The
extensions of the S$^4$G fields shown here (squares with indicated mosaic on top) are 160\,$\times$\,160 arcsec, except for
NGC\,2893 where it is only 100$\times$100 arcsec.}
\label{fig:sample}
\end{figure*}

\begin{table*}
\centering
\caption{{\textsc Galaxy Sample}.~-~(1) Galaxy name, (2) Hubble type, (3, 4)
J2000 coordinates (right ascension, declination), (5) systemic velocity, (6)
inclination, (7) bar length, (8) position angle, (9) bar position angle, (10)
effective radius (\reff). Objects forming part of the Virgo cluster are
marked with a small $v$ next to their name. {\textsc Notes.~-~}All
morphological classifications, coordinates and V$_\mathrm{sys}$ are from the
\textit{NASA Extragalactic Database (NED)}, http://ned.ipac.caltech.edu/. All
inclinations, PAs and effective radii of the galaxy
(\reff) are from the S$^4$G P4 (Salo et al. 2015, accepted to ApJS) while bar lengths and bar PAs are
determined by Herrera-Endoqui et al. (2015, submitted) (here the ones by visual inspection).}
\begin{tabular}{lcccccccccc}
\hline
Galaxy &  Hubble & RA            & Dec           & V$_\mathrm{sys}$ &  Inclination & Bar Length & PA        & PA$_\mathrm{bar}$ & \reff    \\
~      &  Type   & (hh mm ss.ss) & (dd mm ss.ss) & (\kms)           &  (deg)       & (arcsec)   & (degrees) & (degrees)         & (arcsec) \\
(1)    &  (2)    & (3)           & (4)           & (5)              & (6)          & (7)        & (8)       & (9)               & (10)     \\
\hline          
NGC\,1015      & SBa             & 02 38 11.56 &  -01 19 07.3 &  2628           &  30.5  &  21.5  &  $-8.3$  & 101.4           &  19.73  \\
NGC\,2543      & SB(s)b          & 08 12 57.92 &  +36 15 16.7 &  2471           &  59.9  &  14.9  &  37.0    & 105.5           &  26.56  \\
NGC\,2712      & SBb             & 08 59 30.47 &  +44 54 50.0 &  1815           &  60.5  &  20.5  &  3.6     & \phantom{0}22.6 &  25.09  \\
NGC\,2859      & (R)SB0$^+$(r)   & 09 24 18.53 &  +34 30 48.6 &  1687           &  37.2  &  34.4  &  1.8     & 169.6           &  22.41  \\
NGC\,2893      & SB0-a           & 09 30 16.96 &  +29 32 23.9 &  1703           &  17.4  &  12.0  &  88.1    & 164.1           &  \phantom{0}4.58  \\
NGC\,2962      & SB0-a           & 09 40 53.93 &  +05 09 56.9 &  1960           &  49.0  &  30.7  &  8.3     & 172.7           &  20.50  \\
NGC\,3485      & SBb             & 11 00 02.38 &  +14 50 29.7 &  1436           &  20.4  &  21.0  &  $-64.6$ & \phantom{0}40.6 &  26.38  \\
NGC\,3504      & SBab            & 11 03 11.21 &  +27 58 21.0 &  1539           &  12.8  &  37.1  &  $-41.7$ & 148.3           &  11.13  \\
NGC\,4245      & SB0/a(r)        & 12 17 36.77 &  +29 36 28.8 &  \phantom{0}886 &  33.3  &  36.3  &  0.5     & 131.0           &  23.52  \\
NGC\,4262$^v$  & SB0$^-$(s)      & 12 19 30.57 &  +14 52 39.6 &  1359           &  24.5  &  13.4  &  $-6.0$  & \phantom{0}26.5 &  \phantom{0}5.99  \\
NGC\,4267$^v$  & SB0$^-$         & 12 19 45.24 &  +12 47 53.8 &  \phantom{0}983 &  11.9  &  16.9  &  $-27.5$ & \phantom{0}34.0 &  21.07  \\
NGC\,4394$^v$  & (R)SB(r)b       & 12 25 55.53 &  +18 12 50.6 &  \phantom{0}922 &  30.4  &  41.4  &  $-57.6$ & 143.4           &  36.79  \\
NGC\,4643      & SB0/a(rs)       & 12 43 20.14 &  +01 58 41.8 &  1330           &  36.8  &  49.9  &  56.0    & 133.3           &  24.22  \\
NGC\,5350      & SBbc            & 13 53 21.63 &  +40 21 50.2 &  2321           &  50.3  &  15.2  &  7.9     & 120.8           &  28.06  \\
NGC\,5375      & SBab            & 13 56 56.00 &  +29 09 51.7 &  2386           &  29.8  &  27.2  &  $-9.4$  & 171.1           &  24.35  \\
NGC\,5701      & (R)SB0/a(rs)    & 14 39 11.08 &  +05 21 48.5 &  1505           &  15.2  &  39.0  &  52.0    & 174.9           &  25.97  \\
\hline
\end{tabular}
\label{paramsgals}
\end{table*}

\section{Data}
\label{sec:data}

\subsection{Sample selection}

The parent sample from which our target galaxies are drawn is the S$^4$G survey
of nearby galaxies \citep{2010Sheth}. We restricted our initial choice to barred
galaxies with inclinations below 70\dgr\ and brighter than $M$$_B=-18.0$\,mag to
ensure high quality data. As we took our sample from the S$^4$G survey, we were naturally constrained to those
galaxies with $cz\leq3000$\,\kms\ so that important spectral features (H$\beta$,
Mg$b$) remain within the wavelength range probed by SAURON. This instrument is
mounted on the William Herschel Telescope (WHT) in La Palma at the Observatorio
del Roque de los Muchachos. Therefore, galaxies with sky declinations between
$-2^\circ\leq\delta\leq60^\circ$ were chosen to achieve optimal visibility. The entire 
exercise resulted in a large number of galaxies ($\sim$\,100), most
of which were located in the vicinity of or within the Virgo Cluster. Full 2D
spectroscopic analysis of a large sample was beyond our capabilities in terms of
observing time. We thus carefully inspected different sets
of numerical simulations and images of the S$^4$G survey and selected those galaxies with
prominent bars in different apparent stages of their evolution and with
different morphologies. In addition, we selected both early-type and late-type
galaxies and those with inclinations below 60\dgr\ to reduce uncertainties (e.g. in the \Qb\ determination).

Our target sample consists of 16 galaxies (see Fig.~\ref{fig:sample}),
a number that provides a reasonable representation of different types of bars.
The number of SAURON pointings greatly exceeds this number: the dataset for each galaxy is a
mosaic of several pointings (up to seven individual IFU pointings) allowing us
to reach the spatial detail that we aimed for while also covering the bars out to
the beginning of their surrounding discs. The limitations of the size of our
sample are obvious. However, while large integral field surveys such as CALIFA,
SAMI or MaNGA provide large enough samples for statistics, they lack the
detailed spatial sampling provided by this work (e.g. we sample at typically 100~pc, even maintained with our Voronoi-binning (within the bar region) which is in most cases at least a factor 10 better than the larger surveys). Table~\ref{paramsgals} gives the entire list of observed targets and basic properties.

\subsection{Observations}

The observations were carried out in four consecutive runs in March 2012,
January 2013, April 2013 and January 2014 at the WHT in La Palma with the SAURON integral
field unit \citep{2001Bacon}.     In the employed low-resolution mode, this
instrument has a field of view of 33\arcsec\,$\times$\,41\arcsec, spatial
sampling of 0\farcs94\,$\times$\,0\farcs94 per lenslet (1431 in total) and a
spectral resolution of full width at half maximum (FWHM) of 3.9\,\AA. Its
wavelength coverage ranges from 4760\,\AA\ to 5300\,\AA, leading to the above
mentioned redshift limitations chosen to include important emission and stellar
absorption line features. 

We observed up to 7 SAURON pointings per galaxy to build a large mosaic. The
final maps extend along the bars up until the start of the disc, allowing us to
probe radial dependencies within and outside the bar, while also resolving great
spatial detail. This strategy was quite costly in time: for the small sample of
16 galaxies we invested 54 pointings in total, each of 1--2\,hours depending on
the galaxy's surface brightness. Table~\ref{paramsgals2} summarises the number
of pointings and the total exposure times for each galaxy.
Figure~\ref{fig:sample} shows the final extent of the mosaic overlaid on top of
the S$^4$G images. 

Apart from the large pointing offsets, we introduced small dithers within each
pointing of typically 1 to 2\,\arcsec. This helps us to account for a couple of
bad columns in the CCD and to improve our sampling. The orientation of the field
of view (FoV) of SAURON was such that the 146 sky lenslets always pointed away
from the galaxy's centre. They are 1.9\,\arcmin\ from the main FoV and thus
ensure a simultaneous sky exposure during the object exposure.

We took a calibration frame using a Neon lamp before and after each science 
frame. Skyflats were taken at dusk and dawn, as well as continuum lamp 
exposures with a Tungsten lamp. For the flux calibration we observed several 
spectrophotometric standard stars. For further spectral calibration, we also 
observed a broad range of stars with different spectral types from the MILES 
database\footnote{http://miles.iac.es} \citep{2006SB}.

\begin{table}
\centering
\caption[Short\protect\\text]{Summary of the observations: (1) NGC number, (2) Run number, (3) Pointing number, (4) Total exposure time, in seconds.}
\begin{tabular}{cccr}
\hline
Galaxy & Run &  \# P & T$_\mathrm{exp}$ \\
(1) & (2) & (3) & (4) \\
\hline
NGC\,1015  &  2    &  2  & 12\,$\times$\,1800 \\
NGC\,2543  &  1,2  &  3  & 10\,$\times$\,1800 \\
NGC\,2712  &  4    &  4  &  8\,$\times$\,1800 \\
NGC\,2859  &  1    &  4  & 16\,$\times$\,1800 \\
NGC\,2893  &  3    &  1  &  8\,$\times$\,1800 \\
NGC\,2962  &  4    &  3  &  9\,$\times$\,1800 \\
NGC\,3485  &  4    &  2  &  4\,$\times$\,1800 \\
NGC\,3504  &  3    &  4  & 12\,$\times$\,1800 \\
NGC\,4245  &  2    &  4  & 16\,$\times$\,1800 \\
NGC\,4262  &  1    &  2  &  8\,$\times$\,1800 \\
NGC\,4267  &  1    &  3  & 12\,$\times$\,1800 \\
NGC\,4394  &  2    &  6  & 24\,$\times$\,1800 \\
NGC\,4643  &  3    &  7  & 23\,$\times$\,1800 \\
NGC\,5350  &  2    &  2  &  8\,$\times$\,1800 \\
NGC\,5375  &  4    &  4  & 16\,$\times$\,1800 \\
NGC\,5701  &  1    &  3  & 12\,$\times$\,1800 \\
\hline
\end{tabular}
\label{paramsgals2}
\end{table}

\subsection{Data reduction}

The reduction was performed with the available SAURON pipeline XSauron
described in detail in \citet{2001Bacon}. The preprocessing of raw frames 
includes overscan and bias subtraction. The evaluation of dark frames showed that the dark
current is negligible: less than 1\,e$^-$pixel$^{-1}$h$^{-1}$. All frames were 
preprocessed in this same way. After that, a model mask was created to extract the 
spectra. This mask builds a table indicating corresponding positions by 
relating the pixels on the CCD to their associated wavelengths and lenslets. 
The outcome is a set of three-dimensional data cubes ($\alpha$, $\delta$, $\lambda$). 
Wavelength calibration was achieved with the arc (neon) lamp exposures. A 
cross-correlation function between the neon frames taken before and after the 
science exposure and the one of the extraction mask defines potential slight 
offsets between the science frame and the mask. This analysis is based on 11 
emission lines which can be seen in the wavelength range of SAURON. 

The flat-fields were created with a combination of twilight and continuum lamp 
(tungsten) exposures. The former calibrates the spatial component, while 
the latter is responsible for the spectral coordinates. We used for each run a 
representative twilight and continuum flat exposure investigating counts and 
distributions of all flat exposures. On a case-by-case basis, we also chose 
night-dependent flats, but for the vast majority and thanks to our bright 
objects, the former method proved to work well. Cosmic rays were removed before 
the sky subtraction, where the median of the 146 dedicate sky lenslet values 
was computed and subtracted from the science frame spectra. Flux calibration 
was done using the spectrophotometric standards. Their flux correction curve 
was extracted comparing the observed curve with a reference spectrum. The 
resulting correction curve was used to calibrate all science frames. The 
merging and mosaicking of the individual data cubes was achieved with the 
XSAURON software using the integrated intensity contours in comparison with 
those of a $g$-band SDSS image. The entire mosaic was constructed with the 
obtained offsets and scalings between each image.

\subsection{S$^4$G data}

We complement our SAURON mosaics with $Spitzer$ 3.6$\mu$m images from the S$^4$G
\citep{2010Sheth}. As the $Spitzer$ images are very deep, the
outer isophotes are typically close to or beyond the 3.6 = 25.5
mag/arcsec$^2$ in the AB magnitude system, the position angles (PA)
and ellipticities ($\ell$) are taken from the S4G pipeline 4
(Salo et al. 2015, accepted to ApJ). Global galaxy parameters such as the
effective radii of the galaxies (\reff) are from Mu\~noz-Mateos et
al. (2015, submitted) and the barlength measurements are from
Herrera-Endoqui et al. (2015, submitted). In addition, we used the $Spitzer$
images to compute bar strengths for our sample as described in
Section~\ref{photT}. The \Qb\ measurements for the complete S$^4$G are
given in D\'iaz-Garc\'ia et al. (2015, submitted).

\section{Methods}
\label{sec:methods}

A detailed analysis of the stellar and gas kinematics requires a minimum
signal-to-noise ratio (S/N) \citep[e.g.,][]{marel1993}. We adopted the Voronoi
binning scheme of \citet{2003Cappellari} and applied it to our data to reach a
minimum S/N$\approx$40 per pixel for all galaxies. The central spectra remained
unbinned in all cases and exceeded this S/N level (e.g. S/N$>$100) while in the barred regions, we reach a typical S/N of . While a S/N
of 40 ensures high-quality spectra for the extraction of the mean stellar
velocity, velocity dispersion as well as Gauss-Hermite moments h$_3$ and h$_4$,
it is also low enough to preserve the spatial substructures in the galaxies, as seen in the resulting maps (see Appendix~\ref{app:gaskinmaps}).
Before we binned, we also ensured that we would not contaminate our measurements
by poor quality spaxels. Therefore we excluded those spaxels with a S/N below 3
and then limited the data to an isophote with at least this average S/N level.
The resulting extensions of the maps are hence due to the combined mosaic and
this additional S/N minimum threshold.

\subsection{Stellar kinematics}
\label{ppxf}

We extracted the stellar kinematics using the pPXF -- penalized pixel fitting --
code developed by \cite{capems_2004}. The routine fits each galaxy spectrum with
a combination of template spectra from a given library. Here we use a subset of Medium-resolution Isaac Newton Telescope library of empirical spectra (MILES; \citealt{2006SB}) single stellar population (SSP) model spectra \citep{2010Vaz} with a
a range of ages and metallicities of 0.1\,Gyr to 17.8\,Gyr and
-0.40$<$[Z/H]$<$+0.22, respectively. Their mean resolution is of
FWHM\,=\,2.51\,\AA\,\citep{miles} and before the fitting process, we matched the
spectral resolution of the models to that of our data. Throughout this work we
assume a Kroupa initial mass function \citep[IMF,][]{2001Kroupa}. The result of
pPXF is a line-of-sight velocity distribution (LOSVD) described by a
Gauss-Hermite parametrisation \citep{gerhard1993, marel1993} allowing the
measurement of the velocity ($V$), velocity dispersion ($\sigma$) and higher
order Gauss-Hermite moments (h$_3$ and h$_4$). From the stellar velocity and
velocity dispersion, we also calculated the value of the specific stellar
angular momentum $\lambda_{\rm R}$ \citep{2007Emsellem}, radially but also
integrated within 1\,\reff. Using the code developed by \citet{2012Maciejewski},
we furthermore obtained the radial and tangential velocities $V_r$ and $V_t$ for
a subset of our sample (see Section~\ref{kinT} for more details).

\subsection{Emission lines}
\label{gandalf}

We used the Gas AND Absorption Line Fitting ({\tt GANDALF}) package by
\citet{sarzi2006} and \citet{FalconBarroso2006} to determine the ionised-gas
distribution and kinematics. The code treats the emission lines as additional
Gaussian templates on top of the stellar continuum and iteratively searches for
the best match of their velocities and velocity dispersions. The SAURON
wavelength range allows us to measure the emission line of
H$\beta\lambda\lambda$4861 and the doublets [O{\sc iii}]$\lambda\lambda$4959, 5007
and [N{\sc i}]$\lambda\lambda$5200, 5202\,\AA. We tied spectral lines
kinematically to the [O{\sc iii}] doublet to lower the number of free parameters
given to {\tt GANDALF}. We checked that leaving them free resulted in a
consistent outcome. Furthermore, we imposed known relative flux relations to
constrain the freedom of the doublet lines during the fitting process, namely
$F([$O{\sc iii}$]_{4959})=0.350\cdot F([$O{\sc iii}$]_{5007})$.

We will use the results from {\tt GANDALF} to \textit{clean} the spectra and
produce emission-line-free datacubes for our stellar population analysis, to be
presented in our Paper II.

\subsection{Bar strength measurements}
\label{sec:barstr}

Bar strengths have been measured in many different ways (see introduction for
details). For our analysis we will concentrate on the following two methods: (1)
the photometric torque \Qb\ taking advantage of the S$^4$G data and (2) a new
measurement based on the stellar velocity maps which does not include strong
model assumptions (\Qkin).

\subsubsection{Photometric torque using 3.6 $\mu$m $Spitzer$ imaging (\Qb)}
\label{photT}

\begin{figure}
\includegraphics[width=\linewidth]{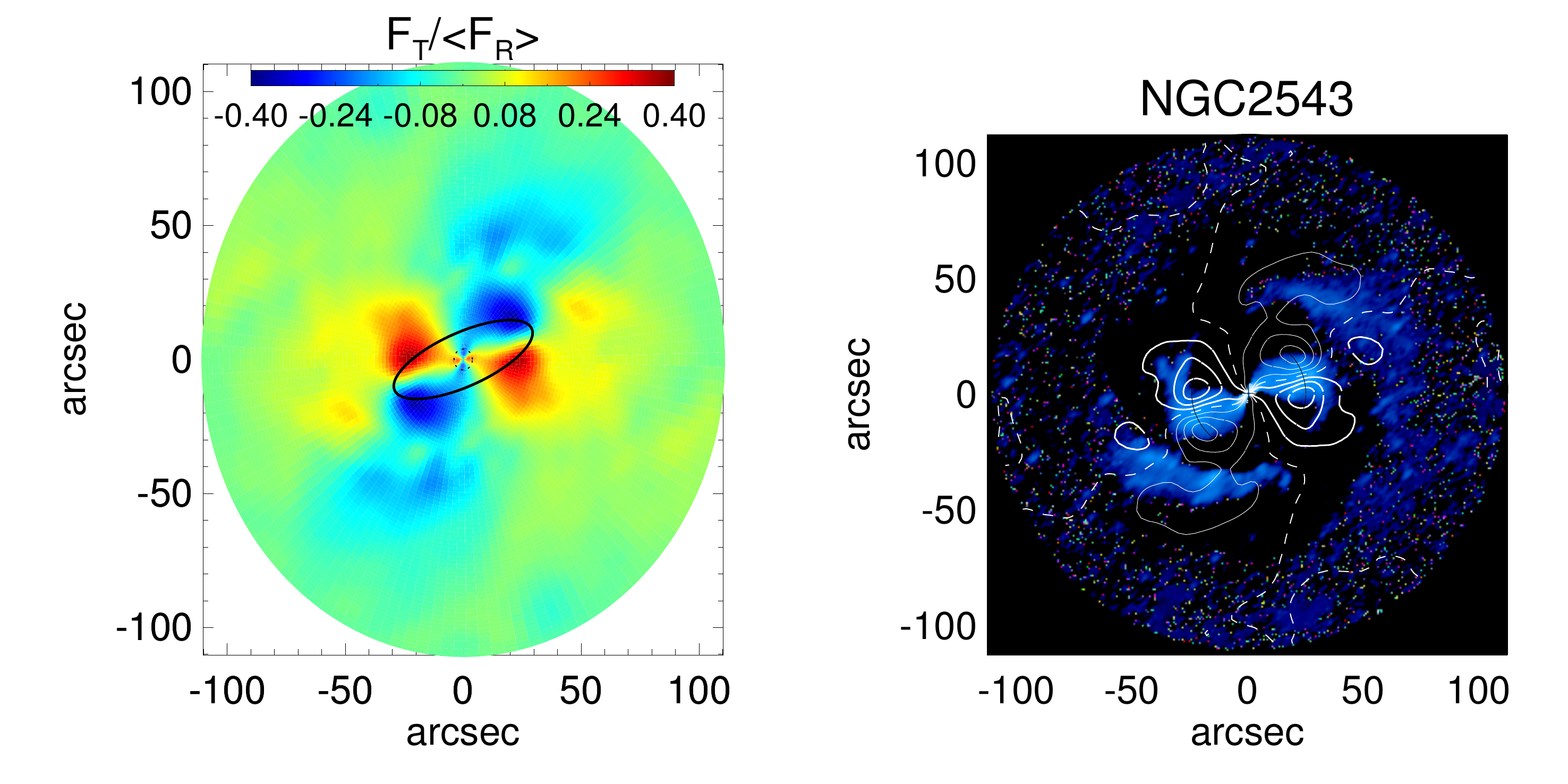}
\caption{\textit{Left panel}: Gravitational torque map of NGC~2543 derived from $Spitzer$ image using only even Fourier components. In the top of the panel
a colour bar shows the maximum and minimum Q$_{\rm G}$; Bar length and
ellipticity are traced with a black solid line; The inner dotted circle
corresponds to the S$^4$G Pipeline 4 \reff\ of the bulge. \textit{Right panel}:
Fourier smoothed density with the axisymmetric component ($m=0$) subtracted;
Contours of equal  Q$_{\rm G}$ are overplotted in white; The dotted lines indicate the
regions where the tangential forces change sign.}
\label{fig:Qb}
\end{figure}

We calculated the gravitational potential of our galaxies from the 3.6 micron images, using the NIRQB code \citep{2002Laurikainen2} based on the polar method developed in \citet{1999Salo}. Before applying the Fourier transformation, the $Spitzer$ images are rectified to face-on. Then, the even Fourier
components (up to 20) of the surface density $I(r,\phi)$ are calculated within a
polar grid. The gravitational potential $\Phi_{m}(r,\phi)$ is then inferred from
the smoothed surface densities by applying a fast Fourier transformation in the azimuthal direction in combination with a direct summation over radial and vertical directions. We use a polar grid with 128 bins in the azimuthal direction, which determines an angle step-size for the azimuthal Fourier transform of $2 \pi/128 = 2.8^{\circ}$.

The calculation of the potential is based on the following assumptions:
\begin{enumerate}
 \item  The mass-to-light ratio is constant.
 \item  The disc vertical scale height $h_{z}$ is constant.
 \item	The disc has an exponential vertical density distribution:

\begin{equation}
\rho_{z}(z)=\displaystyle\frac{1}{2h_{z}} exp(-|z/h_{z}|).  
\label{vert-density}
\end{equation}

 \item  The vertical scale height of the disc scales with the disc size as
h$_{\rm z}$ = 0.1 r$_{\rm k_{20}}$, where r$_{\rm k_{20}}$ is the 2MASS \citep{20062MASS} $K-$band
surface brightness isophote of 20\,mag\,arcsec$^{-2}$.
\end{enumerate}

Tangential forces ($F_{\rm T}(r,\phi)$=$\frac{1}{r}\partial
\Phi(r,\phi)$/$\partial \phi$) and  radial forces ($F_{\rm R}(r,\phi)$=$\partial
\Phi(r,\phi)$/$\partial r$) are obtained via integration. Non-axisymmetric forces in the galaxy are
characterised by the ratio of the tangential force to the mean axisymmetric
radial force field:
\begin{equation}
Q_{\rm G}(r,\phi)=F_{\rm T}(r,\phi)/\langle F_{\rm R}(r)\rangle,
\end{equation}

\noindent where $\langle F_{\rm R}(r)\rangle$ is the azimuthally averaged radial
force at a radial distance $r$. Q$_{\rm G}(r,\phi)$ values are used to construct
the gravitational torque maps of our galaxies (see example in
Fig.~\ref{fig:Qb}). Typically, barred galaxies show a well-defined four-quadrant
Q$_{\rm G}$ map, resembling a \emph{butterfly pattern}, which is roughly
symmetric with respect to the bar major axis. We take the even Fourier
components uniquely (focusing on bi-symmetric structures) and we
symmetrise our maps, reducing in this way the impact of sharp density clumps.

Based on the torque maps, and given a certain radial distance $\rm r$ and
quadrant $q$, one can identify a maximum $\rm Q_{T}(r)^{q}=max(Q_{\rm
G}(r,\phi)^{q})$. We calculate the radial profile of the relative strength of
the non-axisymmetric perturbations throughout the galaxy, $\rm Q_{\rm T}(r)$,
taking the mean of these four maxima. For additional information about the
method, see \citet{2010Salo}. Finally, the
gravitational torque parameter (\Qb) corresponds to the maximum value of $\rm
Q_{\rm T}$ at the bar region. 

The main source of uncertainty ($\approx$15\%) is the poorly known
          vertical thickness: to account for this we have used
          different disc thicknesses in the calculation of the
          gravitational field. A small systematic error is produced by
          the omission of the dark halo contribution on the radial forces,
          but this is likely to be smaller than that associated with
          the vertical thickness (D\'iaz-Garc\'ia et al., submitted.)

\subsubsection{Kinematic torque (\Qkin)} 
\label{kinT}
In order to perform a model-independent measurement and to test the torque measure of \Qb\ we developed a new method solely using the kinematics, resulting in a new parameter which we call the kinematic torque \Qkin.
The basis of this analysis is the stellar velocity field. Using this map, we
extracted the radial and tangential velocities following
\citet{2012Maciejewski}, using their equations 9 and 10. This method is based on assuming a thin disc geometry to obtain the two velocity components in the equatorial plane. Further assumptions in
deriving these two quantities are:

\begin{enumerate}
 \item  A steady state bar, hence not in buckling phases or alike. 
 \item  A symmetric bar with respect to its major axis. 
 \item  A thin galaxy disc resulting in only two velocity components.
\end{enumerate}

\begin{figure}
\includegraphics[width=\linewidth]{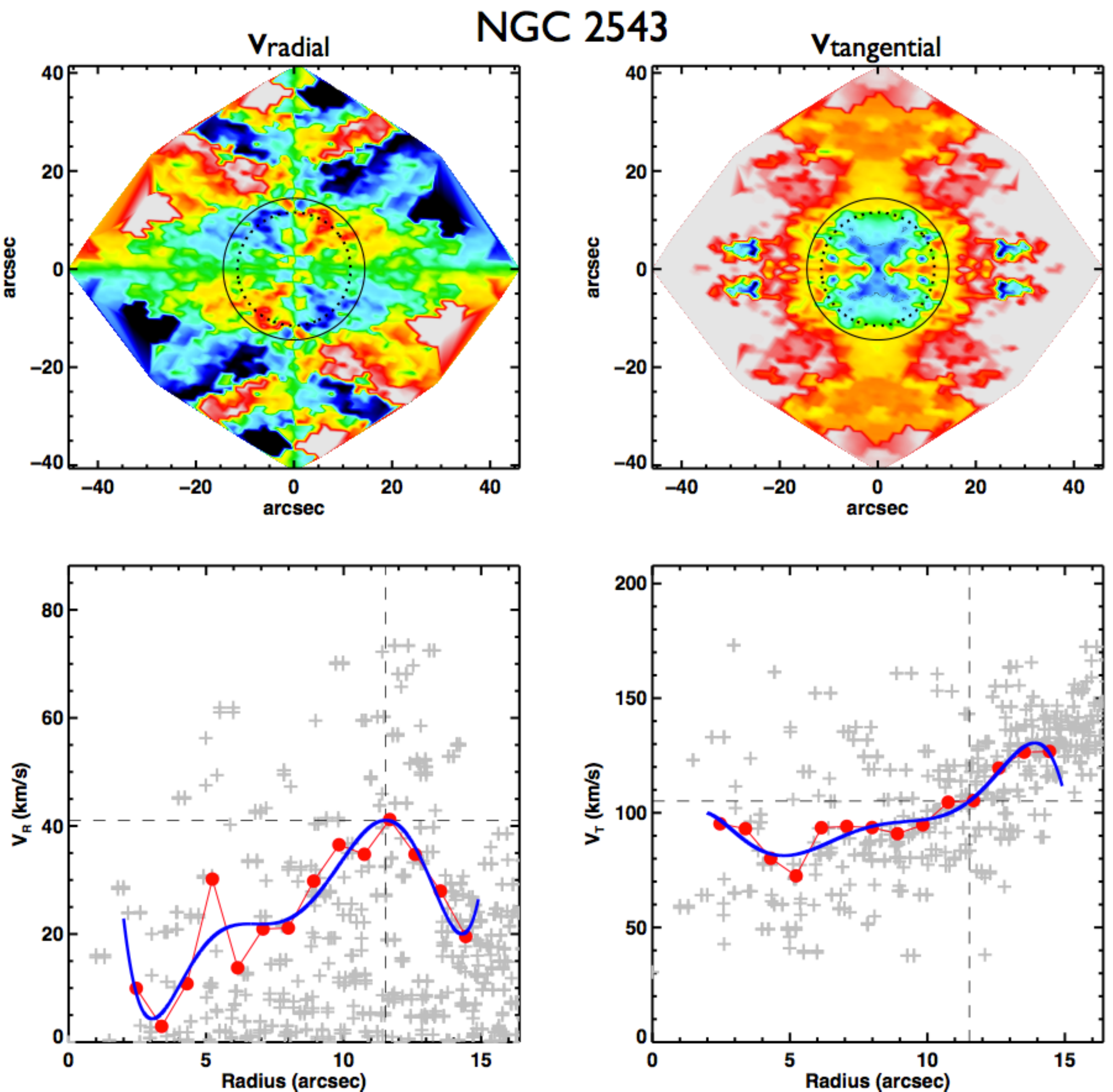}
\caption{Radial and tangential velocities for NGC\,2543. \textit{Upper left:}
radial velocity field, \textit{upper right:} tangential velocity field;  in both
of them, the continuous circle indicates the bar radius and the dotted circle
the radius where we measured the kinematic torque (\Qkin). \textit{Lower left:}
radial velocity along the radius; \textit{lower right:} radial distribution of
the tangential velocity; in both: grey crosses represent individual measurements, 
red points show the obtained modulation (not a fit to the gray points - see text for details) and the blue a smoothed fit to it. The
dashed lines indicate the maximum found in the radial velocity and 
corresponding to the same radius shown for the tangential velocity 
(see text for details). }
\label{fig:Qk}
\end{figure}

As the buckling phase is only a brief evolutionary stage of the bar
\citep[e.g.][]{2004ShenS, 2006MartinezV}, it is much more likely to observe bars in their
steady state. Under the presence of spiral structure, the bar will lose its
symmetry with respect to the major axis, but only at the very edges of the bar.
An aspect to consider is that prominent bulges will break the third assumption
in the inner regions of the bar. Following these limitations, one can conclude
that the most reliable region is within the middle of the radial extension of
the bar, where we expect to measure the strongest radial velocities and
corresponding torques, as outlined below. In addition to these assumptions, the
technique developed by \citet{2012Maciejewski} requires the knowledge of the
systemic velocity, galaxy inclination, position angle of the line-of-nodes, the
bar position angle and bar length. We estimated those from the literature and
close inspection of our own datasets. Furthermore, we rectified our maps to face-on in 
order to apply the technique.

Using the extracted radial and tangential velocities, we defined a new parameter
(\Qkin) that measures the torque directly from the observed kinematics:

\begin{equation}
{\rm Q}_{\rm kin} = \frac{{\rm max}(v_{\rm rad}{\rm (R)})}{ \left<{\rm abs}(v_{{\rm tan, R}})\right>},
\end{equation}

\noindent where we first find the radial position of the maximum value of the
radial velocity ($v_{\rm rad}$), and then determine the corresponding tangential
velocity ($v_{\rm tan}$) as the mean value in a ring around this radius.
This relation is constructed analogous to the calculation of \Qb\ based on the 
fact that $v_{\rm rad}$ is proportional to  $F_{\rm T}/ F_{\rm R} \times v_{\rm rot}$ and $v_{\rm tan}$
roughly equal to $v_{\rm rot}$. Therefore $v_{\rm rad} / v_{\rm tan}$ is expected 
to be proportional to $F_{\rm T}/ F_{\rm R}$ (note that the ratios do not have to be equal, but only proportional). 
Figure~\ref{fig:Qk} shows the radial (left) and tangential (right) velocity maps
(top) and radial (bottom) distribution for NGC\,2543 as an example. The position
of the maximal radial velocity is found by evaluating the radial velocity field
in rings. We expect a certain velocity modulation when tracing a circle through
the four quadrants, i.e., combination of sine and cosine curve when tracing the
radial velocity in a ring. This additional aspect helps us to detect and correct for 
outliers, i.e., unreal peaks or drops of extremely high or low values,
which appear more often in the kinematic data due to higher noise levels. Hence, 
we avoid to simply measure the maximum which would lead to an incorrect result. A smooth version is then obtained by fitting a polynomial.
In the bottom panels, we show the curves obtained when measuring the amplitudes
of the modulation (red and smoothed fit in blue) compared to the individual data points (grey). It is obvious
that there is a significant scatter among the individual points, but
nevertheless a clear maximum can be distinguished in the radial velocity
profile which is well captured with the modulation.

To further constrain the measurement, we evaluated \Qkin\ within the bar region
as determined from the S$^4$G images. This is to avoid choosing areas where high
values appear, either due to higher noise levels towards lower surface
brightness areas, or due to spiral arms. In the example shown in the figure, the
strength of the spiral arms can clearly be seen: in a central $\approx$20 arcsec radius,
we detect the signature of the bar, but further out, the field does not become flat
but shows other maxima and minima due to the torques exerted by the spiral 
arms. Those strong, outer values detected in the radial velocity -- and thus the measured torque -- is not due to
the bar but to the spiral arms in this galaxy. Similar enhancements can also be
seen in Fig.~\ref{fig:Qb} for the computation of \Qb. 

As the value of \Qkin\ depends on the input parameters to determine radial and
tangential velocities, we chose to determine its uncertainty via a set of
Monte-Carlo simulations. For each realisation, we chose a random combination of
initial values of the inclination, line of nodes position angle, bar position
angle and bar length, all within their uncertainties. As inclination is the most
difficult to determine, we allowed an uncertainty of $\pm$10\dgr, whereas we
chose $\pm$5\dgr\ for the other parameters, leading to an overall uncertainty found in 
Tab.~\ref{tab:galqs}. Higher values would result in 
simply higher uncertainties in the measured torques.  

Unfortunately, the determination of \Qkin\ is
only possible when the kinematic major axis and the bar position angle are
neither perpendicular nor parallel (at least 5$^\circ$ off, while an angle of 45$^\circ$ 
would be ideal). It is only under those circumstances that the
method of \citet{2012Maciejewski} can be applied to compute the required $V_{\rm
rad}$ and $V_{\rm tan}$. From the 16 galaxies in our sample, we could only
measure the kinematic torque (\Qkin) on the following 10 systems: NGC\,2543,
NGC\,2712, NGC\,2859,  NGC\,2962, NGC\,3504, NGC\,4245, NGC\,4262, NGC\,4394,
NGC\,5350 and NGC\,5701. Results are summarised in Tab.~\ref{tab:galqs}.

\begin{figure}
\includegraphics[width=1.\linewidth]{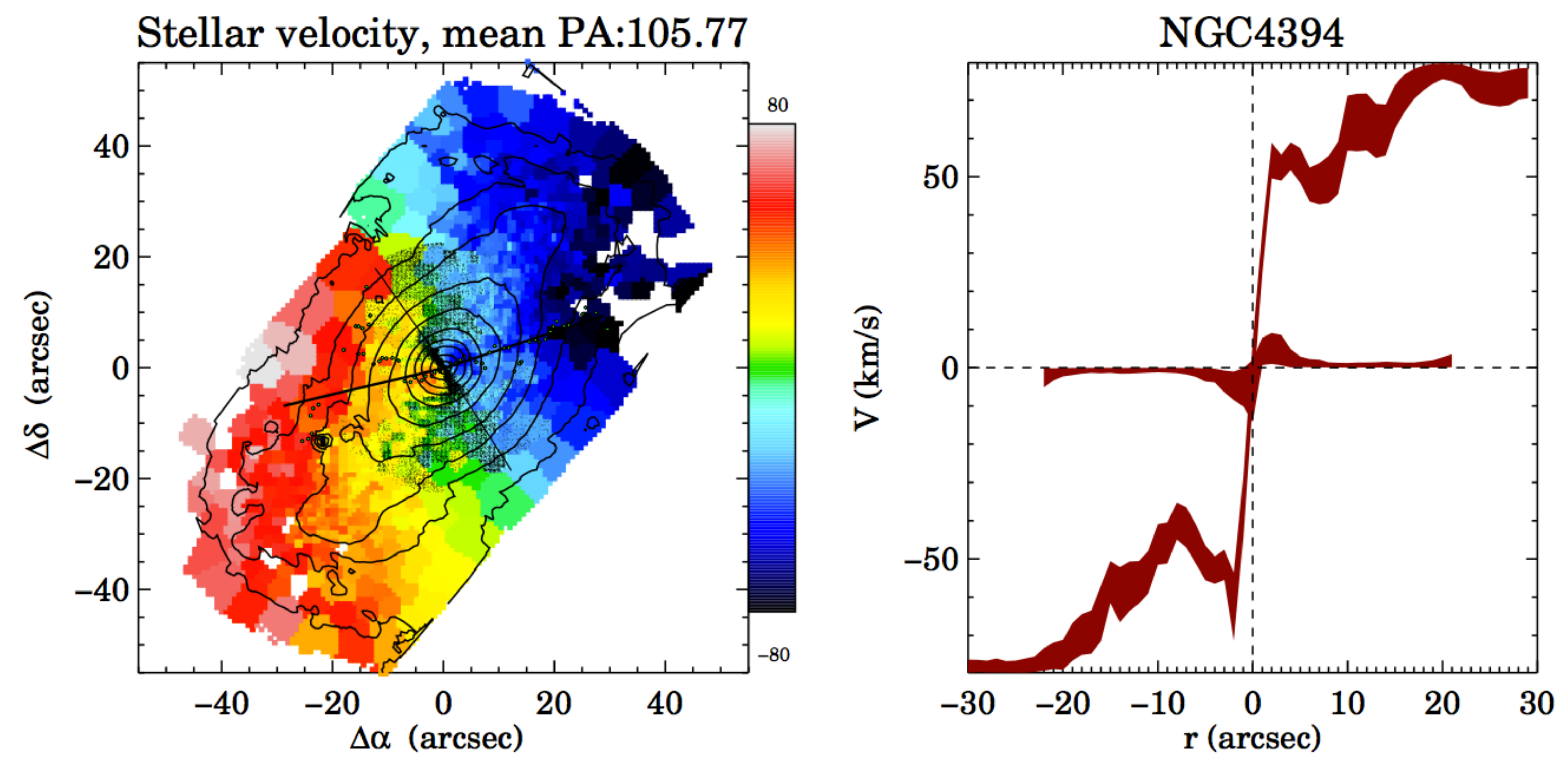}
\includegraphics[width=1.\linewidth]{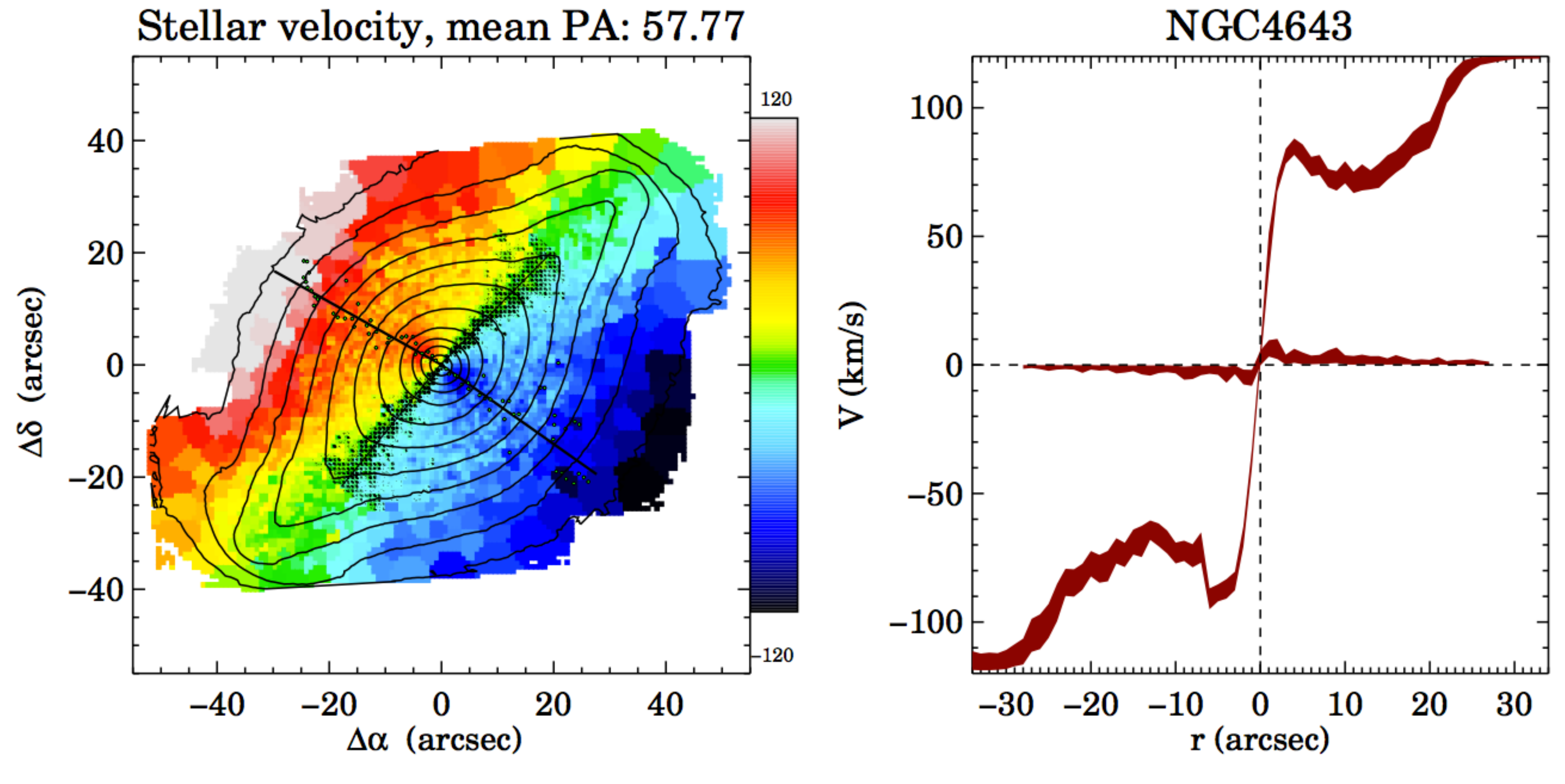}
\caption{Stellar velocity maps and their associated major axis profile for two
galaxies. The colour bars on the side each indicate the range of the
parameter measured. The isophotes shown are derived from the SAURON cube
reconstructed intensities and are equally spaced in steps of 0.5 magnitudes.
The kinematic position angle, based on \citet{2014BarreraB}, is given on the top.
The dots indicate the bins used for determining that angle. The black lines
indicate the photometric position angle, even better seen in the following figures, not demonstrating the method.}
\label{fig:vel}
\end{figure}

\section{Observed kinematic properties}
\label{sec:results}

This section summarises the different parameters extracted from the kinematic
maps of stellar and ionised gas component. Figure~\ref{fig:vel} presents two
examples of absorption-line stellar velocity maps and associated radial profiles
along the major and minor axis for two galaxies in our sample,
NGC\,4643 (early-type) and NGC\,4394 (late-type). The complete set of
kinematic maps, including ionised gas kinematics and Gauss-Hermite moments h$_3$
and h$_4$ are collected in Appendix~\ref{app:gaskinmaps}. Overlaid in all maps,
we show the isophotes of the surface brightness (in mag/arcsec$^2$ with an
arbitrary zero point) reconstructed from the SAURON datacubes and equally spaced
in intervals of 0.5 magnitudes. In this section, we concentrate on an overview
of the general kinematic trends observed in our sample. We also present the bar
strength measurements from these kinematics (\Qkin), in comparison with the ones
derived from the S$^4$G imaging (\Qb).

\begin{figure}
\includegraphics[width=1.\linewidth]{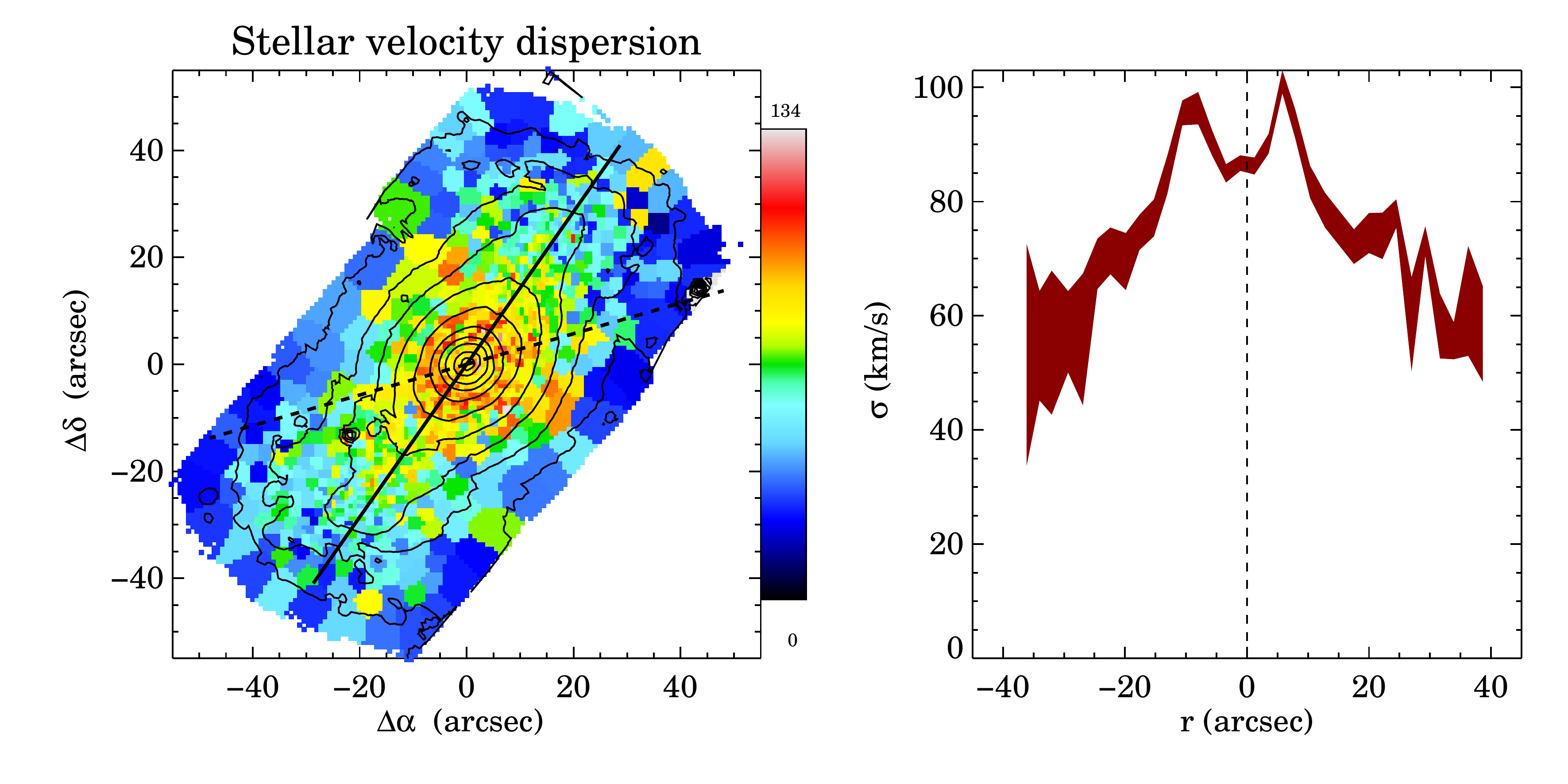}
\includegraphics[width=1\linewidth]{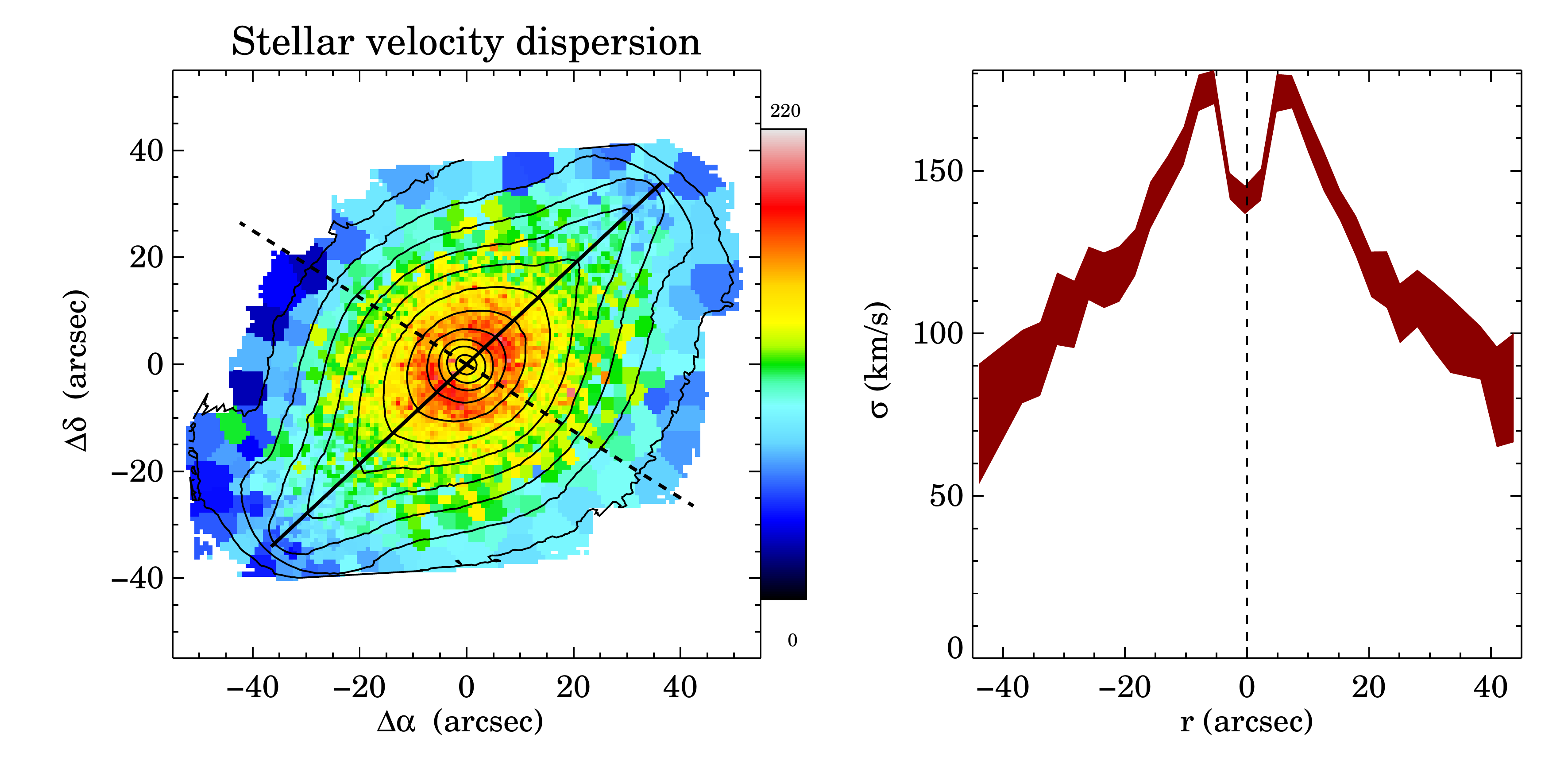}
\caption{Stellar velocity dispersion maps and their associated profile along the
bar major axis for the same two galaxies as in Fig~\ref{fig:vel}. The colour
bars on the side each indicate the range of the parameter measured. The
continuous black line shows the bar major axis, determined from the photometry
of $Spitzer$ images while the dashed line indicates the global photometric
position angle found previously.}
\label{fig:sig}
\end{figure}

\subsection{Stellar and gas kinematics}
\label{kinematics}

We investigate the orientation of the stellar and gas kinematics, comparing them to the bar axis,
 as well as to features that can be linked to bar-driven secular evolution. We use the entire
maps as well as cuts along different axes to better unravel certain features.

A first glance at the maps shown in Appendix~\ref{app:gaskinmaps} reveals that the
overall rotation is not strongly affected by the bar (i.e., the kinematic major axis remains almost constant as a
function of radius as determined using the method by \citealt{2014BarreraB}),
implying that the bar has not changed the global rotation pattern of the
galaxies. We do not detect either large velocity twists in the line-of-nodes (a
kinematic feature observed in simulations). Only NGC\,2712 and NGC\,4394 show 
small deviations. The absence of this feature in our maps may be due to
projection effects or simply to the limited FoV, because the twist is often visible
further out, such as in  NGC\,936 \citep[e.g.,][]{2012Maciejewski}.

Along the kinematic major axis, we do detect in all cases the so-called 
double-hump rotation curve (local inner maximum followed by a slight drop and further rise)
predicted by simulations \citep{2005Bureau}, so far mainly confirmed in
edge-on systems \citep[e.g.,][]{2004Chung}. This feature can be seen in the stellar
velocity maps as an enhanced area of high (low) velocity values on both sides of the nucleus, but is obviously
more apparent in the radial profiles. This double-hump is clearly visible in more
than 60\% of the galaxies in our sample and hints towards the existence of inner
discs or rings. Along the minor kinematic axis, we also find a similar
distortion in the very central parts, visible as a small-scale wiggle in the
minor axis rotation profile. This profile is normally expected to be flat with a
value around zero, but it appears to be present for all cases. This small feature
might indicate a non-perfect estimation of the global photometric position angle
of the galaxies.

\begin{figure}
\includegraphics[width=1\linewidth]{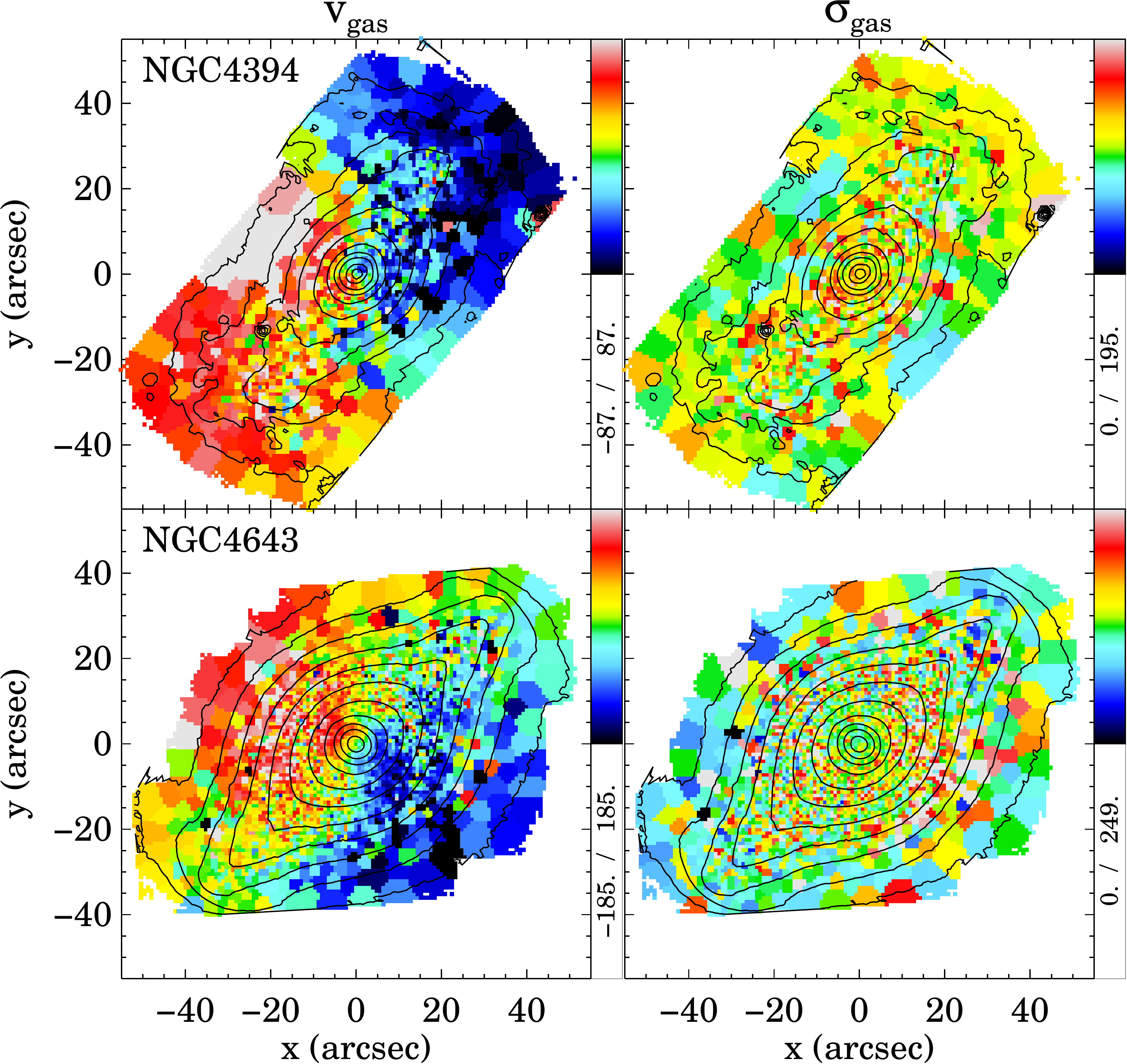}
\caption{Velocity and velocity dispersion maps of the ionised gas component
for the two examples in the BaLROG sample of galaxies.}
\label{fig:gas}
\end{figure}

The stellar velocity dispersion maps, and radial profiles, of the two example
galaxies NGC\,4643 and NGC\,4394 are presented in Fig.~\ref{fig:sig}. The maps
show the presence of $\sigma-$drops, a kinematic feature also predicted by
simulations of barred galaxies \citep[e.g.,][]{2003Wozniak}. The fraction of
galaxies in our sample that show this behaviour is 62.5\%. The early types in our sample show
a larger region of overall higher $\sigma$ than the late types, as expected from
the presence of larger central bulge structures. Peak velocity dispersion values
range between 100\,\kms and 220\,\kms.

The corresponding kinematic maps (velocity and velocity dispersion) for the
ionised gas are shown in Fig.~\ref{fig:gas} for the two example galaxies,
illustrating their difference to the stellar kinematics. The ionised gas maps of
the entire sample appear in general less regular with a more patchy
distribution, reflecting the gas properties. We do not find gas equally
distributed in all the galaxies. The absolute values of minimal and maximal
rotation are generally slightly higher for the gas than for the stars in all
galaxies. The kinematic major axis of the gas velocity field has the same orientation as the stellar velocity field.
Only one galaxy shows a significant change in orientation, NGC\,4262, whereas two others (NGC\,2962 and NGC\,5701) show very mild differences only. 
Previous studies suggest that NGC\,4262 might have been involved in an
interaction \citep{2005Vollmer}. The gas velocity fields also present the
double-hump feature, in some cases significantly more pronounced than in the
stellar maps (e.g. NGC\,3504). This confirms that the gas is more susceptible to
bar-driven processes \citep[e.g.,][]{1981Schwarz,2011Ellison, 2013Atha}. 
The ionised gas velocity dispersion differs significantly from the stellar 
$\sigma$. It does not show a central elevation (tracing the bulge component) but exhibits slightly higher values throughout the area inside the bar isophotes. However, the pattern is extremely patchy and we lack enough coverage of the disc for a fair comparison between the morphological components. Especially in the
late-type systems, some regions display higher velocity dispersion in the gas
than in the stars. These are typically associated with spiral arms (e.g.
NGC\,3504, NGC\,4394).

We will link the investigated features to the strength of the bars in 
Sec.~\ref{sec:compq}.

\begin{figure}
\includegraphics[width=1\linewidth]{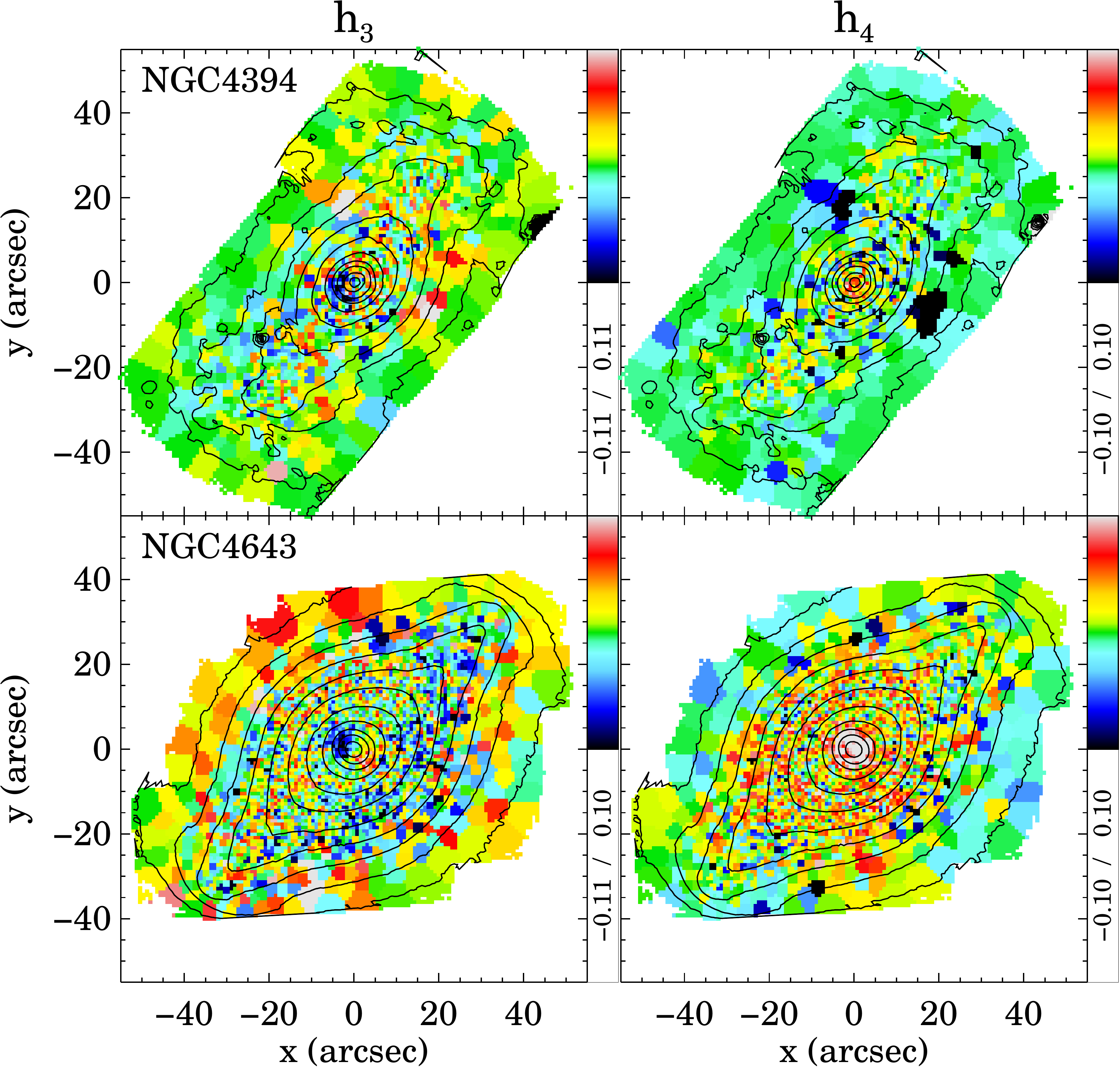}
\caption{Gauss-Hermite maps of the stellar component for the two examples in the
BaLROG sample of galaxies.}
\label{fig:h3h4}
\end{figure}

\begin{figure*}
\includegraphics[width=1.0\linewidth]{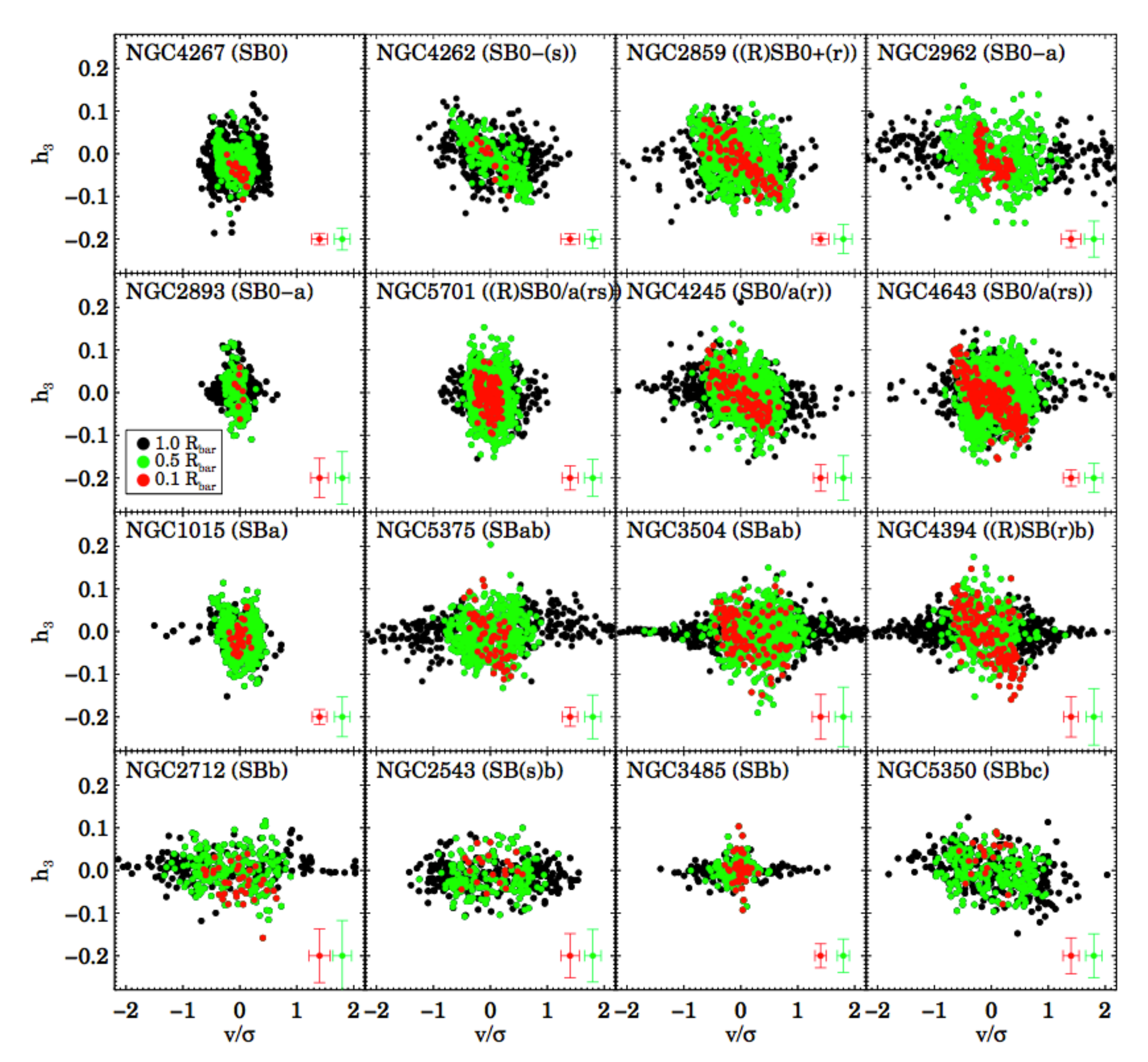}
\caption{Individual Voronoi-binned values of h$_3$ versus the stellar velocity over
the stellar velocity dispersion, within 1\,$R_{bar}$, 0.5\,$R_{bar}$ and
0.1\,$R_{bar}$. Representative error bars for the red and green regions are indicated in the 
lower left corner of each panel.}
\label{fig:h3rbar}
\end{figure*}

\begin{figure*}
\includegraphics[width=1.0\linewidth]{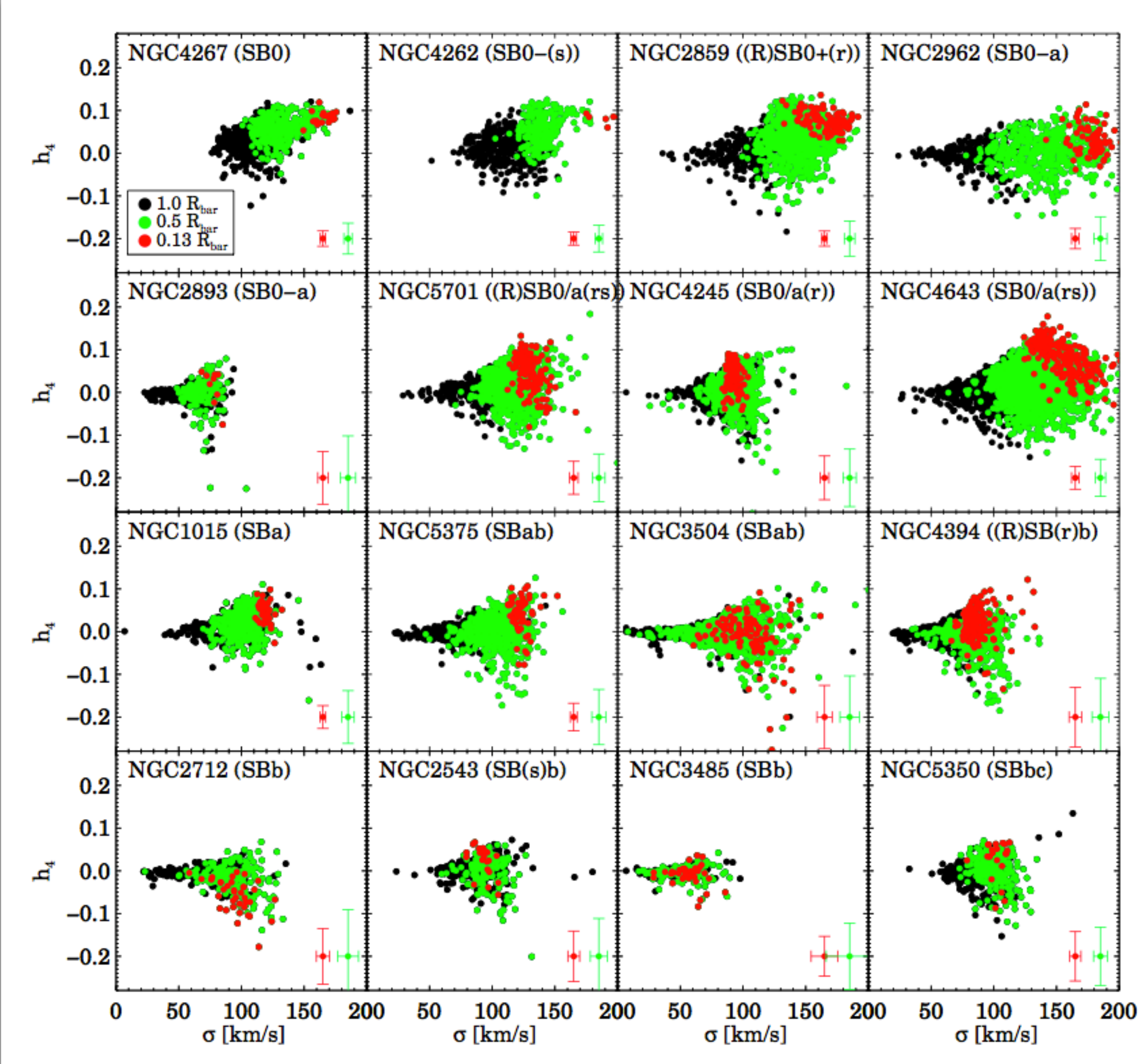}
\caption{Individual Voronoi-binned values of h$_4$ versus the stellar velocity
dispersion, within 1\,$R_{bar}$, 0.5\,$R_{bar}$ and 0.1\,$R_{bar}$. Representative error bars 
for the red and green regions are indicated in the 
lower left corner of each panel.}
\label{fig:h4}
\end{figure*}

\subsection{Gauss-Hermite moments: h$_3$ and h$_4$}

In addition to the first and second moment of the LOSVD distribution ($V$,
$\sigma$), we measured the h$_3$ and h$_4$ Gauss-Hermite moments.
They help to understand the distribution of orbits along the line-of-sight and 
can be used to distinguish dynamically distinct regions and thus indicate whether
bars influence their formation.

Mathematically, h$_3$ measures the skewness of the LOSVD, i.e., wings on either side of the peak deviating from the
otherwise Gaussian profile, while h$_4$ is a measurement of the kurtosis
\citep[e.g.][]{marel1993}. Within our sample, we find a large variety in those
maps and just by visual inspection, we cannot identify a systematic pattern
which could be attributed to the bars of their host galaxies.

Figure~\ref{fig:h3h4} shows the two Gauss-Hermite moments h$_3$ and h$_4$ for
NGC\,4394 and NGC\,4643. In the first galaxy, the maps appear more uniform than in
the second one, which does exhibit low-level structure. However, the majority
of the maps of h$_3$ and h$_4$ (see more in the appendix) show very low values
(below 0.1) and are rather flat throughout the FoV. In several cases a slight
anti-correlation between the h$_3$ moment and the stellar velocity can be seen
(NGC\,1015, NGC\,2959, NGC\,2962, NGC\,4245, NGC\,4262, NGC\,4643 and
NGC\,5350), while h$_4$ moments correlate in most cases with the velocity
dispersion. We investigated the h$_4$ profiles along the bar major axis for
kinematic signatures of peanut-shaped bulges \citep[e.g.,][]{2005Debattista}, but
could not detect any clear evidence. This is a property of mainly very low
inclined galaxies with strong peanut shapes seen in the photometry, so our
sample is not an ideal selection for the detection of this characteristic. So
far, only a few studies \citep[e.g.,][]{2008MendezAbreu, 2014MendezA} have confirmed this prediction.

In the literature, the h$_3$ moment is often related to the stellar velocity
and both correlations and anti-correlations are found
\citep[e.g.,][]{2004Bureau}. For a more robust measure we chose to correlate h3
with V/$\sigma$, shown in Fig.~\ref{fig:h3rbar}, in order to compensate for
different masses. However, we checked the relation correlating with the stellar
velocity alone and did find very similar results. In the figure, Voronoi-binned 
values within the bar length are shown in black, within 0.5 times the bar length
in green and within 0.1 times the bar length in red. There is no clear
(anti-)correlation for the full extent of the bar length (nor for the effective
radius, which we tested for comparison, but which is not shown here). However, the
more central the aperture, the stronger the anti-correlation for about 50\% of
our sample, clearly depicted by the red points. The other $\sim$50\% of our
sample do not show any (anti-)correlation at all, but a simple spread of h$_3$
values around zero. This behaviour is consistent across both measures -- bar
length or the effective radius -- with the exception of NGC\,4262. The effective
radius for this galaxy is significantly smaller and captures the inner part
only. Hence, as smaller apertures decreased the scatter if an anti-correlation
was present, it stands out better for the effective radius measurement. In many
of the other galaxies the effective radius is comparable to the bar length or at
least not less than half its size. Overall, we cannot detect a tendency between
late or early types, because both late types (e.g. NGC\,4394) and early types (e.g.
NGC\,2859) show the above described behaviour, although with a mild bias towards
earlier types showing stronger correlations. In the figure, we ordered the
galaxies according to their Hubble type (SB0 top left to SBbc bottom right) and
one can appreciate the larger scatter also amongst the innermost (red) points in
the bottom row (latest types of our sample). The fact that there is no striking
difference though might mean that the Hubble type is not the crucial factor,
neither the bar, but the presence of significant substructures. More than 50\%
of those with strong central anti-correlations have confirmed substructures such
as nuclear rings, nuclear lenses or nuclear ring or bar lenses (see classifications by
\citealt{2011Laurikainen} and \citealt{2015Buta}).
 
The influence of bars on building up a central component is supported by this
h$_3$--$V$ anti-correlation in the centres of about 50\% of our sample of
galaxies. Earlier studies have found a correlation as well as anti-correlation
between the stellar velocity and h$_3$ moment, depending on the area and type of
galaxy sampled. In edge-on barred galaxies, \citet{2004Bureau} and
\citet{2005Bureau} detected an h$_3$--$V$ correlation over the projected bar
length, expected for a thick bar. In the centres, however, they also found an
anti-correlation in more than 60\% of the galaxies. This can indicate the
presence of multiple components with different kinematics. Hence a significant
number of barred galaxies, not only in edge-on systems but also in our sample of
different inclinations, show the presence of cold and dense
(quasi-)axisymmetric central stellar discs. This supports the scenario of the
bar driving gas towards the centre and nourishing star formation, resulting in
this additional central component. The coincidence of a steep central light
profile and star-forming ionised gas discs in these same regions
\citep[e.g.][]{1999Bureau} supports this theory further. \citet{FalconBarroso2006} also 
found a link between more intense star forming regions and lower gas dispersion values.
However, due to our limited wavelength range, we could not account consistently for dust correction which could influence the values of h$_3$, being sensitive to dust. Dust is particularly important in late-type galaxies, but the majority of our sample are early-types and no obvious dust lanes are detected in the later types. 

Figure~\ref{fig:h4} shows similar measurements for h$_4$. The galaxies are
again ordered by Hubble type. This parameter measures the symmetric
deviation from a Gaussian profile, indicating a velocity distribution which is
less (or more) peaked (negative or positive values). It is expected to
correlate with the velocity dispersion \citep{marel1993}, so we chose to plot it against
$\sigma$. Overall, but especially for the early types, we observe elevated
values in the central regions, most probably associated with the bulge. The
presence of higher h$_4$ values together with an occasional $\sigma-$drop hints
at the presence of components with more recent star formation within a classical
elliptical-like bulge component \citep{2003Wozniak, 2005Bureau}. In
Fig.~\ref{fig:h4}, this behaviour stands out even more clearly: a higher velocity
dispersion is found in the centres where the red points (corresponding to most
central values, i.e., within 0.1 bar lengths) also show elevated h$_4$ values.
Later-type galaxies are clearly different. Not only does the velocity dispersion cover a
larger range of values, also the h$_4$ values are not particularly high in the
centre or even go significantly below zero (e.g. NGC\,2712, NGC\,3504,
NGC\,3485, NGC\,5350, NGC\,5875). The bulge in the later-type galaxies is significantly
less pronounced, therefore the mixture of more components could cause this
spread of h$_4$ values. The strongest central concentration of elevated h$_4$
values, at an almost constant $\sigma$ and h$_4$, are found in NGC\,1015,
NGC\,2859, NGC\,2962, NGC\,4262 and NGC\,4267. These are not the
galaxies showing the h$_3$--velocity central anti-correlation, but those to have
 a prominent bulge component. NGC\,5701 also has a large
bulge, but contains confirmed nuclear spiral structure \citep{Erwin2004},
which could contribute to the higher spread in h$_4$. 

In conclusion, the analysis of Gauss-Hermite moments suggests
that the centers of barred galaxies (within at least half a bar length, and 
even more obvious within 0.1 bar lengths) 
host dynamically distinct components. These could have been altered 
by bar-driven evolution.

\subsection{Angular momentum: $\lambda_{\rm R}$}

We calculate $\lambda_{\rm R}$ as a measure of rotational versus pressure 
support following the prescription given in \citet{2007Emsellem}, as bars are 
meant to work as engines redistributing angular momentum amongst 
the different components of a galaxy. This 
parameter is based on the first two stellar velocity moments and the 
corresponding flux and is defined as
\begin{equation}
\lambda_{\rm R} = \frac{\sum_{i=1}^{N_p} F_i R_i | V_i |}{\sum_{i=1}^{N_p} F_i 
R_i \sqrt{ V_i^2 + \sigma_i^2}}
\end{equation}
\noindent for two-dimensional spectroscopy, where $F_i$ denotes the flux, $R_i$ the
circular radius, $V_i$ the velocity and $\sigma_i$ the velocity dispersion of
the $i$th spatial bin (going to $N_p$ bins). As outlined in
\citet{2011Emsellem}, it improves the characterisation of the dynamical state of
a galaxy compared to the simple measure of $V/\sigma$. It shows a clear
difference, especially for non-regular rotators with irregularities in their
velocity fields, whilst at the same time being correlated to the specific
angular momentum of the stars.

\begin{figure}
\includegraphics[width=1.0\linewidth]{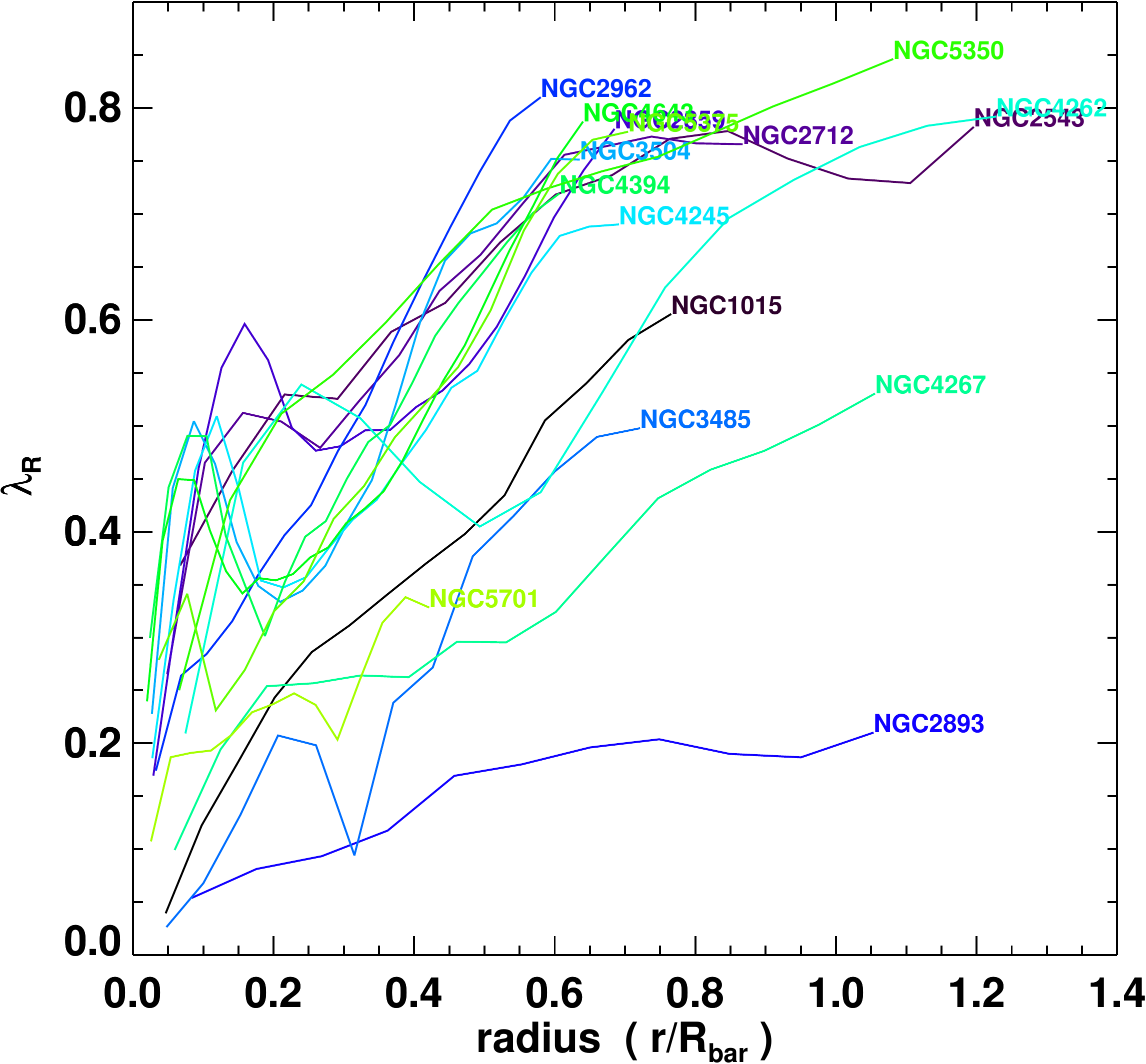}
\caption{Radial profiles of $\lambda_{\rm R}$ for all galaxies, normalised to
the radius of the bar. A rise and consecutive dip is observed at a similar position
in the profiles of the majority of galaxies, around 0.2$\pm$0.1 R$_{\rm bar}$. }
\label{fig:lamR}
\end{figure}

We calculated $\lambda_{\rm R}$ in our sample both radially (see
Fig.~\ref{fig:lamR}) as well as within one \reff\ ($\lambda_{\rm Re}$, given in
Tab.~\ref{tab:galqs}). We normalised the radial profiles to the bar radius to test the influence of the bar on
the shape of the profile. Considering the small sample size and large variety of
Hubble types and bar types, it is not surprising that we recover a variety of
profiles. Yet more than 70\% show a dip in $\lambda_{\rm R}$ at around
0.2$\pm$0.1 R$_{\rm bar}$. The only galaxy which shows a clear offset of this
dip is NGC\,4262, where the stellar and gas velocity fields are
clearly misaligned.  

This feature appears to be related to the double-hump in the velocity profile
\citep{2005Bureau}, in combination with the rise in $\sigma$ after the
$\sigma-$drop. Three galaxies (NGC\,1015, NGC\,2893, and NGC\,5350) do not exhibit
a double-hump nor a strong  $\sigma-$drop. In other galaxies (NGC\,2712, NGC\,4267,
NGC\,4262, NGC\,5375), the hump feature in the stellar velocity alone seems to
be strong enough to produce the drop in the $\lambda_{\rm R}$ profile. In other
cases, in particular NGC\,4245, NGC\,3485 and NGC\,5701, the peak of the hump in
the velocity profile coincides with the peak of the velocity dispersion profile
(after the central drop). Thus, despite the distinct morphologies and
inclinations, we observe a common behaviour and influence on the stellar
kinematics. This is likely produced by the bar, since these features seem to
accumulate around a similar radius related to the bar length. This feature could
probably be associated to the inner Lindblad resonance (ILR)
\citep[e.g.][]{1994Elmegreen, 1990Pfenniger}. In fact, in several of our
galaxies, nuclear rings have been detected at those locations: NGC\,2859 \citep{2002Erwin}, NGC\,3504
\citep{1993ButaC, 1997Elmegreen}, NGC\,4262 and NGC\,4245 \citep{2010Comeron}.

We also tested the location of the dip as a function of the effective radius of
the bulge and the disc scale-length (derived from the S$^4$G) but did not find any
correlation. This supports our suspicion that this feature is related
to the bar. Similar studies have already related outer ring radii to the bar
sizes \citep[e.g.,][]{2012Perez} based on earlier studies and simulations
\citep[e.g.][]{1994Byrd, 1995Buta}. \citet{2010Comeron} estimate that the
maximal possible extension of a nuclear ring should be located at 0.25 bar
lengths.

\section{Contrasting bar strength measurements}

\begin{table}
\centering
\caption[Short\protect\\text]{Summary of the values obtained for $\lambda_{\rm Re}$ and the bar strength measurements: (1) NGC number, (2) $\lambda_{\rm Re}$, (3) photometric torque, (4) error in the photometric torque, (5) kinematic torque, (6) error in the kinematic torque, (7) kinematic torque of the gas component, (8) error in the kinematic torque of the gas component. As the error of the measure for  $\lambda_{\rm Re}$ is negligible, we do not list it here. }
\setlength{\tabcolsep}{0.5em}
\begin{tabular}{llllllll}
\hline
Galaxy & $\lambda_{\rm Re}$  &  \Qb & $\Delta$\Qb &  \Qkin & $\Delta$\Qkin &  \Qkin$_{\rm,g}$ & $\Delta$\Qkin$_{\rm,g}$\\
(1) & (2) & (3) & (4) & (5) & (6)  & (7) & (8) \\
\hline
NGC\,1015  &   0.25  &   0.26 &    0.074 &    -      &        -    &       -   &    -      \\
NGC\,2543  &   0.62  &   0.36 &    0.070  &   0.39   &      0.069  &    1.1    &   0.23     \\
NGC\,2712  &   0.65  &   0.28 &    0.044  &   0.37   &      0.044  &    0.46   &   0.07    \\
NGC\,2859  &   0.37  &   0.17 &    0.025  &   0.22   &      0.044  &    0.94   &   0.40     \\
NGC\,2893  &   0.06  &   0.16 &    0.020  &    -     &        -    &      -    &     -    \\
NGC\,2962  &   0.44  &   0.14 &    0.024  &   0.080  &      0.040  &    0.62   &   0.44    \\
NGC\,3485  &   0.52  &   0.38 &    0.064  &   0.33   &      0.084  &    0.84   &   0.43    \\
NGC\,3504  &   0.29  &   0.26 &    0.044  &   0.24   &      0.082  &    1.5    &   0.99     \\
NGC\,4245  &   0.33  &   0.18 &    0.020  &   0.10   &      0.033  &    0.20   &   0.19    \\
NGC\,4262  &   0.33  &   0.07 &    0.012 &   0.14   &      0.048  &    1.8    &   0.25     \\
NGC\,4267  &   0.24  &   0.04 &    0.013 &    -     &        -    &       -   &    -       \\
NGC\,4394  &   0.46  &   0.23 &    0.036 &   0.23   &      0.12   &    0.52   &   0.12     \\
NGC\,4643  &   0.28  &   0.28 &    0.069  &    -     &        -    &      -    &   -        \\
NGC\,5350  &   0.62  &   0.44 &    0.076  &    -     &        -    &      -    &   -         \\
NGC\,5375  &   0.47  &   0.23 &    0.044  &    -     &        -    &      -    &   -         \\
NGC\,5701  &   0.20  &   0.18 &    0.022  &   0.08   &      0.076  &    0.42   &   0.66    \\
\hline
\end{tabular}
\label{tab:galqs}
\end{table}

\begin{figure}
\includegraphics[width=1.\linewidth]{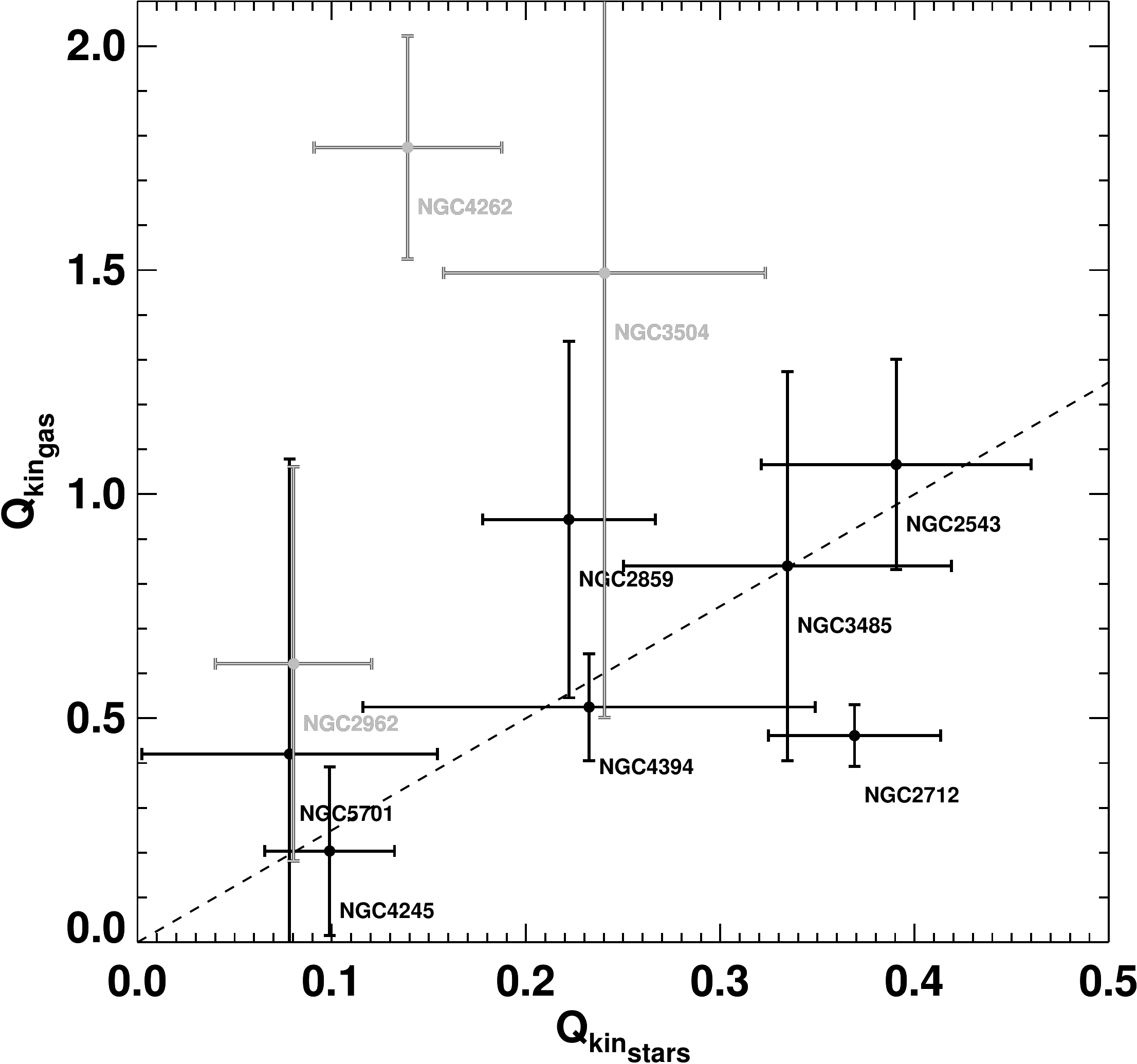}
\caption{Kinematic gas torque versus kinematic stellar torque. The dotted line
indicates a 2.5-correlation. Grey galaxies show obvious offsets between gas and
stellar kinematics.}
\label{fig:qbqkgas}
\end{figure}

In this section we compare bar strength measurements determined according
to the descriptions given in Section~\ref{sec:barstr}. For the case of \Qkin\ we
additionally measure it for both the stellar and ionised-gas components. This
comparison will allow us to establish the ruler that will be used in
Sect.~\ref{sec:discuss} to evaluate the impact of bars of different strengths on
different kinematic properties of our galaxies.

\subsection{\Qkin\ from stars and ionised gas}

We calculated the kinematic torque (\Qkin) from both our stellar and ionised-gas
velocity maps (values given in Tab.~\ref{tab:galqs}). The comparison is shown in Fig.~\ref{fig:qbqkgas}. The gas is
expected to respond more strongly to the bar than the stars. Overall the torque
derived from the gas velocity fields is significantly higher than the stellar
one. For almost all galaxies it is about 2.5 times greater than the value
derived from the stellar velocity maps (dotted line). The grey points mark
significantly higher values. Inspecting their gas velocity maps, we detect
clear differences from the stellar velocity fields. In particular NGC\,4262
shows the highest value of Q$_{\rm kin,gas}$. The gas velocity field is 
counter-rotating with respect to the stellar velocity field. Therefore, a
significant impact due to another process (e.g. galaxy interaction or close
encounter) might be at work in this galaxy (see \citealt{2005Vollmer}). In
NGC\,3504, the gas velocity field shows extreme enhancements in the central
regions which are not present in the stellar velocity field. It is not clear at
this stage what is causing this difference. Overall, we find that the stars seem
to be more stable and therefore the stellar kinematic torque agrees better with the
photometric torque (see \S~\ref{sec:compq}), whereas the gas is more susceptible to
other processes, leading to a larger number of outliers.

\subsection{Kinematic vs photometric bar strengths}
\label{sec:compq}

\begin{figure}
\includegraphics[width=1.\linewidth]{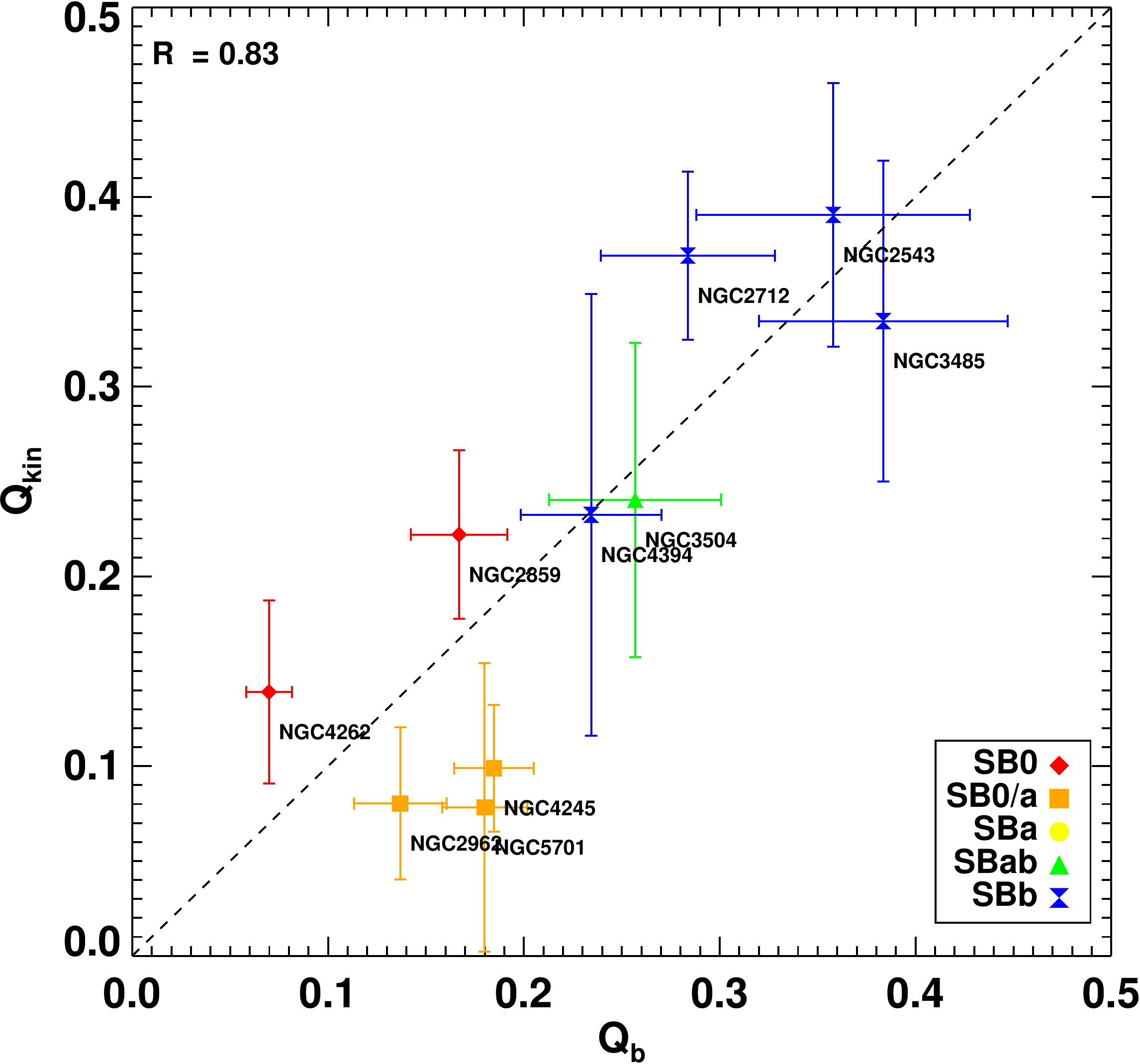}
\includegraphics[width=1.0\linewidth]{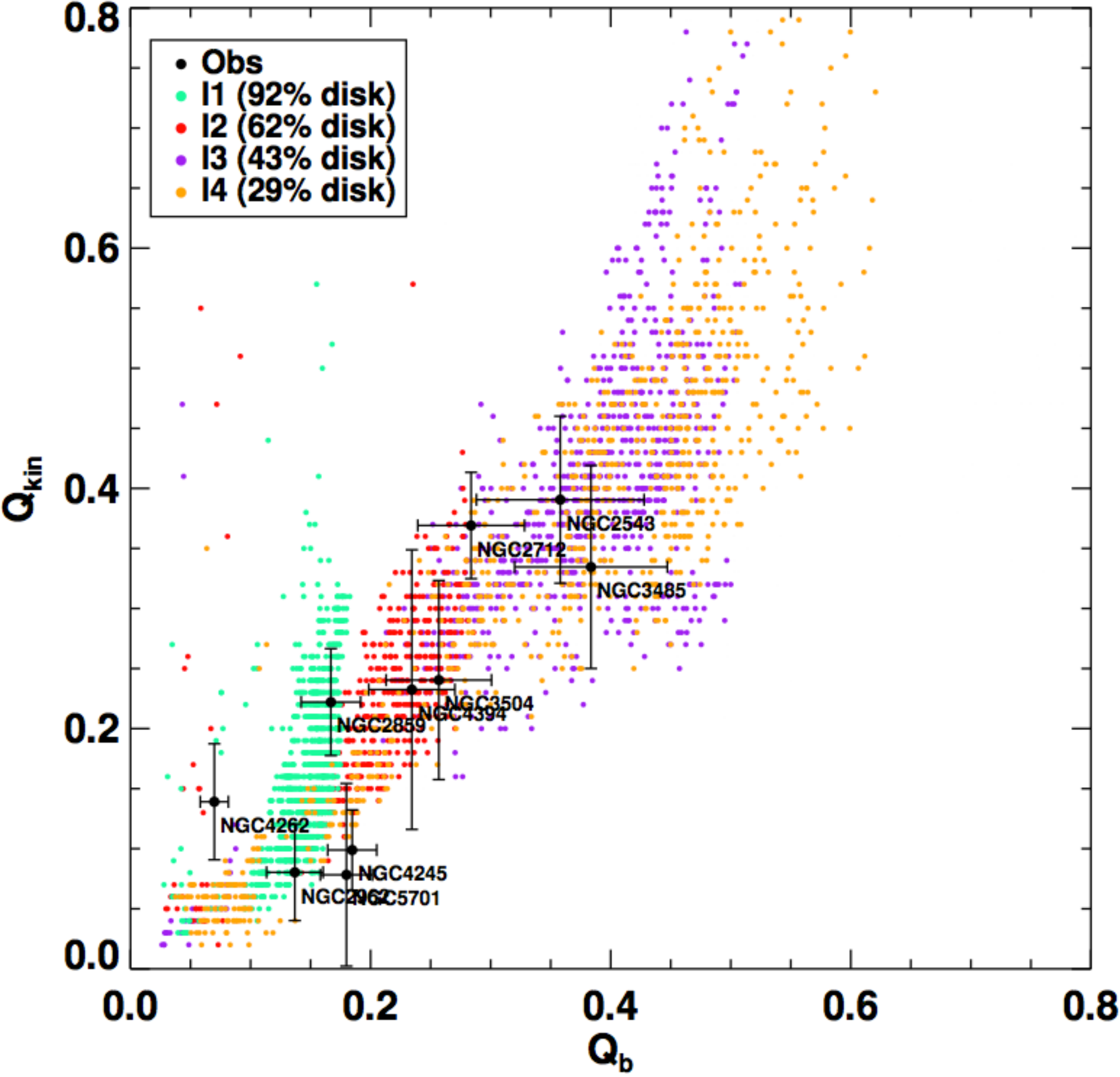}
\caption{(\textit{Top panel}) Kinematic versus photometric torque (\Qkin\ vs \Qb).
The dotted line indicates a one-to-one correlation. Symbols are colour-coded
according to Hubble type. The relation has a linear Pearson correlation coefficient of R=0.83.
(Bottom panel) Comparison of \Qkin\ and \Qb\ for the observations (black) and
the four sets of numerical simulations ($I_1$, $I_2$, $I_3$ and $I_4$).
We only plot stages of the simulations that exclude the buckling phases.}
\label{fig:qbqk}
\end{figure}

Figure~\ref{fig:qbqk} (top panel) compares the kinematic (\Qkin) versus the
photometric (\Qb) torque measurements for the subsample of 10 galaxies, where
the kinematic method was possible. Despite large uncertainties, the correlation
between the two parameters is obvious. This is confirmed by a measured
linear Pearson correlation coefficient of 0.83. The distribution of existing bar strengths
within our limited sample is representative of larger samples of nearby galaxies
\citep{2004bLaurikainen}. Overall, early-type galaxies have lower values than the later
types, confirming earlier results by e.g. \citet{2005Buta,2007Laurikainen}. This could be a
result of different factors: i) the influence of stronger spiral arms that still
alter the motions within the bar region (although we tried to avoid them in
our analysis), ii) the presence of more gas in later types which is more
responsive to the bar could also influence the stellar motions, iii) discs in
earlier types are simply hotter leading to more random versus ordered rotational motion. 
The dotted line in the figure indicates a one-to-one correlation.

To investigate further the relation between stellar \Qkin\ and \Qb, we
have produced an extensive set of numerical simulations of barred galaxies
following those in \citet{2006MartinezV} and \citet{2011MartinezVG}. Here we use
four simulations series, $I_1$, $I_2$, $I_3$ and $I_4$, each one with a
different disc-to-total ratio: 0.92, 0.62, 0.43, 0.29 respectively.
This setup allows us to explore the effect of distinct dark matter haloes on the
torque parameters. We analysed 1800 snapshots taken at different points in time
of the bar evolution. In addition,  we also varied the inclination and the
position angle of the bar relative to the galaxy's position angle to
have different viewing angles and thus assess the influence of these parameters
(see Appendix~\ref{sec:inclPAinf} for more details). The bar strength
measurements of the simulations are presented in Fig.~\ref{fig:qbqk} (bottom
panel).

We analysed the simulations in the same way as the observations. We
calculated \Qb\ from their simulated intensity distribution and \Qkin\ from
their associated stellar velocity maps (see Appendix~\ref{sec:decsim} for
details). Due to the lack of r$_{\rm K20}$ (k-band photometric parameter) to infer the scale height $h_z$ (for the calculation of \Qb) \citep{2008Spelt}, we applied the \citet{1998DeGrijs} relation for intermediate type galaxies which links the scale-height to the scale-length, assuming an exponential disc without truncations. The overall trend found is consistent with what we find with the
observations. Nonetheless, distinct simulation series behave systematically
differently; the figure shows that higher disc fractions consistently lead to
lower bar strengths, both in \Qb\ and \Qkin\ . Each simulation series exhibits
low bar strengths, which correspond to snapshots in very early times in the bar
formation. While $I_1$ soon seems to saturate and cannot grow stronger bars, the
others do and saturate at later stages such that the strongest bars are found in
the simulation series $I_4$, the one with the highest dark matter fraction. 

Given the good agreement between \Qb\ and \Qkin\ for our subset
of galaxies along with the large number of simulations, we will use the photometric
values determined from the S$^4$G images for the bar strength values, because these are available for our entire
sample. Our study also serves for verifying the technique and results of \Qb.

\section{The effect of bar strength on galaxy properties}
\label{sec:discuss}

In this section we try to understand if stronger bars affect the properties of
the host galaxy in a systematic way, focusing on whether it leads to stronger or
weaker kinematic features. 

\subsection{Relation with Hubble type}

\begin{figure}
 \includegraphics[width=1.0\linewidth]{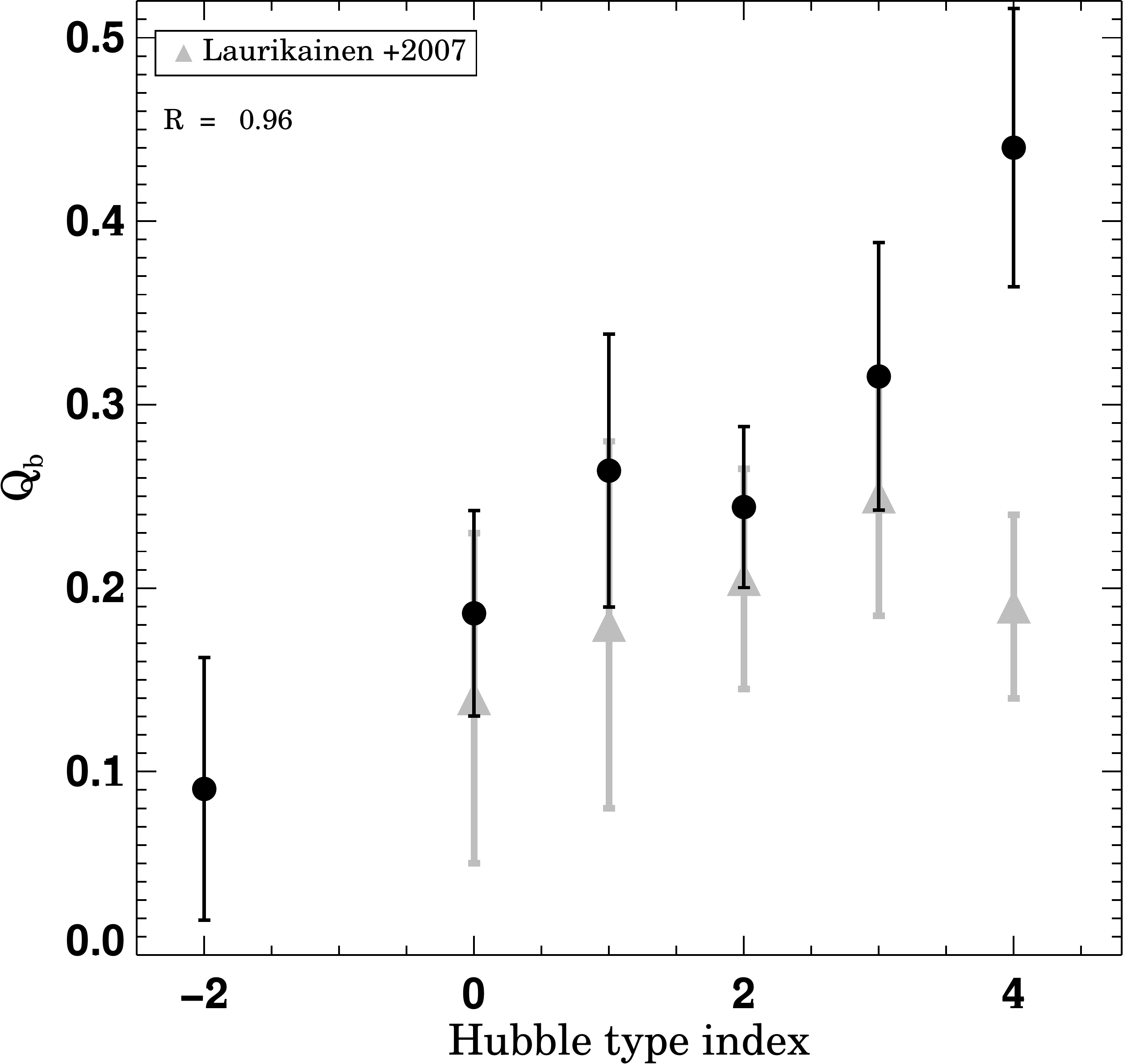}
 \caption{Bar strength as a function Hubble type. We represent our sample averaged per Hubble type in comparison with the sample analysed in \citet{2007Laurikainen}.}
 \label{fig:qbtype}
 \end{figure}

Figure~\ref{fig:qbtype} illustrates the already observed trend of \Qb\ with Hubble type \citep[e.g.,][]{2007Laurikainen}, resulting in a linear Pearson correlation coefficient of $R=0.96$ for our sample (averaged values per Hubble type bin). In comparison with the observed trend found in \citet{2007Laurikainen}, we cover slightly stronger bars throughout but conserving the trend, making our sample representative of rather strongly barred galaxies. For a wider study of \Qb\ as a function of Hubble type based on the S$^4$G sample, please refer to D\'iaz-Garc\'ia (2015, submitted).

\begin{figure}
 \includegraphics[width=1.\linewidth]{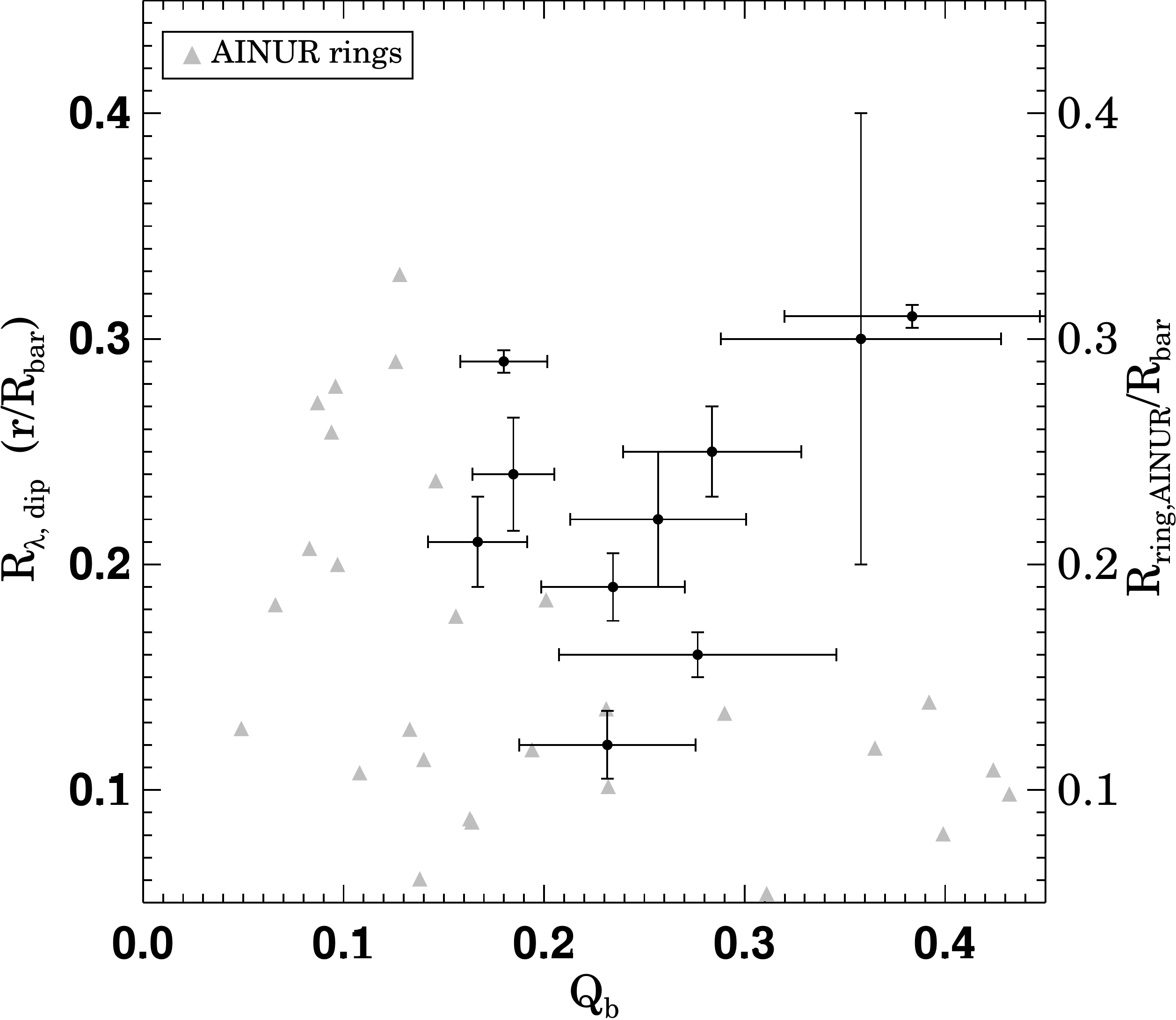}
 \caption{Position of the dip in the $\lambda_{\rm R}$ profile as a function of
 bar strength based on the profiles obtained in Fig.\ref{fig:lamR}. Grey points indicate ring positions (as bar length
 fractions) also as functions of the bar strength, both as measured in the AINUR
 sample \citep{2010Comeron}.}
 \label{fig:qbLh}
 \end{figure}

In Fig.~\ref{fig:qbLh} we depict the position of the dip in the $\lambda_{\rm R}$
 profile, depicted in Fig.\ref{fig:lamR}, as a function of bar strength. The dip is not observed in all galaxies, therefore we only
 show those, which exhibit this feature. Is this dip feature related to inner structures such as nuclear rings?
We plot in gray positions of rings from the AINUR sample \citep{2010Comeron}, including three galaxies (NGC~2859, NGC~3504 and NGC~4245) that we share.
Evidently, ring and dip positions do not correlate and we do not find any mathematically significant trend for either sample.

\begin{figure}
 \includegraphics[width=1.0\linewidth]{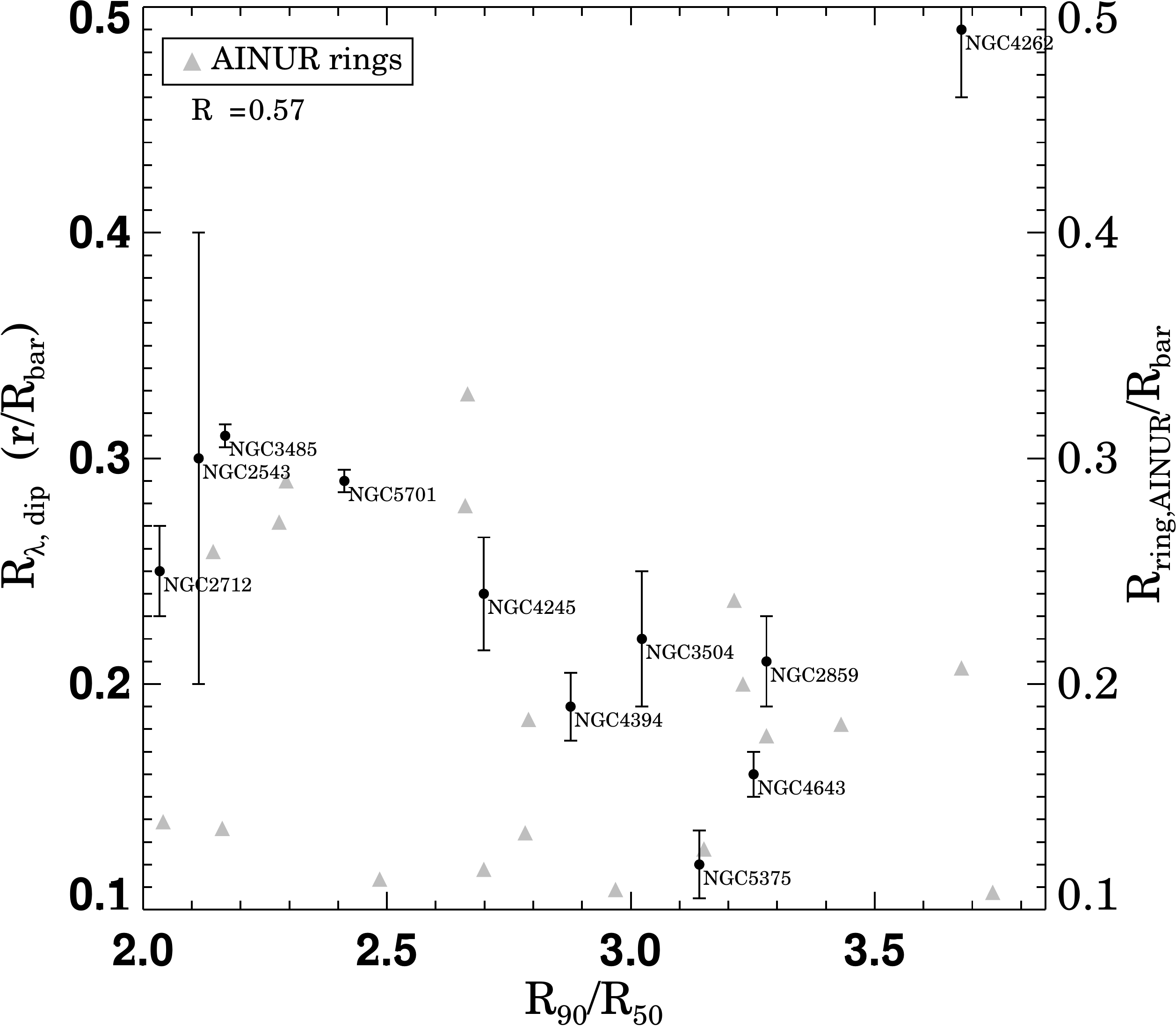}
 \caption{Position of the dip in the $\lambda_{\rm R}$ profile as a function of
 light concentration R$_{90}$/R$_{50}$ based on the profiles obtained in
 Fig.\ref{fig:lamR}. Grey points indicate ring positions (as bar length
 fractions) also as functions of the concentration, derived in the same way as for our sample.}
 \label{fig:concLh}
 \end{figure}

Fig.~\ref{fig:concLh} shows again the position of the dip in the $\lambda_{\rm R}$
 profile, now as a function of light concentration
 R$_{90}$/R$_{50}$. Apart from NGC~4262 - already found
 not to follow other observed trends, probably due to a recent interaction - the
 galaxies seem to follow a downward trend: the more concentrated the bulge, the
 closer is the dip feature towards the center. This could be directly related to
 the bulge: in our simulations with more concentrated bulges, we also find that the
 ILRs are located closer to the center. It could also mean that these features
 are more evolved in time, supporting the prediction of the migration of nuclear
 rings towards the center \citep[e.g.,][]{1995Knapen, 2000Fukuda,2003Regan,2009Ven}, also
 recently observed by \citet{2014PinolF}.
 
This trend is only mildly observed for the values of the AINUR sample, taking 
 their measured ring radii as a comparison, because no $\lambda_{\rm
 R}$ profiles are available for that data. We further determined the position of iILR 
 and oILR by a simple linear approximation analysing  $\Omega$-curves and estimates for the bar pattern speed. 
 Neither the position of the dip in the $\lambda_{\rm R}$
 profile, nor the ring radius are found at the exact same position as these resonances (in a forthcoming
 paper, we will verify this by a more robust calculation of bar pattern speeds).

\subsection{Influence of the bar strength on the global position angle}

\begin{figure}
\includegraphics[width=1.0\linewidth]{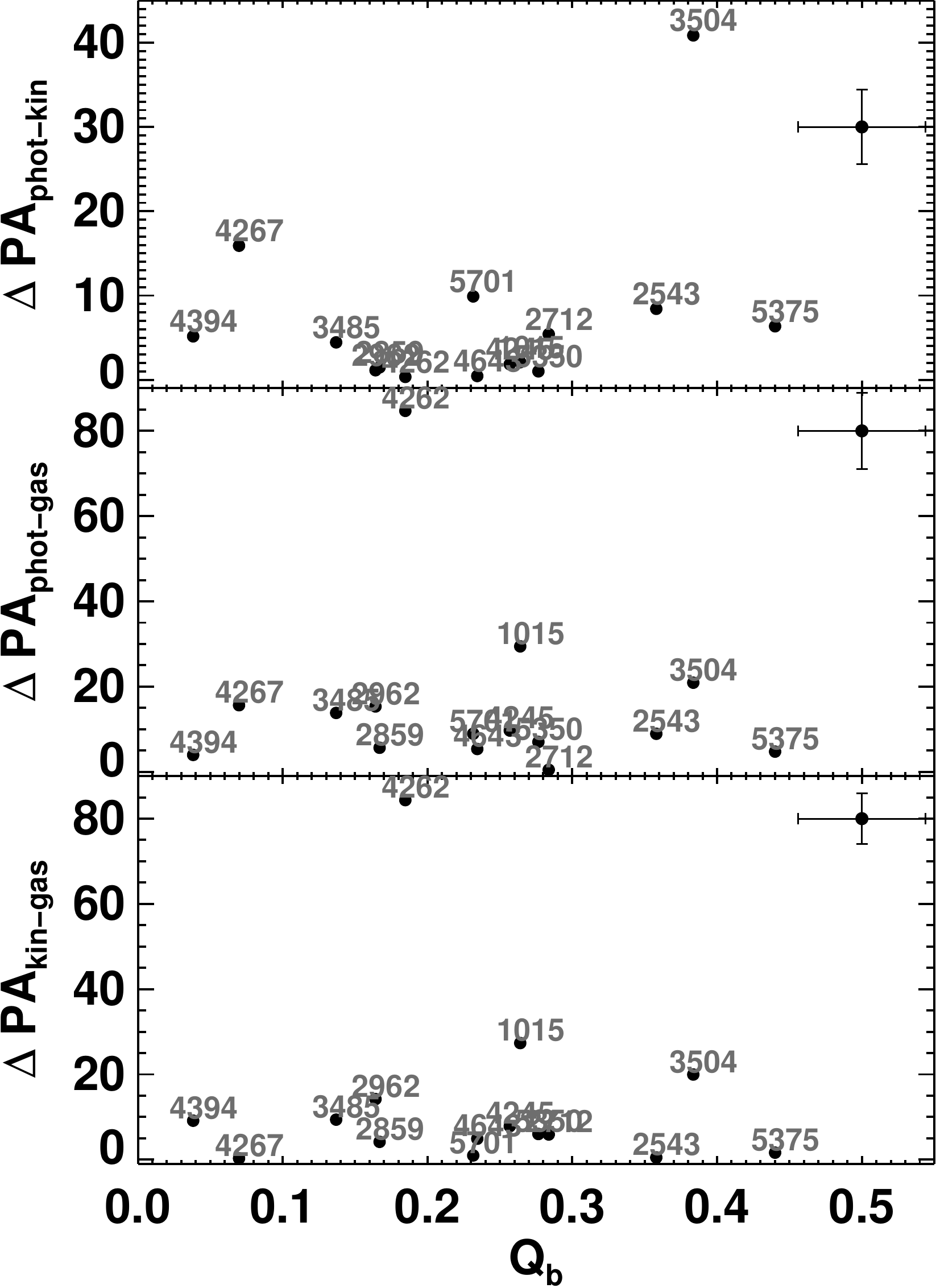}
\caption{Photometric and kinematic position angle differences between stars and
ionised gas as a function of bar strength. \textit{Top:} difference between
global photometric and stellar kinematic PA. \textit{Middle:} difference of
photometric to kinematic gas PA. \textit{Bottom:} difference between the stellar
kinematic and gas kinematic PA. Representative error bars are indicated in the
top right corners.}
\label{fig:QPA}
\end{figure}

To quantify the influence of the bar on the global velocity map, we
analysed the difference between the photometric and kinematic position angles as
a function of the bar strength. We compared the photometric PA with the stellar
and gas kinematic PA, as well as the difference between the stellar and gaseous
components. This is shown in Fig.~\ref{fig:QPA}. We measured the kinematic PAs
directly from the velocity maps following \citet{2014BarreraB}.

No trends are observed with bar strength in any of the three cases. In previous
studies including barred galaxies \citep[e.g.,][]{FalconBarroso2006,
2009Fathi,2011Krajnovic, 2014BarreraB}, the observed misalignments were neither
strong nor systematic. Misalignments were found mainly in systems with complex
kinematics (non-regular rotators), systems in interaction and only in some
barred galaxies, but their amplitudes largely depended on the FoV. The detailed
study by \citet{2014BarreraB} further concludes that morphological substructures
only influence the redistribution of angular momentum, but the global kinematics
such as rotation are driven by the overall disc mass. Here, we also only
detected one galaxy, NGC\,3504, with a larger difference between the photometric
and kinematic PAs. The ionised gas is known to react strongly to the bar
producing a twist in the zero-velocity curve \citep[e.g.,][]{1980PetersonH,
2006Ems}. However, for the gas too, we only detected large misalignments in
NGC\,4262. The overall misalignment is slightly larger than that for the stellar
kinematic/photometric PA difference, but not significantly. Finally, the
difference between the two kinematic PAs (stellar and ionised gas PA) is found
to be equally small and not correlated with the bar strength, confirming the
results from earlier studies (as described above).

\subsection{Stellar kinematic features related to bar strengths}
\label{kinBS}

Bars can be depicted as engines that, on the one hand, drive gas towards the
central regions and consequently nourish star formation
\citep[e.g.][]{1985ElmegreenE, 1989ElmegreenE,2005Erwin, 2011Ellison} and, on the other hand, support radial motions of
stars \citep[e.g.,][]{2010Minchev, 2011Brunetti}. Due to these factors, they are
natural triggers of changes in the centre of galaxies, and in the stellar
velocity dispersion in particular. We do not find, however, any trend between
the central velocity dispersion and the bar strength. This reinforces the
picture that the central stellar velocity dispersion is determined by global
galaxy properties. At least, it does not vary significantly due to morphological
substructures in a systematic way, except for the occasional central
$\sigma-$drop.

While simulations predict and find a significant influence of the bar on the
host galaxy in various ways \citep[e.g.,][]{2006MartinezV, 2013Atha,
2014Sellwood}, we only find mild signatures on the kinematic maps in our sample,
such as the proposed double-hump rotation curve and occasional $\sigma-$drops.
Despite this lack of major, bar-induced alterations in global galaxy kinematic
parameters, we detect some relation between those subtle kinematic features and
the bar strength. It is thus logical to assume a connection between those features
and the bars.

Since double-humps and $\sigma-$drops exist commonly
among barred galaxies, we tested their amplitude in relation to the bar
strength. In other words, would stronger bars produce stronger humps or deeper
drops? We quantified the strength of the hump by the difference of its inner
peak and consecutive drop, calling this parameter $\Delta$V. We further
normalised this value by the maximum rotation -- corrected for inclination --
that we could detect for each galaxy. We are aware that asymmetric drift could
attenuate this signal slightly, but do not expect a major change for the trend observed.

For the velocity dispersion we determined
the amplitude of the central $\sigma-$drop if present, again normalised by the
maximum velocity dispersion (following \citealt{2012Peletier}). We chose to
compute these quantities at the position angle where the signal was stronger.
Since the hump in the velocity is seen strongest along the major axis, we took
the profile along that axis. We chose to take the velocity dispersion profiles
along the bar major axis, because the drop is most pronounced along that direction. 

\begin{figure}
\includegraphics[width=\linewidth]{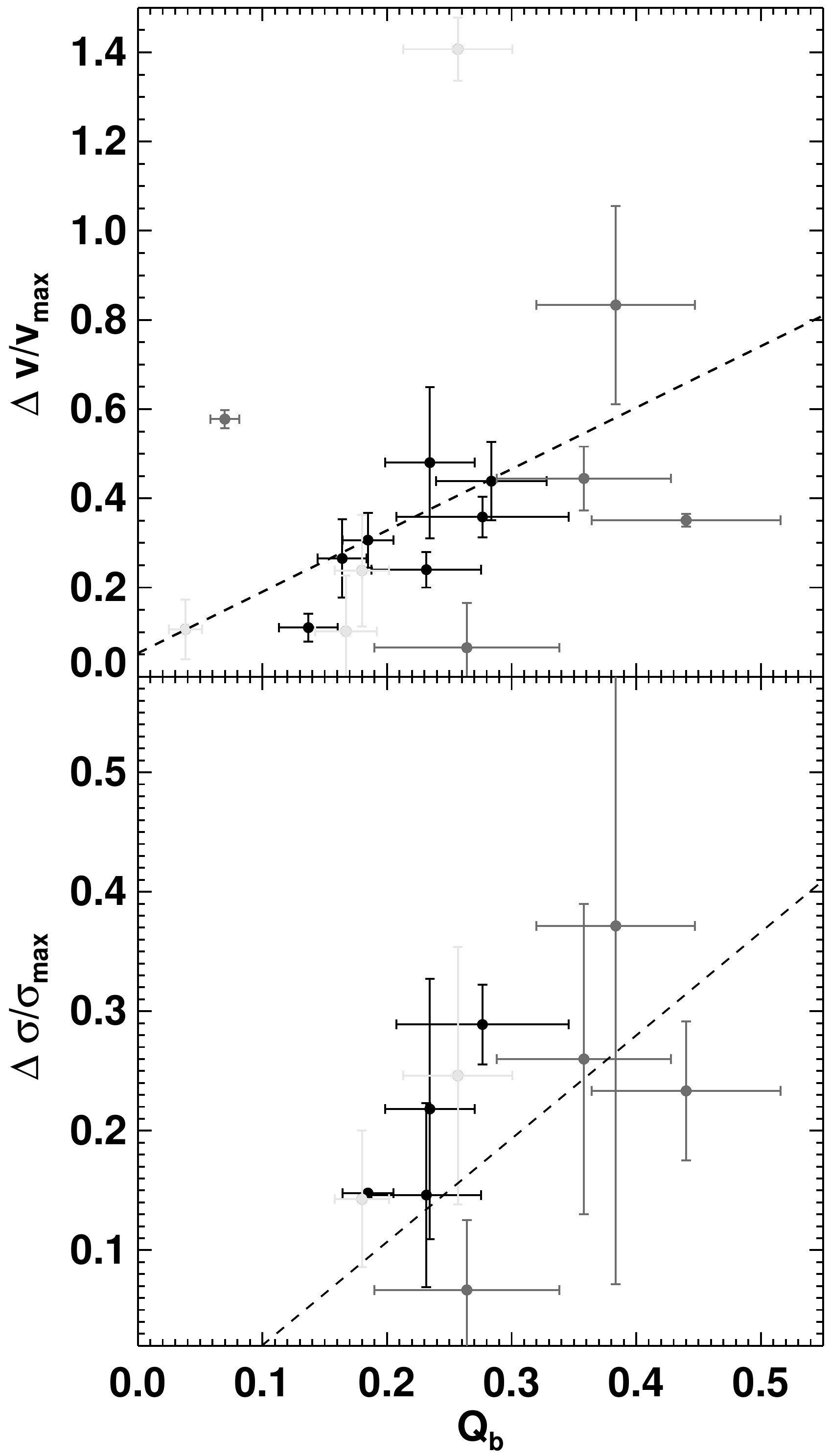}
\caption{Different parameters detected in the kinematic maps of barred galaxies
as a function of bar strength. \textit{Top:} Magnitude of the velocity hump 
along the major axis rotation curve, normalised by the maximum rotation velocity
and corrected for inclination (see text for details). \textit{Bottom:}
Magnitude of the $\sigma-$drop, normalised by the maximum dispersion (i.e.
difference between central drop and highest surrounding velocity dispersion). 
In both panels, the dashed line indicates a linear fit to the black and dark gray points.}
\label{fig:Qkingrads}
\end{figure}

Figure~\ref{fig:Qkingrads} shows the results. We identified galaxies with very
low inclinations in light grey, galaxies with intermediate but still low
inclinations or larger uncertainties in their velocity fields in grey, and
reliable points in black. As our sample is very small, we did not 
discard any points, but indicate that we are conscious about the bias introduced
by measuring at different inclinations. As inclination effects in the velocity
dispersion are very difficult to characterise (i.e., it depends on the projection
of the velocity ellipsoid being probed and anisotropy), we did not attempt any
correction.

We find tentative evidence that stronger bars produce stronger humps in the
velocity profile. After the
inclination correction, the low-inclination galaxies also follow this trend and we
obtain a linear Pearson correlation coefficient of R\,=\,0.57. Discarding unreliable galaxies (light gray points),
the correlation coefficient increases to R\,=\,0.76. As the hump could sometimes be
distinguished better in the ionised gas, we also determined these parameter in
the gas velocity profile (not shown here). The results follow the same trend. In
the bottom row of Fig.~\ref{fig:Qkingrads}, we show the measurement of the
magnitude of the $\sigma-$drop. Stronger bars produce a stronger
$\sigma-$drop features. We obtained an overall linear Pearson correlation coefficient of
R\,=\,0.74. All except one of the galaxies without a central dispersion drop
have \Qb\ values below 0.15. The lack of this drop feature seems to be most
evident in galaxies with weaker bars.

\subsection{Stellar angular momentum as a function of bar strength}

\begin{figure}
\includegraphics[width=\linewidth]{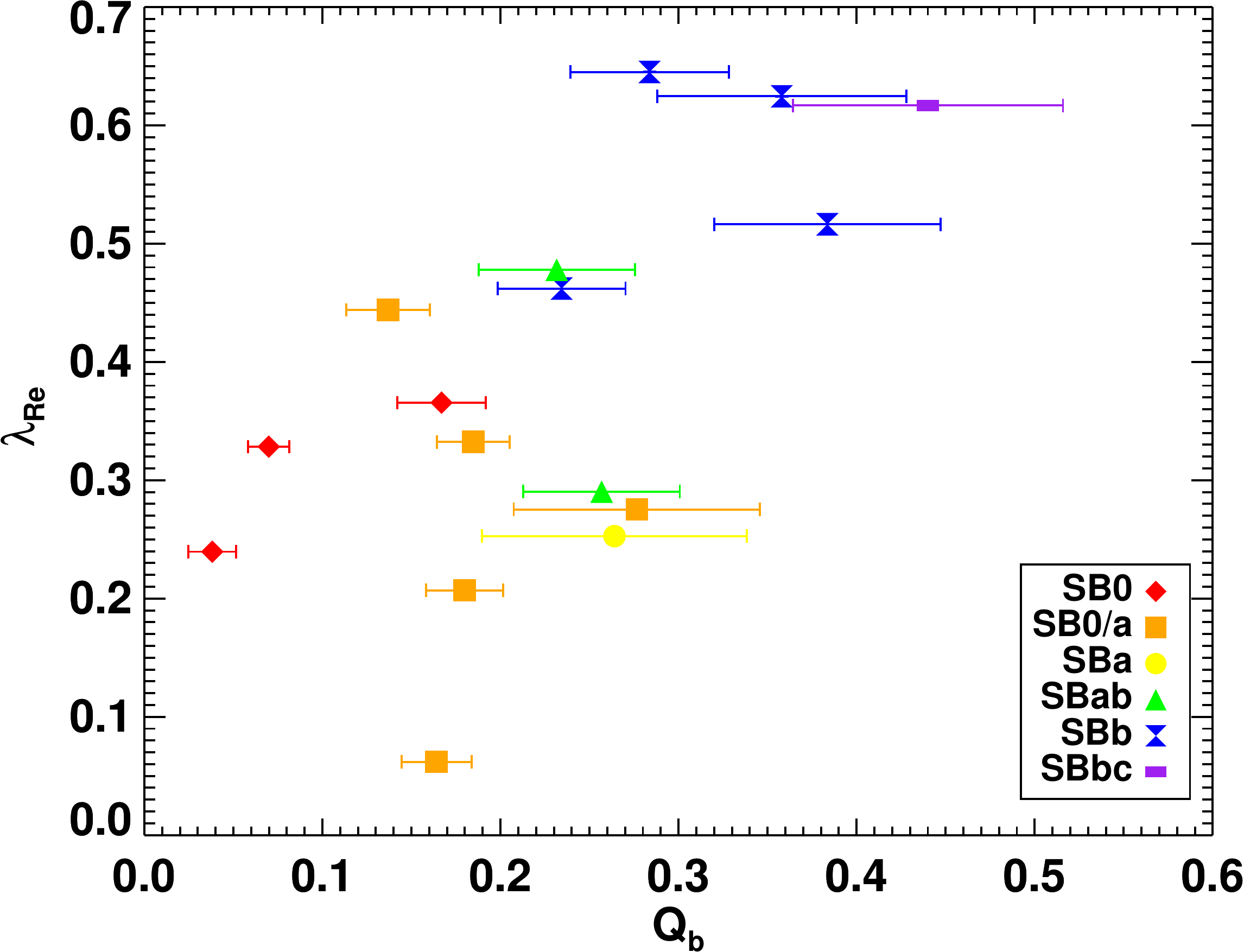}
\caption{Apparent stellar angular momentum within one effective radius
\citep{2007Emsellem} as a function of bar strength. The different Hubble types
are shown in different colours.}
\label{fig:LamQ}
\end{figure}

We now inspect the influence of the bar on the integrated angular momentum
within one effective radii ($\lambda_{\rm Re}$). This is shown in
Fig.~\ref{fig:LamQ}. The values obtained for $\lambda_{\rm Re}$ are consistent
with the values found in the literature for barred galaxies
\citep[e.g.][]{1994Bender, 2008Krajnovic, 2011Krajnovic}. These studies,
however, do not include bar strength measurements. We observe an increasing
value of $\lambda_{\rm Re}$ with bar strength. This is somewhat connected to
Hubble type, because the later-types in our sample display the largest $\lambda_{\rm Re}$ values.
The high $\lambda_{\rm Re}$ values observed in the later-type galaxies are likely 
due to the higher fraction of the disc, and thus high rotation, included
within the one effective radius aperture.

As there is angular momentum transfer between the bar, disc and outer halo \citep[e.g.,][]{1993Combes, 2006MartinezV}, the value of $\lambda_{\rm Re}$ should be higher in barred galaxies
compared to their non-barred counterparts. \citet{2003Atha} showed, however, that while
angular momentum is transferred to the disc, the bar also slows down and
therefore contributes to a decrease in $\lambda_{\rm Re}$. The current available
$\lambda_{\rm Re}$ values in the ATLAS3D \citep{2011Emsellem} or CALIFA
samples \citep{2014FB} do not indicate distinct values for barred
and non-barred galaxies.

\subsection{Bars as drivers of radial motions}
\label{sec:radm}

Bars have been studied as a major driver of radial mixing for a long time
\citep[e.g.][]{1993FriedliB}, but spiral arms \citep[e.g.,][]{2002SellwoodB} or
the combination of their resonance overlap \citep[e.g.,][]{2010Minchev,
2011Shevchenko} are also held responsible for an increase of this.
Investigating the latter, \citet{2011Brunetti} found that kinematically hot
discs are not as efficient environments and exhibit less radial mixing than
kinematically colder ones. We investigate the level of radial motions induced by
bars in our sample by studying the radial gradients of the stellar velocity
dispersion. The expectation is that barred galaxies would display flatter
gradients than those measured in non-barred systems.

\begin{figure*}
\includegraphics[width=1\linewidth]{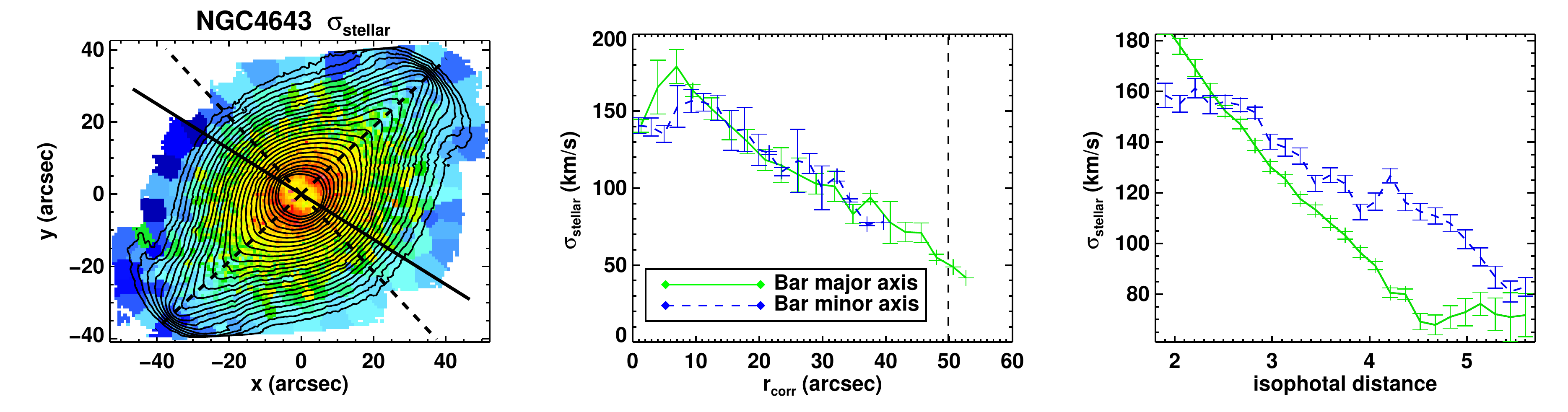}
\caption{Stellar velocity dispersion map of NGC\,4643 and associated profiles.
(\textit{Left}) Stellar velocity dispersion map with bar major and minor
axes indicated with dashed lines. (\textit{Middle}) Averaged profiles, corrected for inclination. Dashed line indicates the bar length.
(\textit{Right}) Averaged
profiles along isophotes (first point corresponds to first isophote seen on the
top left image). }
\label{fig:sprof}
\end{figure*}

We start by comparing velocity dispersion profiles along the major and minor
axes of the bar. In earlier literature studies, velocity dispersion profiles
were investigated typically only along the major axis of the bar
\citep[e.g.][]{2009Perez}. For our sample, this is shown in Fig.~\ref{fig:sprof}
for one galaxy as an example (similar plots for other galaxies are presented in
Appendix~\ref{app:gaskinmaps}). We show the overall stellar velocity dispersion
map for reference and, next to it, the radial profiles along the major and minor
axes (minor axis radii are corrected for inclination) of the bar extracted from this map. The profiles along the axes are
generally overlapping, hence we do not observe any increase along the bar major
axis. In one third of our sample, however, we observe a mild difference around
the central parts (also seen in the example), with the major axis showing a
higher dispersion. This is probably linked to the aforementioned kinematic
substructures such as inner discs or rings, possibly a result of barred secular
evolution. The major axis profiles observed in \citet{2009Perez} show a
similar behavior to ours, but no minor axis measurements have been performed in that work.

As bars are structures seen in the photometry, we decided to also trace the
profile comparing their points along the same isophotes. The isophotal profiles
reveal a larger overall $\sigma$ along of the major axis than the minor axis,
and not just in the central parts. The fact that the velocity dispersion further
out is higher along the minor axis, compared with the same isophote on the major
axis, shows that the dispersion of the bulge -- traced by the minor axis --
dominates strongly, regardless of the prominent bar seen in the photometry. It
shows nonetheless that the kinematics of the bar is significantly different than
the bulge and it is more similar to the disc, because at the outer end of the
profiles, reaching the disc, values along the major and minor axes start
coinciding again.

\begin{figure}
\includegraphics[width=\linewidth]{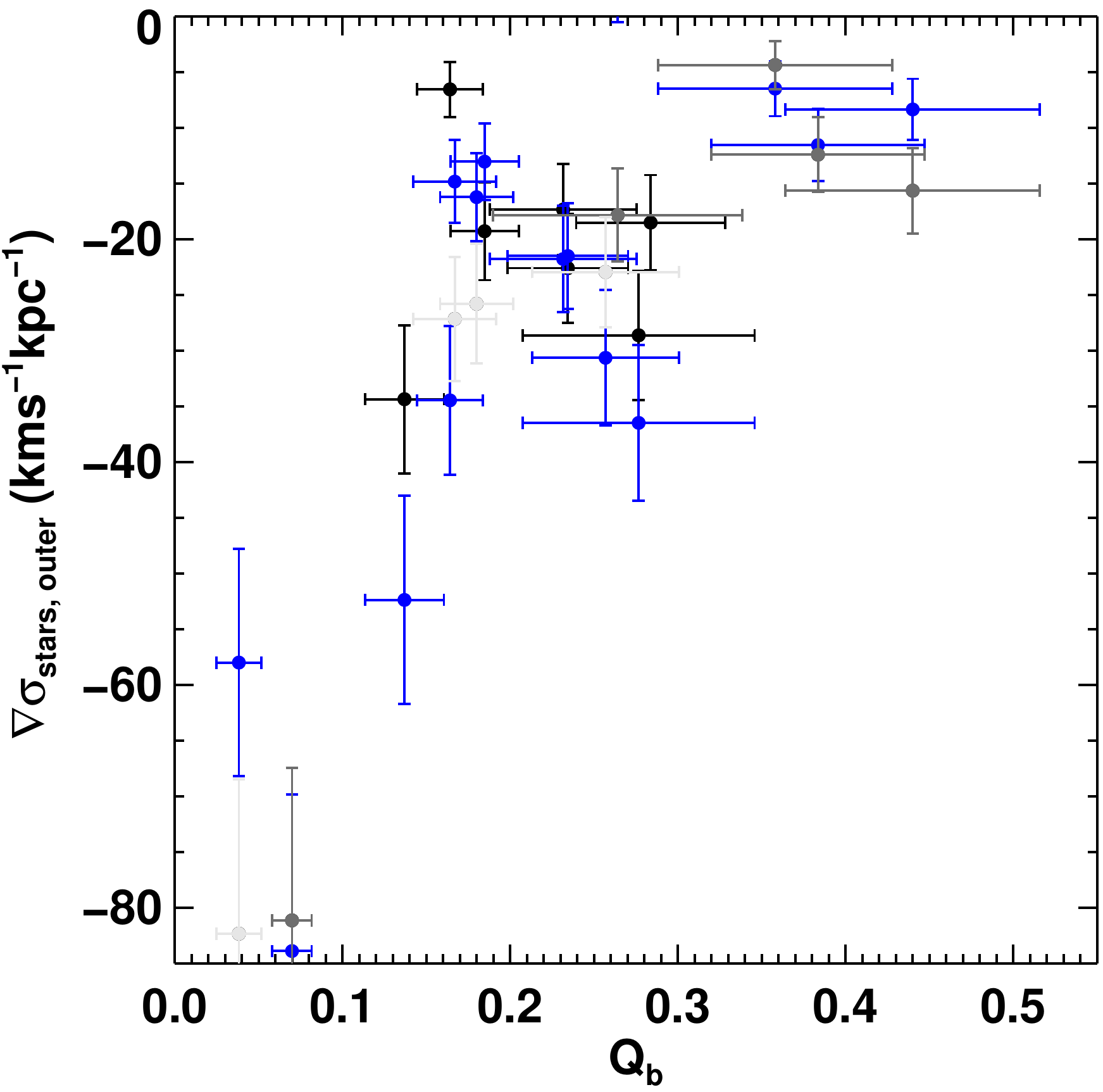}
\caption{Gradient of the outer velocity dispersion, ignoring the central regions, as a function of bar strength. Blue points indicate profiles taken along the bar minor axis, while black points show the ones along the bar major axis.  }
\label{fig:Qkingrads2}  
\end{figure}

Figure~\ref{fig:Qkingrads2} shows a relation between the outer gradient of the stellar velocity 
dispersion and the bar strength. The gradient is shallower for stronger 
bars. Nevertheless, the 
trend is based on only very few points, in particular in the low bar strength 
regime. 
The presence of a bar can cause enhanced radial motions which perturb the system. Thus the orbital mixing increases which in turn can lead to higher dispersion and shallower gradients. This could be the reason for the observed flattening of the gradients with higher \Qb.
Additionally, we measured the gradient 
not only along the major axis of the galaxy (black points), but also along 
the minor axis (blue points). If the bar would significantly flatten the gradient along the 
major axis, the minor axis values would be expected to show steeper gradients. 
The results, however, show a scatter of shallower as well as steeper gradients 
along the minor axes compared to the points measured along the
major axes and we cannot identify a systematic behavior.

The measurement of the maximal radial velocities that we use to calculate the 
kinematic torques also indicate the average radial displacement. As 
mentioned in previous sections, the average radial motion is much higher in the 
gas than for the stars. Stars move at velocities between 10 and 60 km/s radially, 
corresponding to about 10-60 kpc/Gyr or 0.09 - 0.88 when normalised with the 
rotation at $\approx$\,\reff\,, a value which is similar in
magnitudes to what we find with our own simulations (e.g. in $I_4$ it is around 0.2)
whereas the gas move at 40 - 100 km/s and in extreme cases such as NGC~4262 at more 
than 300 km/s. The latter is most probably due to an outer influence such as an
interaction \citep{2005Vollmer}. Nevertheless, the stronger effect on the gas 
than on the stars has been seen in numerous simulations \citep[e.g.]{2013Atha, 
2013Kubryk}. \citet{2014arXiv1412.0585K} find a particular influence of the 
bar-induced radial inflow on the gaseous profile. Furthermore, 
\citet{2012Maciejewski} obtain values which are in the range of the ones we 
recover using the same method. The recent work of \citet{2015Goz} analyzing two 
simulations of barred galaxies resulting from N-body+SPH cosmological 
simulations, shows one case with significantly higher radial motions (around 
150 km/s) whereas in the other case the magnitude is comparable to what we 
observe (around 30 km/s). 

In a large number of simulations \citep[e.g.,][]{2012Minchev} bars are found to 
be the most efficient driver of radial migration, in particular through their 
corotation resonance. We have not yet determined the radius of corotation for 
our sample, but plan to compute this in a forthcoming paper. This parameter, 
together with the stellar populations will further complete the picture of bar-induced 
mixing. In particular in Paper II of this series, we will assess the impact 
of the radial motions determined here, on the stellar population
properties, which will allow us to shed light onto radial migration 
effects \citep[e.g.,][]{1994Friedli, haywood2008,Roediger2012, 2013Kubryk}.

\section{Summary and conclusions}
\label{sec:summ}

We present the BaLROG sample of 16 barred galaxies of different Hubble types, spanning
the typical bar strengths found in the local Universe. Our large mosaics with
the integral field unit SAURON cover the bars out to the radius where the disc begins to dominate, at a spatial 
resolution of typically 100\,pc. 
For every galaxy we also use $Spitzer$ observations from the S$^4$G
survey of nearby galaxies \citep{2010Sheth} to determine several photometric
parameters, as well as to derive the bar strength \Qb. From the velocity maps, we 
calculate radial and tangential velocities to compute the bar strength
based on the kinematics, \Qkin. Our aim is to establish a reliable yardstick,
namely bar strength, to probe the influence of the bars on different parameters
of the host galaxies.  

In this paper we focus on the kinematics of the galaxies,
deriving stellar and gas velocities and velocity dispersions, h$_3$ and h$_4$
Gauss-Hermite moments and the stellar angular momentum $\lambda_{\rm R}$ and carefully comparing to a large set of N-body simulations.
The analysis of our observations leads to the following results and conclusions:

\begin{itemize}

\item Bars do not strongly influence the global kinematics of their host
galaxies, regardless of their strength. Our work confirms previous studies
\citep[e.g.][]{FalconBarroso2006, 2009Fathi,2011Krajnovic, 2014BarreraB} and
shows the lack of strong kinematic misalignments between the galaxies'
photometric and kinematic axes.

\item Bars do have an influence on more subtle kinematic features, especially
in the inner regions of galaxies. We detect double-hump velocity profiles and
velocity dispersion drops \citep[e.g.,][]{2005Bureau}, which increase
in intensity with increasing bar strength. 

\item We find evidence for the presence of inner structures such as inner
rings or discs in about 50\% of our sample. These features are detected from the
anti-correlation between h$_3$ and and V/$\sigma$ within the effective radius of
the galaxies ($\approx$0.1 bar lengths).

\item The derived $\lambda_{\rm R}$ profiles show a dip at 0.2$\pm$0.1
R$_{\rm bar}$, which we suggest is connected to the presence of inner
substructures.

\item We also derived the integrated angular momentum within one effective radius
($\lambda_{\rm Re}$) and find that galaxies with stronger bars exhibit a higher
$\lambda_{\rm Re}$ value. This may be a secondary effect of late-type galaxies, because they
are more rotationally supported and thus also host stronger bars.

\item We developed a new method to determine the bar strength from stellar
or ionised gas velocity maps (\Qkin). This method relies on the extraction of
the ratio of radial and tangential velocities using the technique developed by \citet{2012Maciejewski}.
Values of this parameter agree well with independent measurements obtained from
imaging, \Qb, e.g., \citet{2002Laurikainen2}, and predictions from numerical N-body
simulations.

\item Bar strength values measured from ionised-gas kinematic maps are a factor
$\sim$2.5 larger than those determined from the stellar kinematic maps.

\item We observe a flattening of the outer stellar velocity dispersion profiles 
with increasing bar strength.

\end{itemize}
 
These results suggest a complex influence of bars in nearby galaxies, especially
affecting central regions. We do not observe a significant influence on global properties, 
but bars seem to affect only on small scales. The gas is clearly more strongly 
affected, reflected in higher gaseous than stellar torques. In our sample we detect a difference 
between bars in early and late-type galaxies hinting towards a different mechanism, 
maybe due to the presence of higher and lower gas
fractions. To better answer these questions and determine time scales, we will
investigate the stellar populations of these galaxies in detail in BaLROG II 
(Seidel et al., in prep) and also determine their corotation radii and pattern speeds in
relation with their dark matter fractions (BaLROG III) .

\section*{Acknowledgments}

We would like to thank Alfonso Aguerri for stimulating comments and discussion and
Martin Stringer for editorial help. MKS and JFB wish to express their gratitude
to the Roque de los Muchachos Observatory on La Palma and the different
operators of the William Herschel Telescope during our numerous runs, as well as
observing support by Agnieszka Rys, Carolin Wittman and Thorsten Lisker. We also
acknowledge support from grant AYA2013-48226-C3-1-P from the Spanish Ministry of
Economy and Competitiveness (MINECO). IMV acknowledges support from grant
AYA2009-11137.  JHK acknowledges the support from grant AYA2013-41243-P. MKS acknowledges the support of the Instituto de Astrof\'isica
de Canarias via an Astrophysicist Resident fellowship. We also acknowledge
support from the FP7 Marie Curie Actions of the European Commission, via the
Initial Training Network DAGAL under REA grant agreement number 289313. 
This research made use of Montage, funded by the National Aeronautics and Space Administration's Earth Science Technology Office, Computational Technnologies Project, under Cooperative Agreement Number NCC5-626 between NASA and the California Institute of Technology. The code is maintained by the NASA/IPAC Infrared Science Archive. This research has made use of the NASA/IPAC Extragalactic Database (NED) which is operated by the Jet Propulsion Laboratory, California Institute of Technology, under contract with the National Aeronautics and Space Administration. The paper is based on observations obtained at the William Herschel Telescope, operated by the Isaac Newton Group in the Spanish Observatorio del Roque de los Muchachos of the Instituto de Astrof\'isica de Canarias. 

\bibliographystyle{mn2e_fixed.bst}
\bibliography{refs_jabref_v4}

\newpage

\appendix{}

\section{Description of the simulations}
\label{sec:decsim}

\begin{figure}
 \includegraphics[width=1.\linewidth]{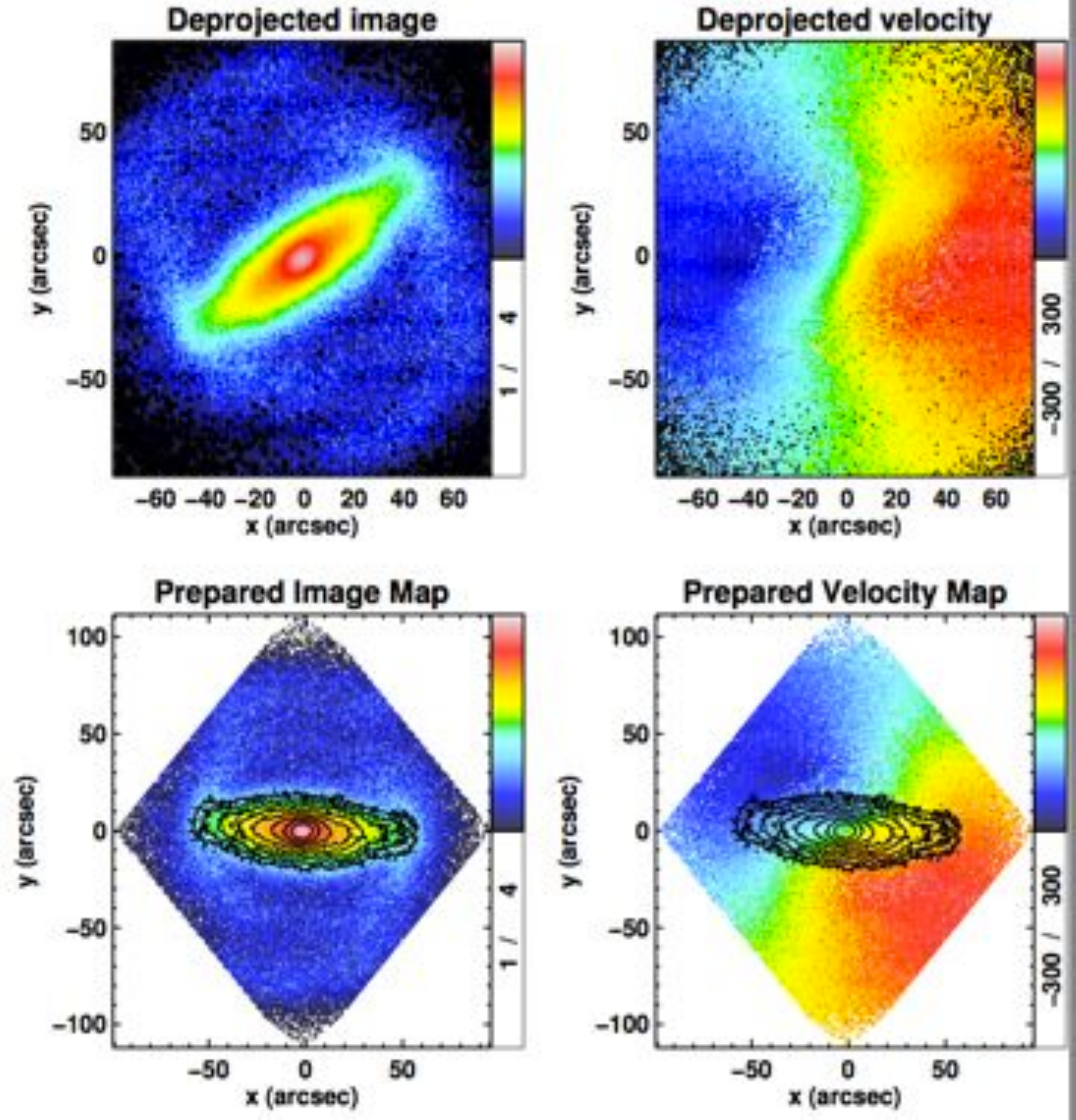}
  \caption{Intensity (in logarithmic units) and velocity maps 
(translated in km/s) of one of the snapshots of the $I_3$ simulation at an 
intermediate point of the bar evolution t=150 at PA = 35 and inclination = 30. The bottom panels show the 
prepared maps which can be symmetrised by the code, with the bar in a 
horizontal position and adjusted field extensions. In those, we overlay the isophotes.}
  \label{I1}
  \end{figure}

\begin{figure}
 
\includegraphics[width=1.\linewidth]{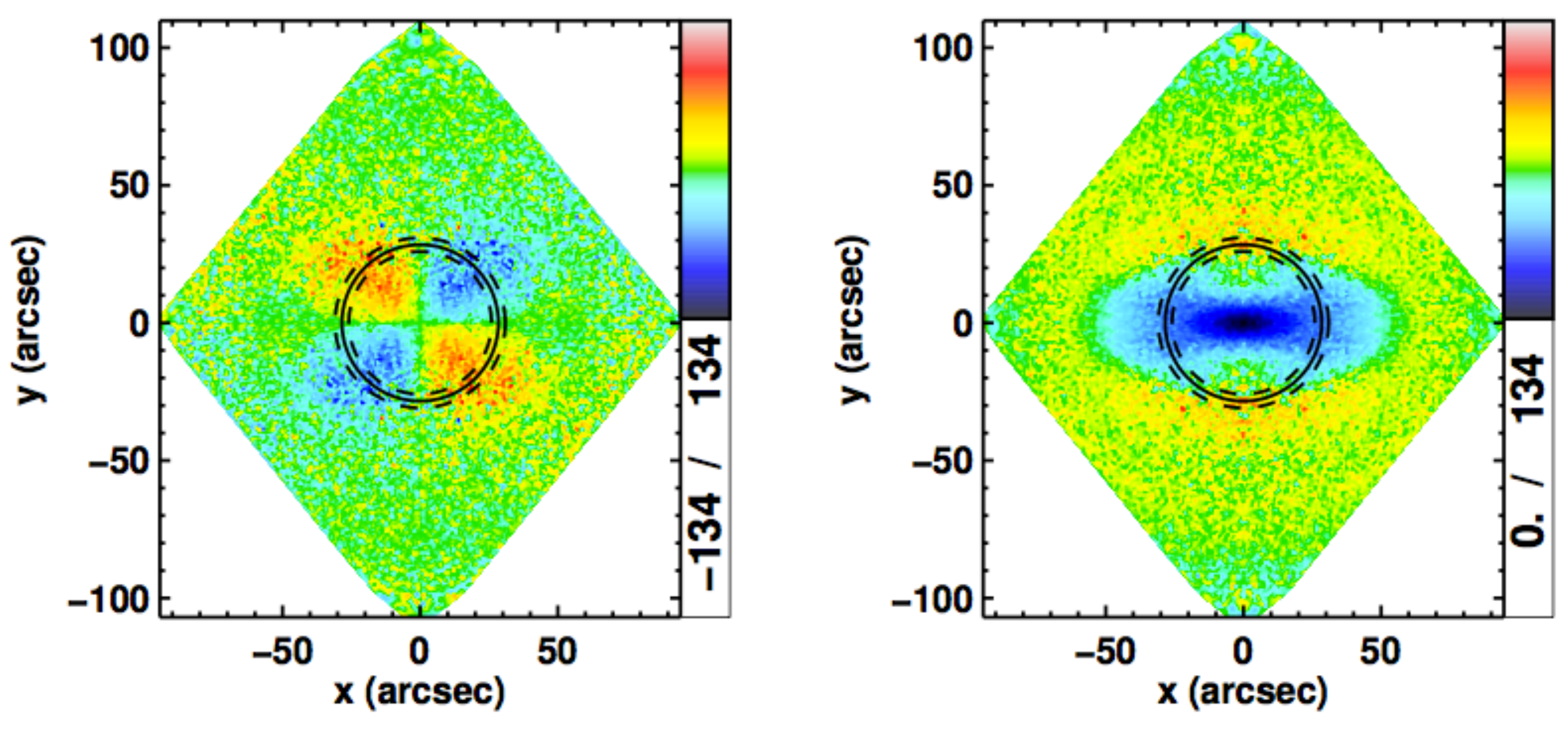}
  \caption{The obtained radial and tangential velocities for the simulated 
galaxy above, but now symmetrised.}
  \label{I1-2}
  \end{figure}

\begin{figure}
 
\includegraphics[width=1.05\linewidth]{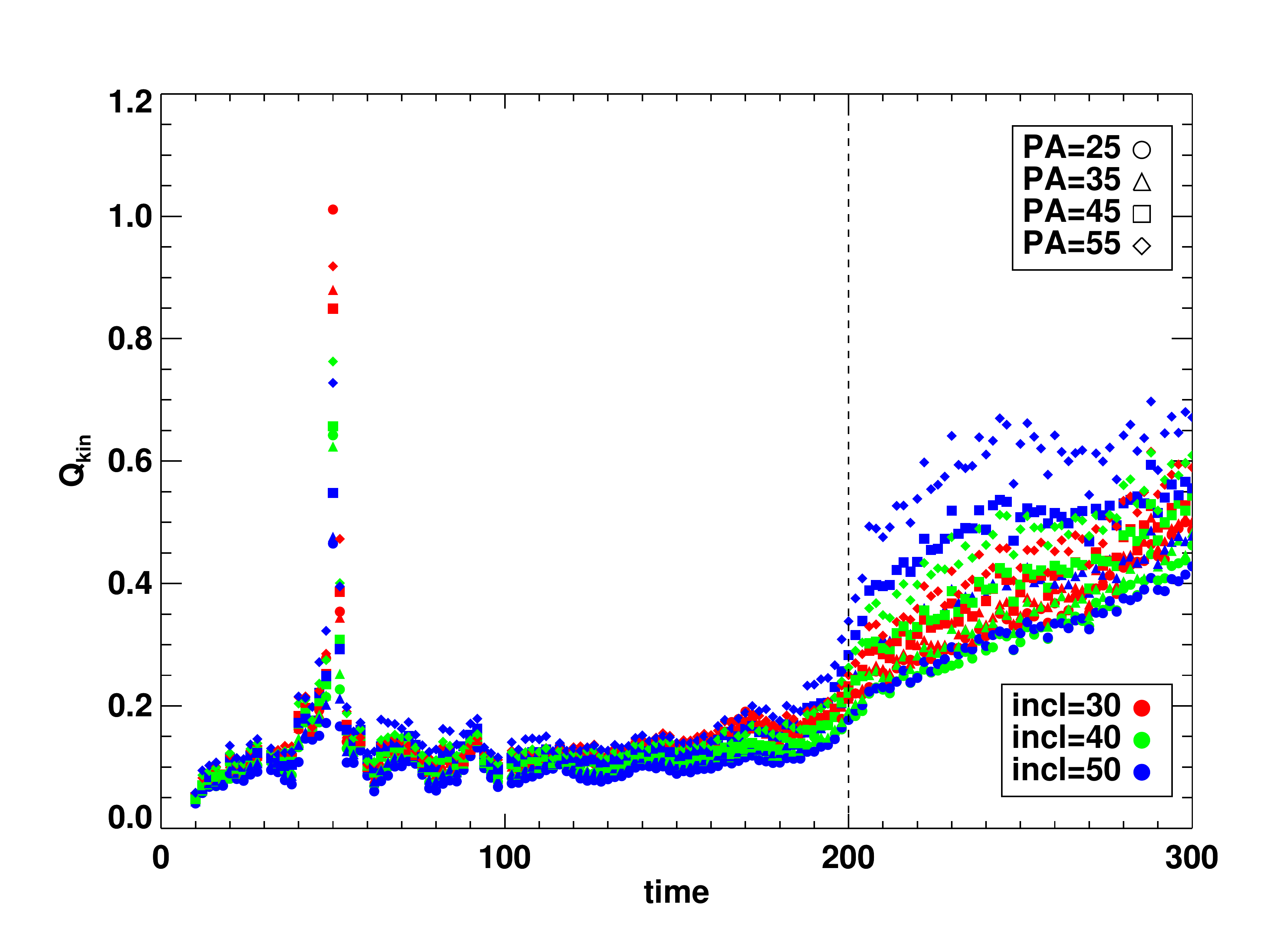}
  \caption{The kinematical torque evolution for one complete simulation, 
here $I_3$ as an example, indicating different inclinations and PAs. The dotted 
line indicates the point in time where the bar has reached its full strength 
and further evolution is not reliable, hence points to the right of 
it will not be considered in our analyses.}
  \label{sim1time}
  \end{figure}

In the following we illustrate the give a more detailed description of the simulations and how we used to obtain the torque measures. 
Figure~\ref{I1} shows as an example the intensity and velocity maps of one of the snapshots of the $I_3$
simulation at an intermediate time step, at an inclination of 30 degree and a 
bar-to-line-of-nodes-PA of 55 degree. Figure~\ref{I1-2} shows the corresponding 
obtained radial and tangential velocities. In Fig.~\ref{sim1time}, the full 
time series (all 1800 snapshots) of this particular simulation is shown. The 
early very strong peak is associated with the first buckling of this bar (not 
stable in time). The shape of the obtained curve indicating the kinematical 
torque values resembles the curve of the A2 value measured during the 
simulation and also gives an indication on the bar strength. For the other 
simulations series, these measurements along with the shape of the final curve 
differs significantly.

The variations of $\rm {Q}_{\rm kin}$ with inclination and PA (shown with 
different colors and symbols respectively in Fig.~\ref{sim1time} are due to 
the fact that the assumptions of a thin disc and stable bar are not 100\% 
correct. This is the case especially during the buckling event of the bar 
and during the later stages of the evolution, which is to be expected due to 
the thickening of the bar. If it was a perfect measurement, all points should 
overlap vertically, since we simply rotate the simulated galaxy slightly in 
order to achieve its different orientation. Since the buckling event is a short 
moment during bar evolution, we can safely exclude these points from our 
comparison as the likelihood to find a bar in the buckling phase is rather low. 
Without these points and below t=200, the measurements agree rather well. A 
more extensive test can also be found in the Appendix~\ref{sec:inclPAinf}.

To perform the analysis for the comparison of the two torque measurements, we calculated the values 
of $\rm {Q}_b$ (in addition to those of $\rm {Q}_{\rm kin}$) for all sets of 
simulations in the same way as done for the observations. In Fig.~\ref{fig:qbqk} we presented the final results showing the observational 
measurements overlaid on the four simulations, $I_1$, $I_2$, $I_3$ and $I_4$ 
(level of disc-to-total ratios: 0.92, 0.62, 0.43, 0.29 respectively).

As seen in Fig.~\ref{fig:qbqk}, none of the simulation series coincides perfectly with 
all the data points, but instead form a continuum. It is reassuring however that 
the simulation series fall onto the same relation found as for the observations.

Simulation $I_1$ (92\% disk, shown by the green points) best represents the lower bar strengths. With 
decreasing disc content, the both the photometric and 
kinematic bar strengths increase. We also distinguish a particular behavior for $I_4$: 
compared to the other sets, this series seems to adapt to all ranges starting 
from very weak to very strong bars. This is the simulation with the lowest disc 
percentage within the bar region initially and also after halo relaxation and 
is at every point in time submaximal. Overall, we find a trend of stronger bar 
development with increasing dark matter halo fraction in our simulations.
\\ \\
 Comparing the velocity maps of the three different simulations with the 
observations, we find good agreement, especially in the case of $I_1$. 
Simulations $I_3$ and $I_4$ in particular develop strong distortions 
in the stellar velocity fields which are not as pronounced in our 
observations. However, the bar strength of $I_1$ never reaches a higher value 
than about 0.2-0.3, unlike the observations. To reach higher values, we needed to increase 
the halo fraction.

\section{Influence of the inclination and PA on the bar strength measurements}
\label{sec:inclPAinf}
\begin{figure*}
\begin{center}
 \includegraphics[width=1.\textwidth]{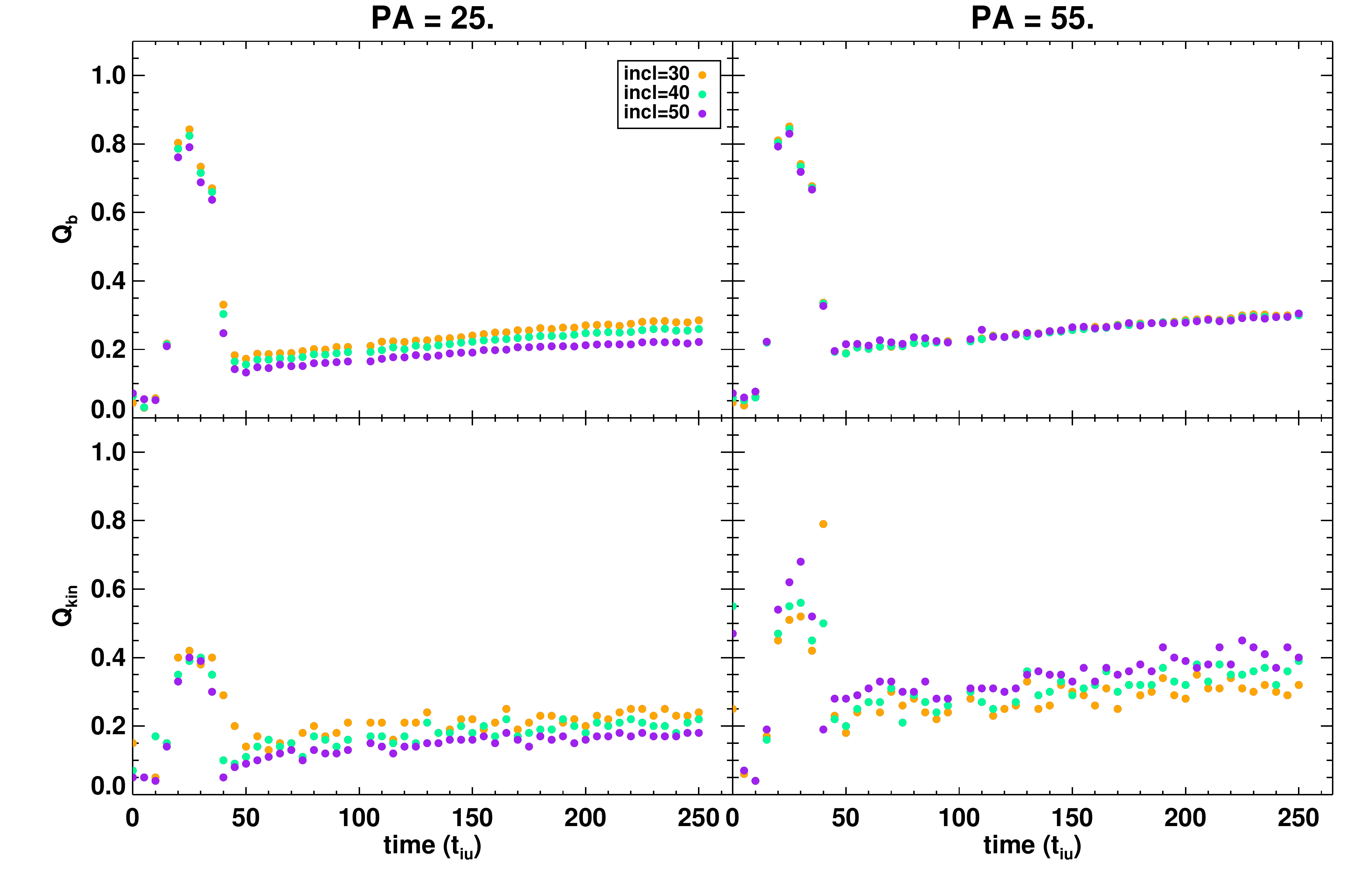}
  \caption{For a simulation with 60\% disc we plot $\rm {Q}_b$ and $\rm {Q}_{\rm kin}$ 
versus time for different inclinations and PAs: left: PA=25, right: PA=55. Top: 
 $\rm {Q}_b$ and bottom: $\rm {Q}_{\rm kin}$. The elevation at early times is 
due to the buckling event in the bar evolution.}
  \label{PAinc}
 \end{center}
\end{figure*}

We used four simulation sets ($I_1$, $I_2$, $I_3$ and $I_4$) to test 
the influence of PA and inclination on the bar strength measurements, namely 
$\rm {Q}_b$, $\rm {Q}_{\rm kin}$ and A2, on a large enough sample. 
Figure~\ref{PAinc} illustrates those tests with the example of the $I_2$ simulation.

In the majority of cases the influence of these two parameters causes a consistent change in any 
of the different bar strength measurement methods. The PA influence always shows 
a clear trend: larger PAs result in higher strength values overall. 
Furthermore it causes less spread within the distinct inclinations, especially 
for $\rm {Q}_b$ and A2. 

The effect of inclination is two-fold: in the simulations with higher 
disc-percentages ($I_1$ and $I_2$), its effect is reversed for low and high PAs. 
For low PAs, we detect that a lower inclination results in higher values in all 
three parameters, whereas for higher PAs, high inclinations result in higher 
values overall. The spread here is less. 

For the other two simulations with higher DM content ($I_3$ and $I_4$), the 
effect of inclination is always the same despite the distinct PAs: a lower 
inclination results in higher values in the three measured parameters. Again, 
the spread is less at higher PAs (except for $\rm {Q}_{\rm kin}$ in $I_4$). 

The fact that the influences are similar in spite of the different measurement 
methods probably helps to produce the observed relation between them. It is 
important to bear the influence of these parameters in mind when checking the 
observations: low inclinations might lead to higher values, and, depending on 
the PA and DM fraction, high inclinations can also lead to lower values. 

We also compared $\rm {Q}_b$ with the A2 values directly from the simulations and  
find that $\rm {Q}_b$ resembles A2 very well.  

Overall, values of $\rm {Q}_{b}$ are expected to be higher at lower 
inclinations whereas values of $\rm {Q}_{\rm kin}$ should be higher at higher 
inclinations as motions can be better measured with increased inclination. Our 
tests, however, suggest that in almost all cases, the chosen methods indicate a 
lower limit for the bar strength, in particular in the case of $\rm {Q}_{b}$.

\section{Complete kinematic maps for stars and ionised gas}
\label{app:gaskinmaps}
\begin{figure*}
\includegraphics[width=0.32\linewidth]{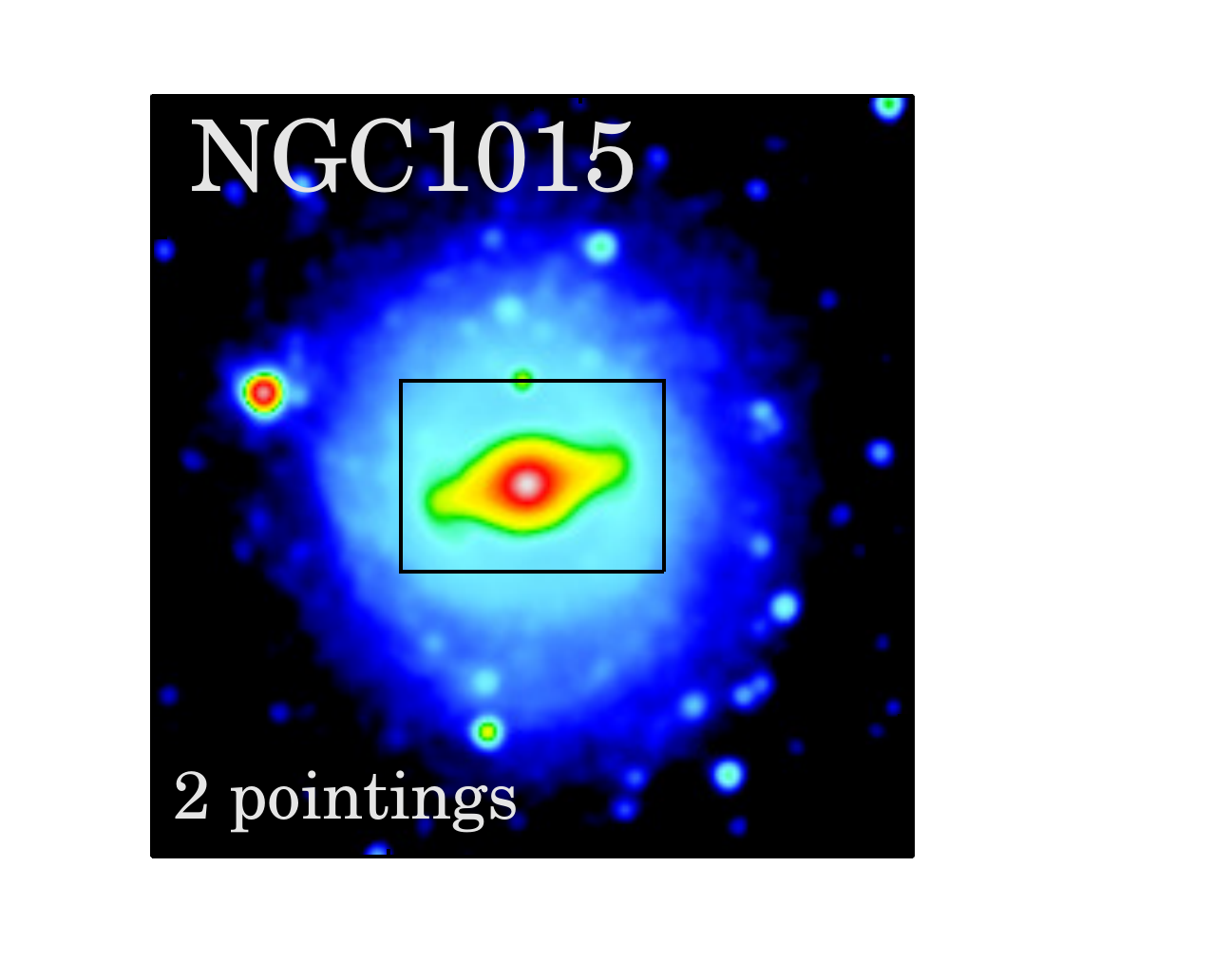}
\includegraphics[width=0.32\linewidth]{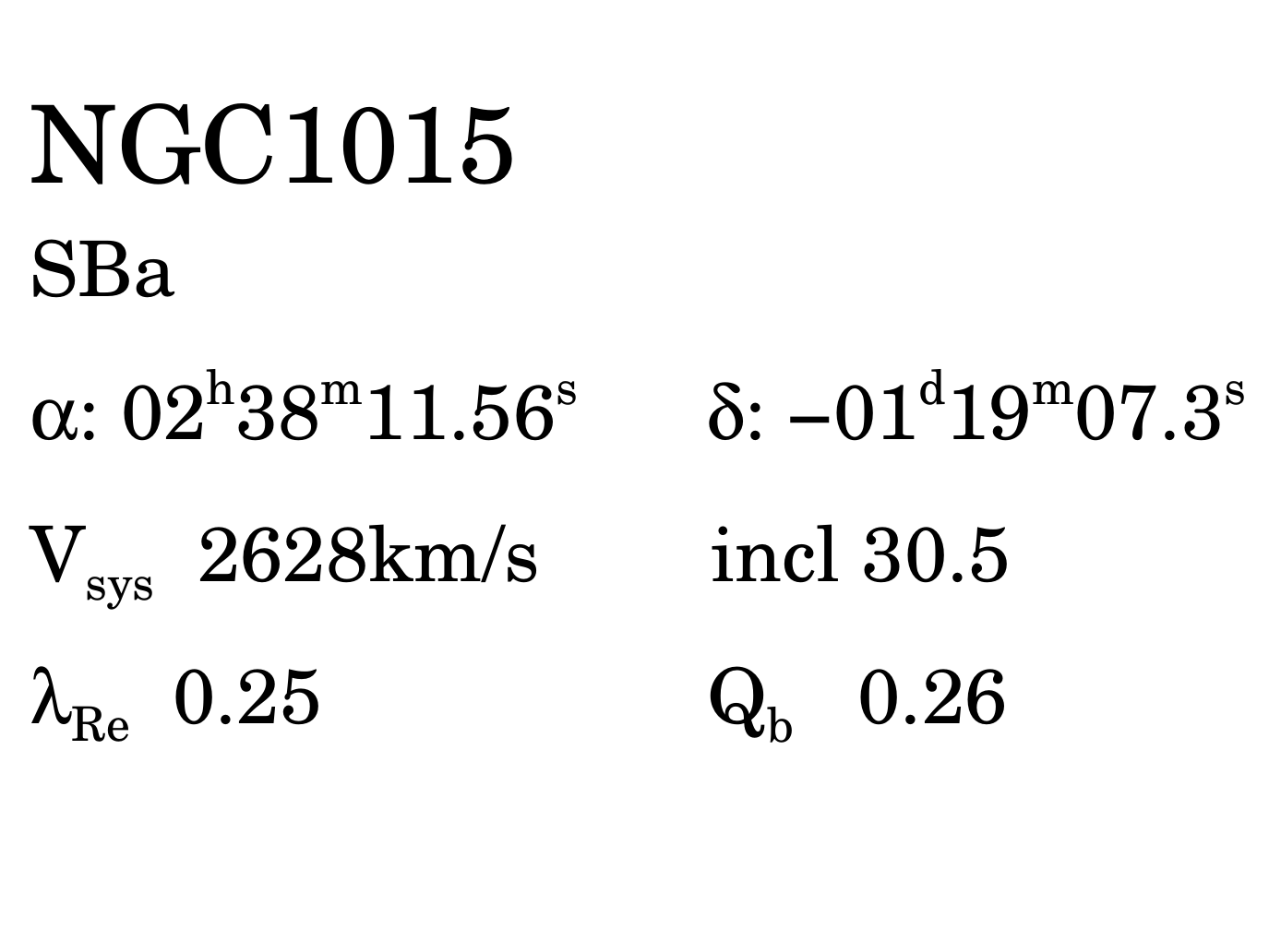}
\includegraphics[width=0.88\linewidth]{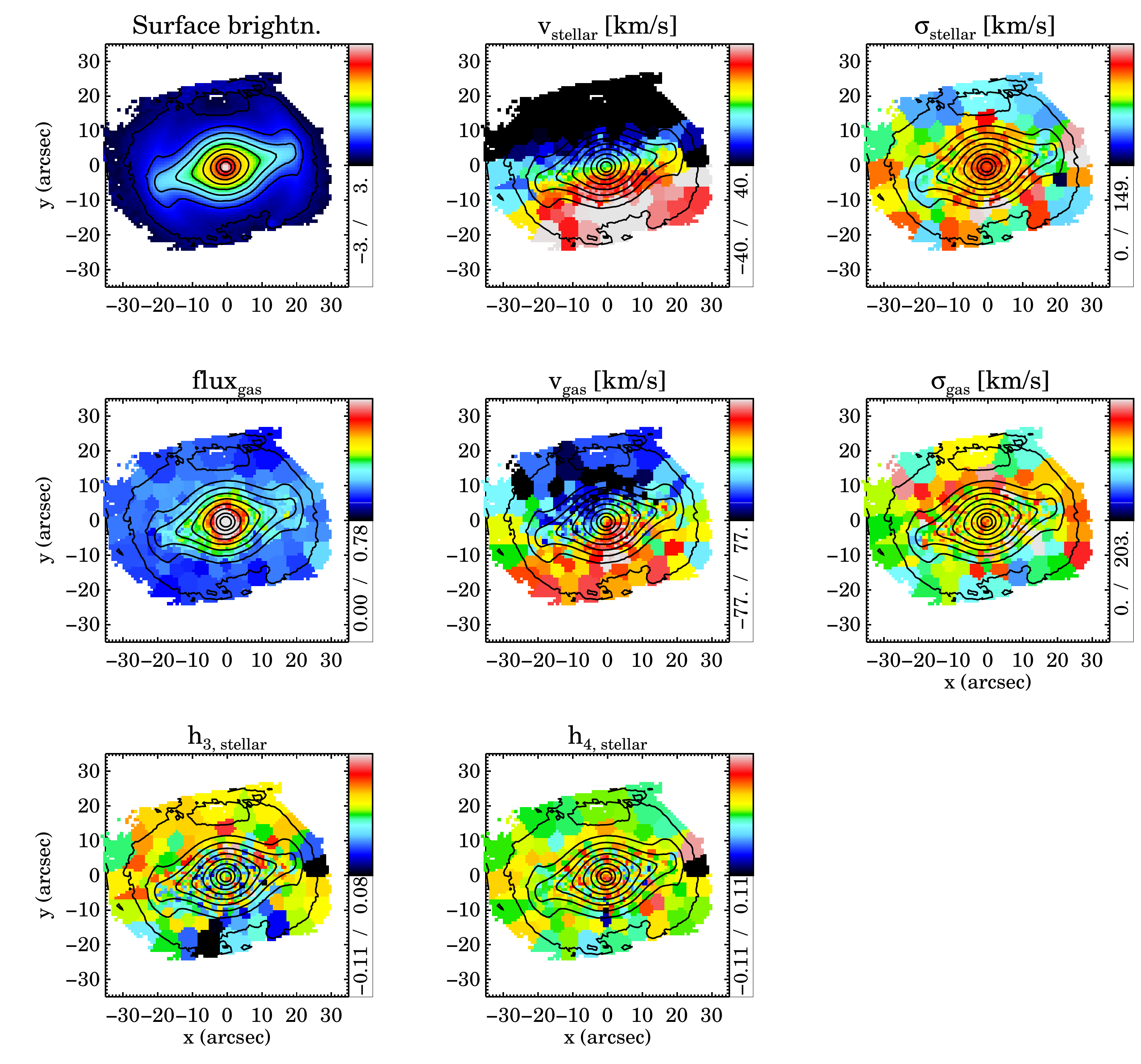}
\includegraphics[width=0.27\linewidth]{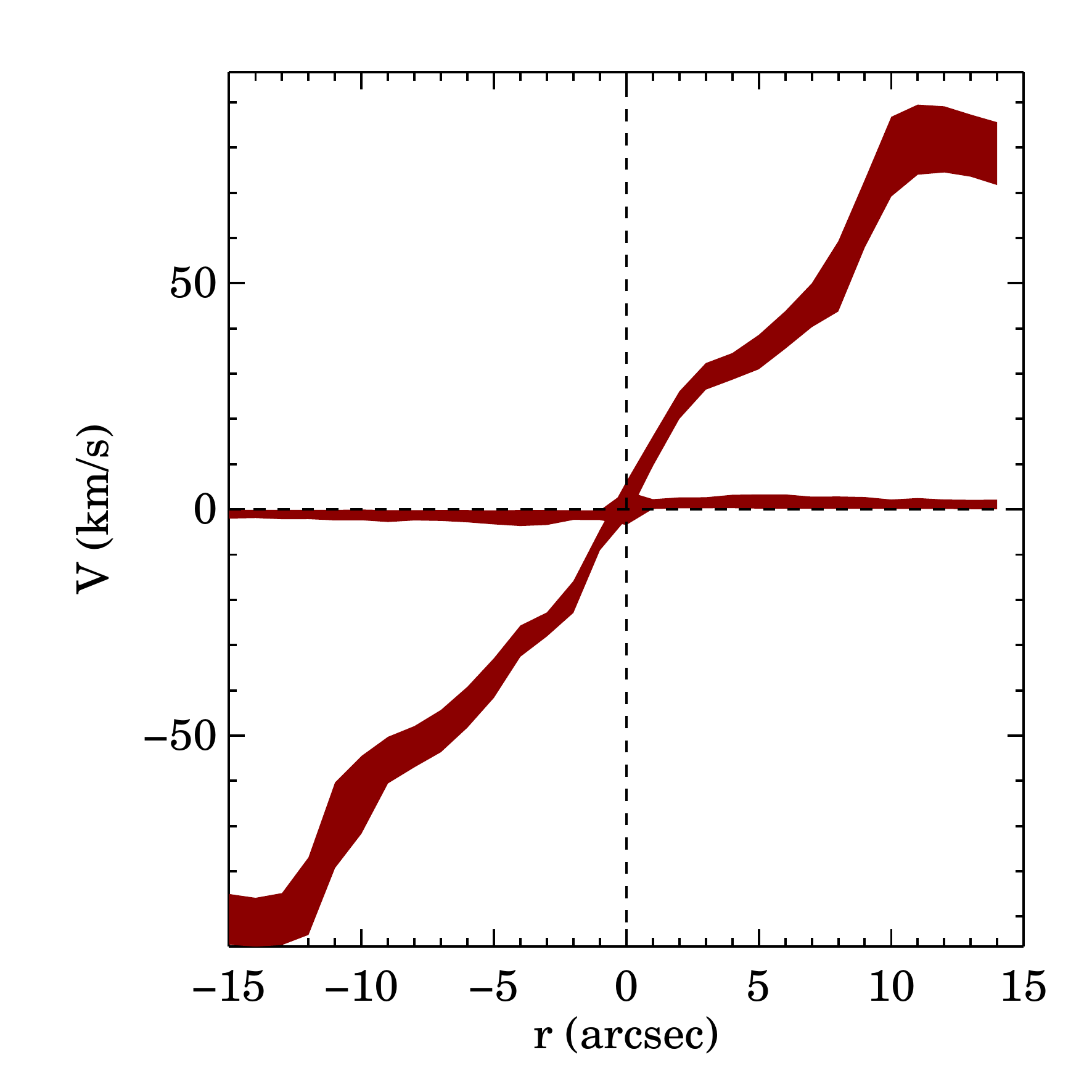}
\includegraphics[width=0.58\linewidth]{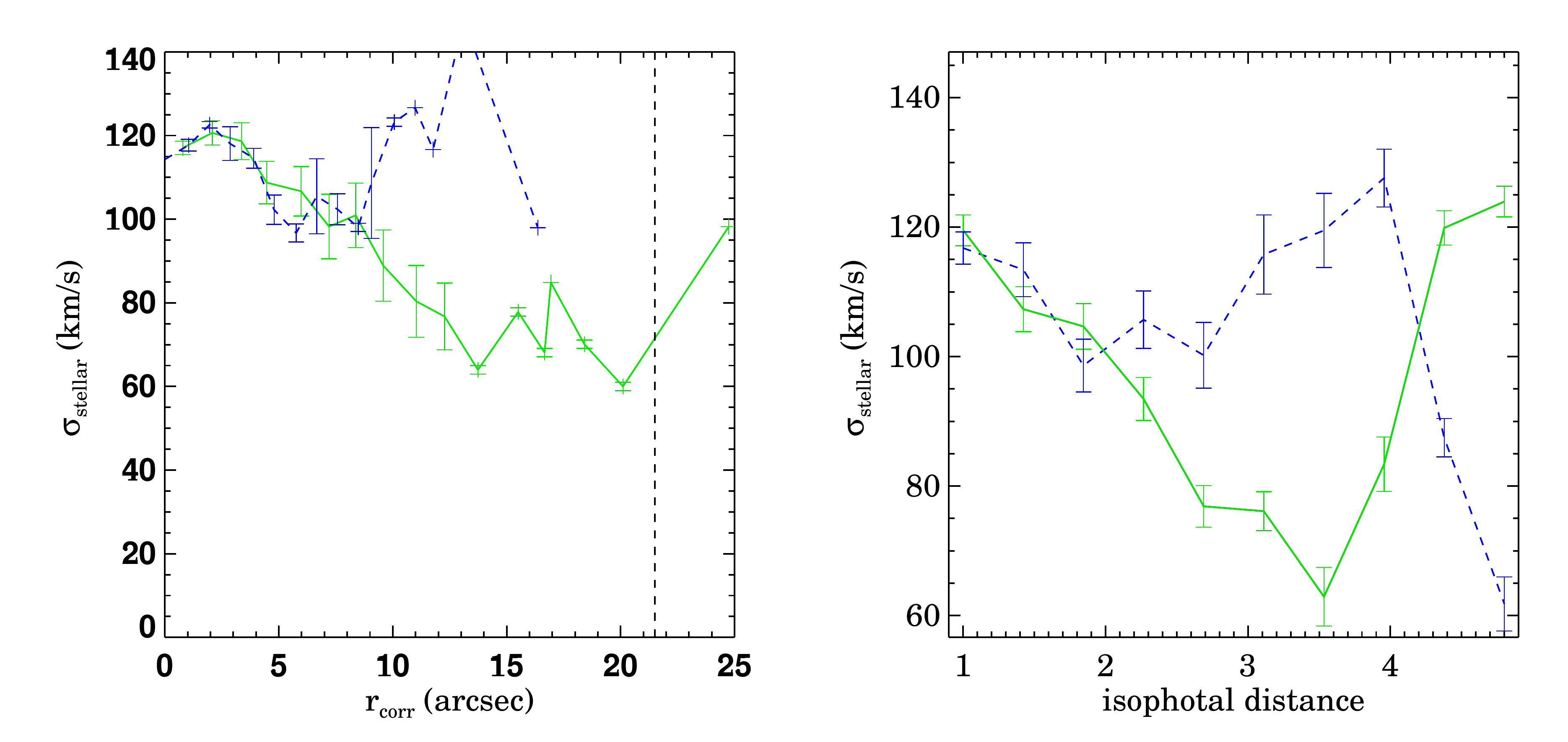}
\caption{Summary of the kinematic maps for stars and ionised gas for each galaxy.}
\label{fig:gasvel}
\end{figure*}

\begin{figure*}
\includegraphics[width=0.32\linewidth]{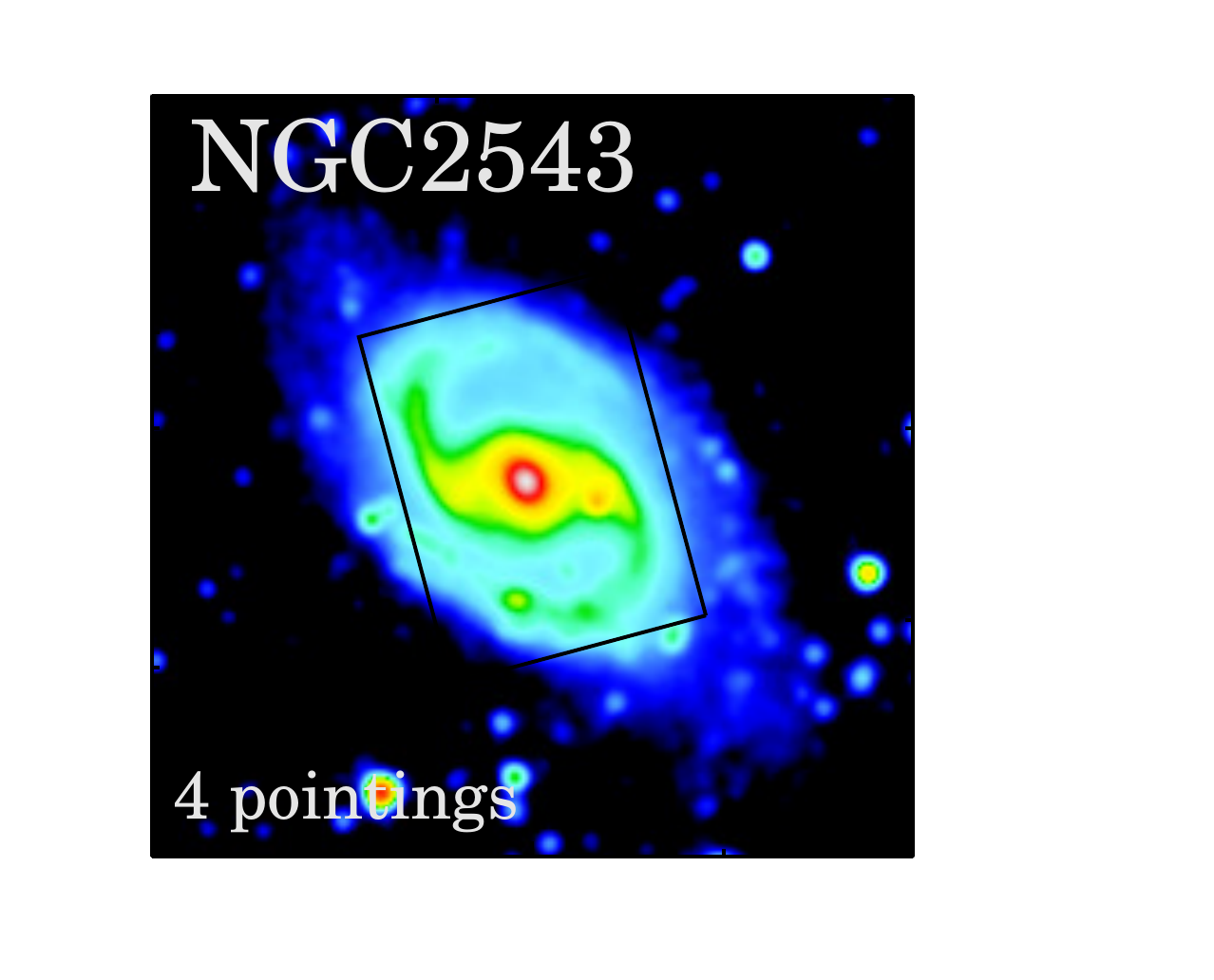}
\includegraphics[width=0.32\linewidth]{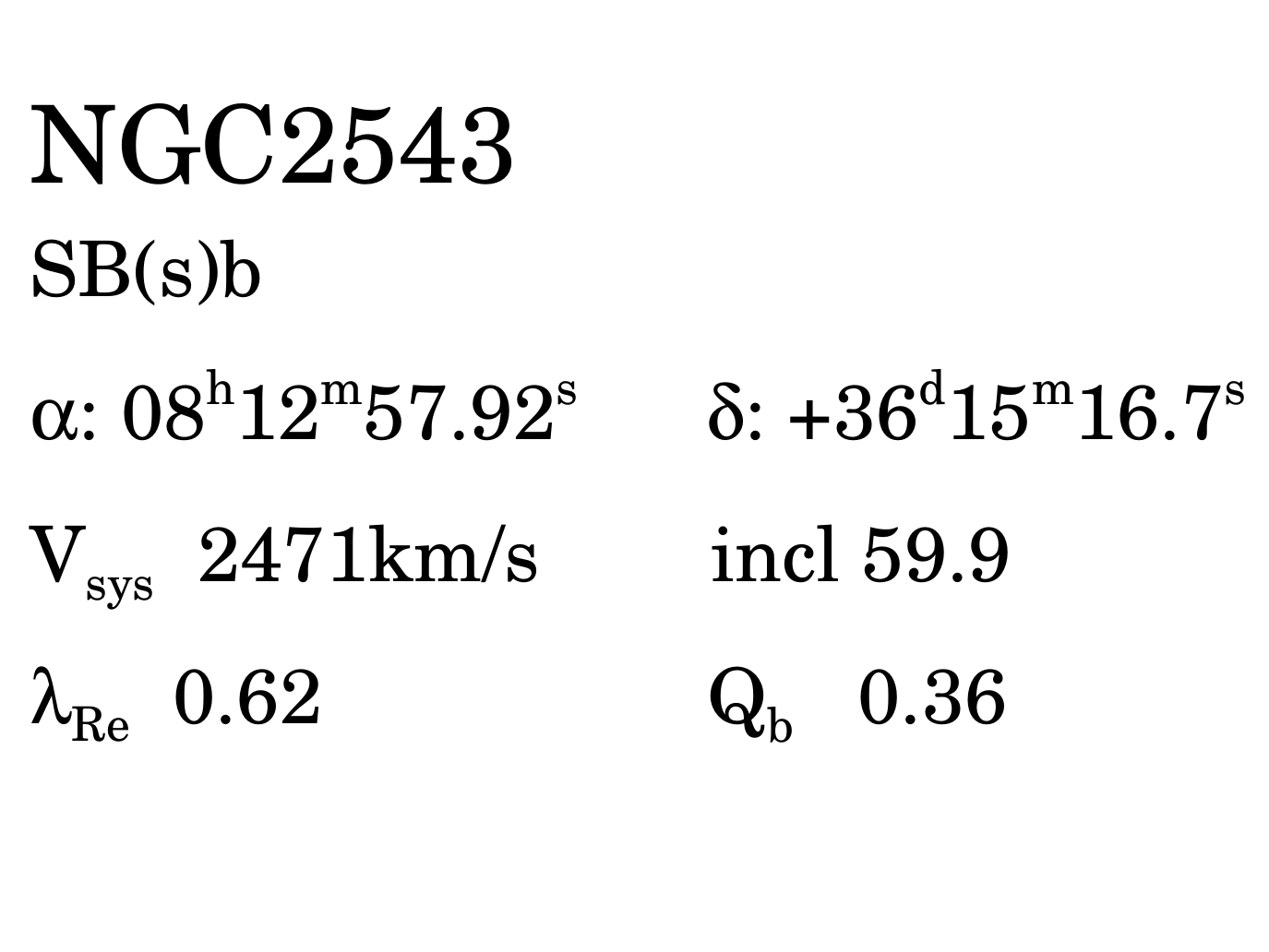}
\includegraphics[width=0.88\linewidth]{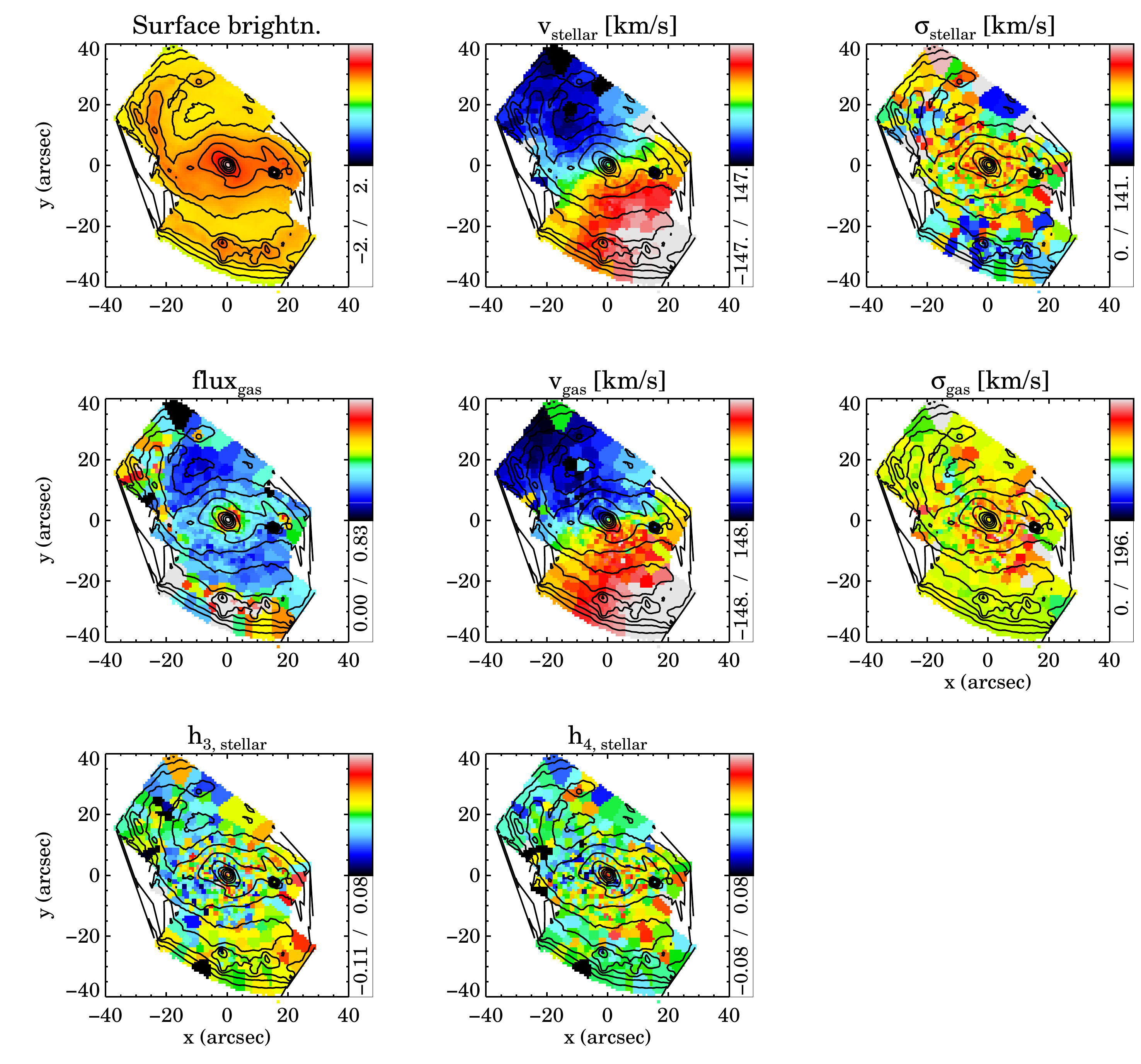}
\includegraphics[width=0.27\linewidth]{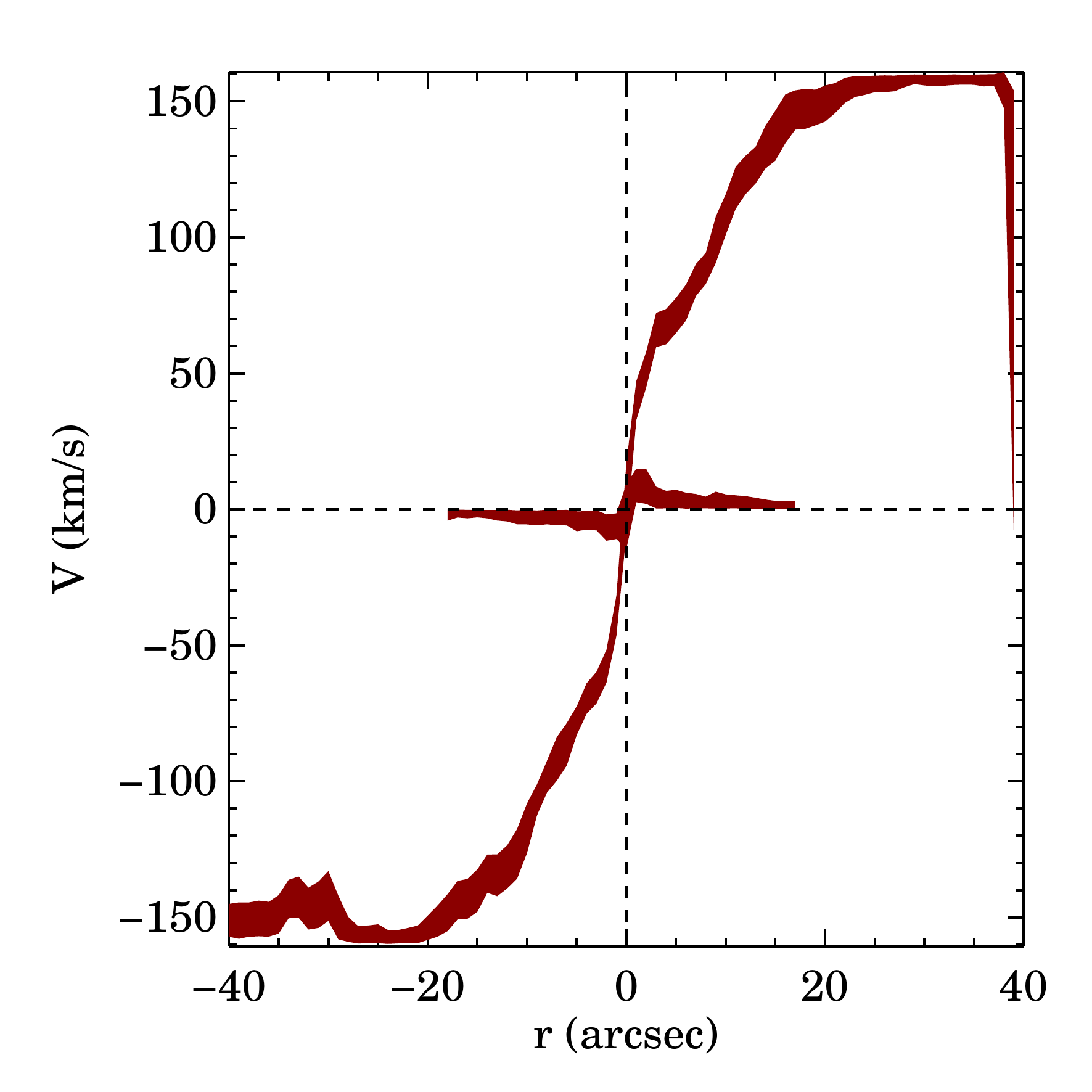}
\includegraphics[width=0.58\linewidth]{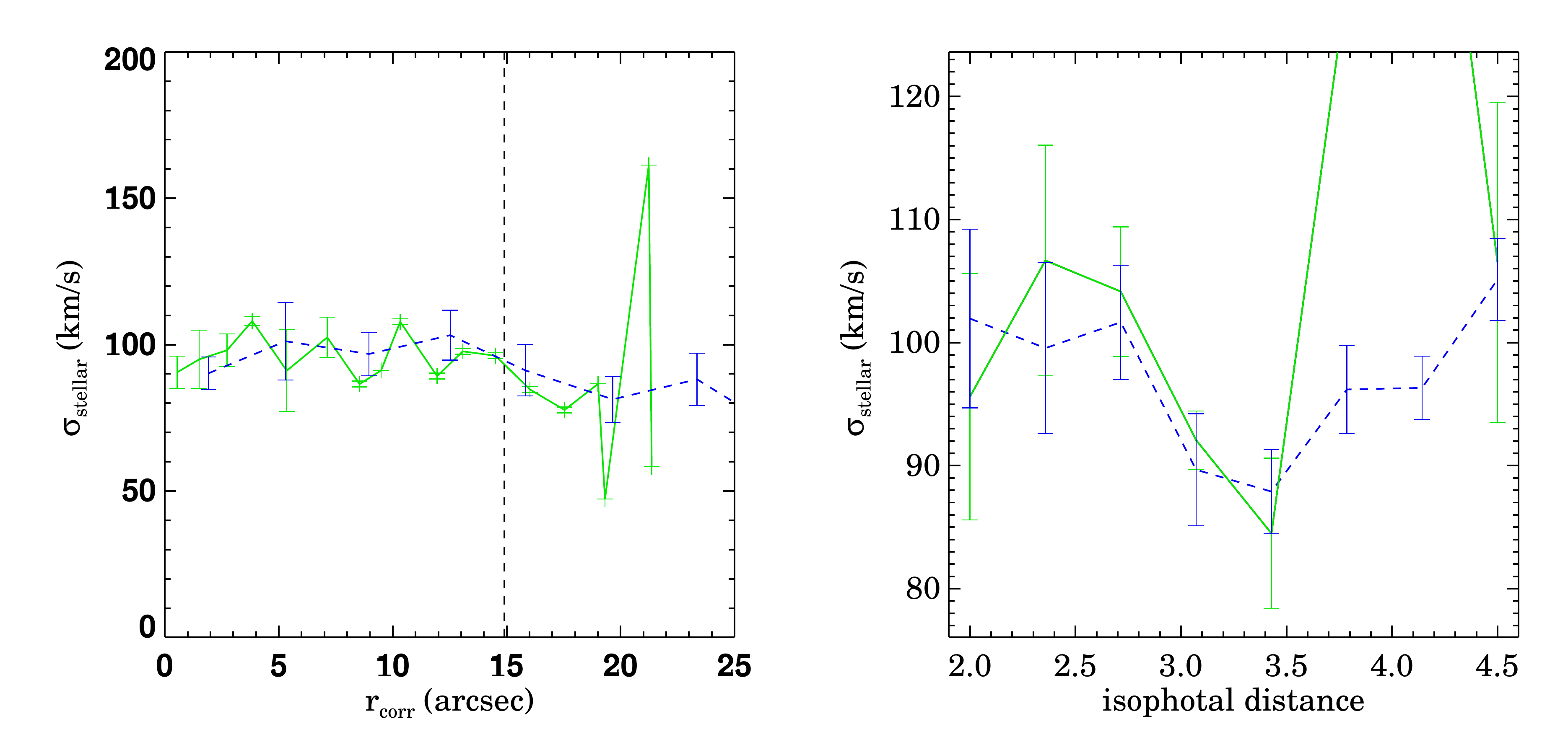}
\caption{Figure \ref{fig:gasvel} continued.}
\label{fig:2543}
\end{figure*}

\begin{figure*}
\includegraphics[width=0.32\linewidth]{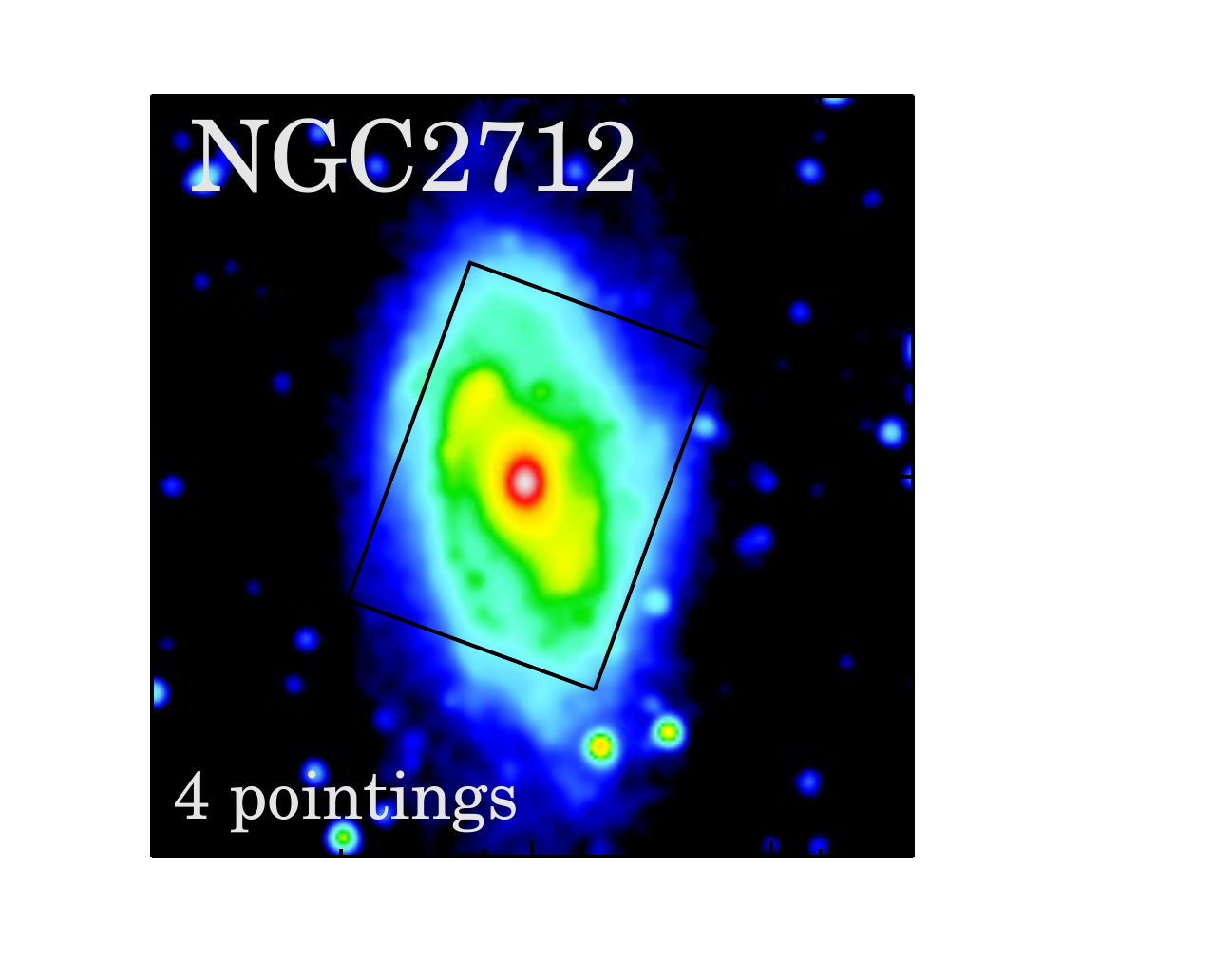}
\includegraphics[width=0.32\linewidth]{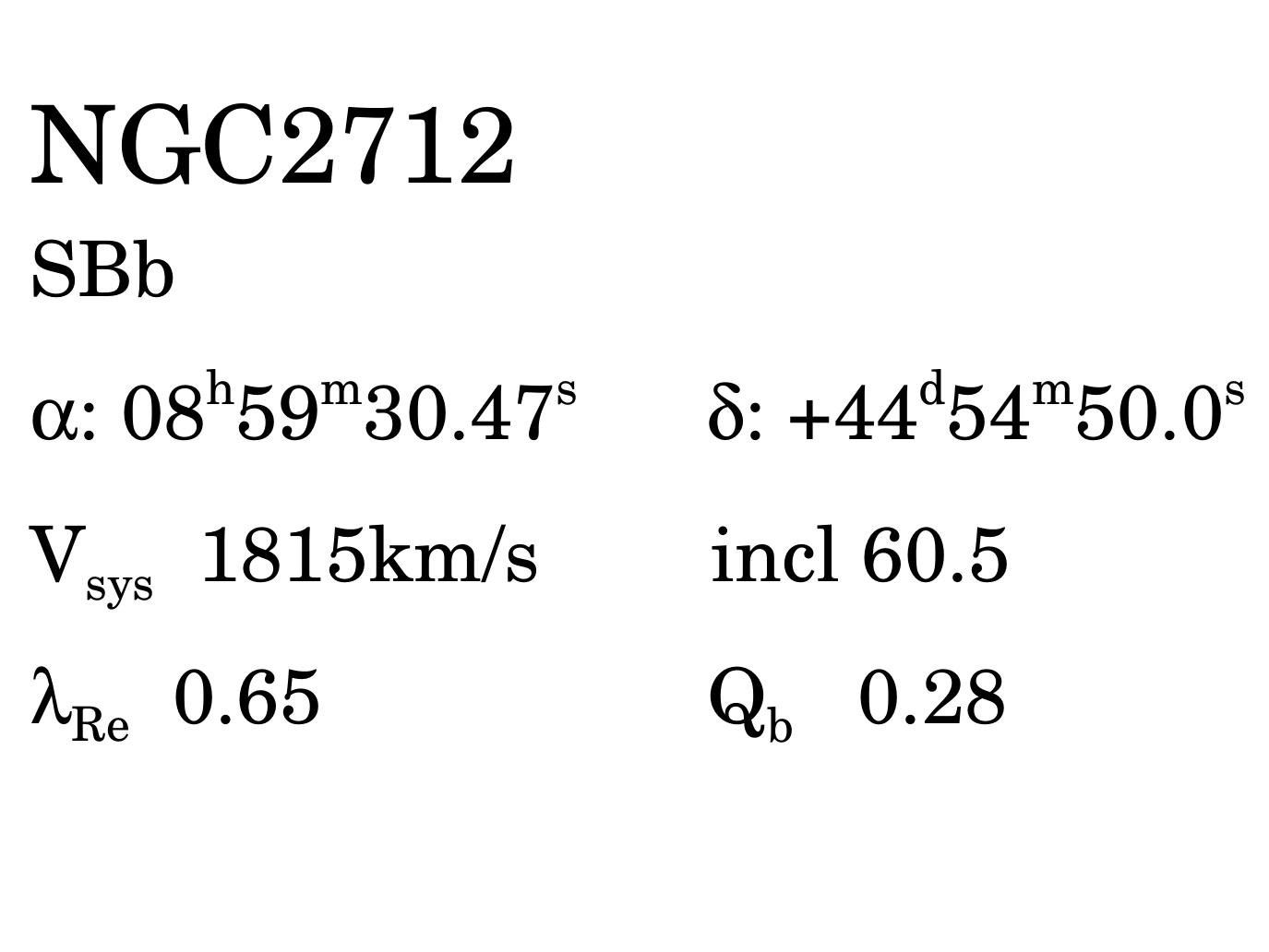}
\includegraphics[width=0.88\linewidth]{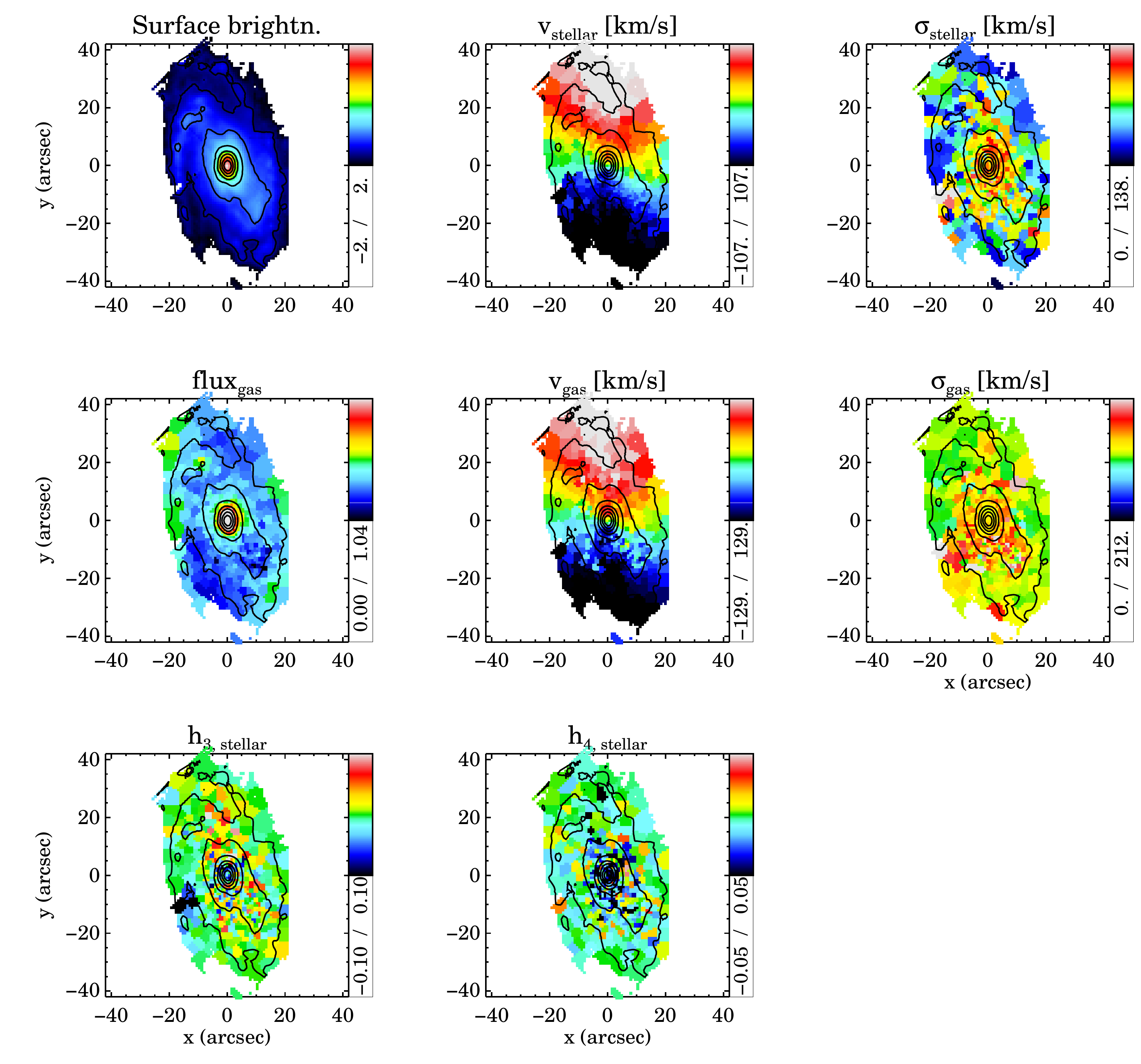}
\includegraphics[width=0.27\linewidth]{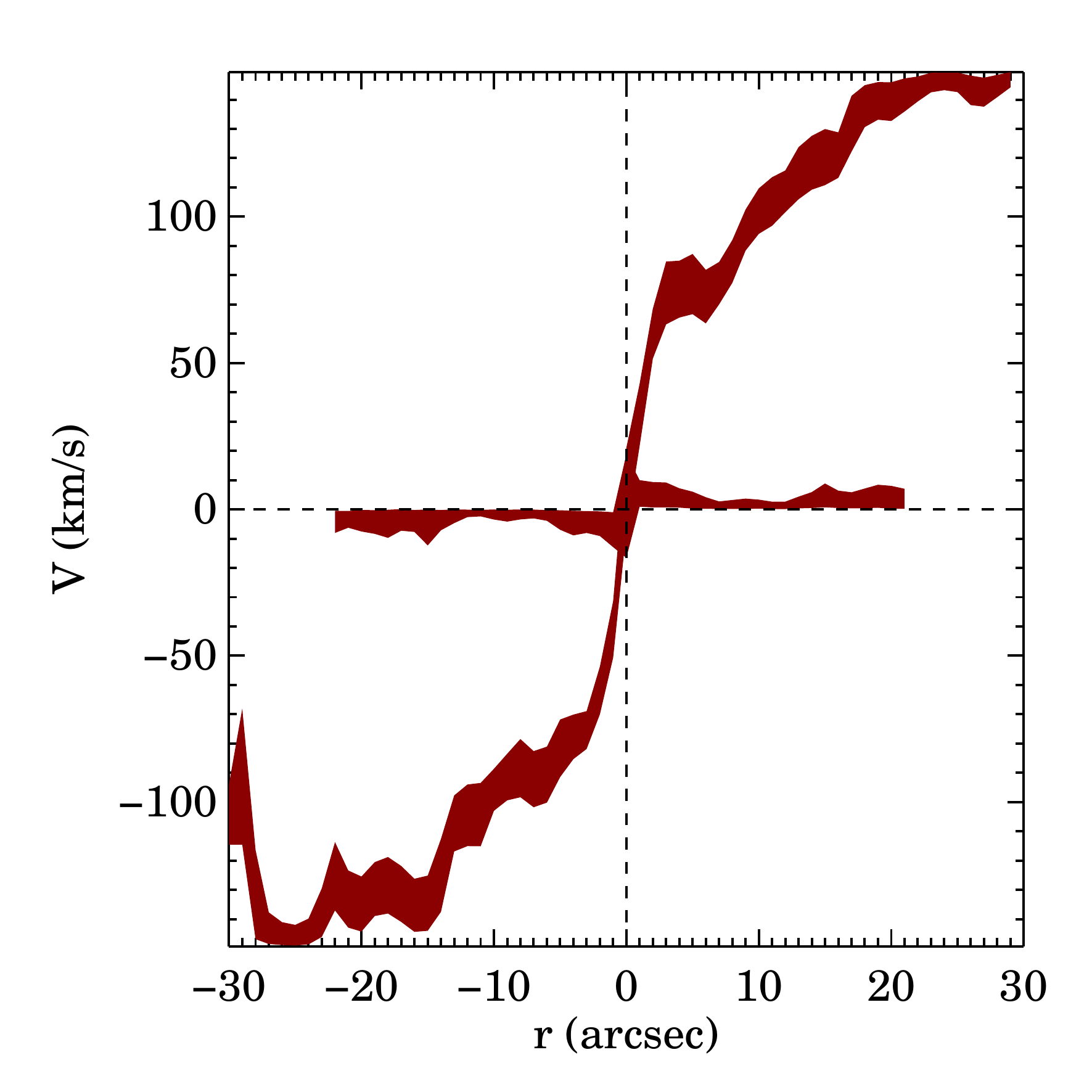}
\includegraphics[width=0.58\linewidth]{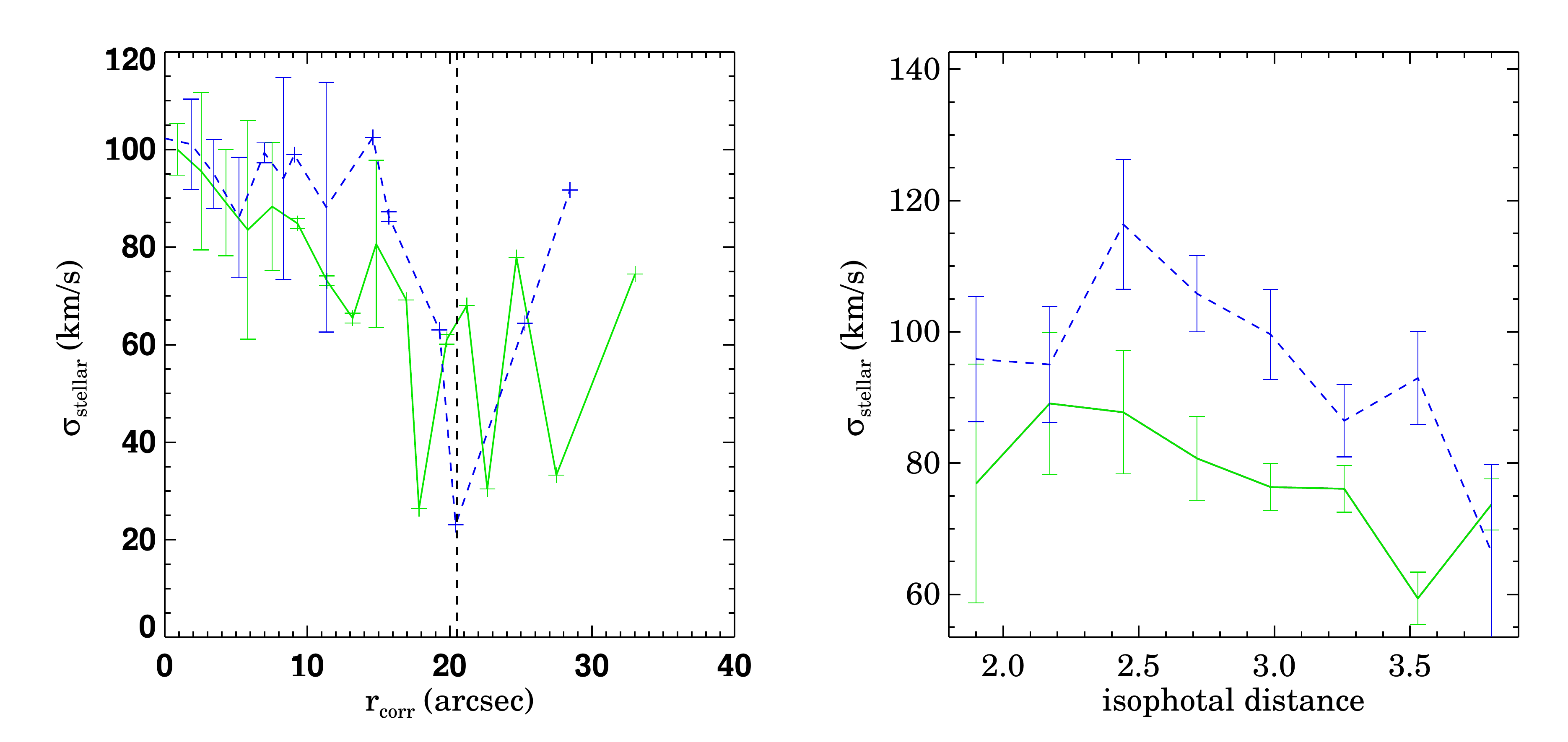}
\caption{Figure \ref{fig:gasvel} continued.}
\label{fig:2712}
\end{figure*}

\begin{figure*}
\includegraphics[width=0.32\linewidth]{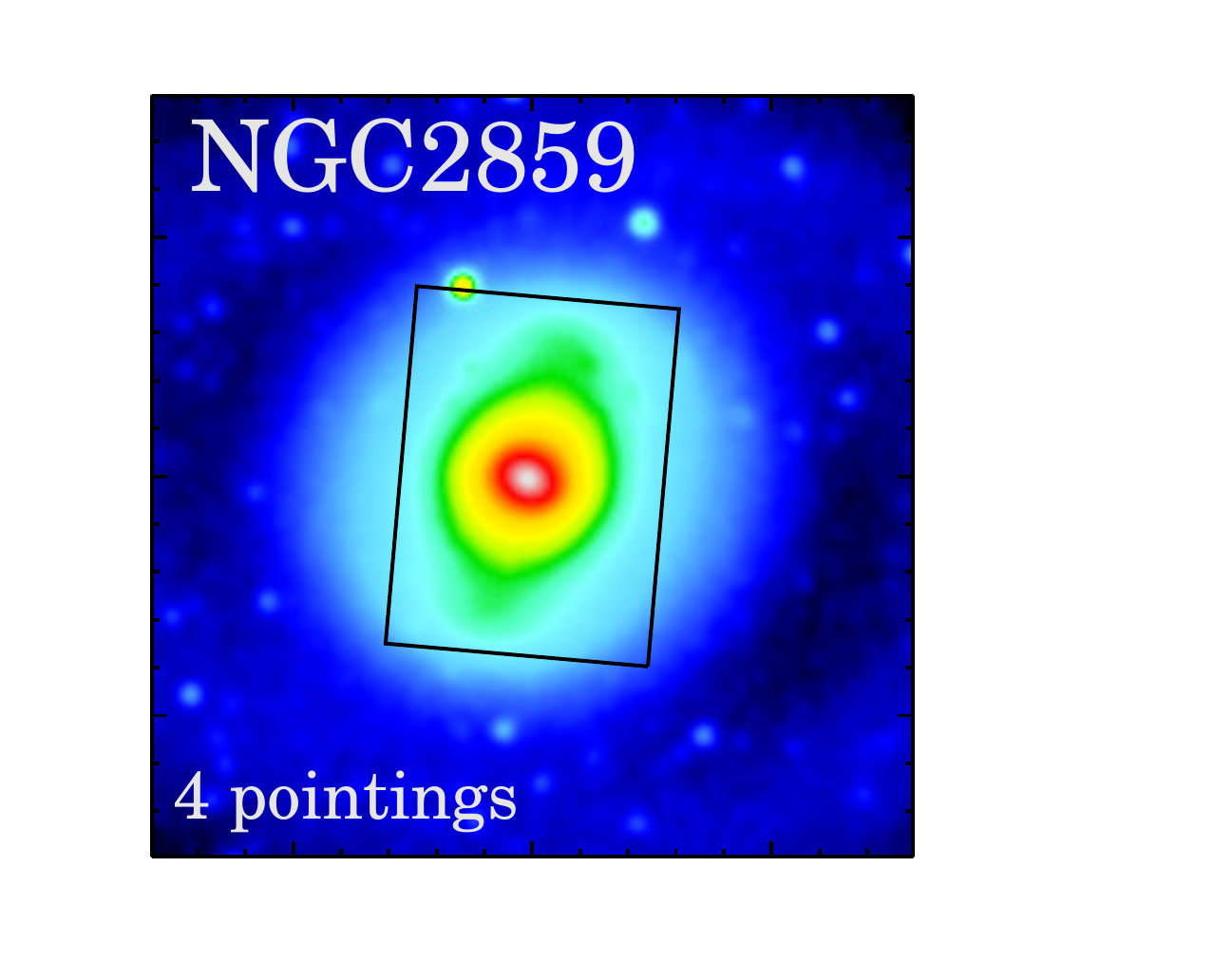}
\includegraphics[width=0.32\linewidth]{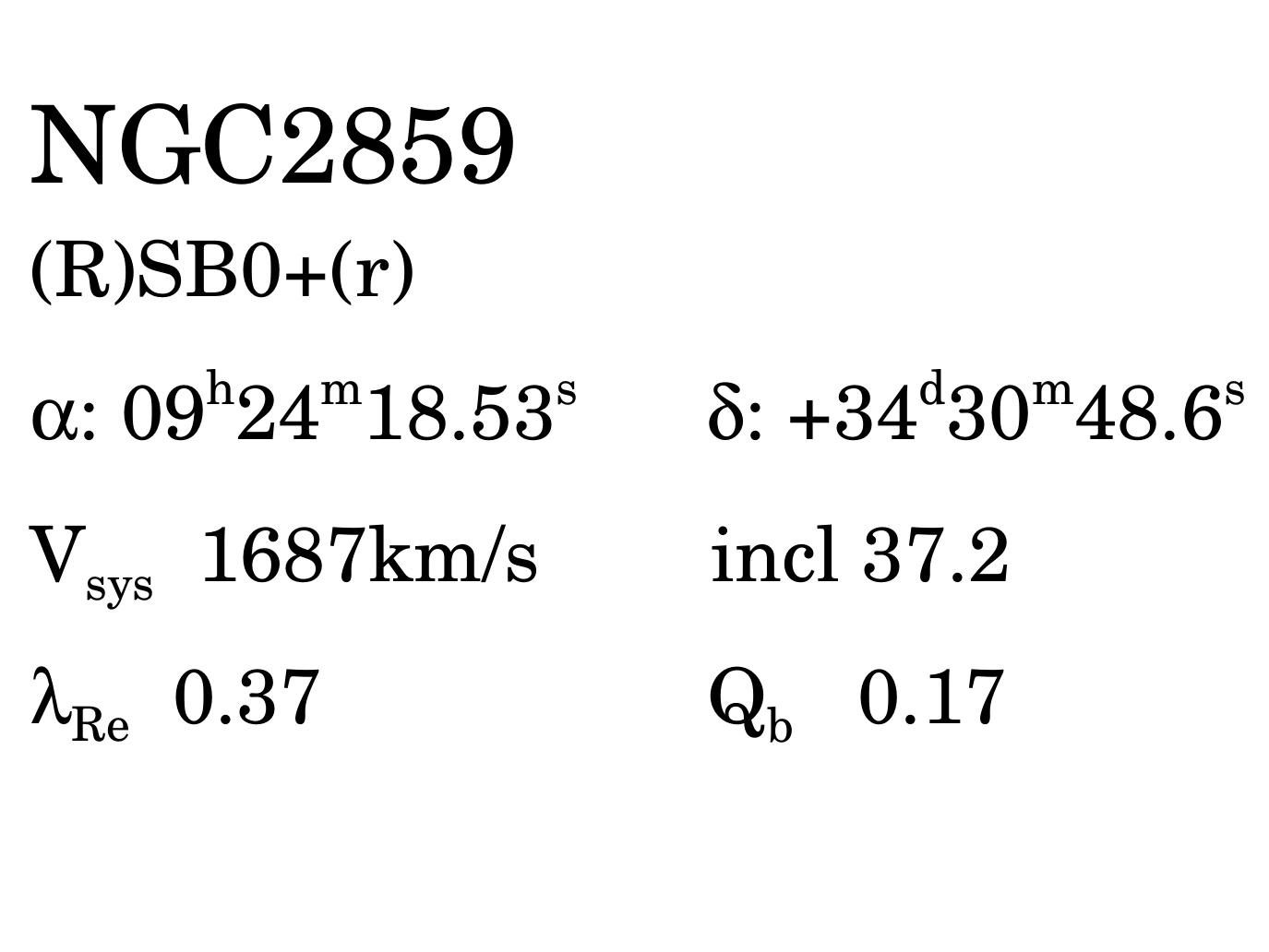}
\includegraphics[width=0.88\linewidth]{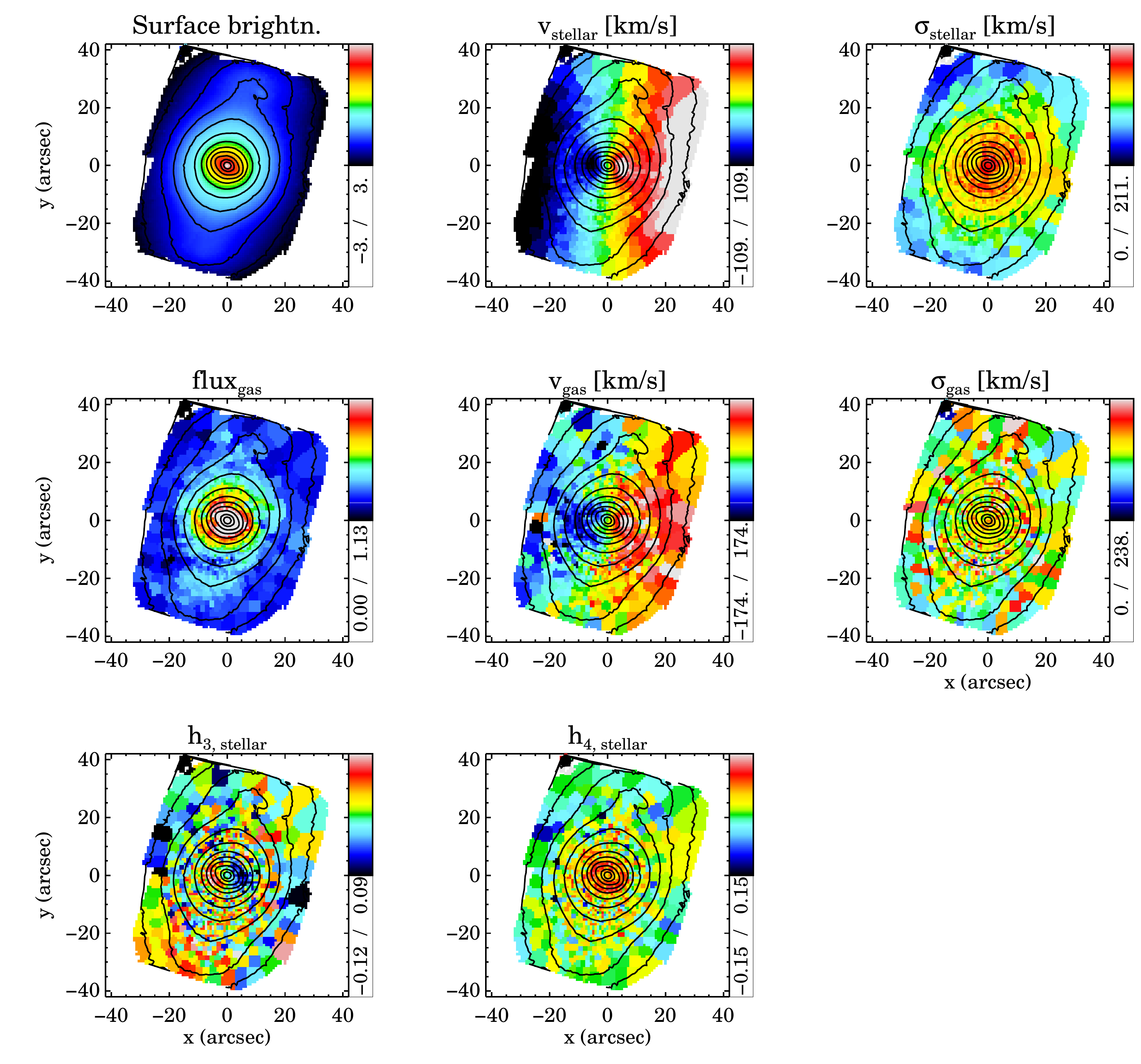}
\includegraphics[width=0.27\linewidth]{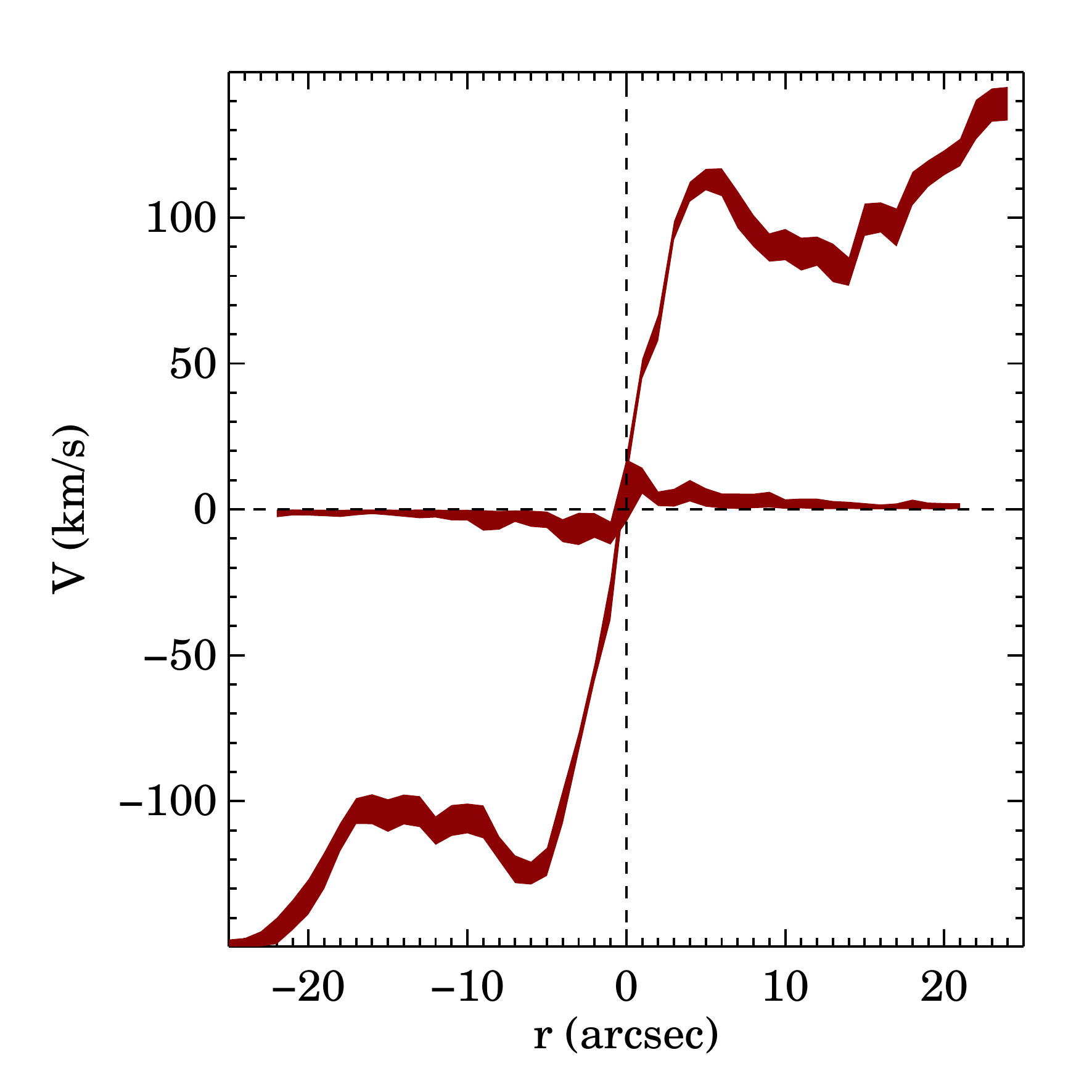}
\includegraphics[width=0.58\linewidth]{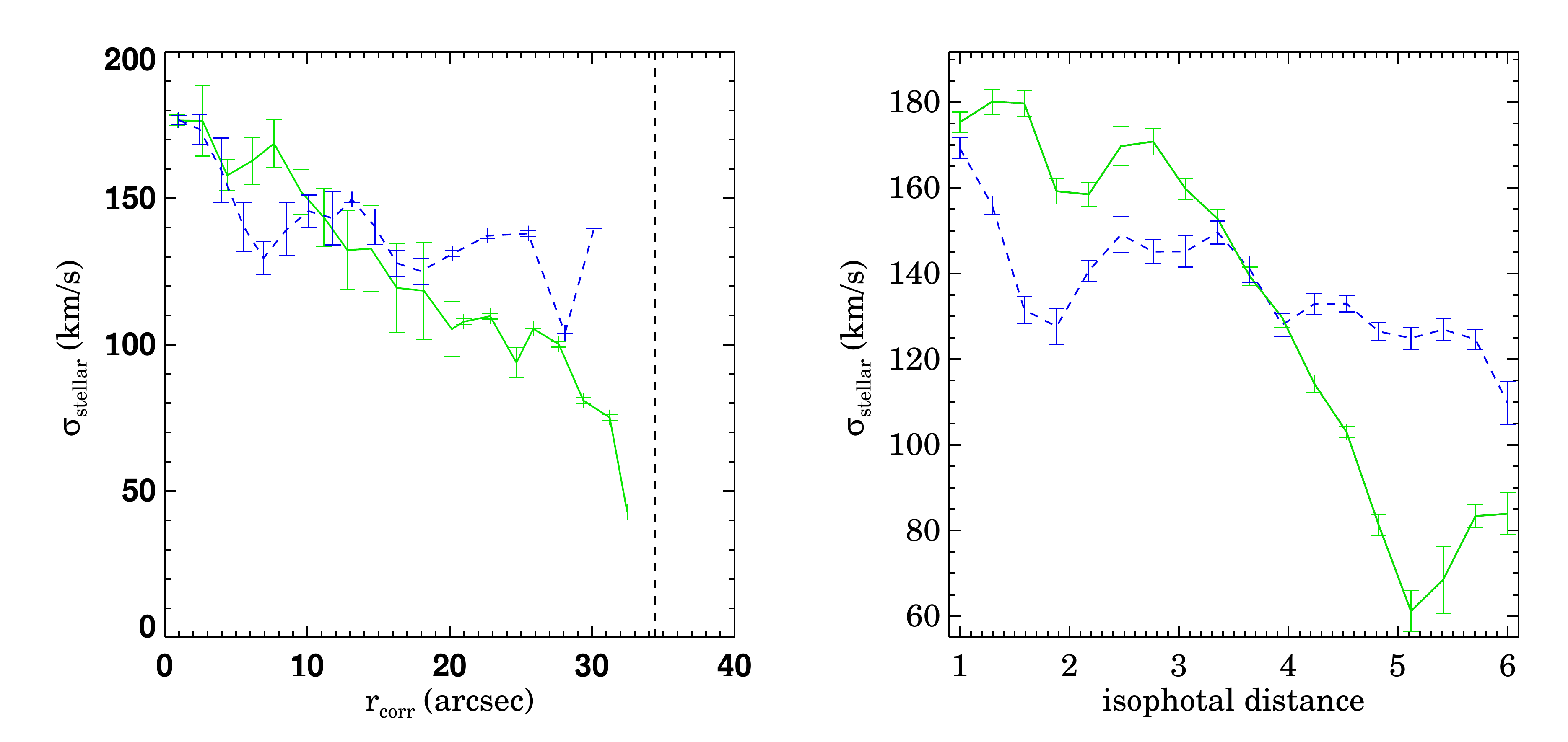}
\caption{Figure \ref{fig:gasvel} continued.}
\label{fig:2859}
\end{figure*}

\begin{figure*}
\includegraphics[width=0.32\linewidth]{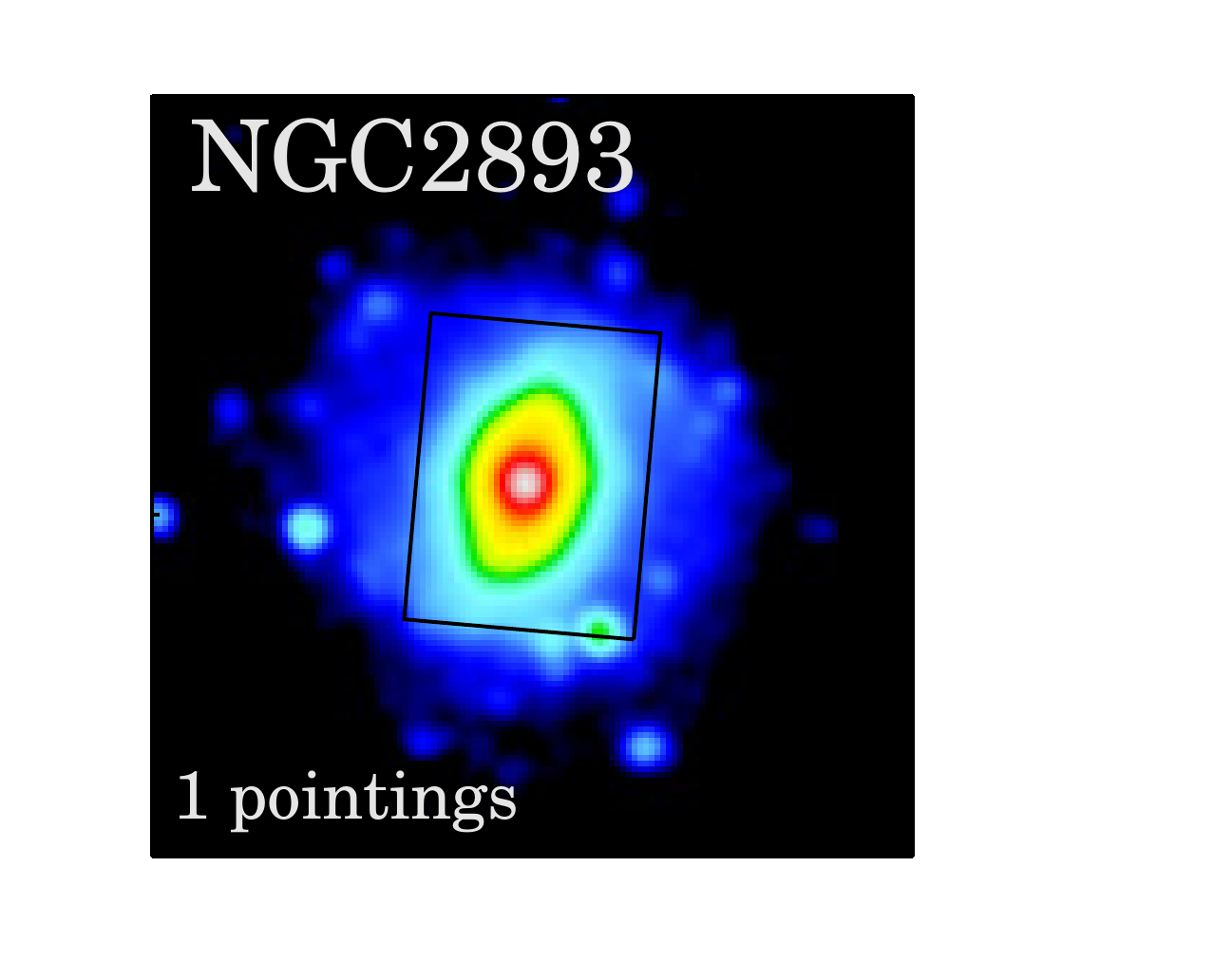}
\includegraphics[width=0.32\linewidth]{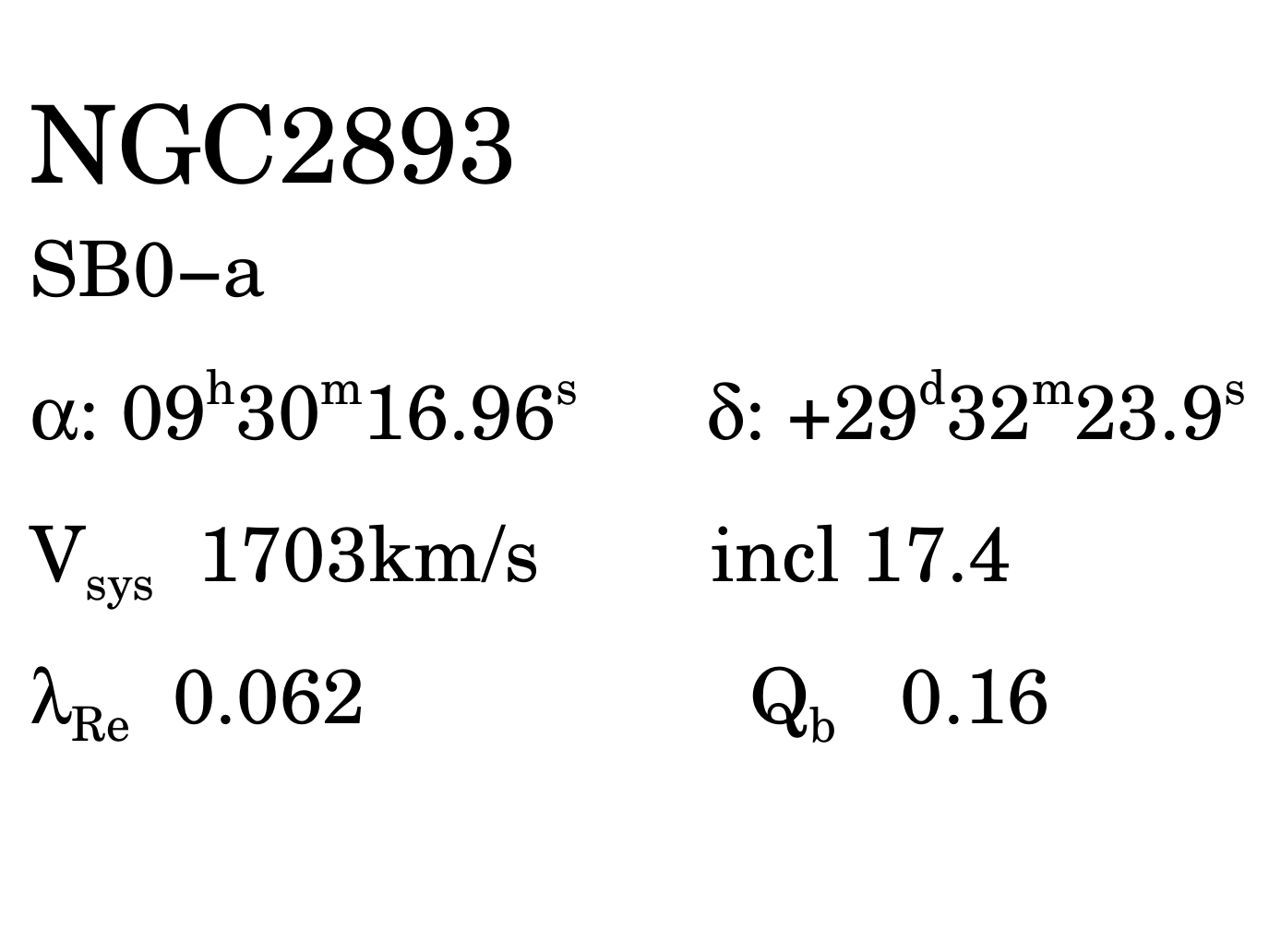}
\includegraphics[width=0.88\linewidth]{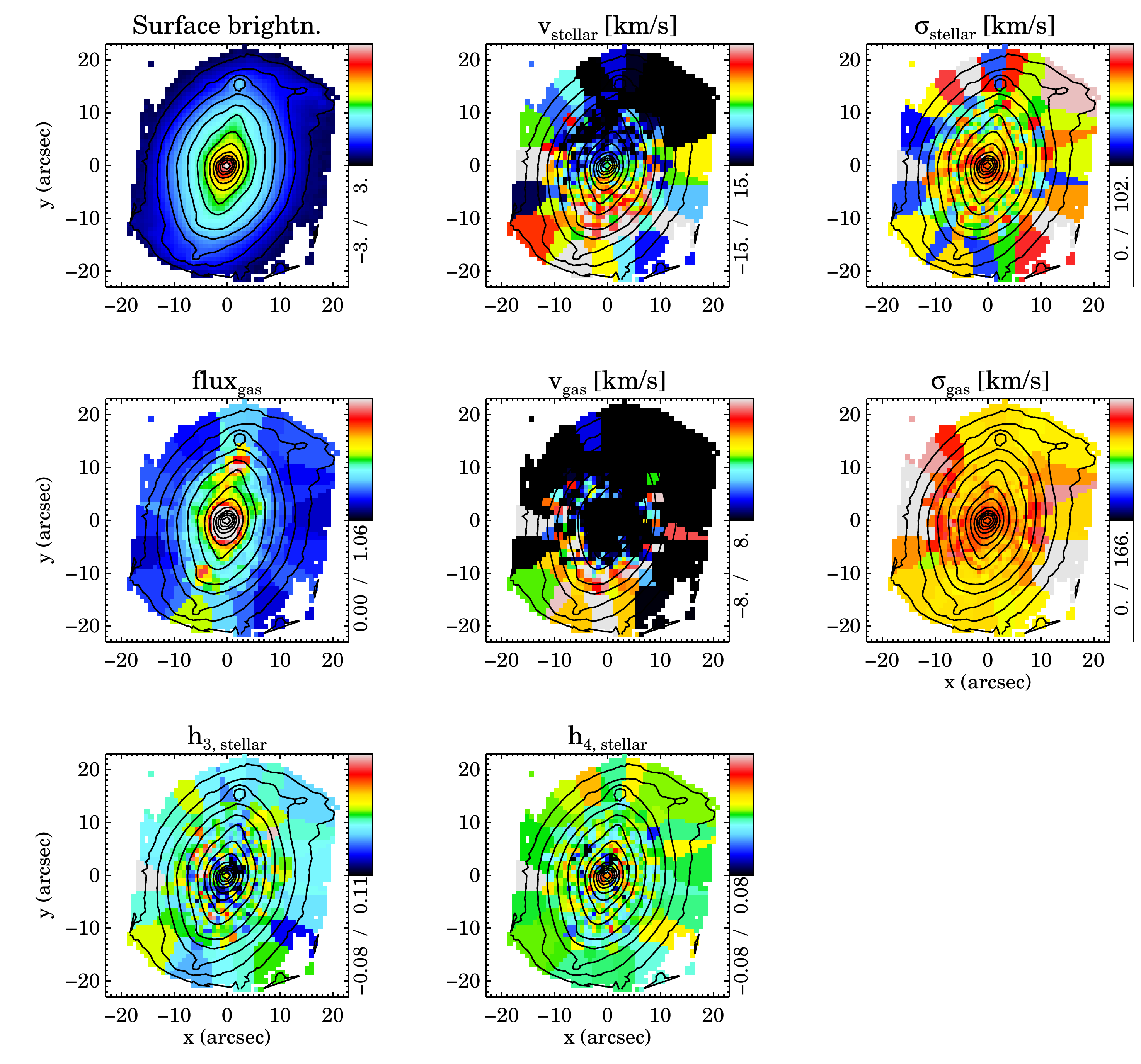}
\includegraphics[width=0.27\linewidth]{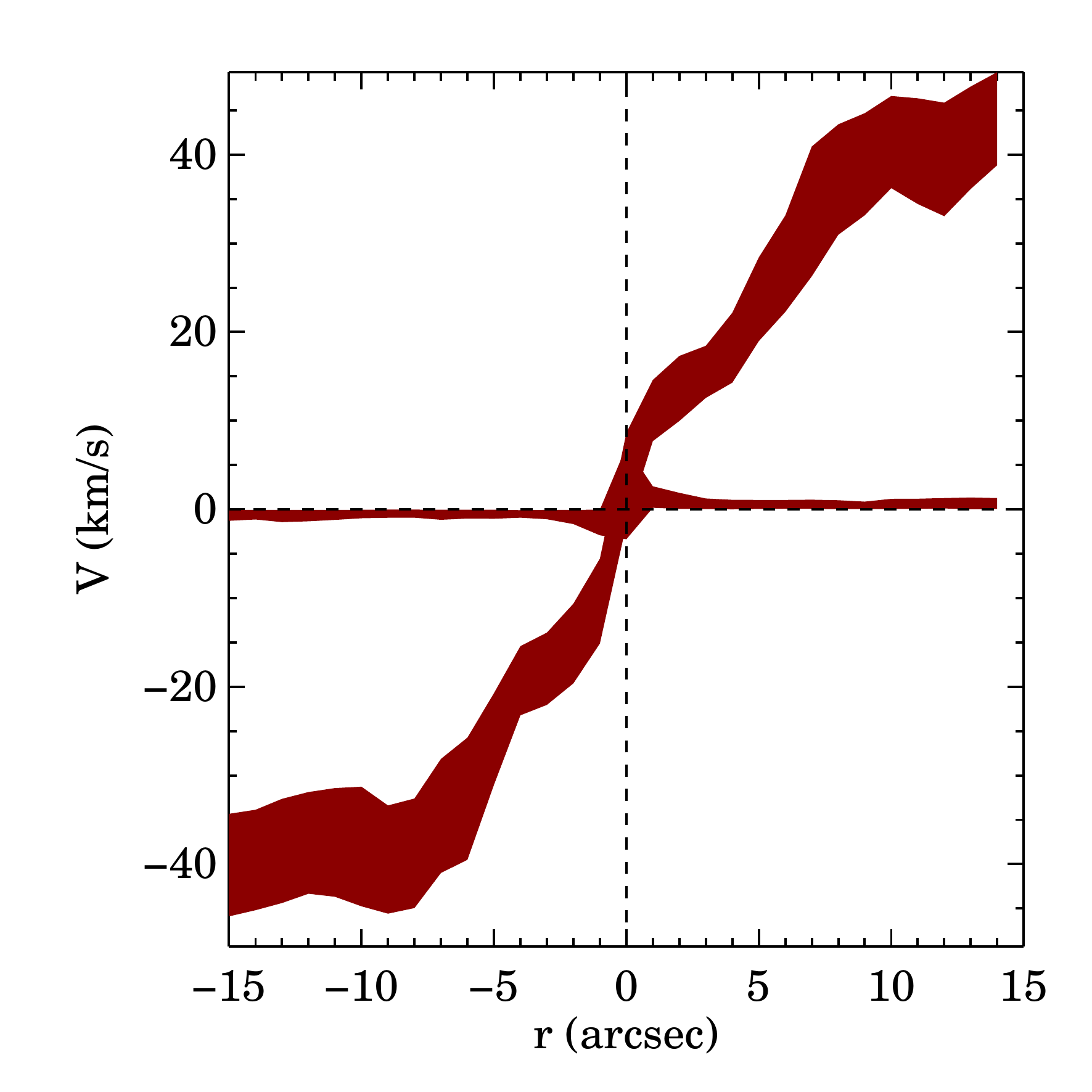}
\includegraphics[width=0.58\linewidth]{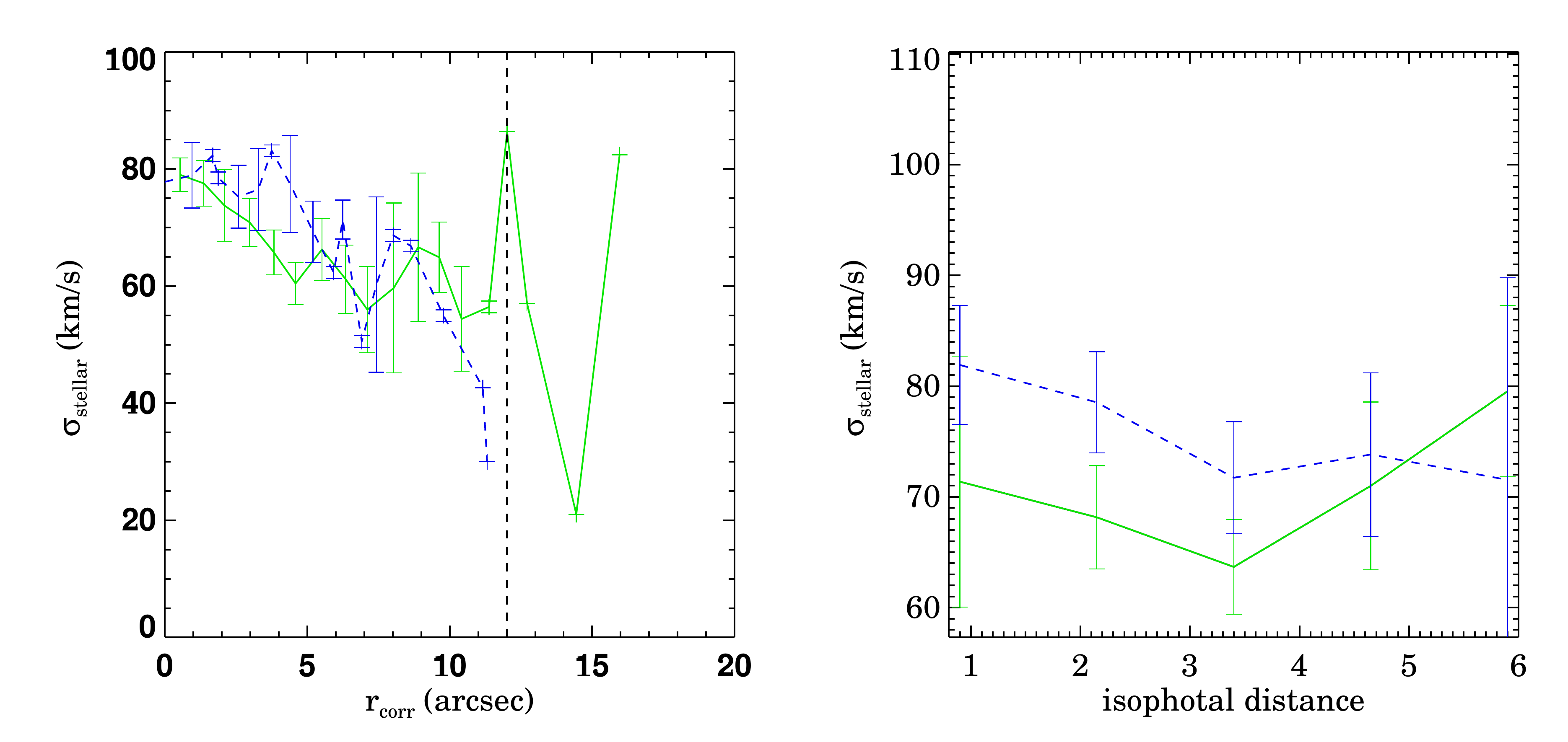}
\caption{Figure \ref{fig:gasvel} continued.}
\label{fig:2893}
\end{figure*}

\begin{figure*}
\includegraphics[width=0.32\linewidth]{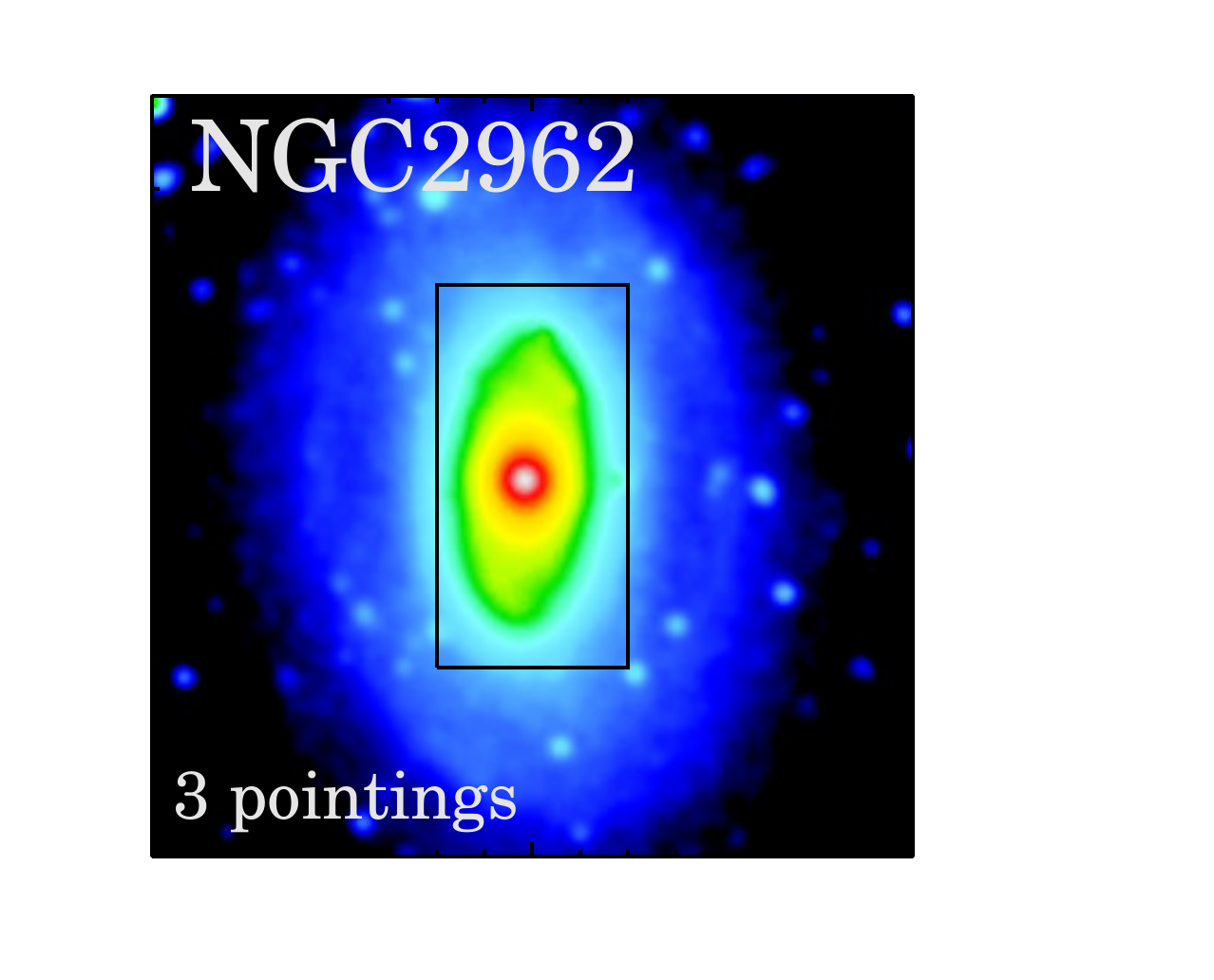}
\includegraphics[width=0.32\linewidth]{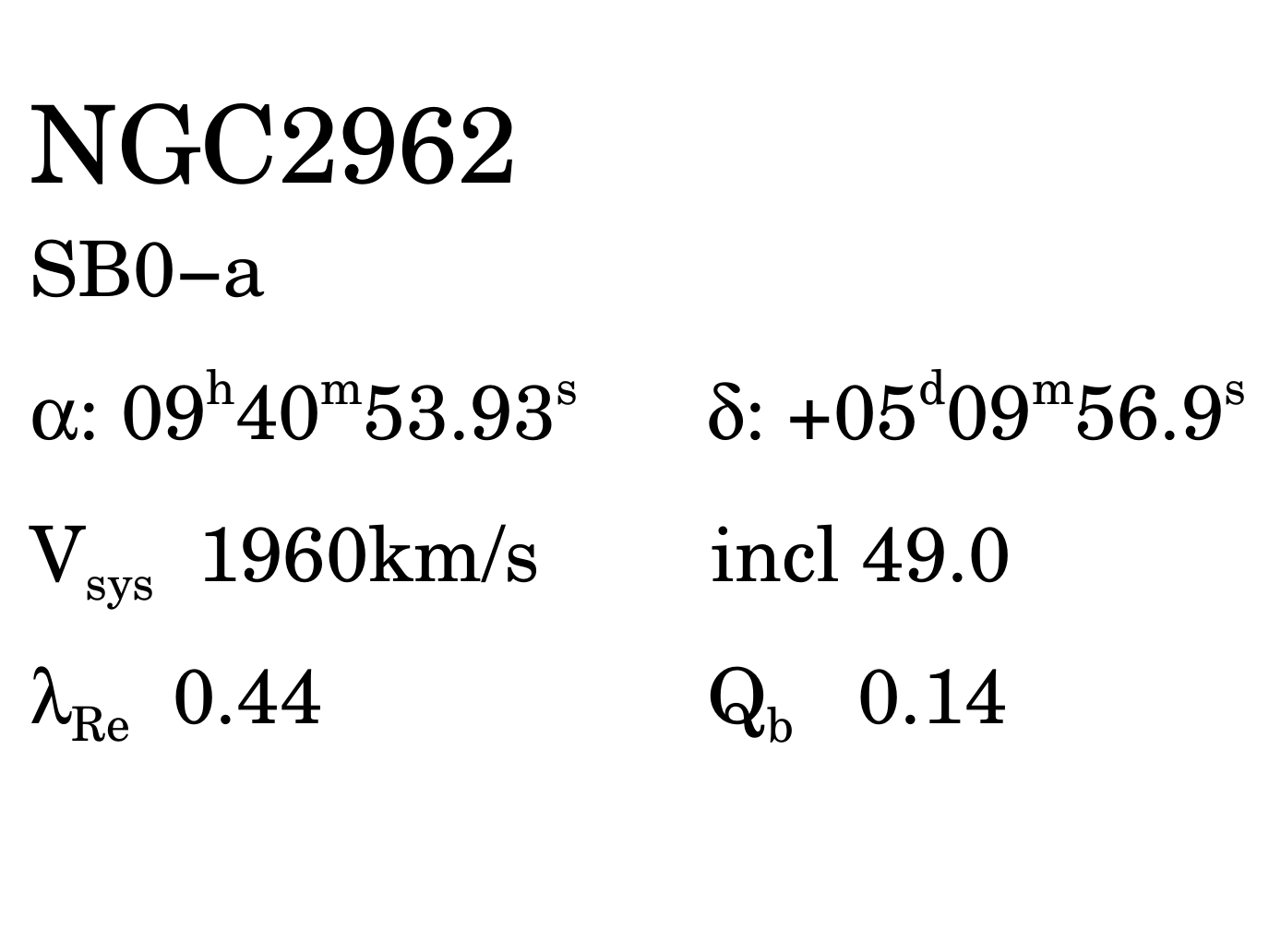}
\includegraphics[width=0.88\linewidth]{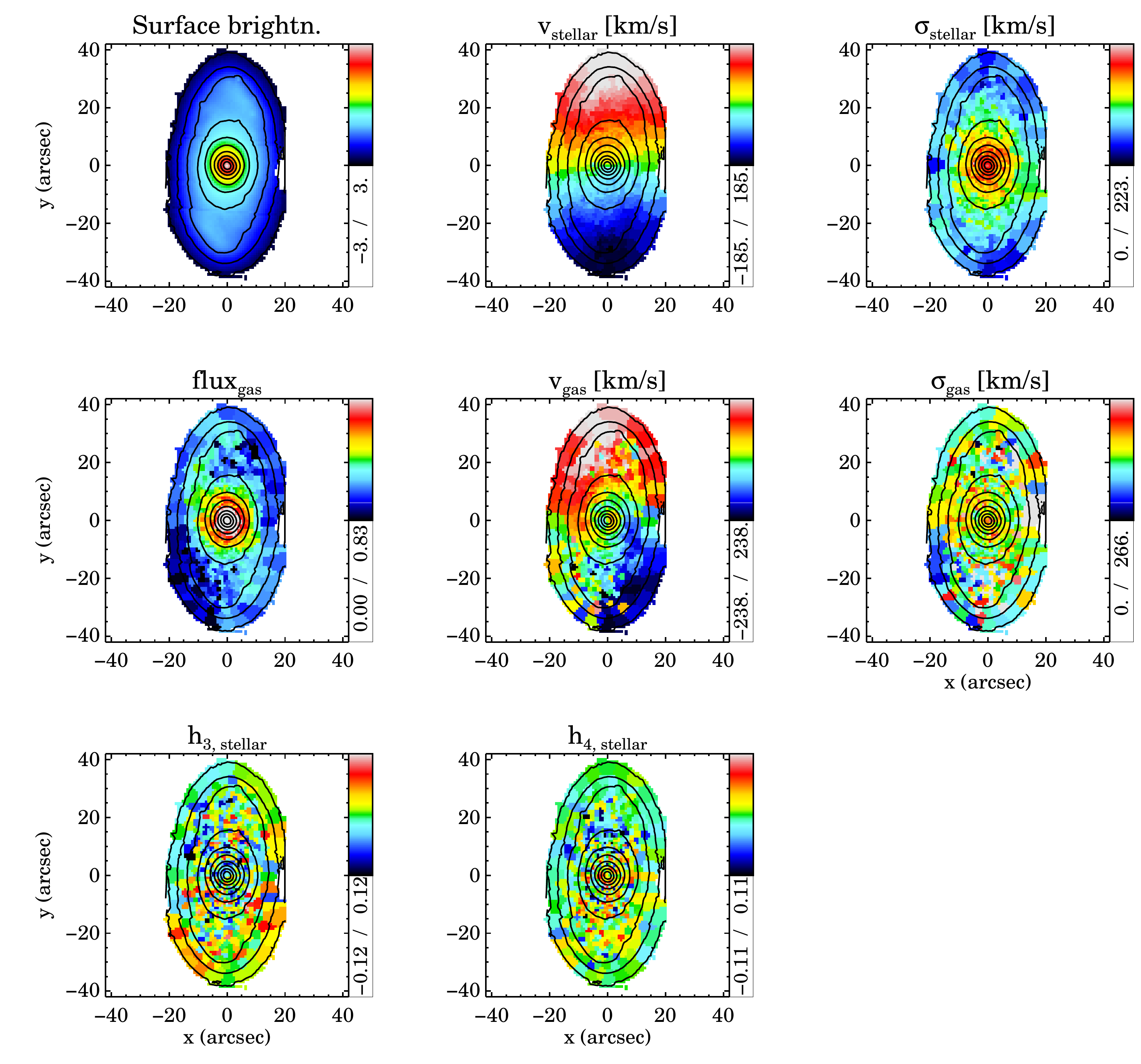}
\includegraphics[width=0.27\linewidth]{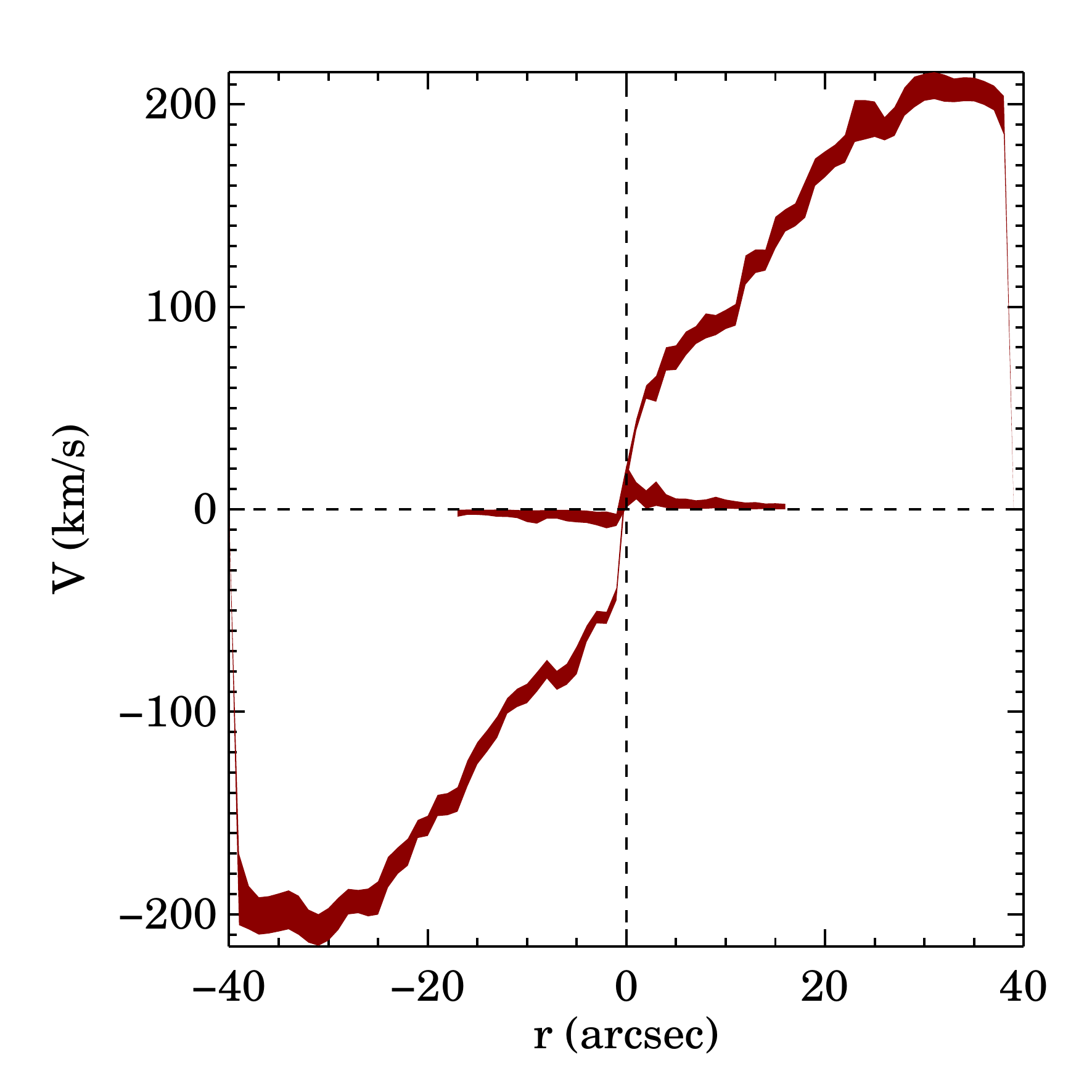}
\includegraphics[width=0.58\linewidth]{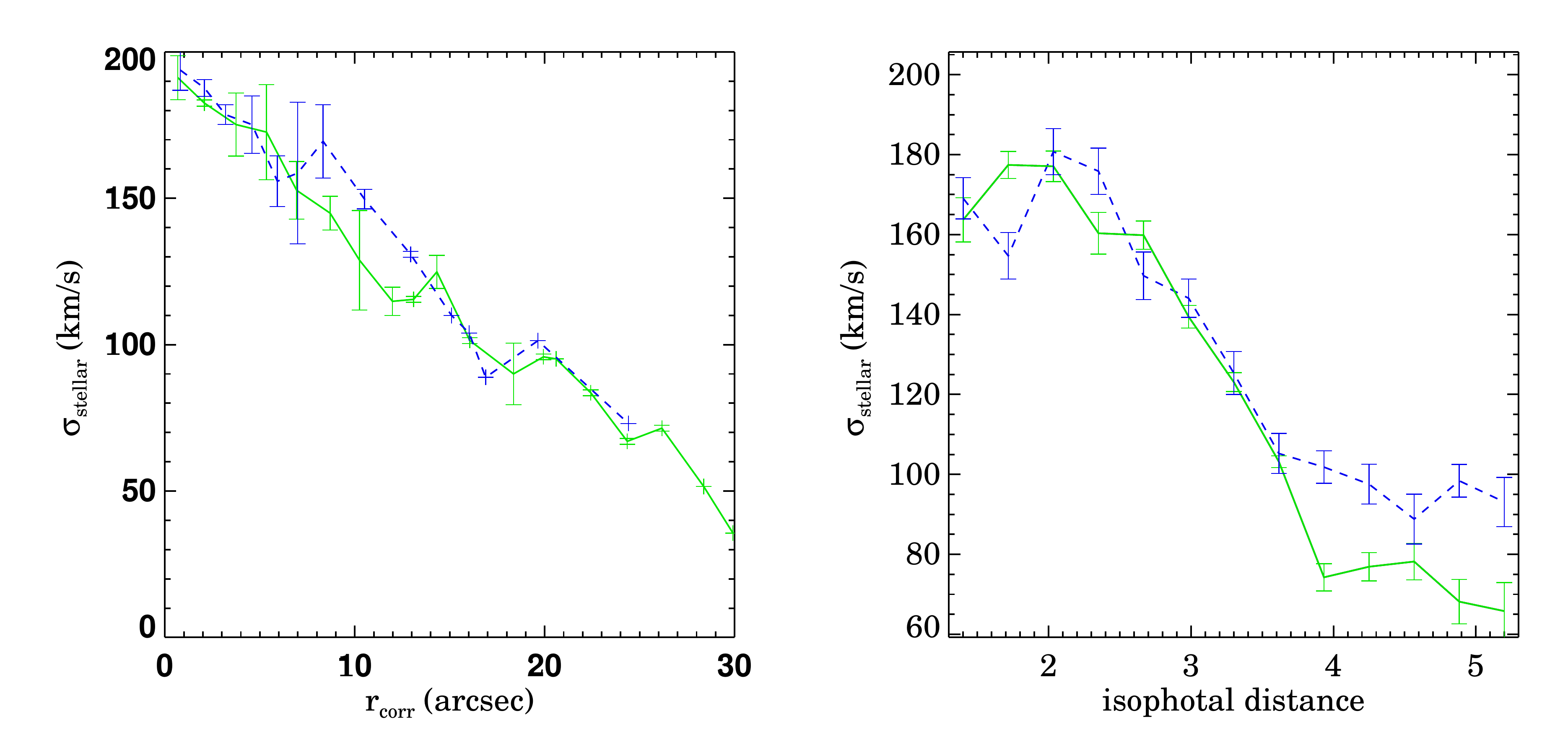}
\caption{Figure \ref{fig:gasvel} continued.}
\label{fig:2962}
\end{figure*}

\begin{figure*}
\includegraphics[width=0.32\linewidth]{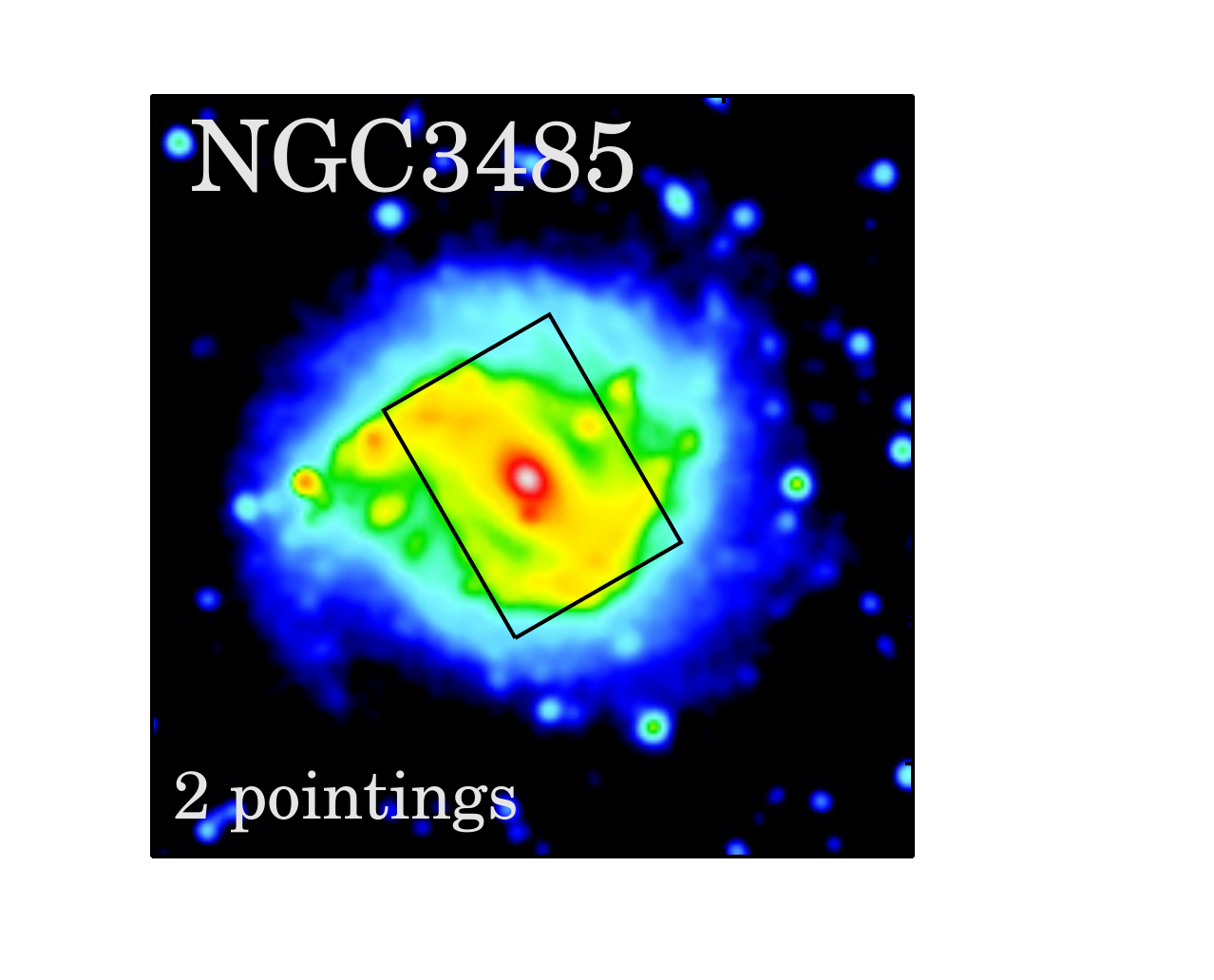}
\includegraphics[width=0.32\linewidth]{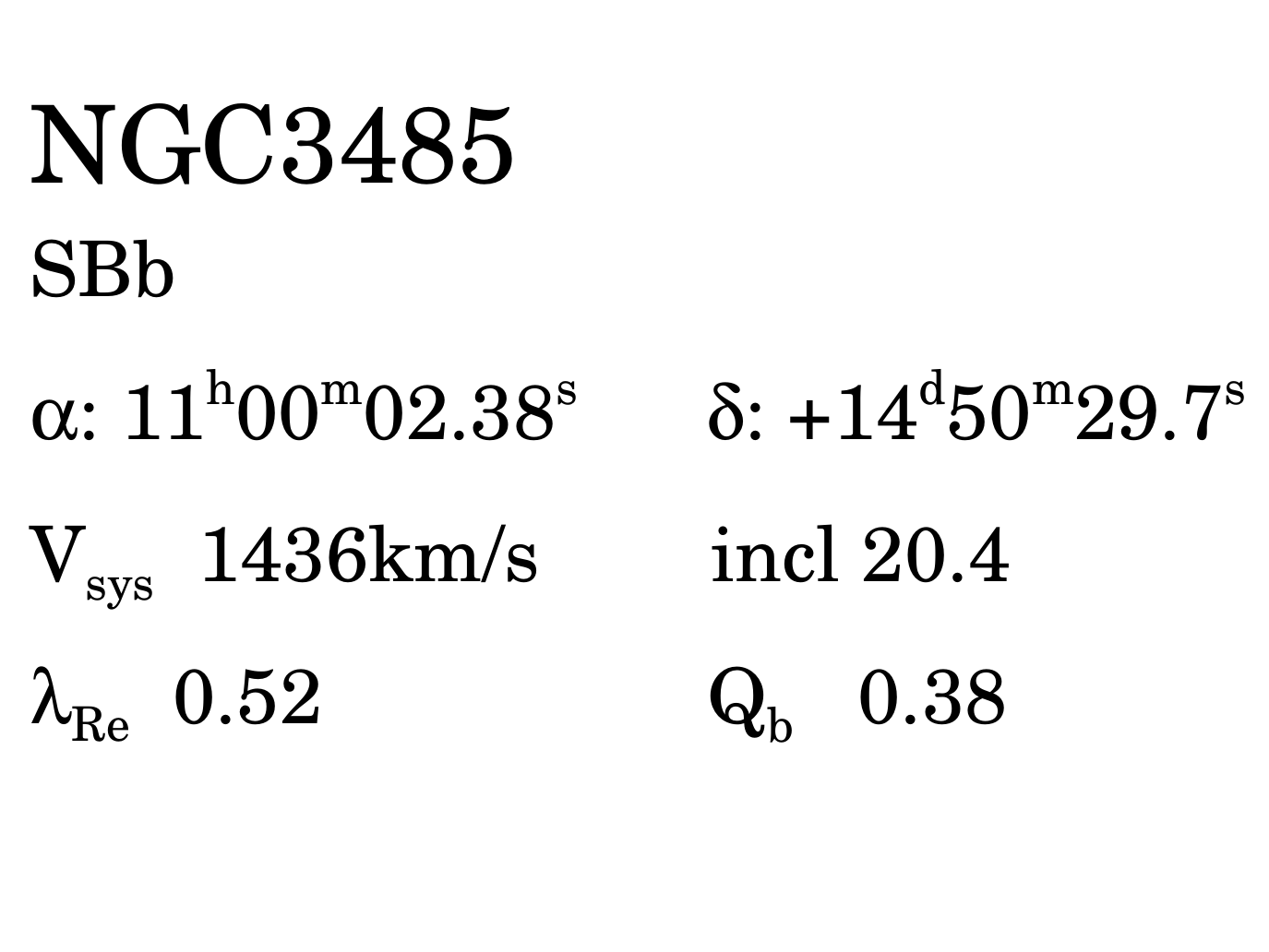}
\includegraphics[width=0.88\linewidth]{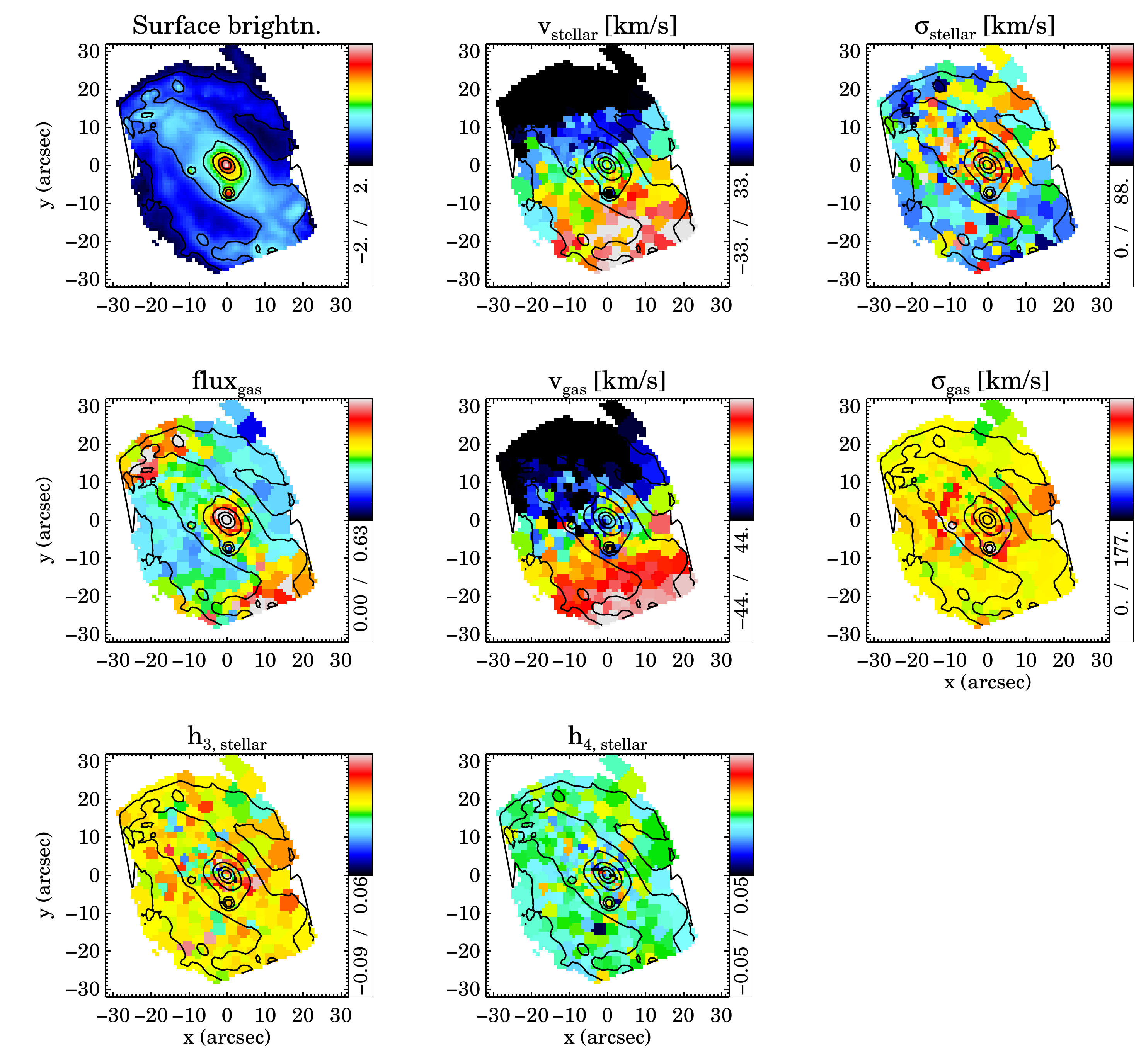}
\includegraphics[width=0.27\linewidth]{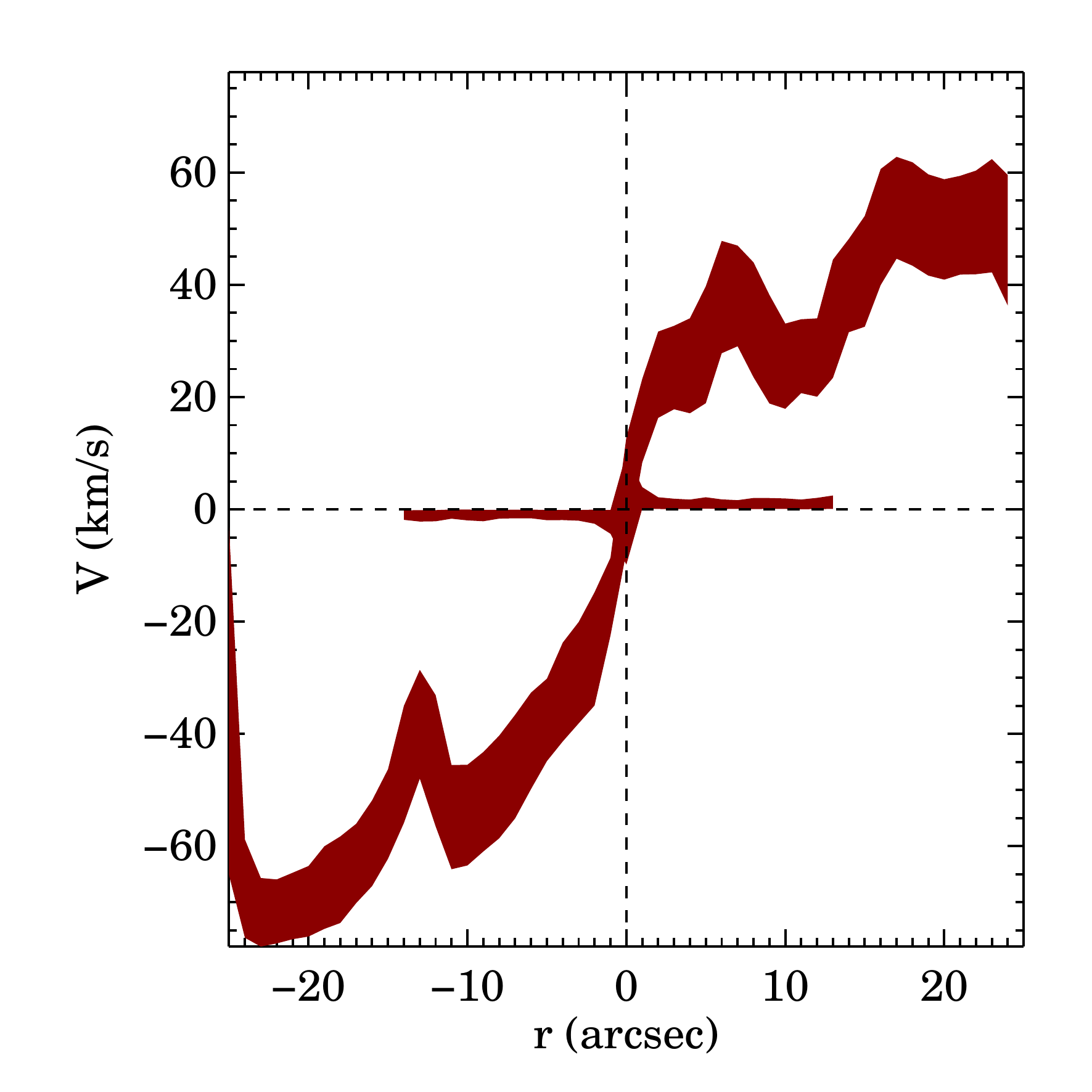}
\includegraphics[width=0.58\linewidth]{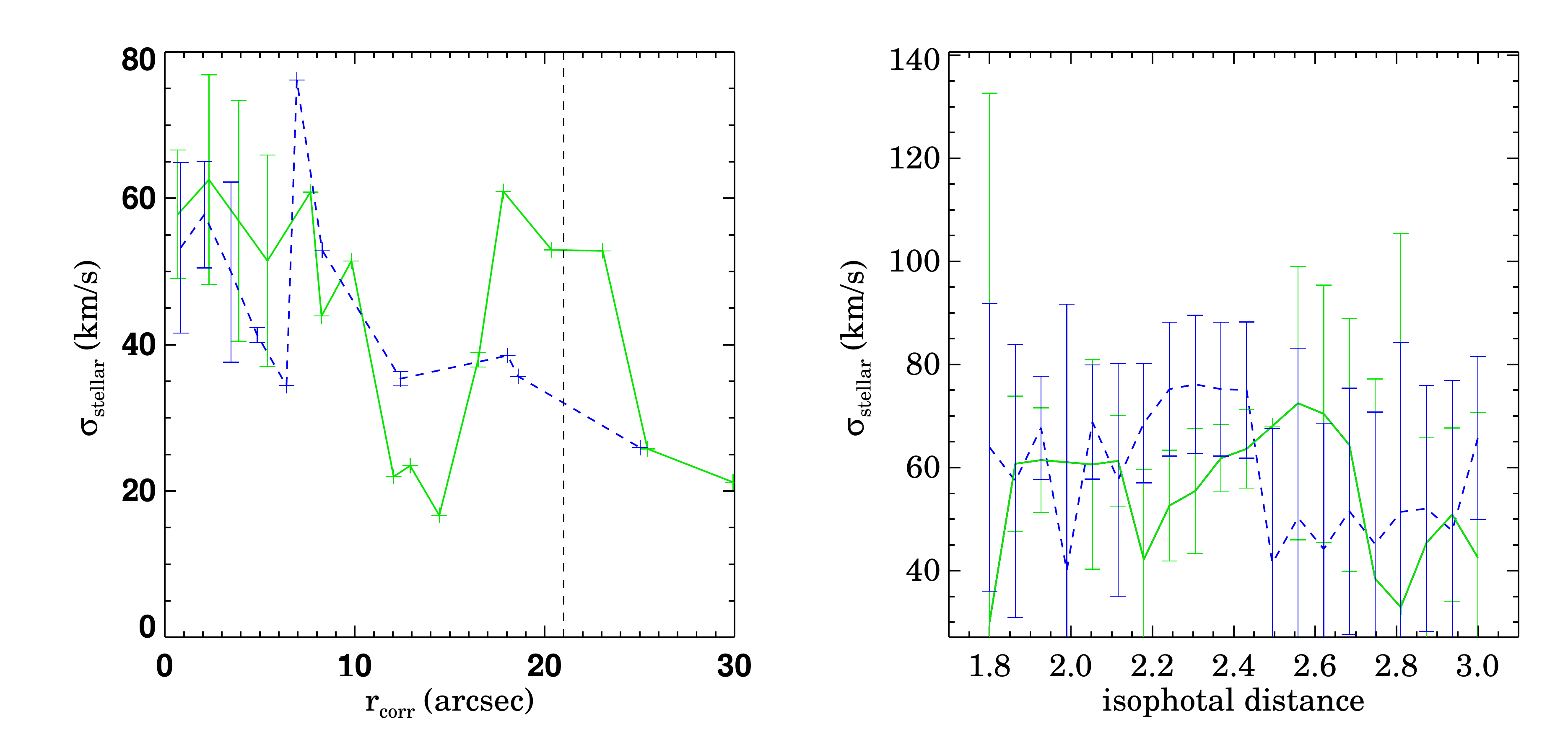}
\caption{Figure \ref{fig:gasvel} continued.}
\label{fig:3485}
\end{figure*}

\begin{figure*}
\includegraphics[width=0.32\linewidth]{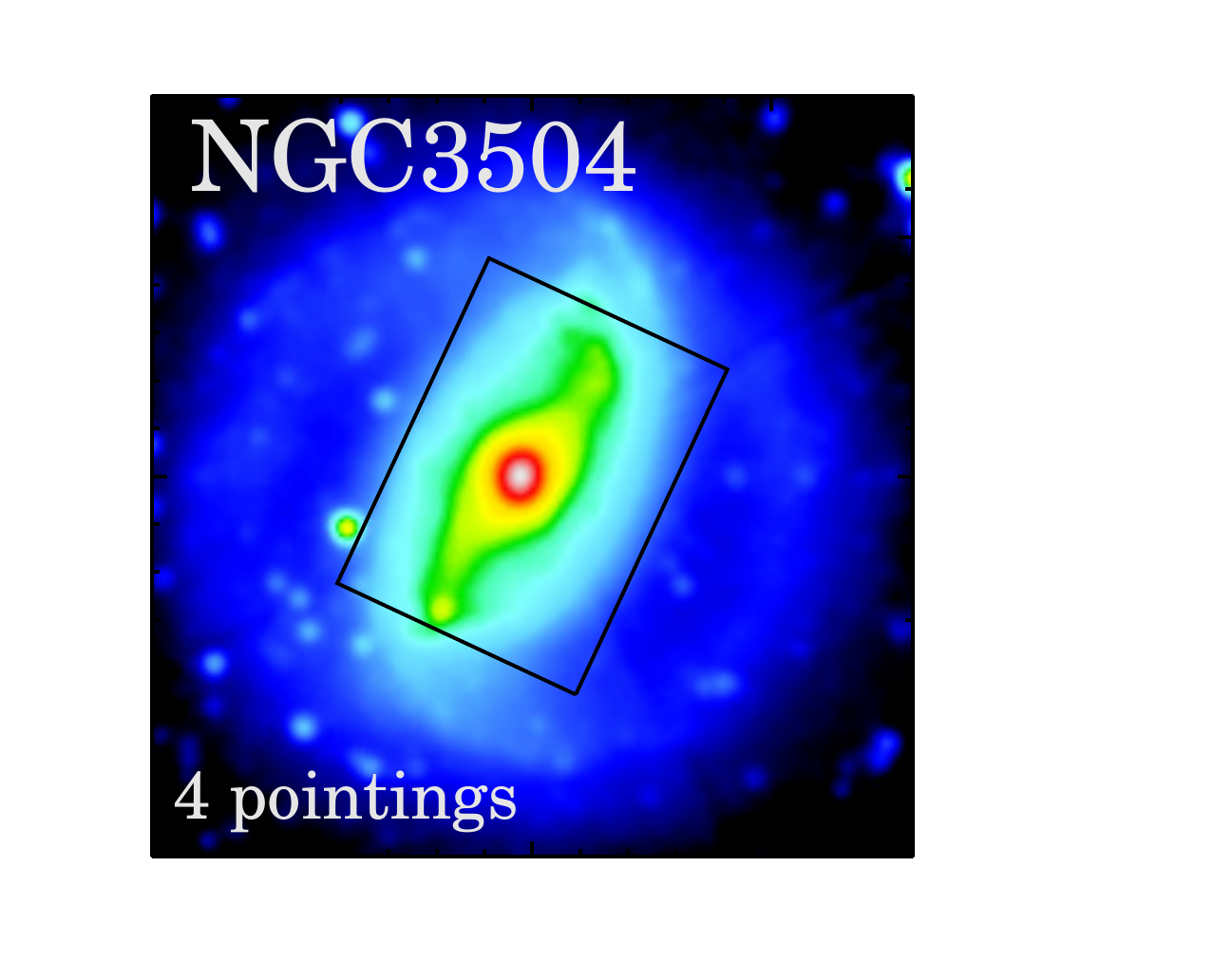}
\includegraphics[width=0.32\linewidth]{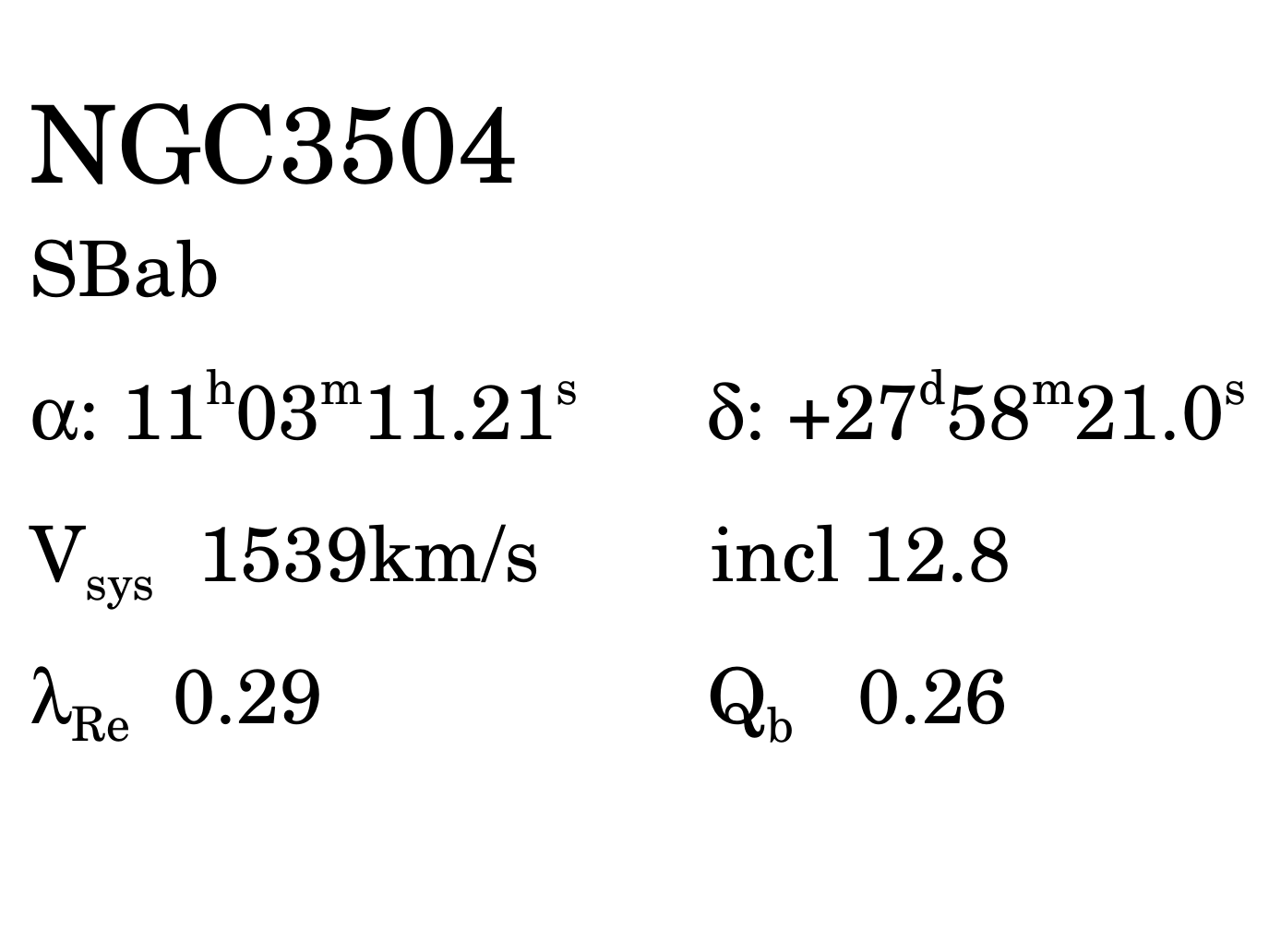}
\includegraphics[width=0.88\linewidth]{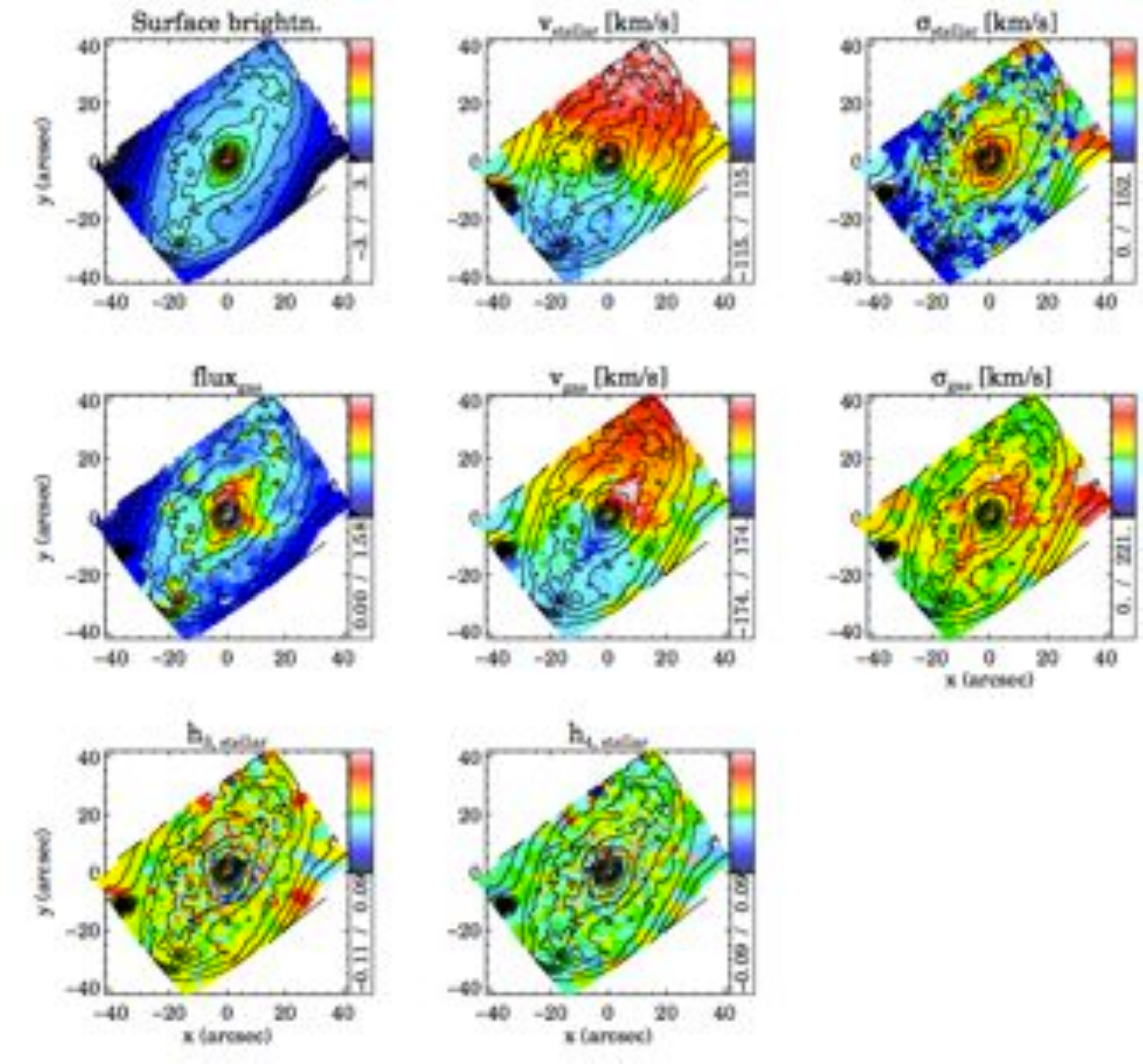}
\includegraphics[width=0.27\linewidth]{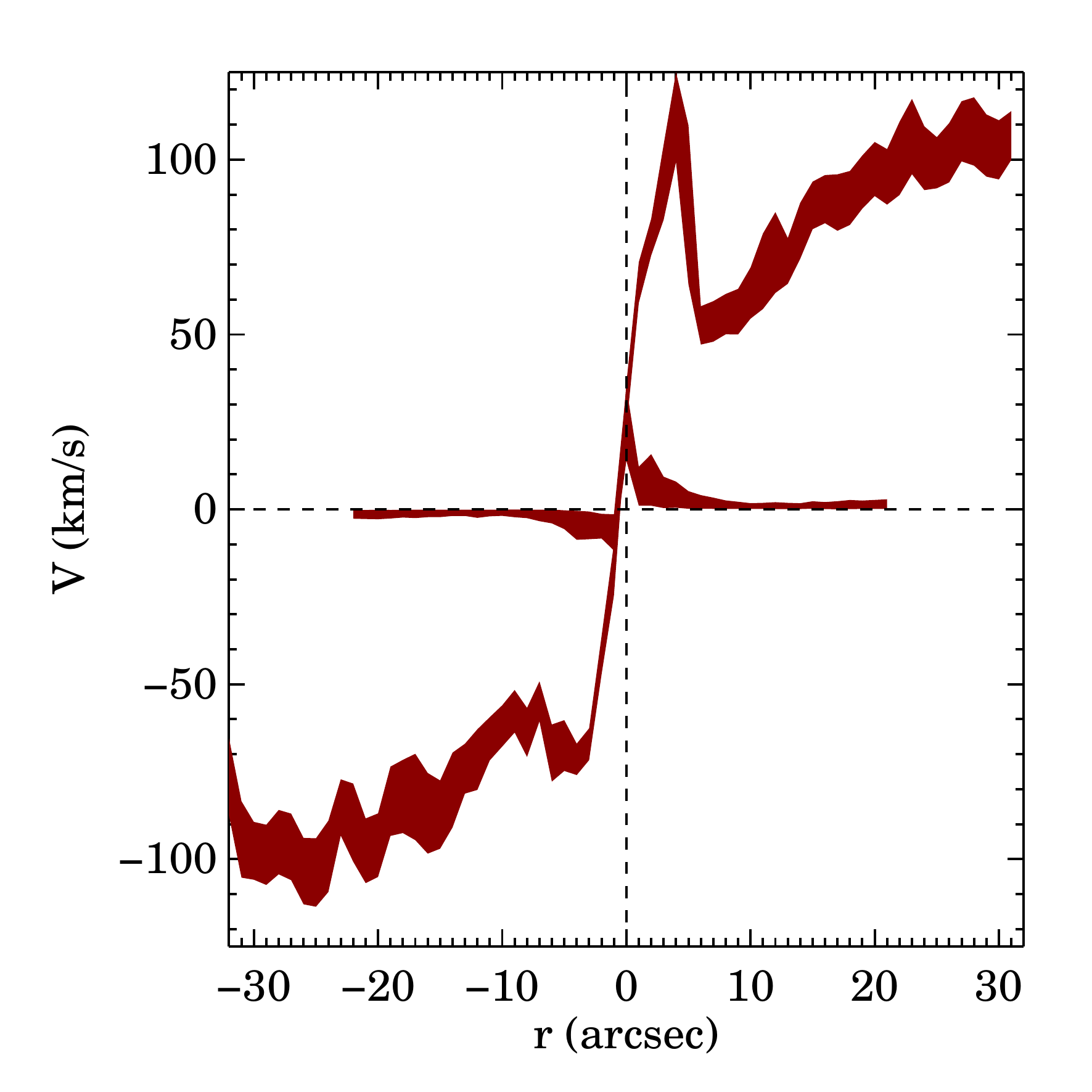}
\includegraphics[width=0.58\linewidth]{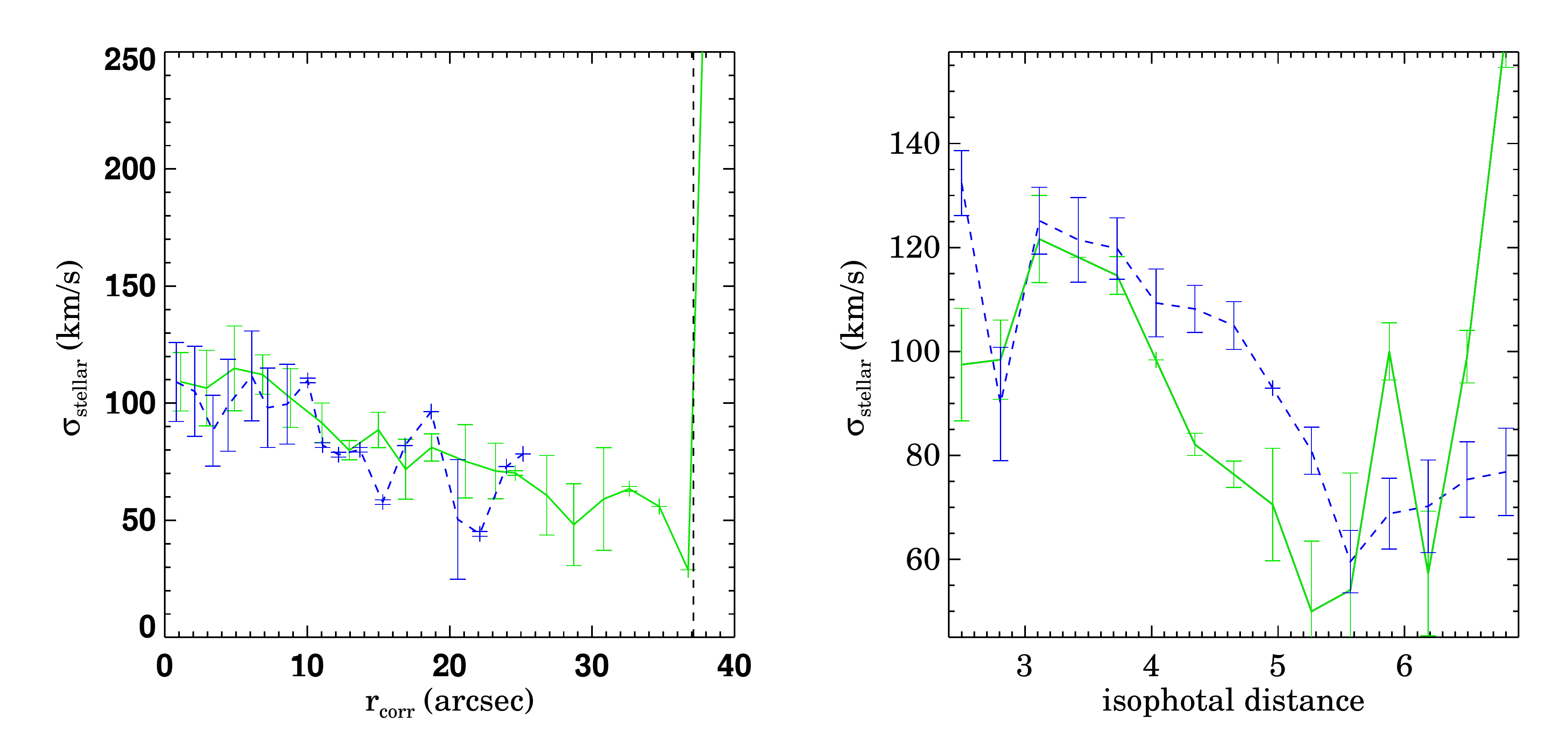}
\caption{Figure \ref{fig:gasvel} continued.}
\label{fig:3504}
\end{figure*}

\begin{figure*}
\includegraphics[width=0.32\linewidth]{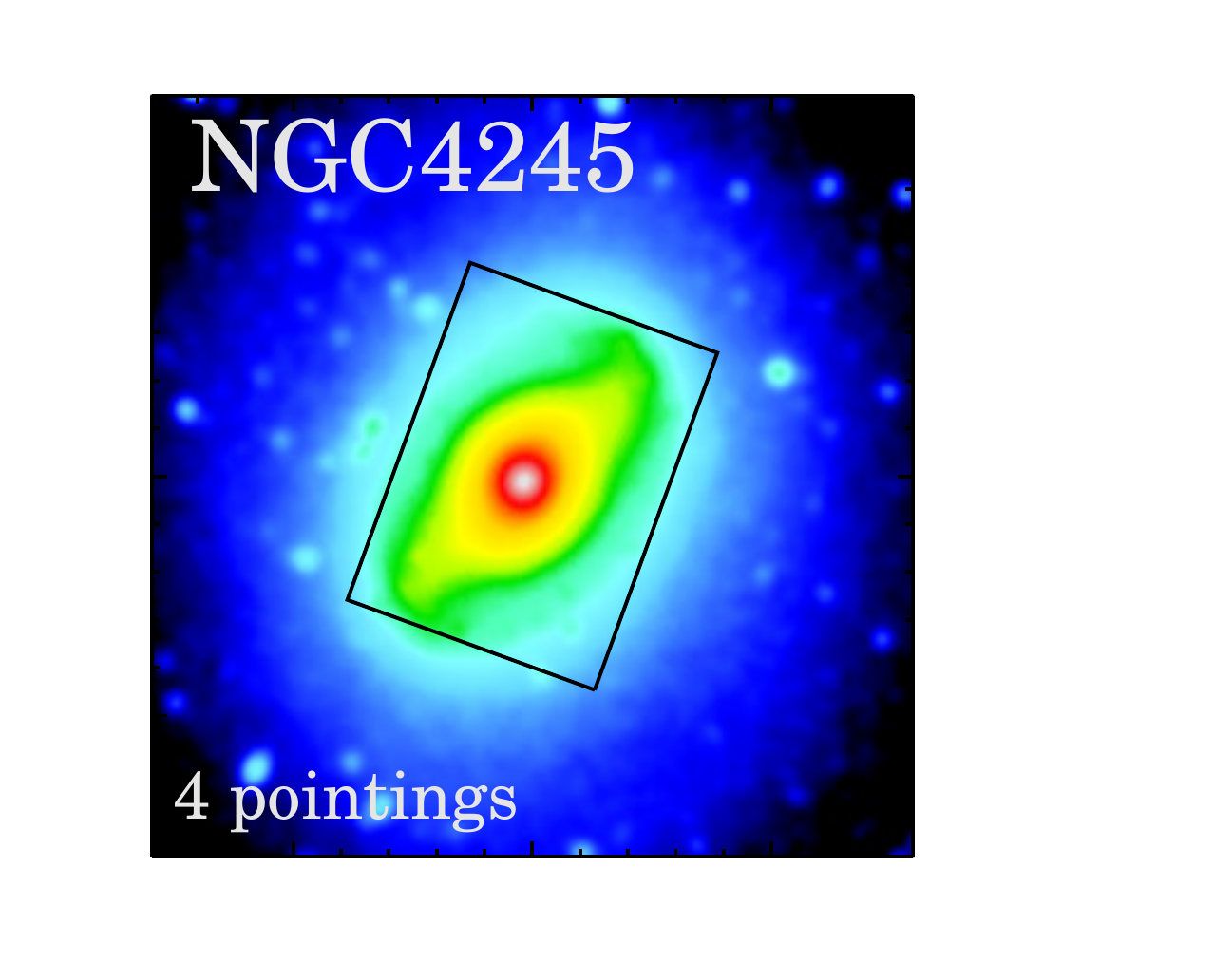}
\includegraphics[width=0.32\linewidth]{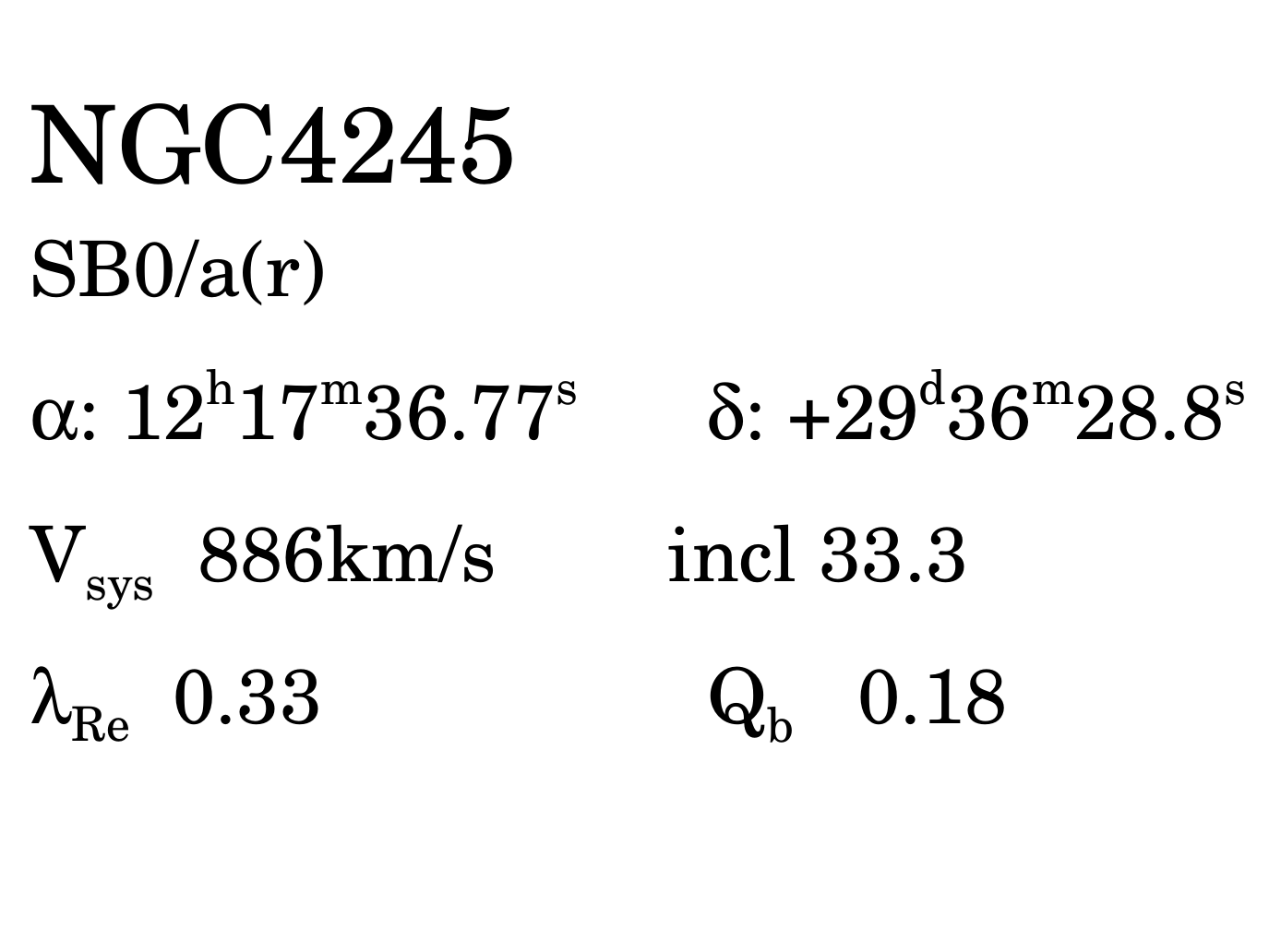}
\includegraphics[width=0.88\linewidth]{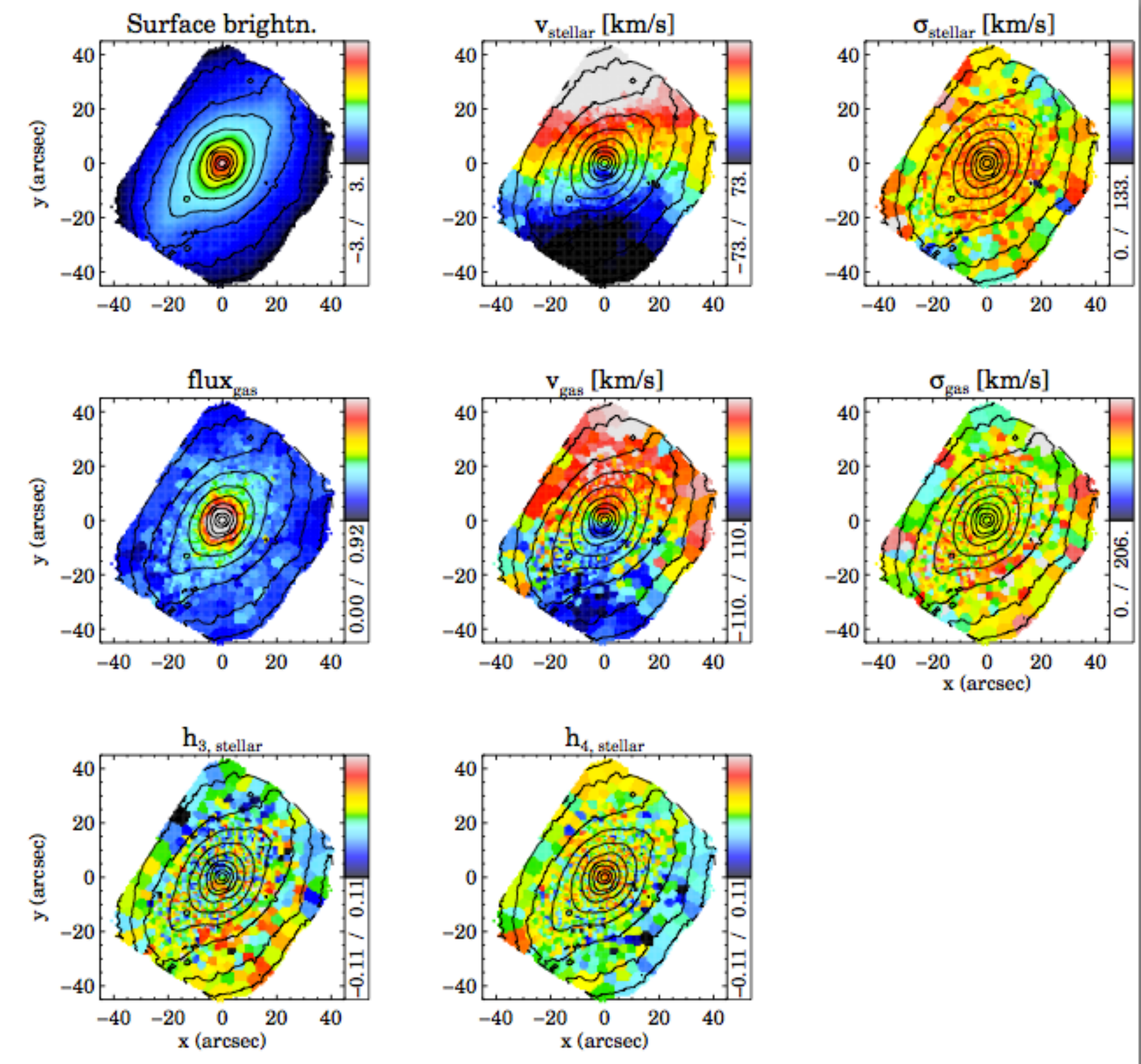}
\includegraphics[width=0.27\linewidth]{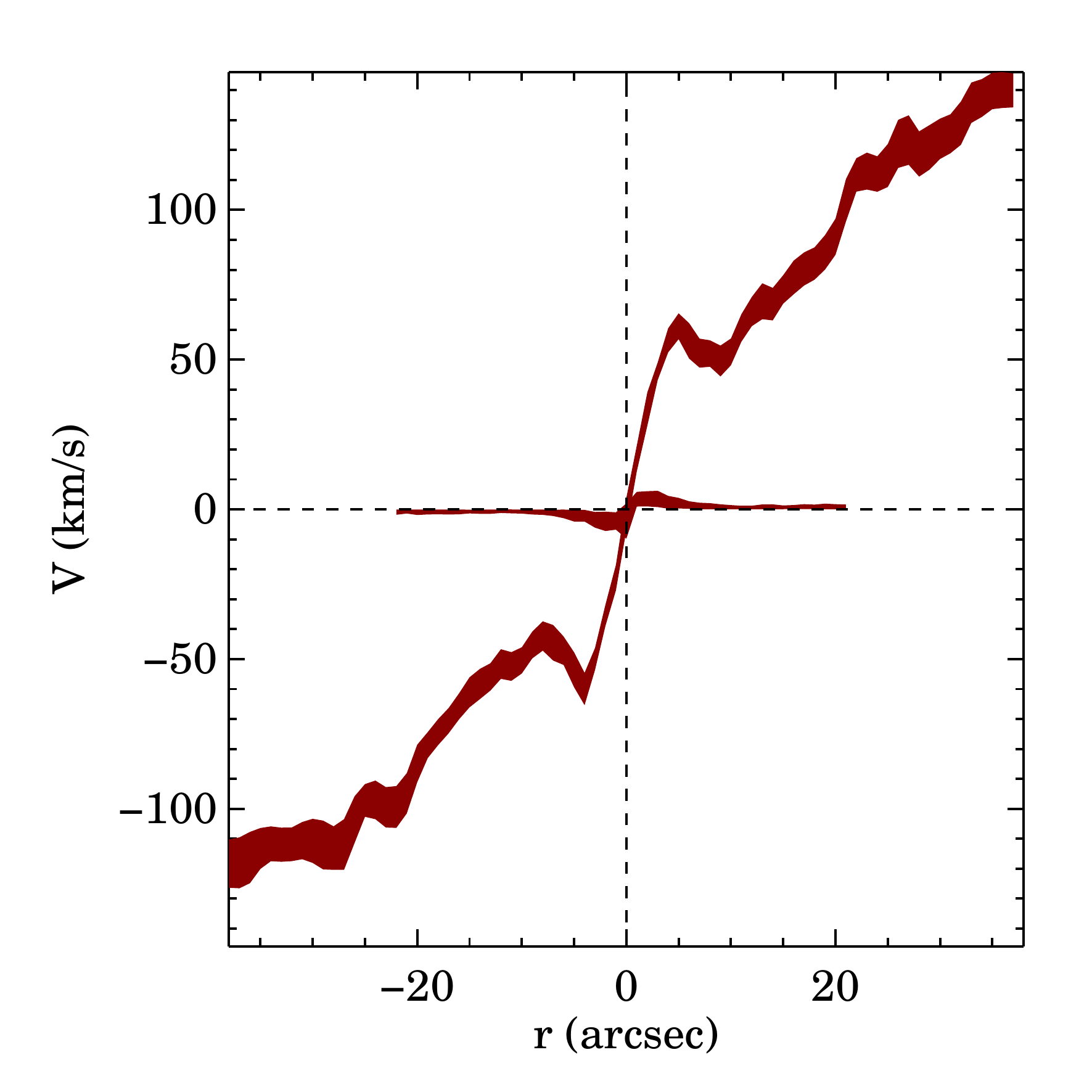}
\includegraphics[width=0.58\linewidth]{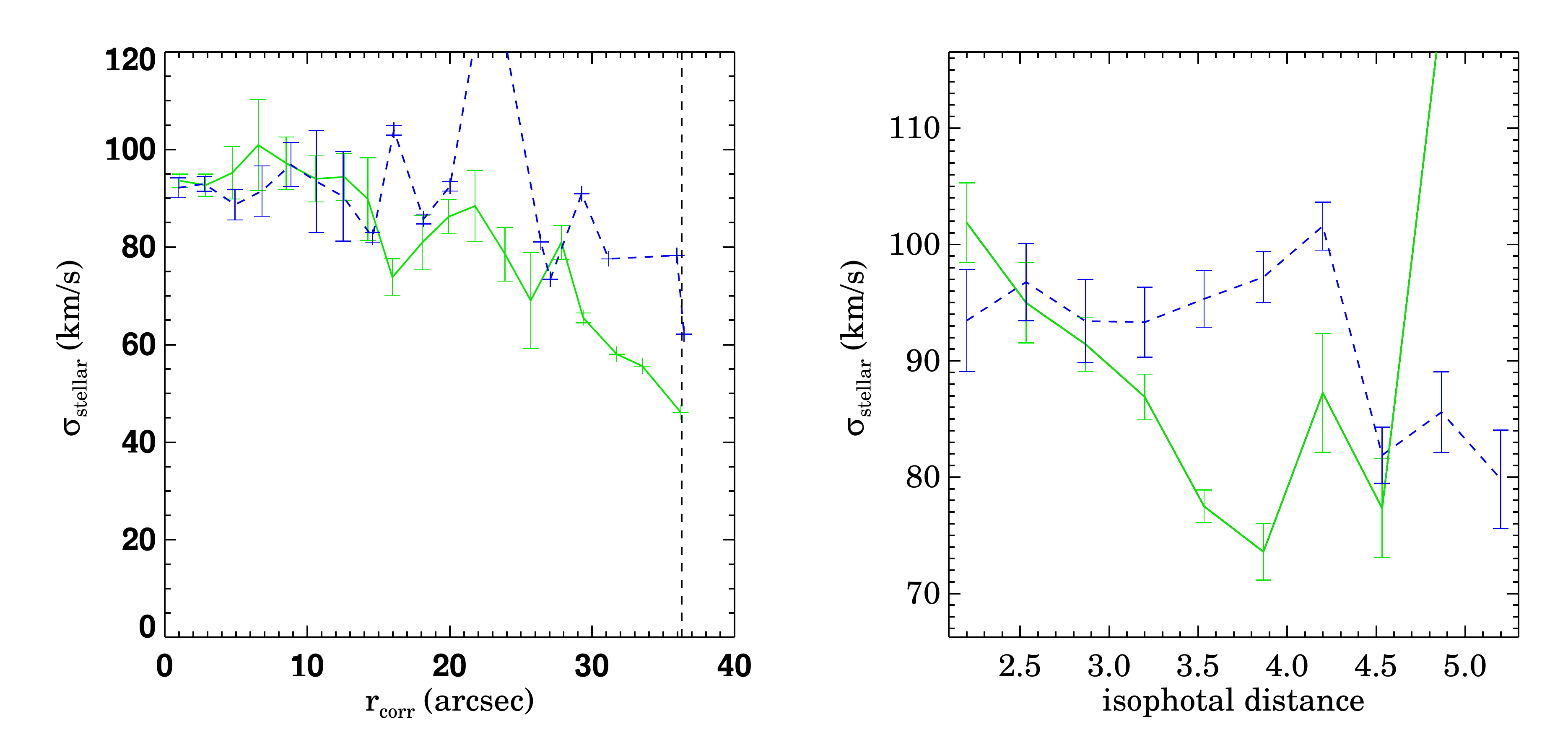}
\caption{Figure \ref{fig:gasvel} continued.}
\label{fig:4245}
\end{figure*}

\begin{figure*}
\includegraphics[width=0.32\linewidth]{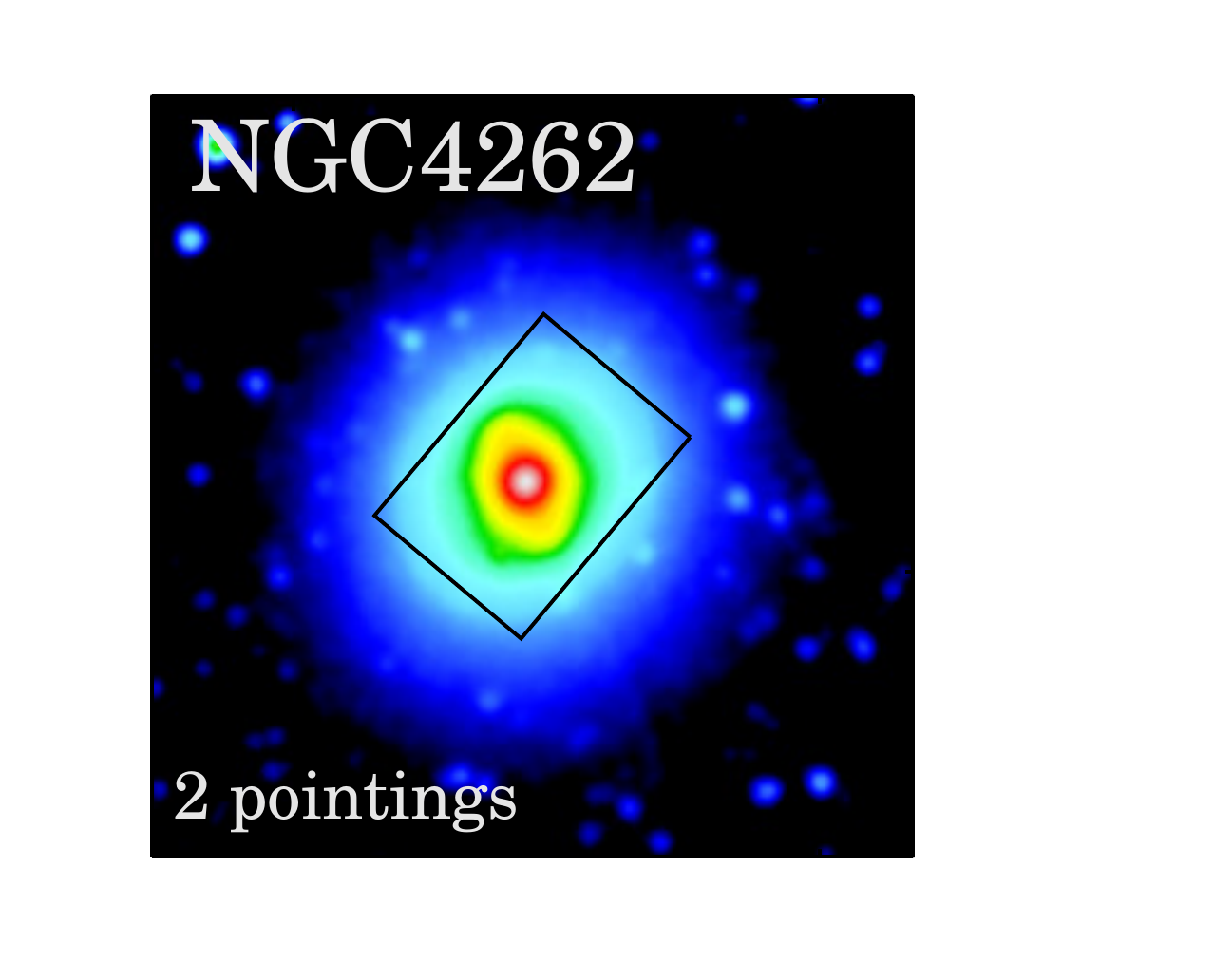}
\includegraphics[width=0.32\linewidth]{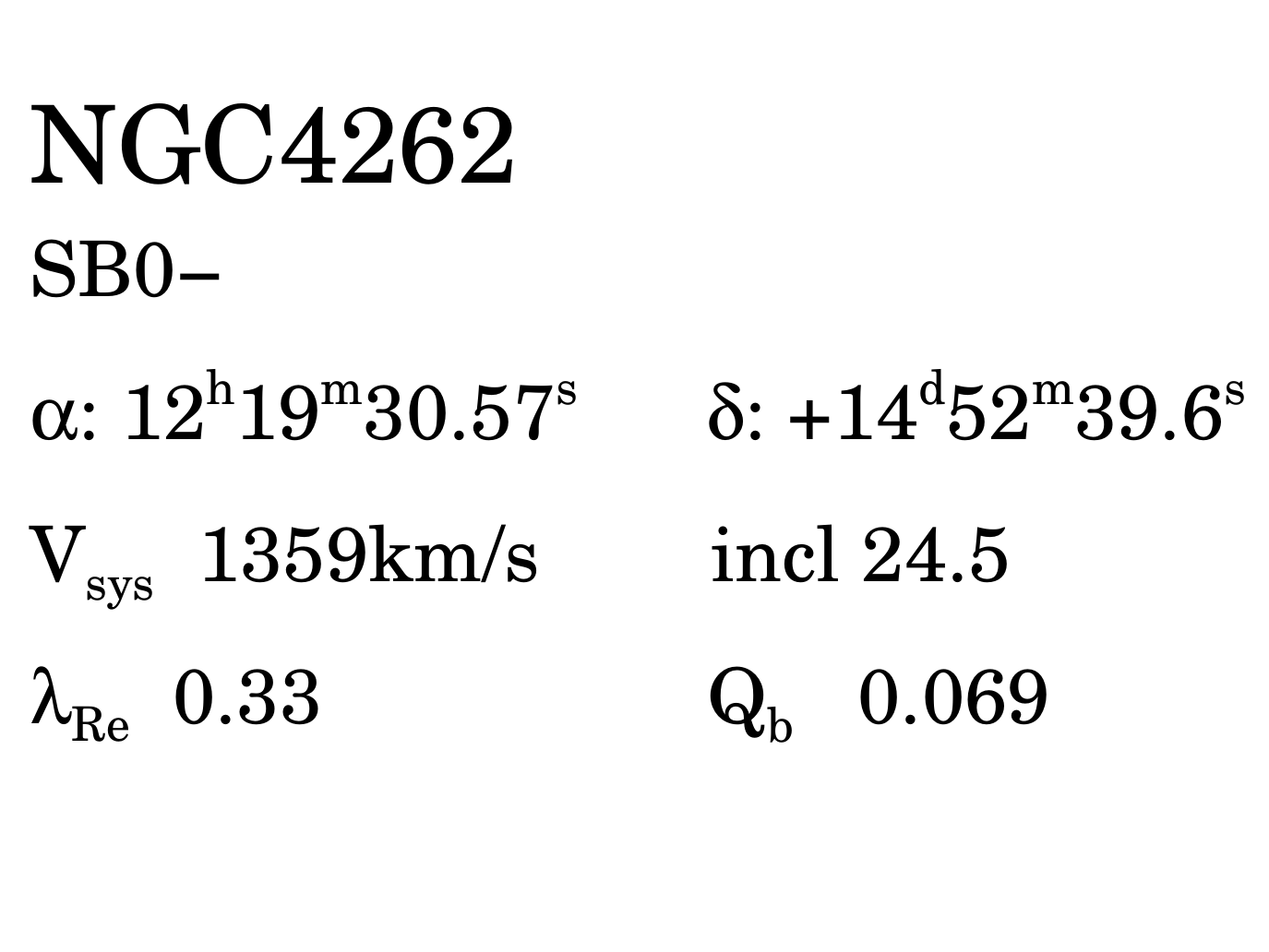}
\includegraphics[width=0.88\linewidth]{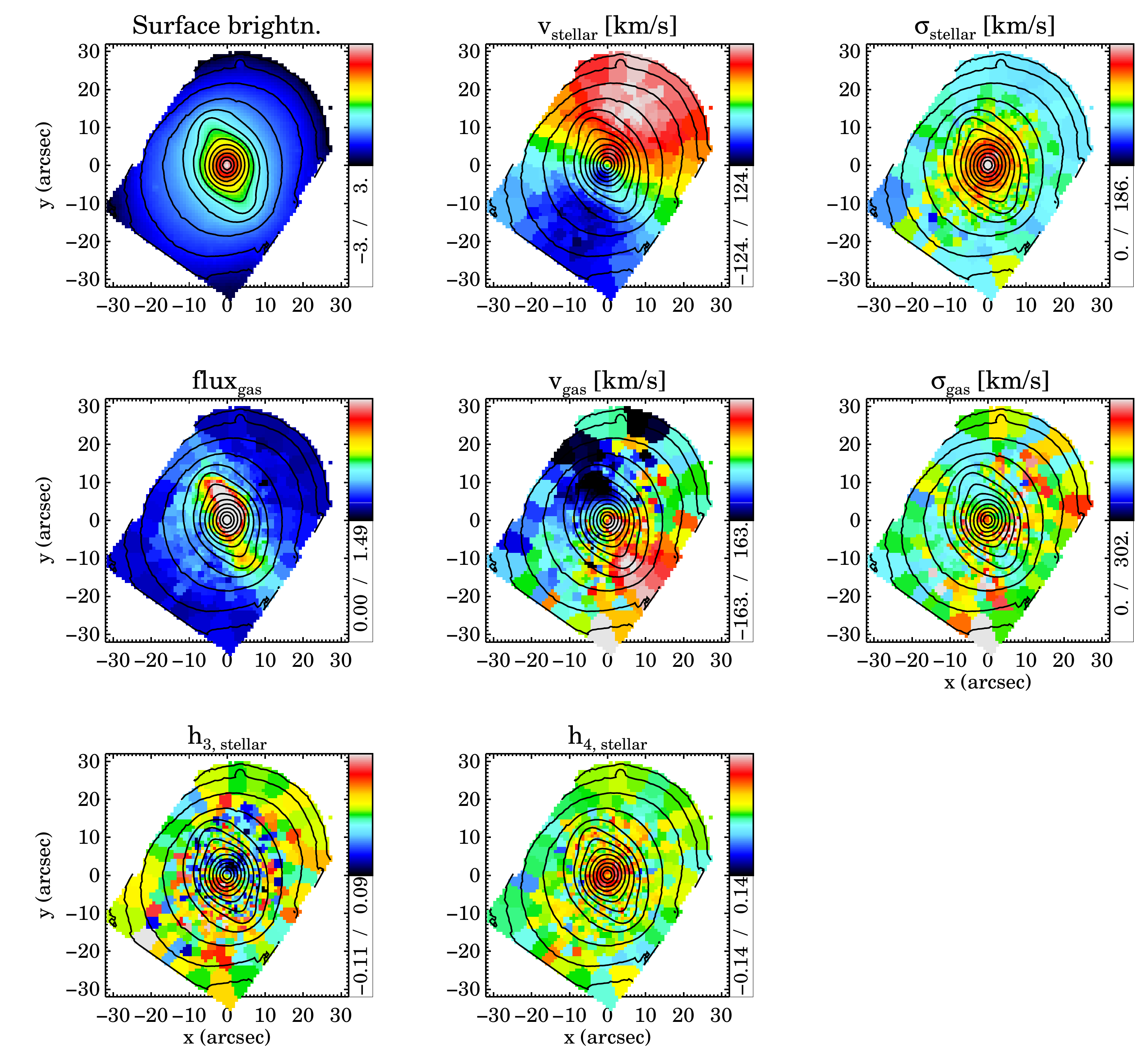}
\includegraphics[width=0.27\linewidth]{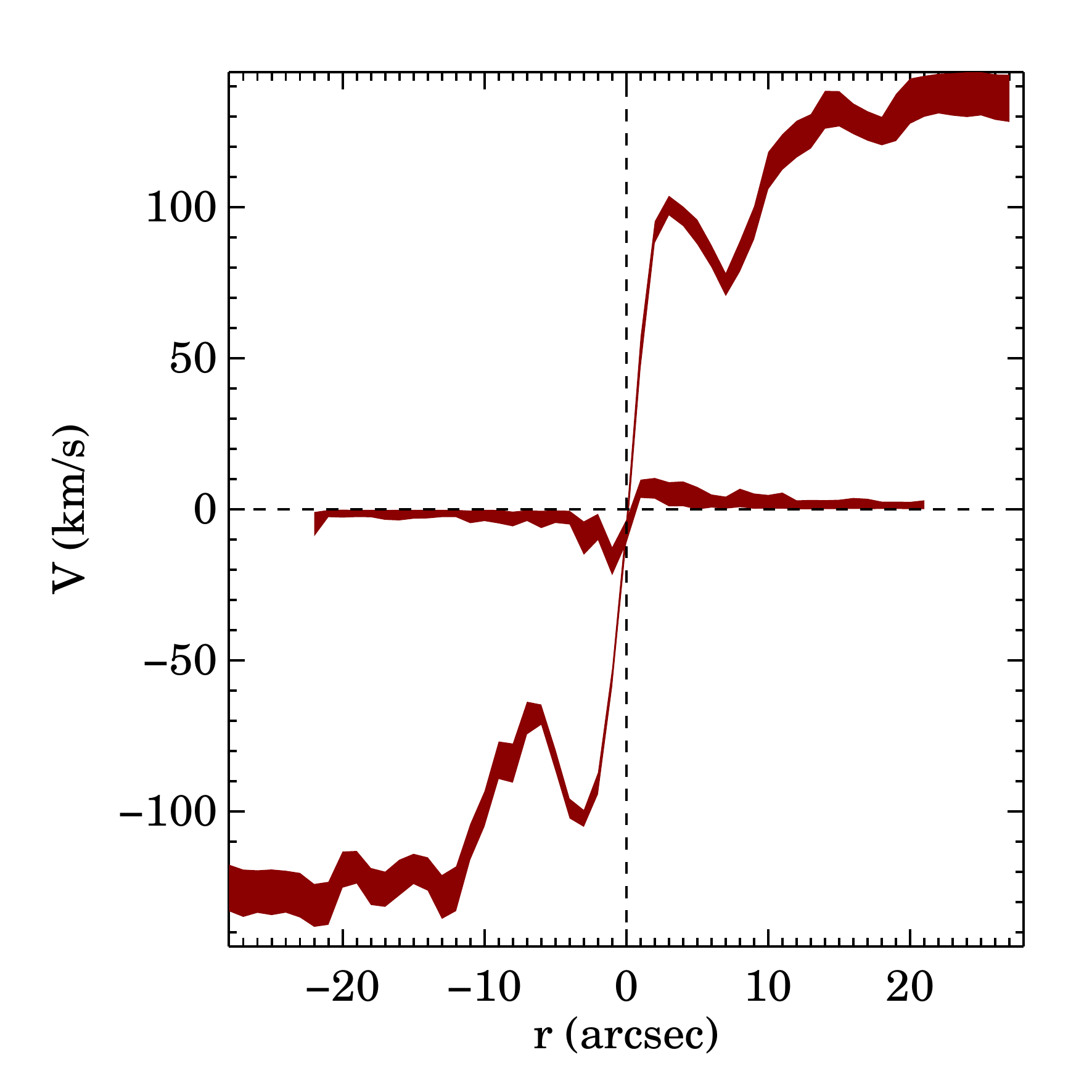}
\includegraphics[width=0.58\linewidth]{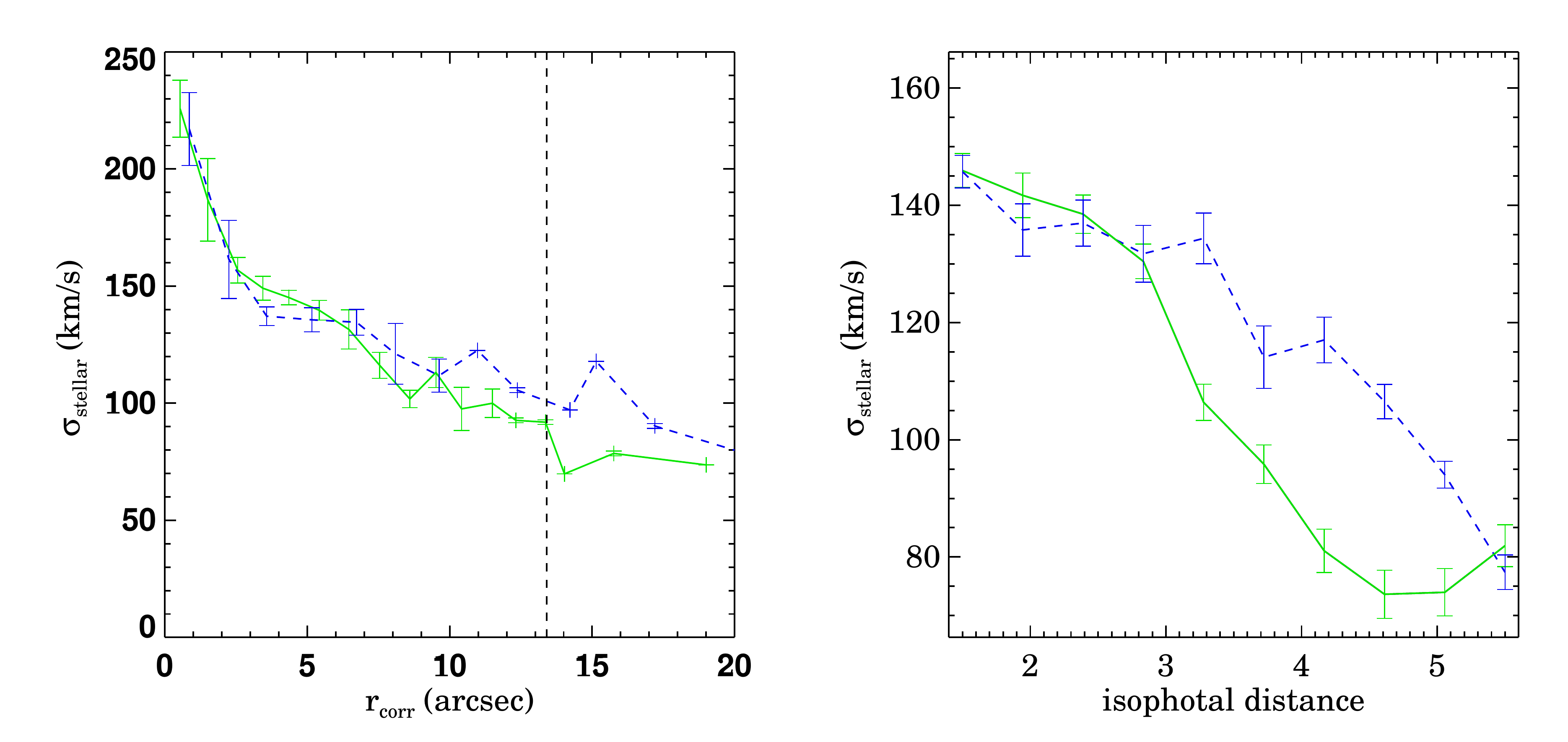}
\caption{Figure \ref{fig:gasvel} continued.}
\label{fig:4262}
\end{figure*}

\begin{figure*}
\includegraphics[width=0.32\linewidth]{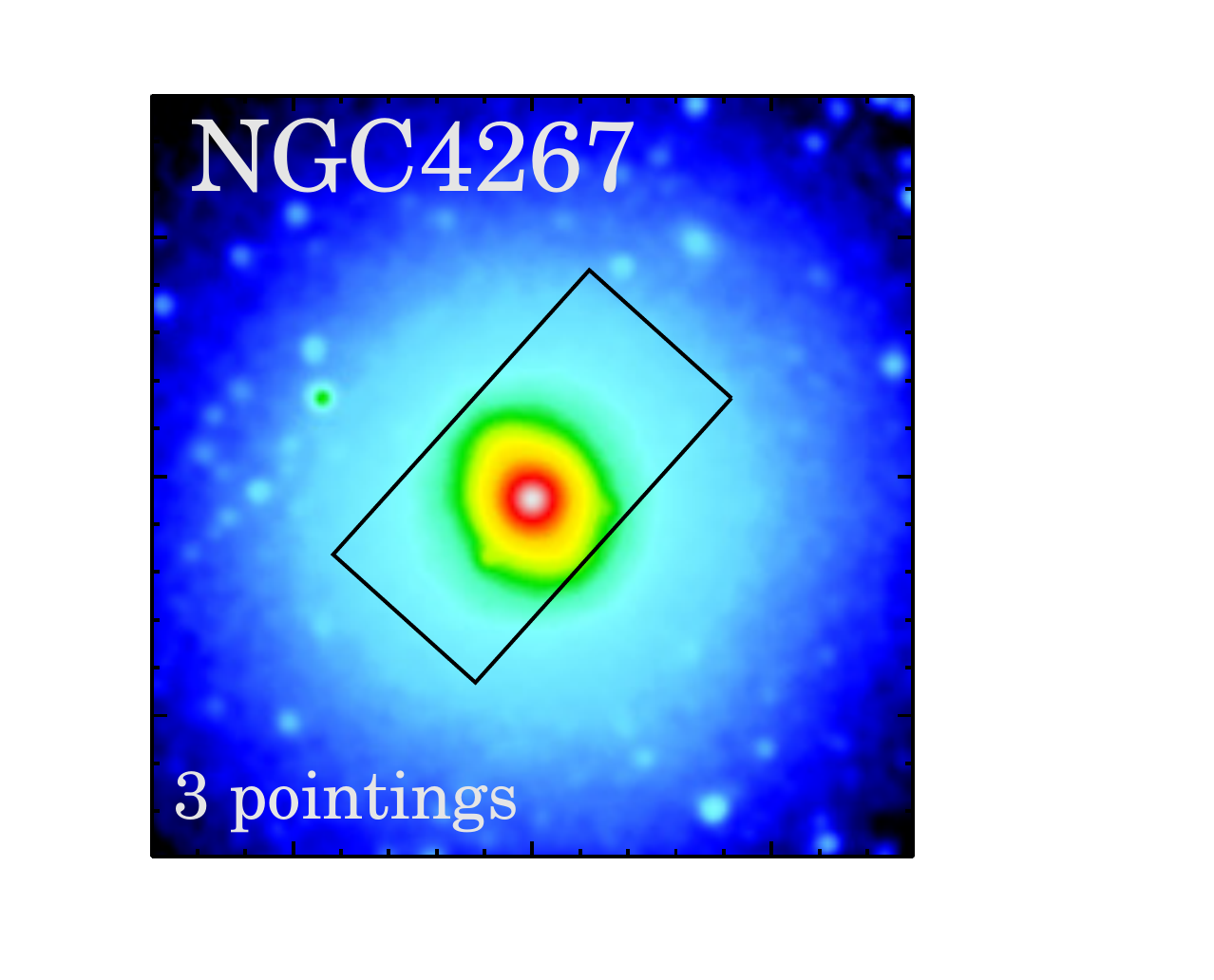}
\includegraphics[width=0.32\linewidth]{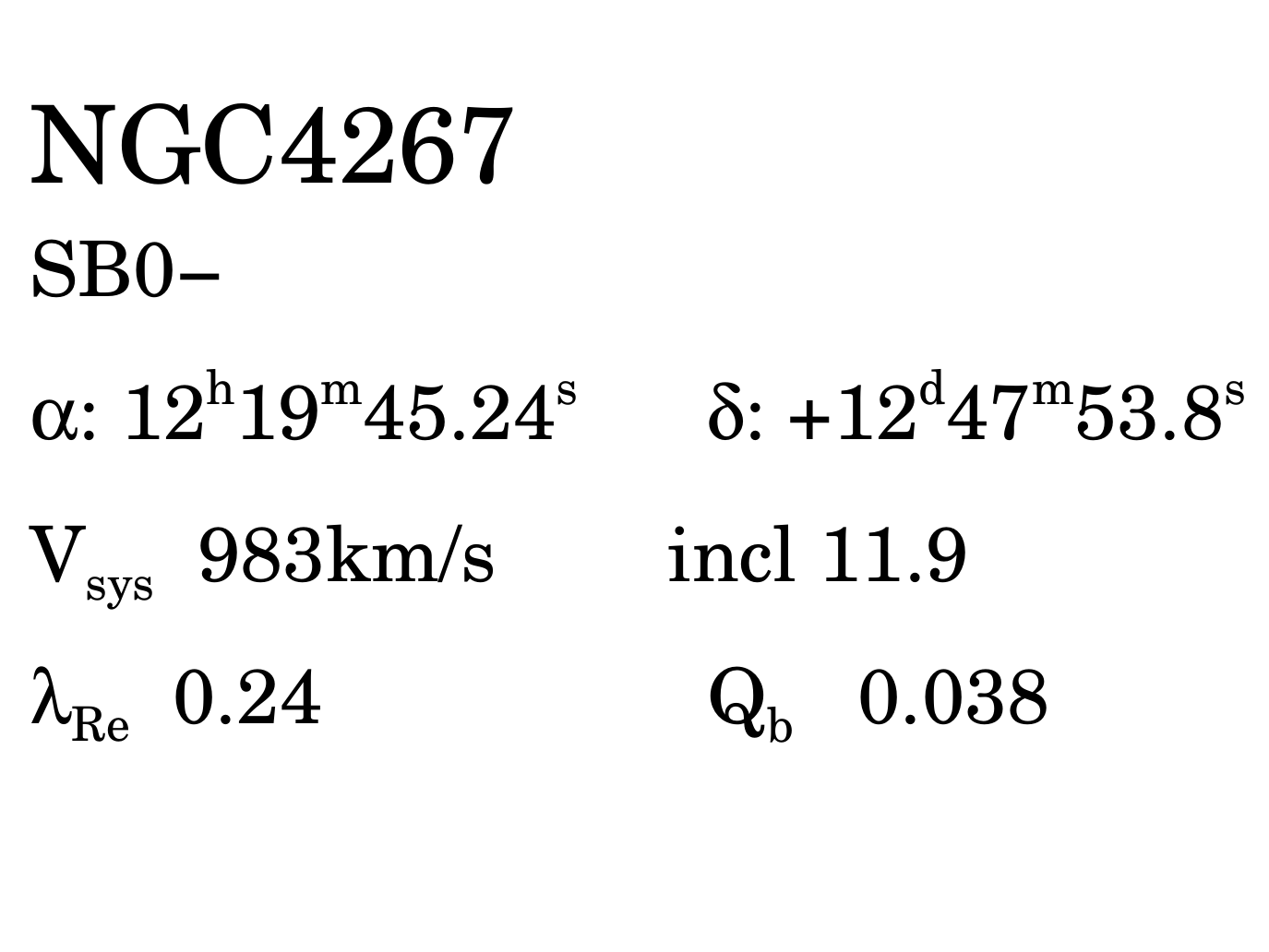}
\includegraphics[width=0.88\linewidth]{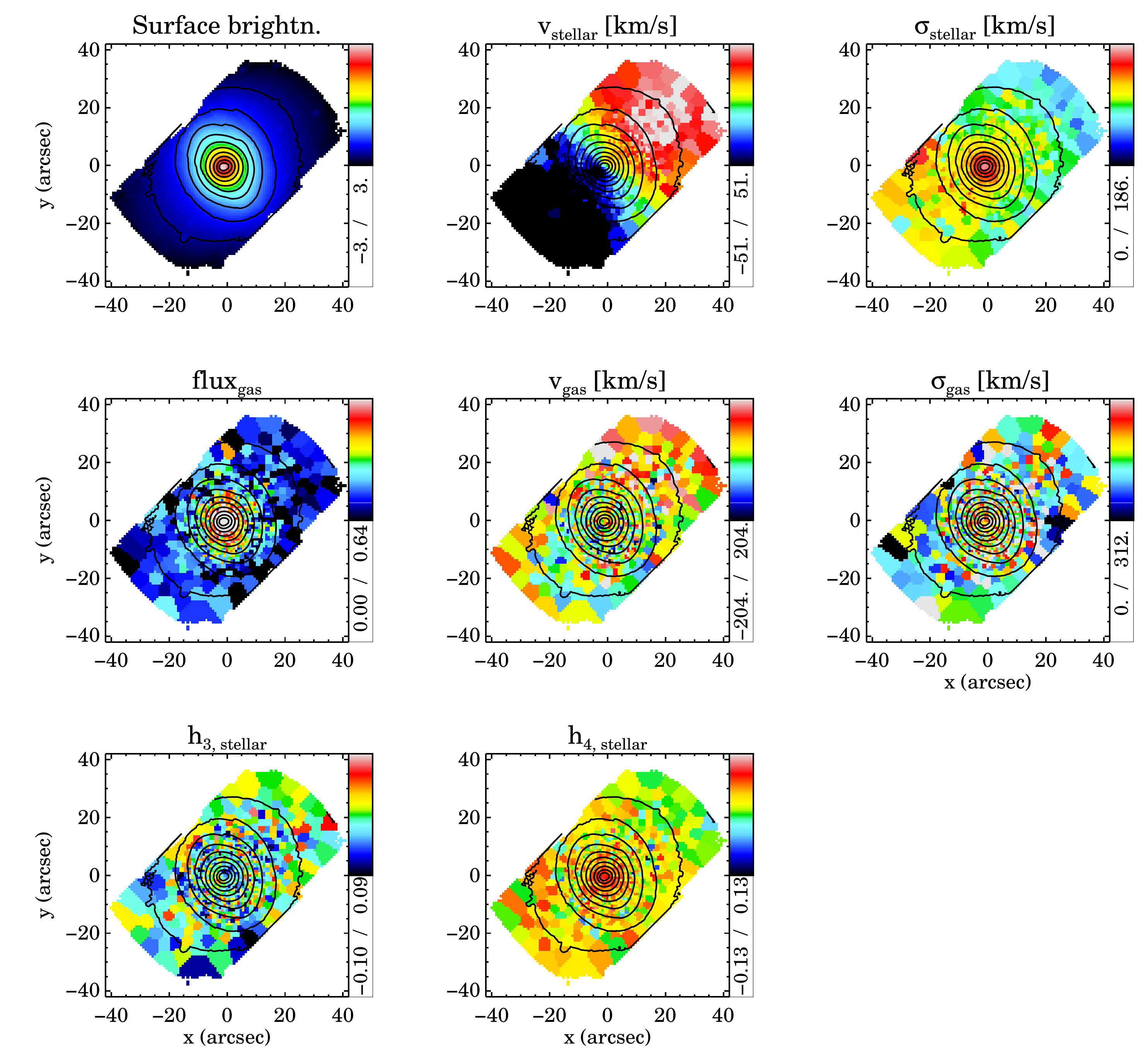}
\includegraphics[width=0.27\linewidth]{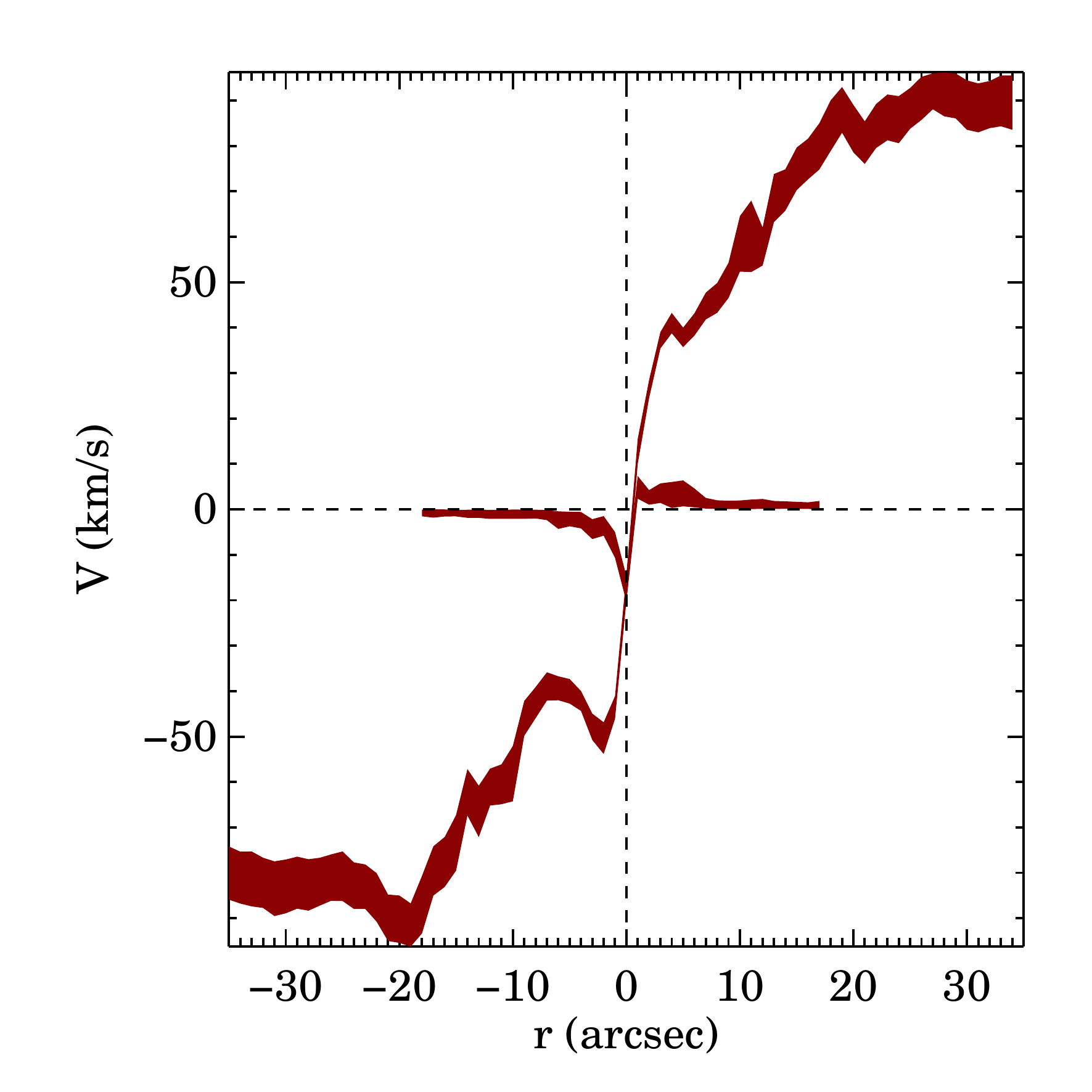}
\includegraphics[width=0.58\linewidth]{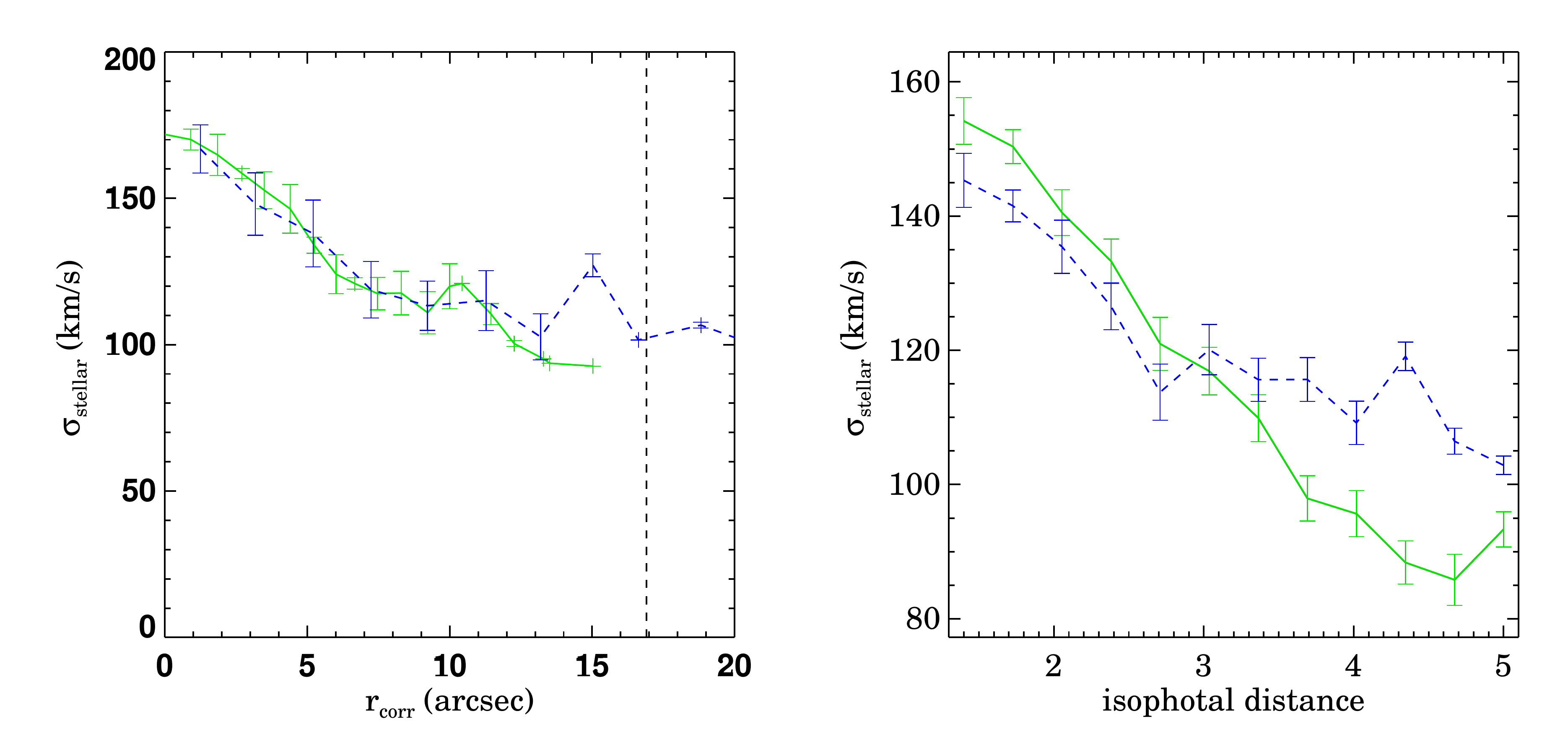}
\caption{Figure \ref{fig:gasvel} continued.}
\label{fig:4267}
\end{figure*}

\begin{figure*}
\includegraphics[width=0.32\linewidth]{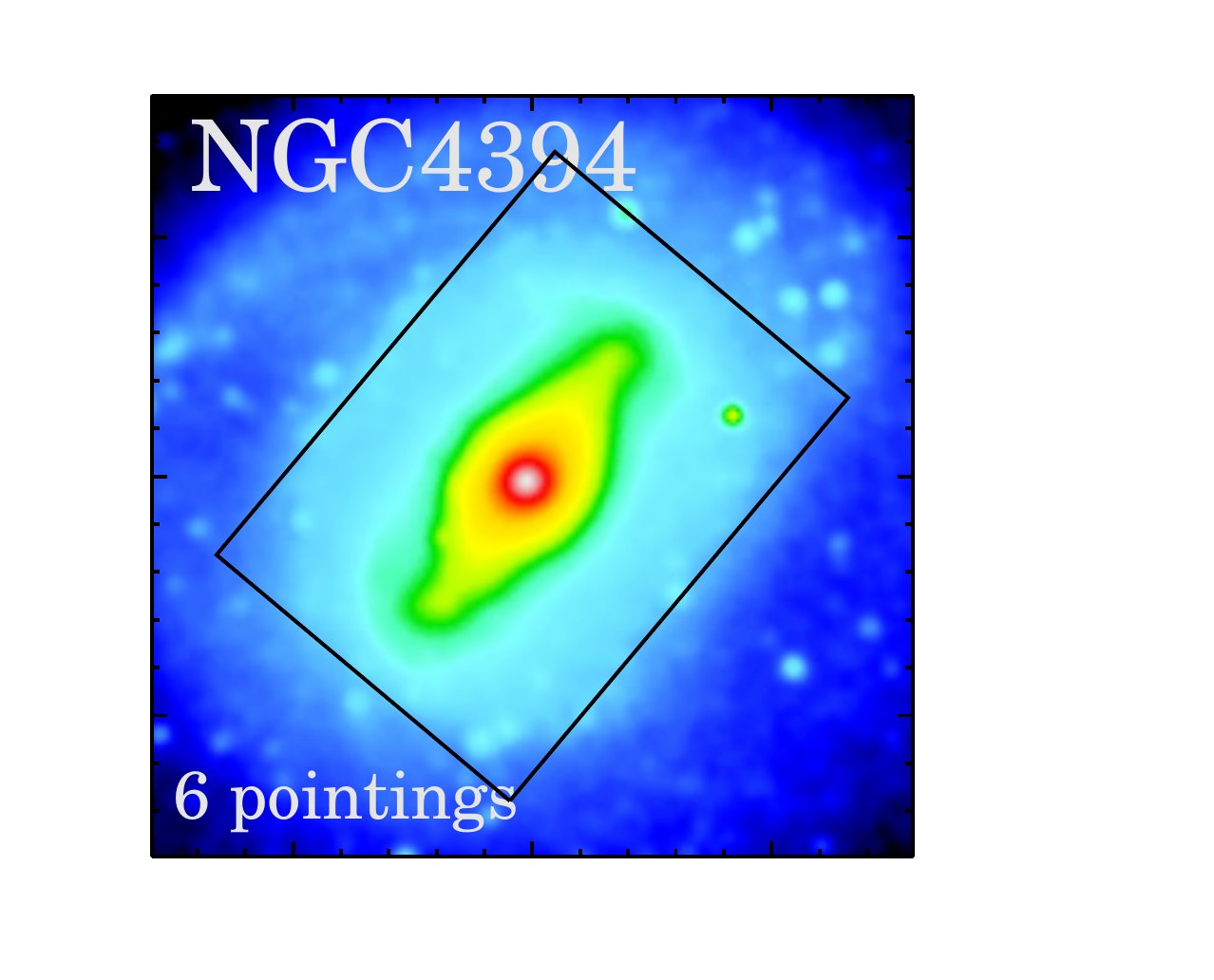}
\includegraphics[width=0.32\linewidth]{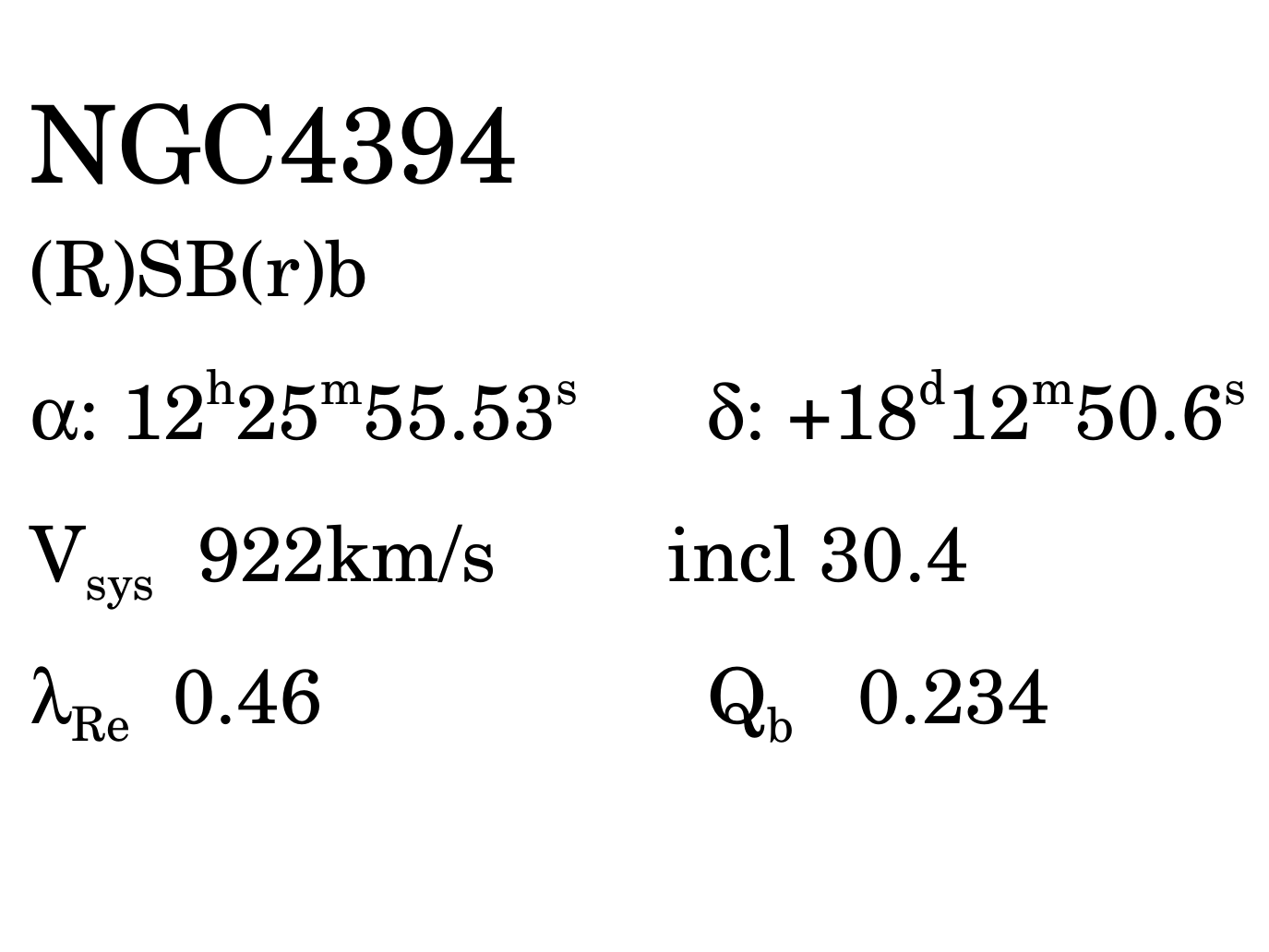}
\includegraphics[width=0.88\linewidth]{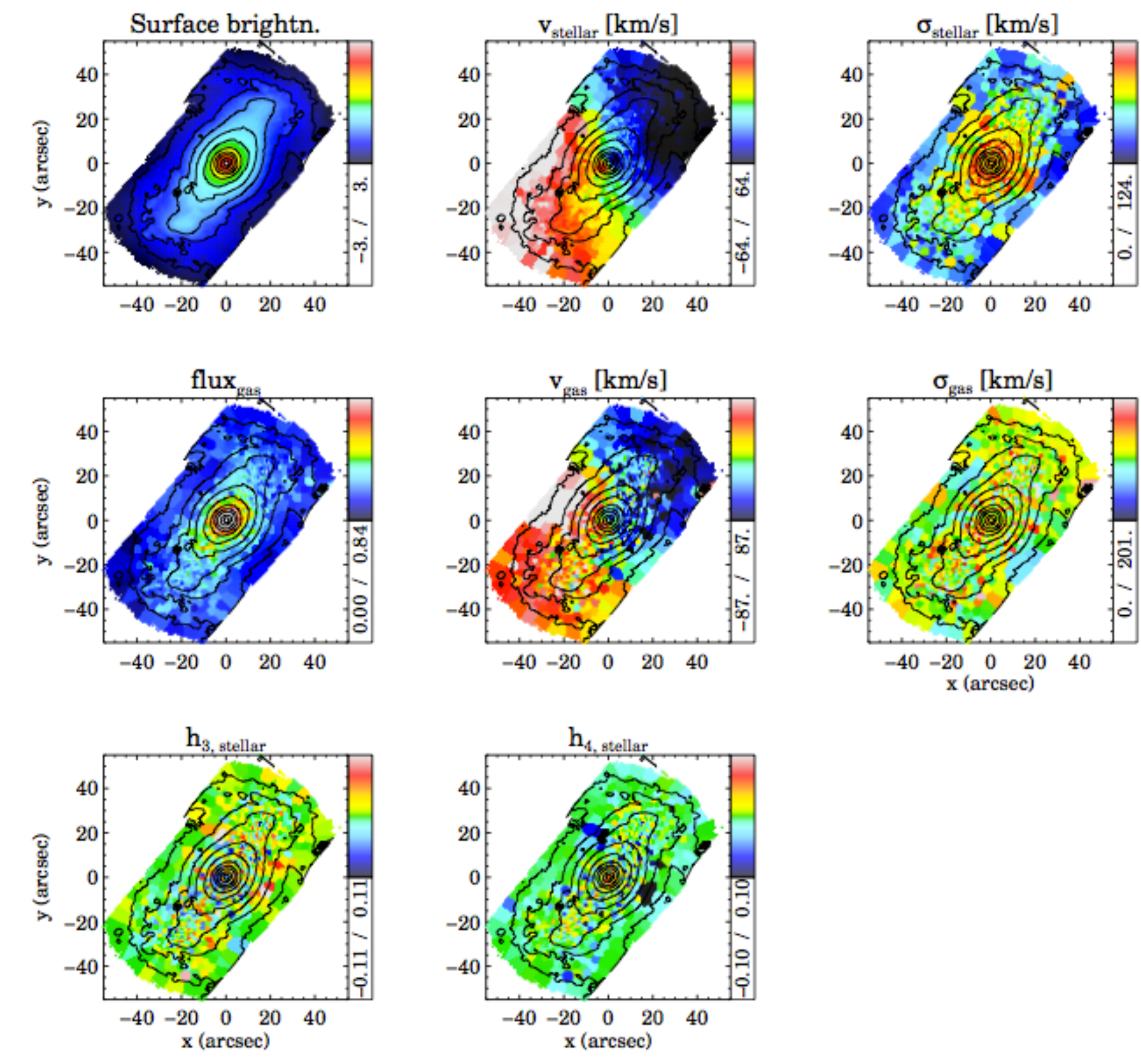}
\includegraphics[width=0.27\linewidth]{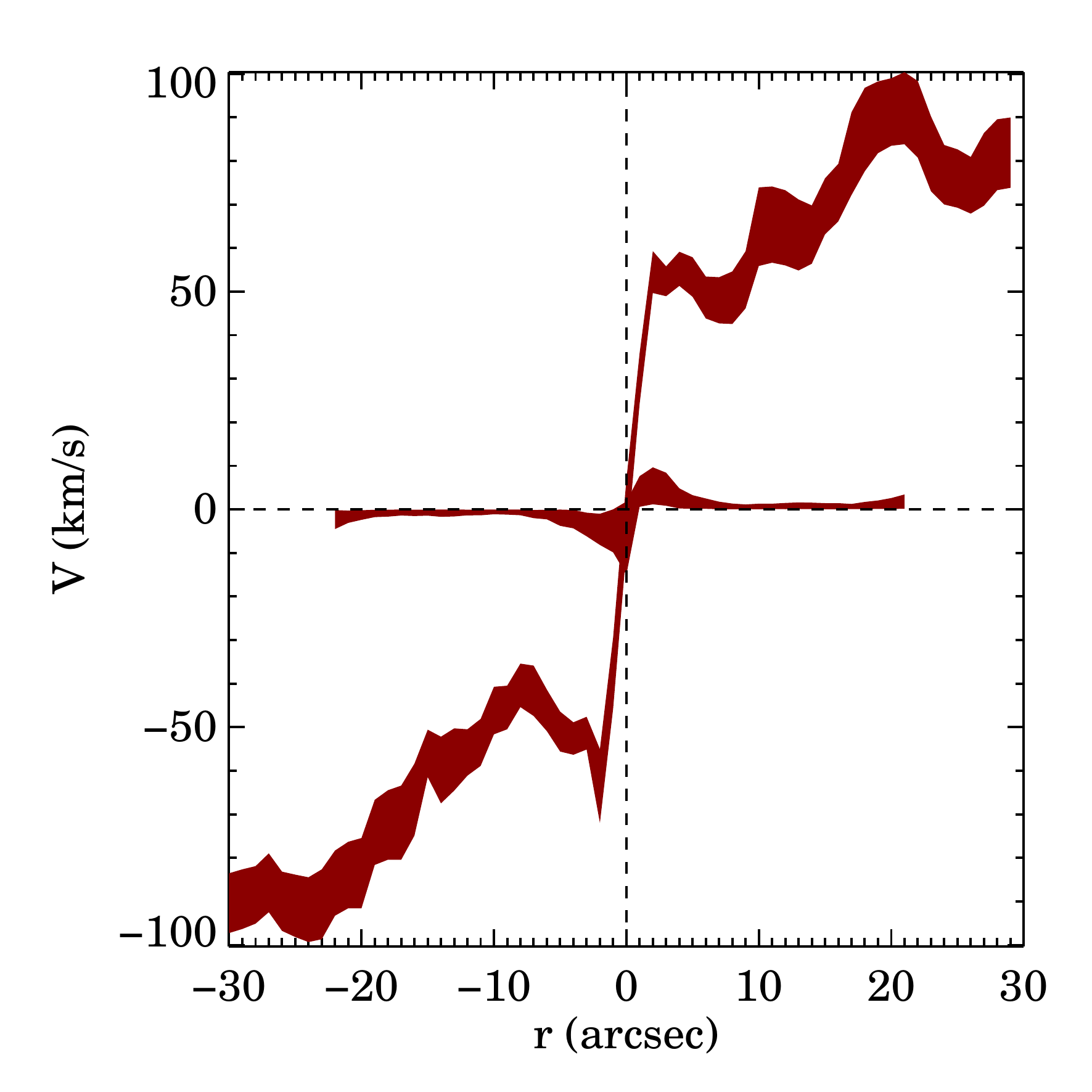}
\includegraphics[width=0.58\linewidth]{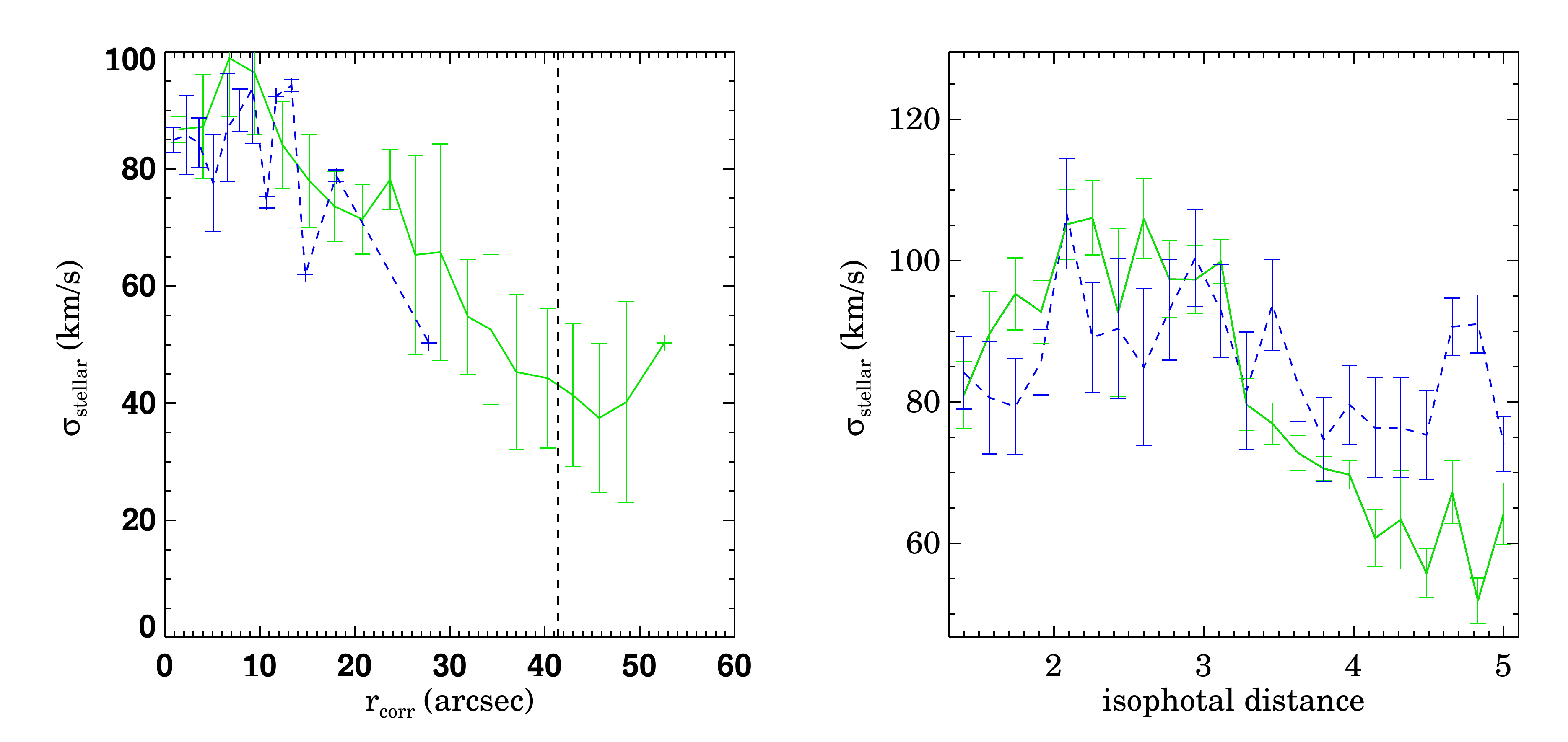}
\caption{Figure \ref{fig:gasvel} continued.}
\label{fig:4394}
\end{figure*}

\begin{figure*}
\includegraphics[width=0.32\linewidth]{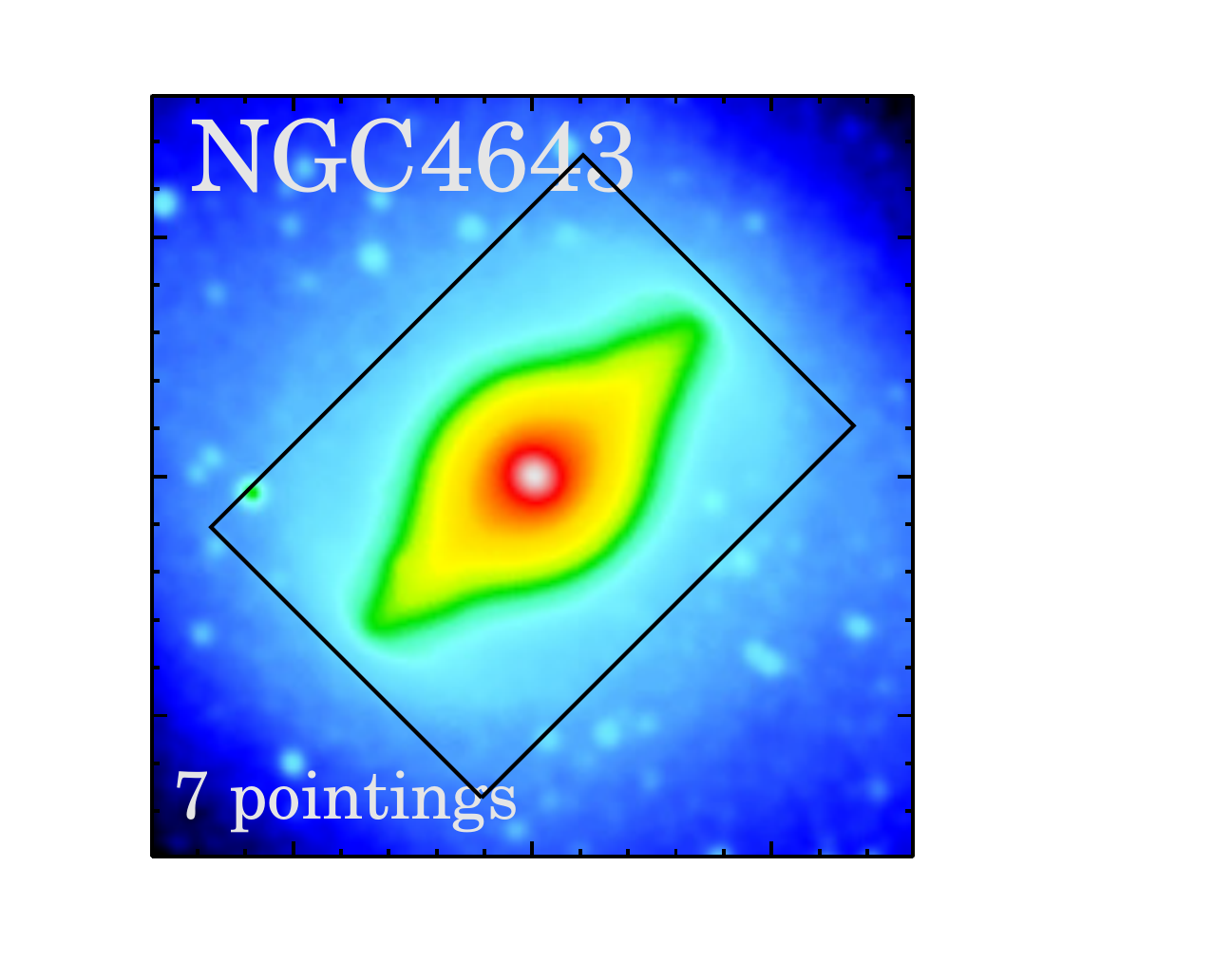}
\includegraphics[width=0.32\linewidth]{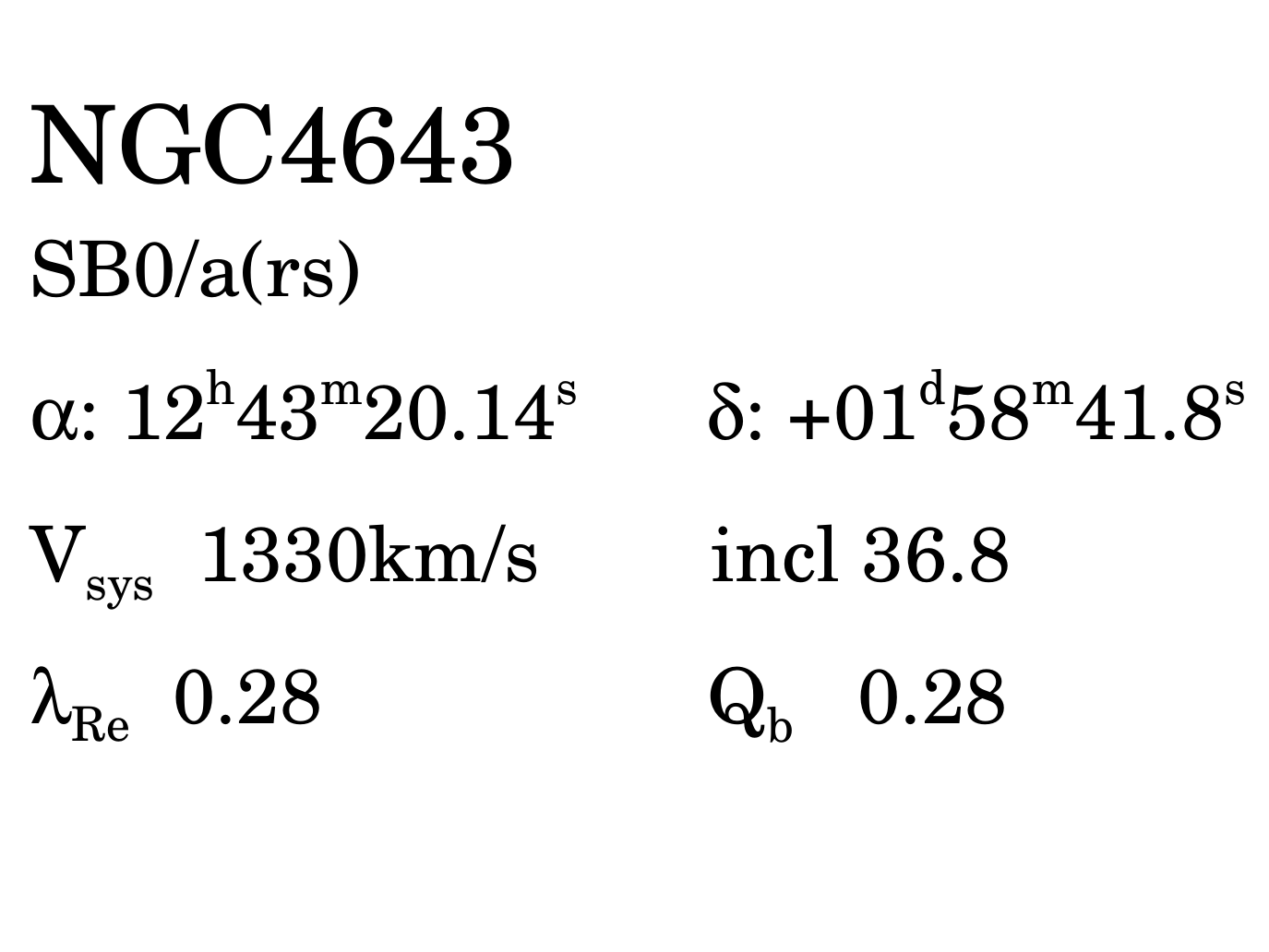}
\includegraphics[width=0.88\linewidth]{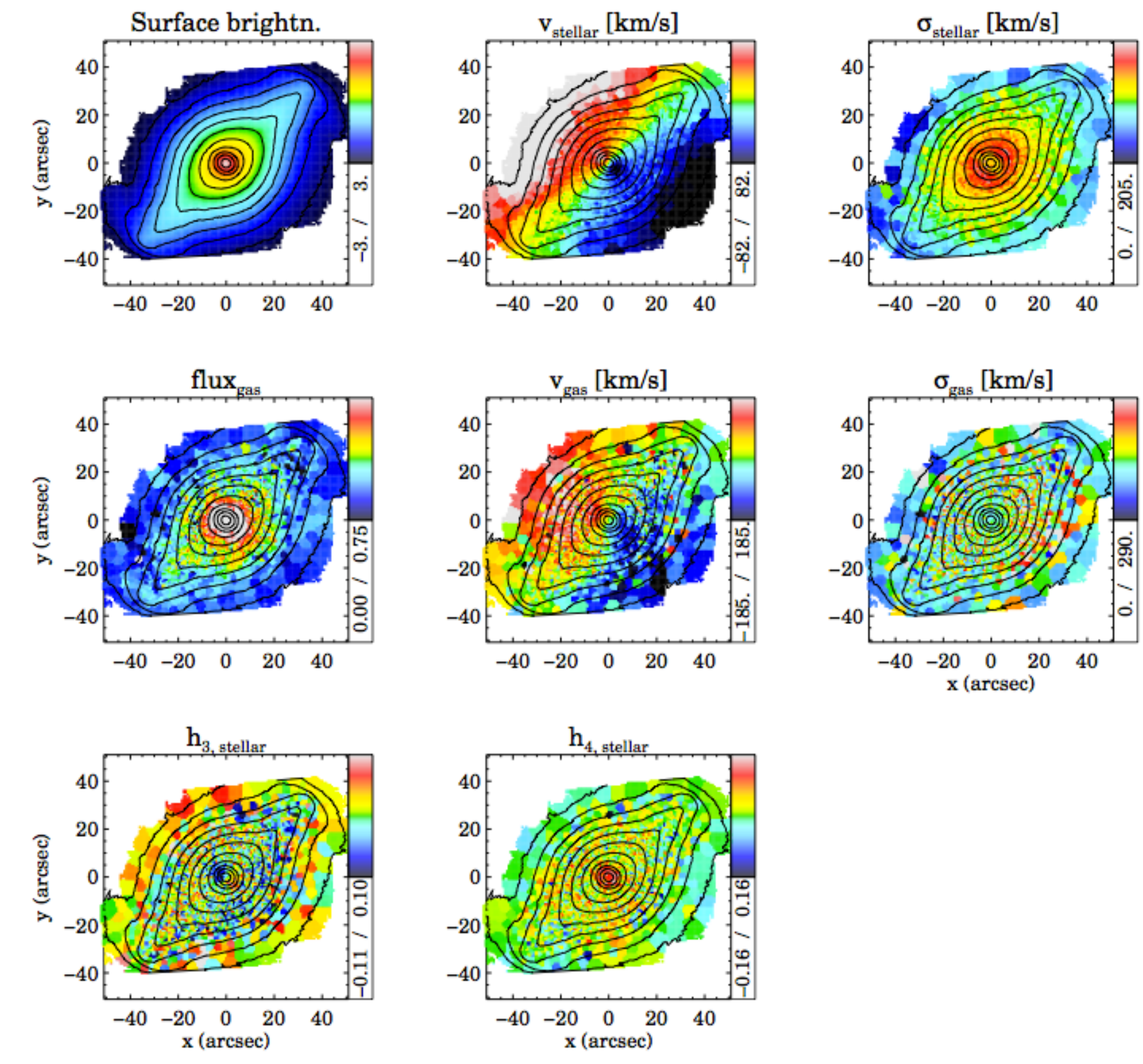}
\includegraphics[width=0.27\linewidth]{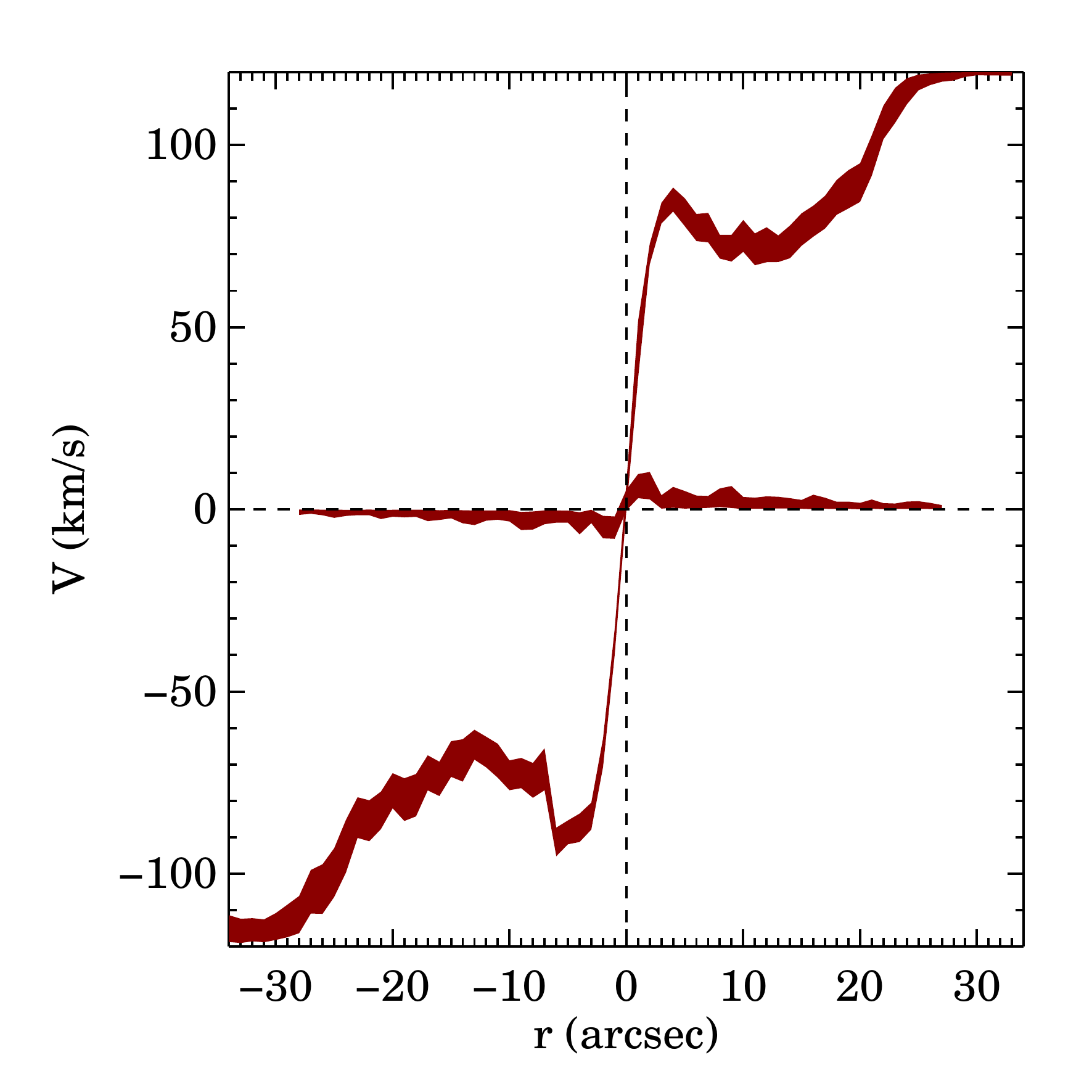}
\includegraphics[width=0.58\linewidth]{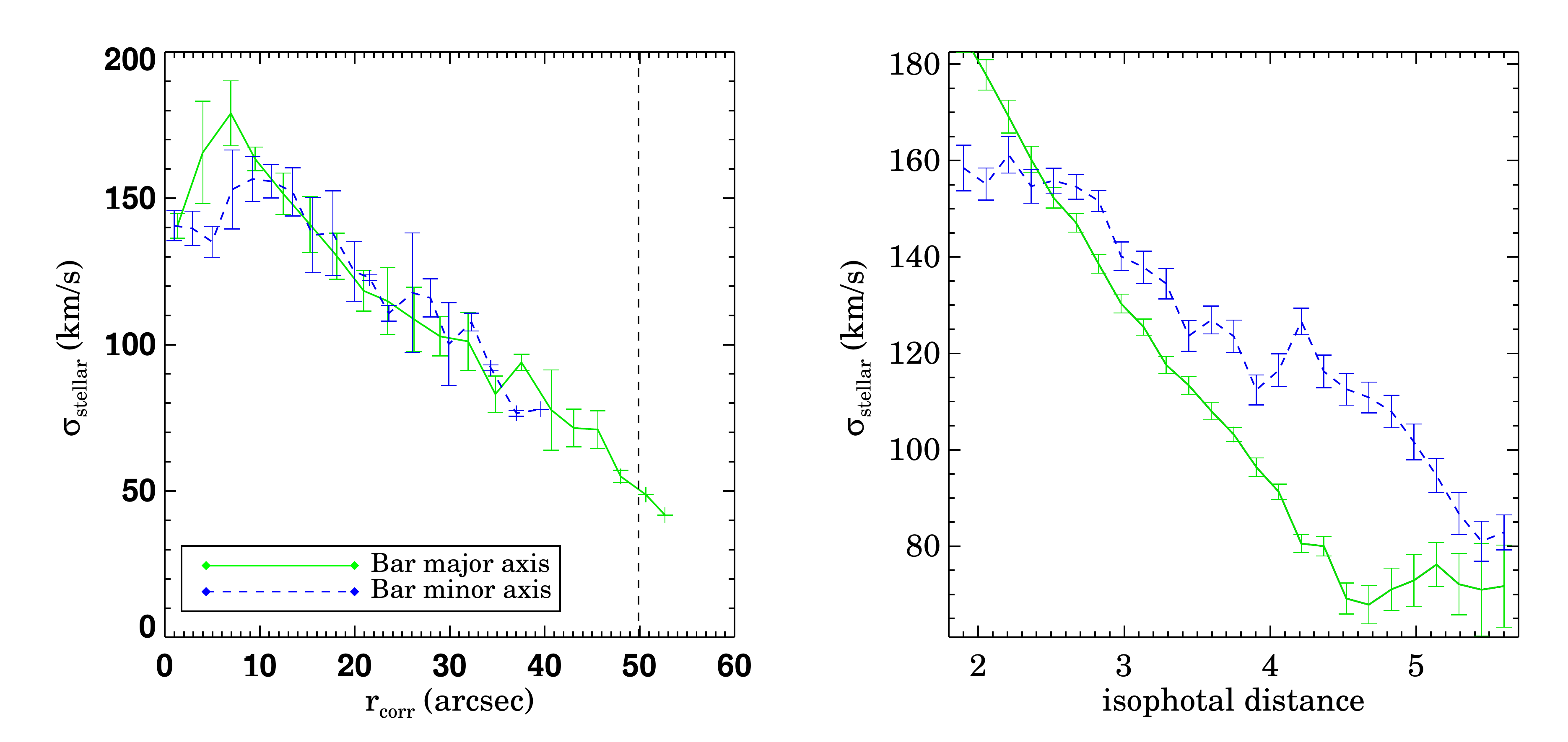}
\caption{Figure \ref{fig:gasvel} continued.}
\label{fig:4643}
\end{figure*}

\begin{figure*}
\includegraphics[width=0.32\linewidth]{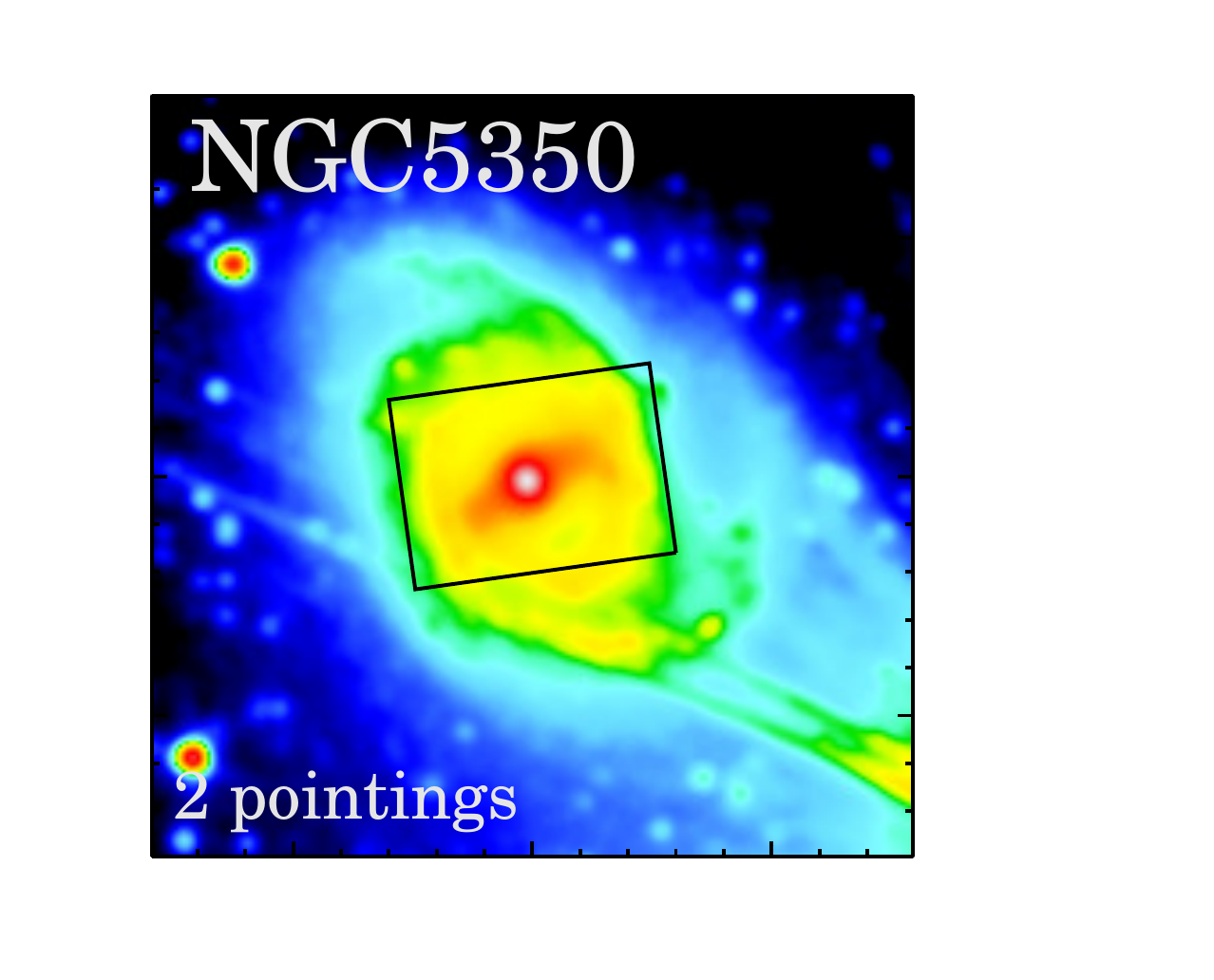}
\includegraphics[width=0.32\linewidth]{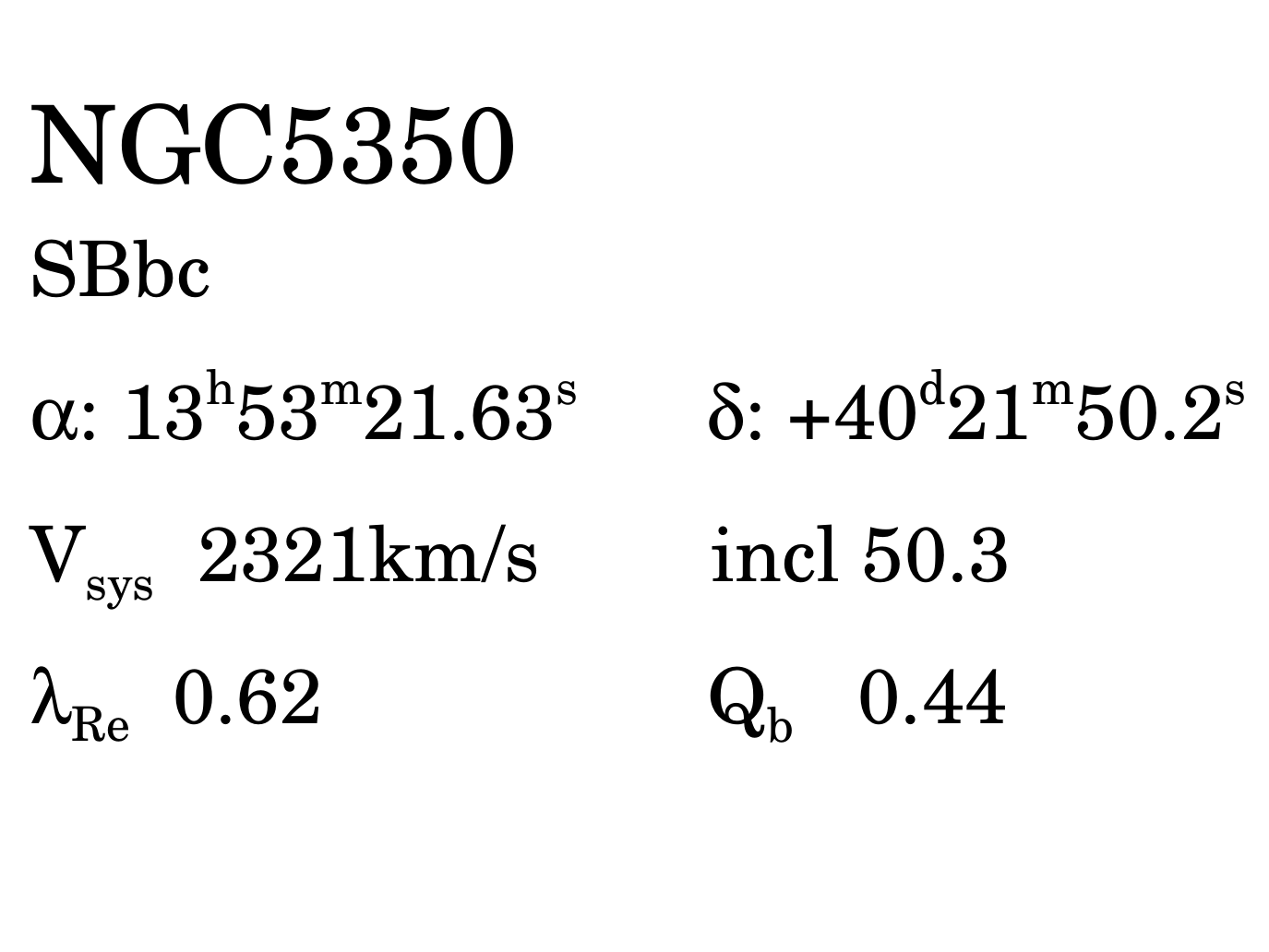}
\includegraphics[width=0.88\linewidth]{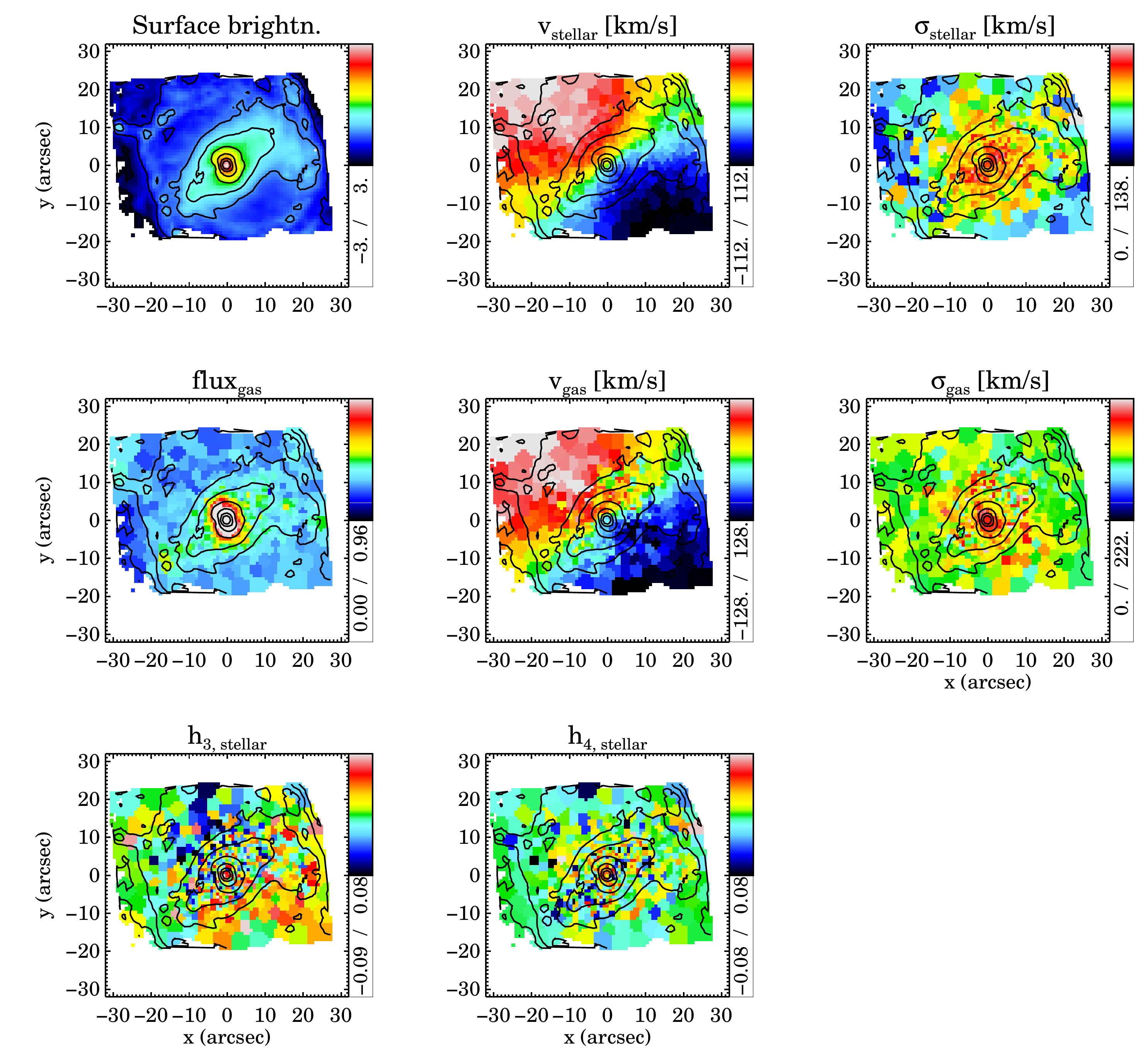}
\includegraphics[width=0.27\linewidth]{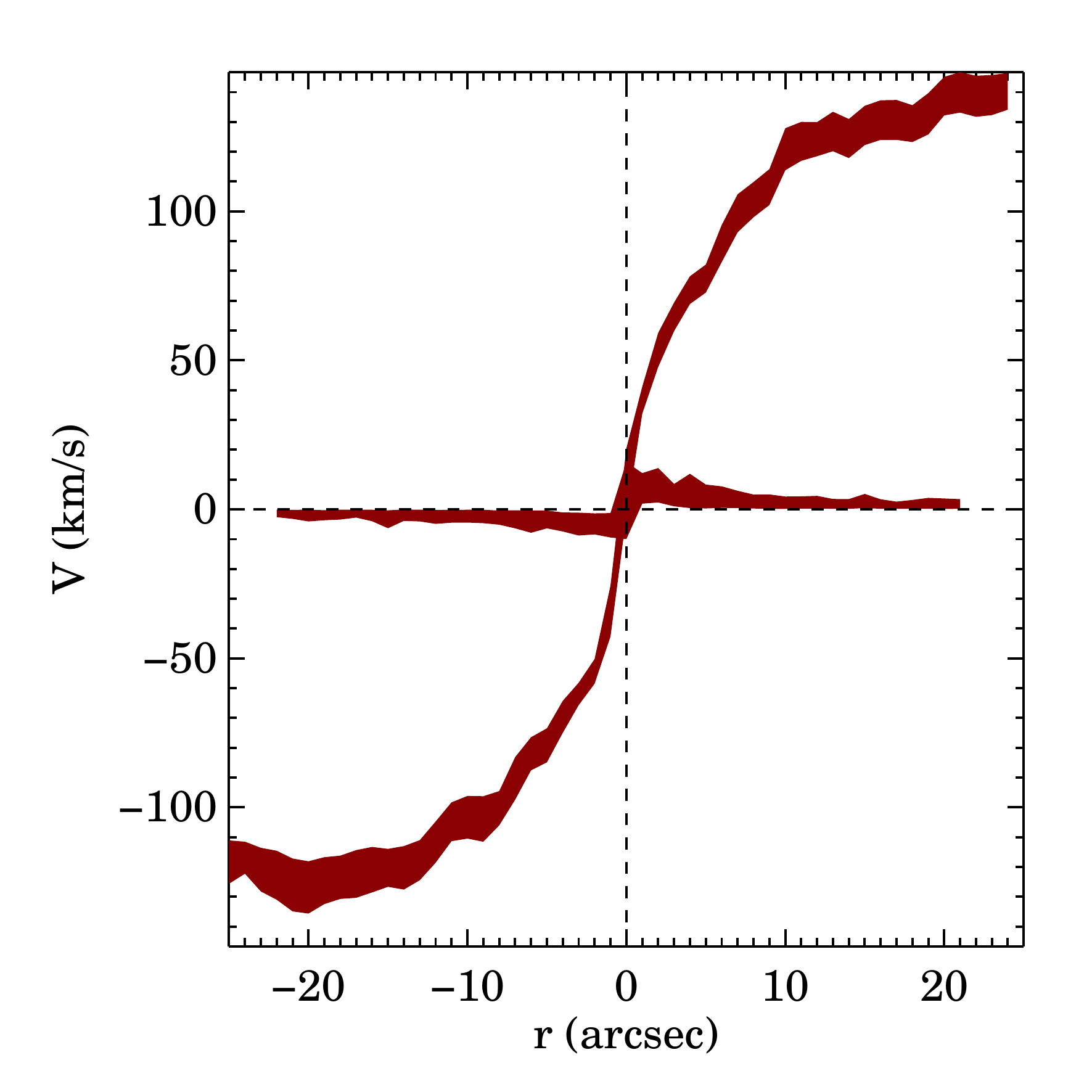}
\includegraphics[width=0.58\linewidth]{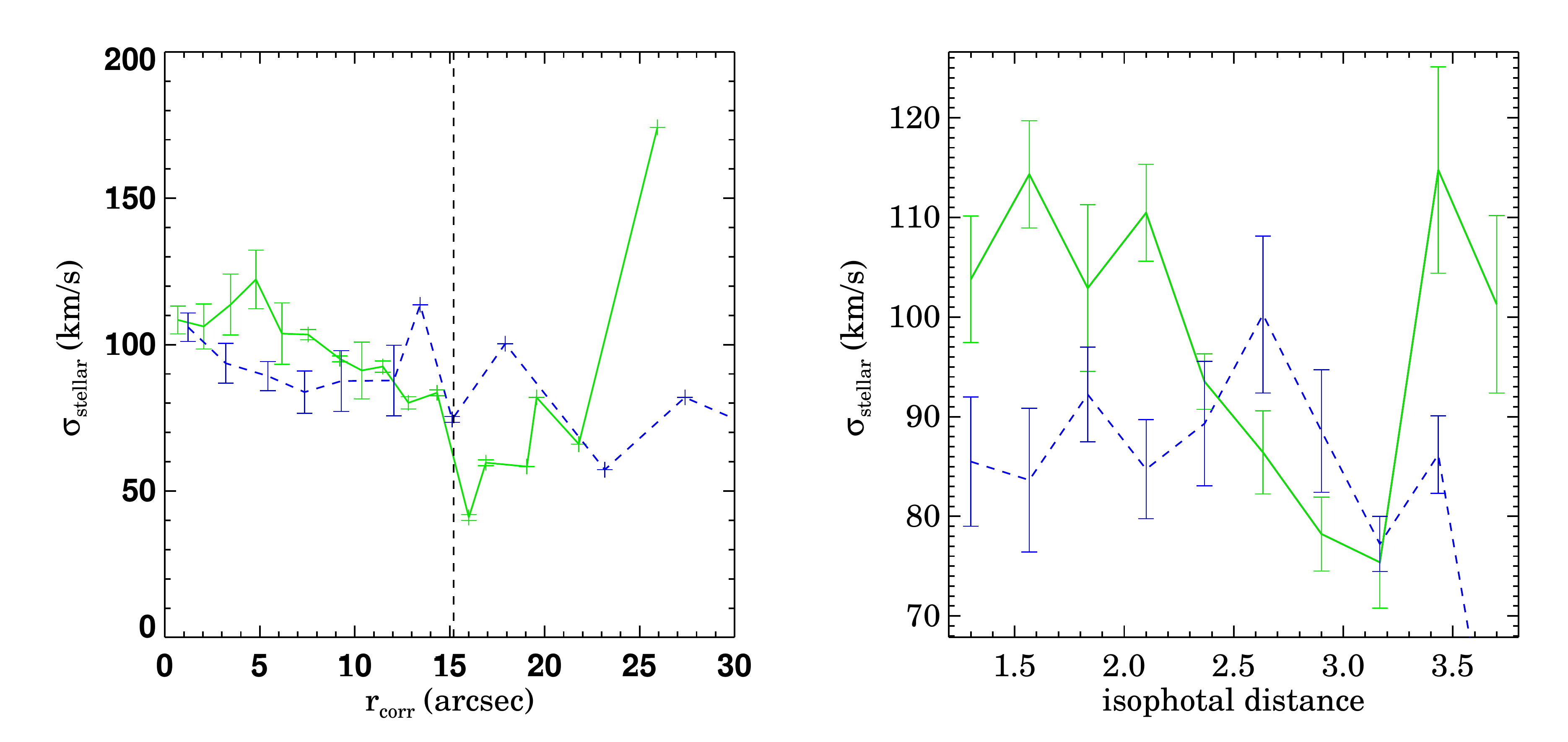}
\caption{Figure \ref{fig:gasvel} continued.}
\label{fig:5350}
\end{figure*}

\begin{figure*}
\includegraphics[width=0.32\linewidth]{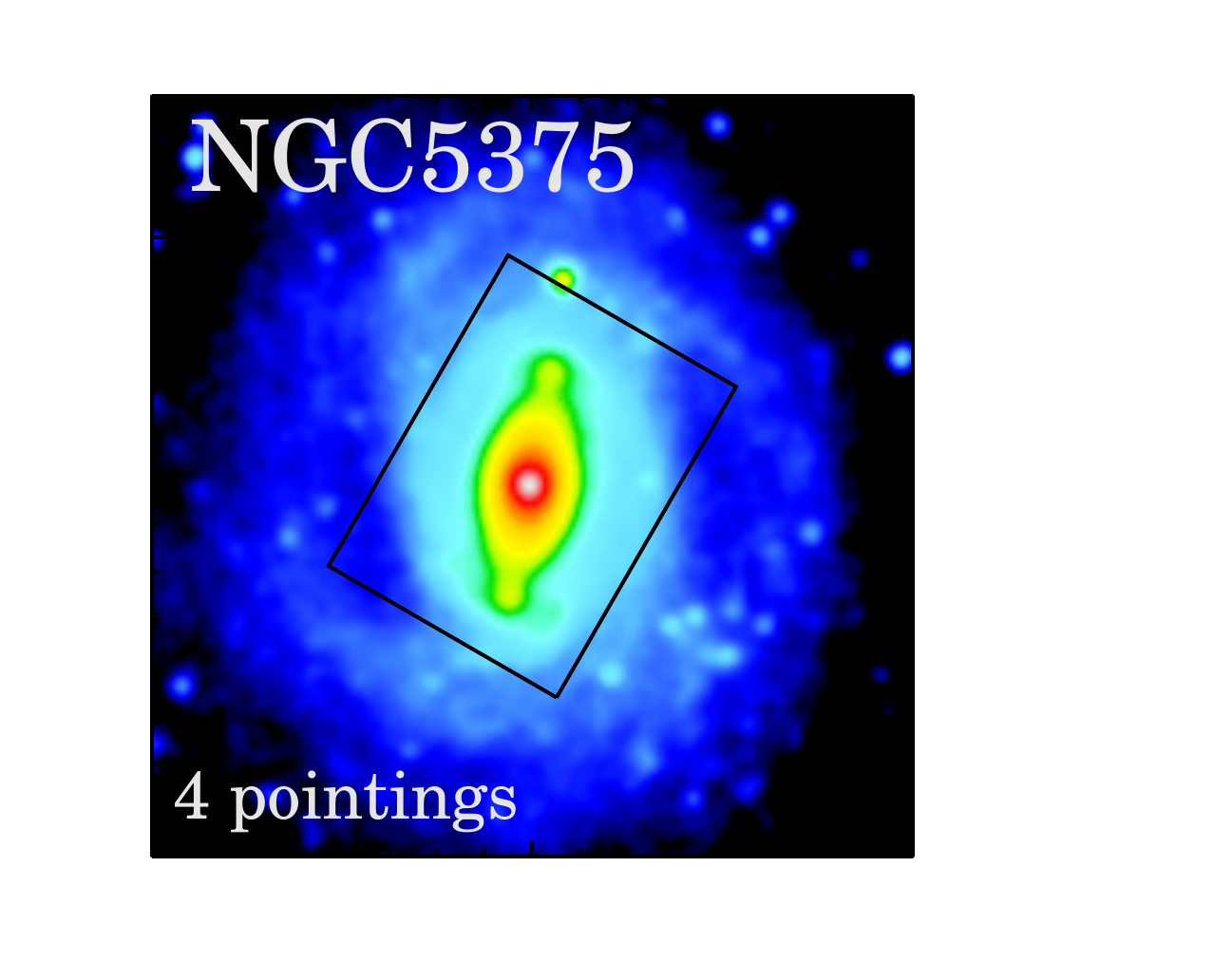}
\includegraphics[width=0.32\linewidth]{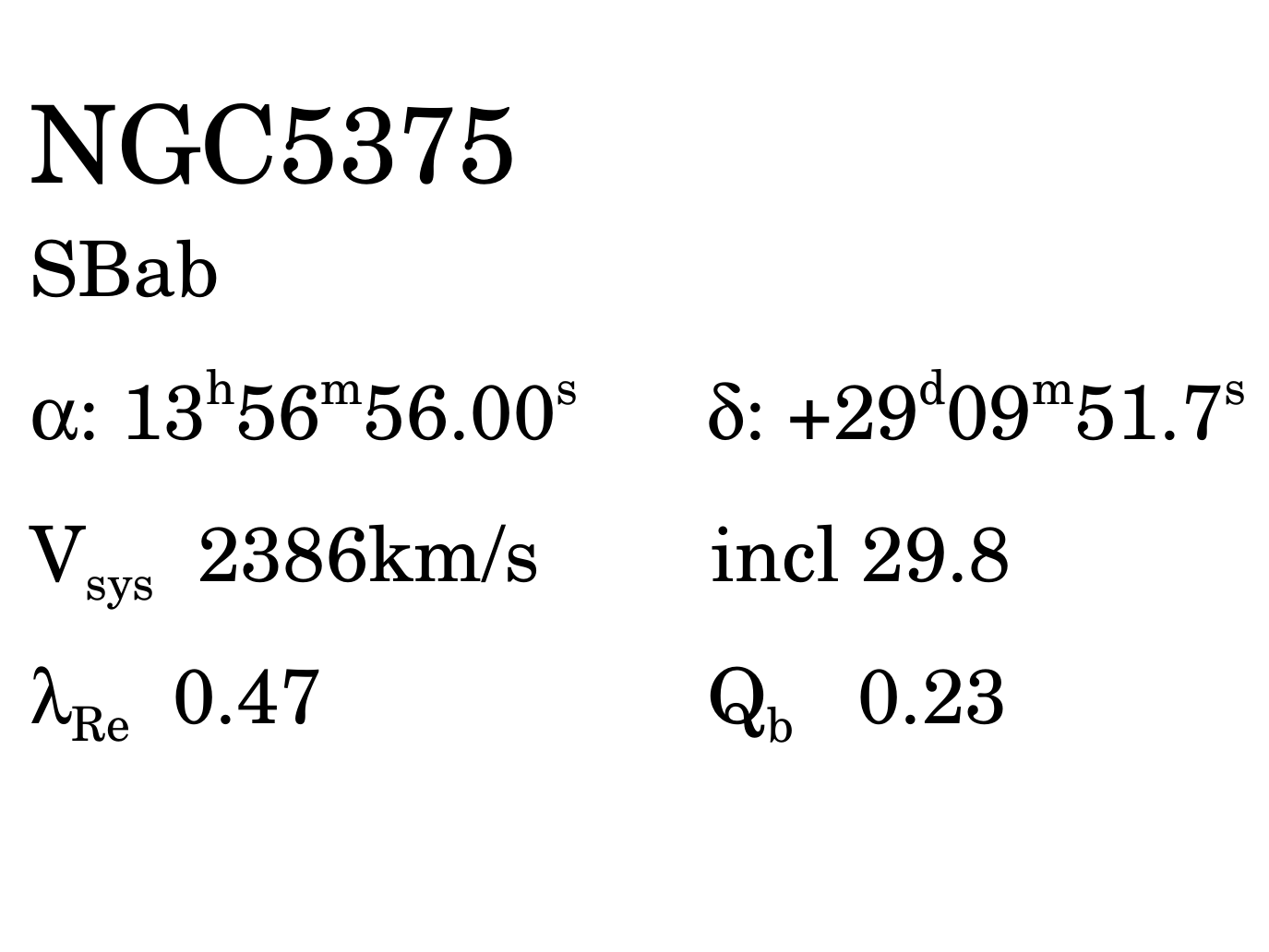}
\includegraphics[width=0.88\linewidth]{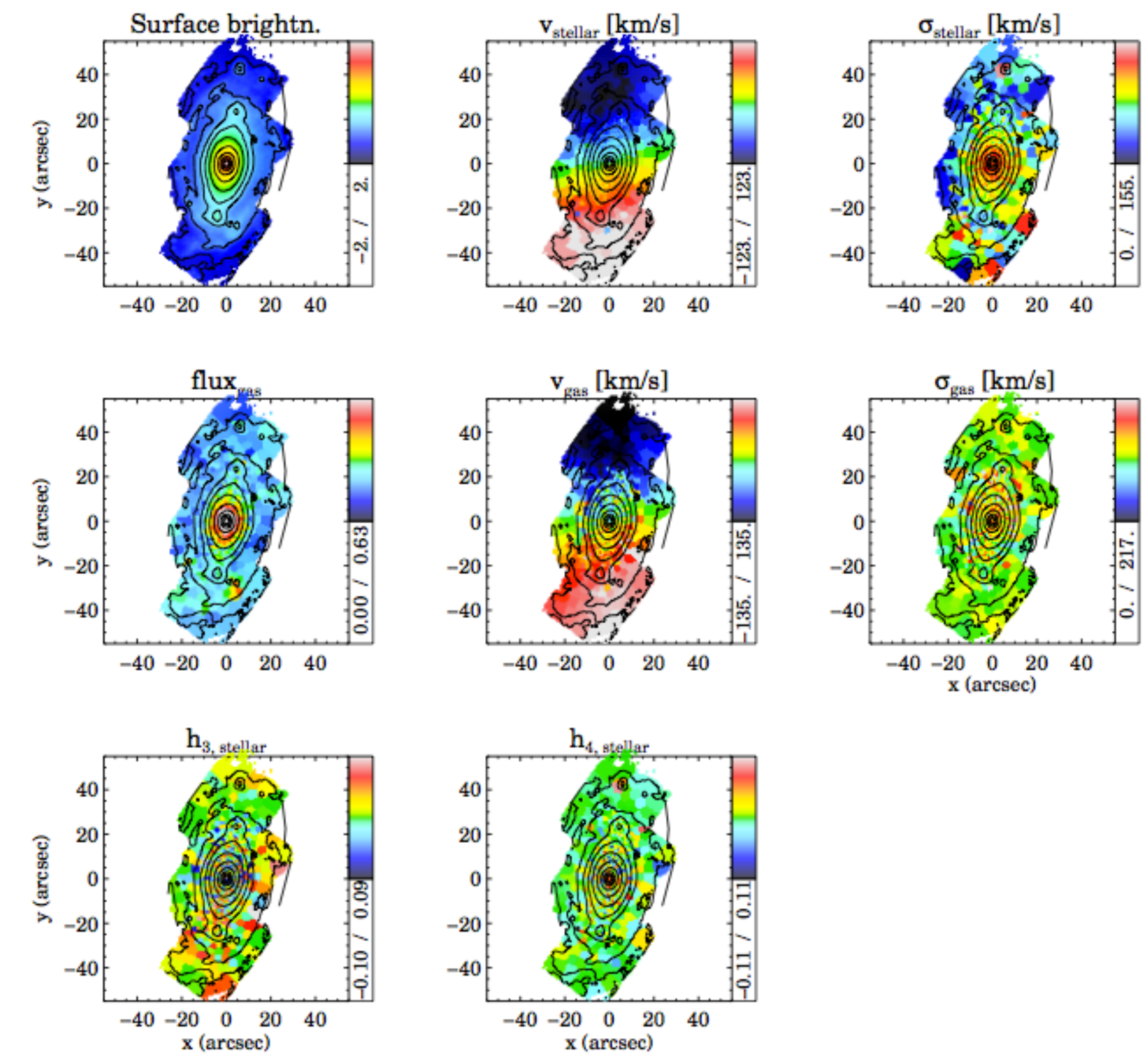}
\includegraphics[width=0.27\linewidth]{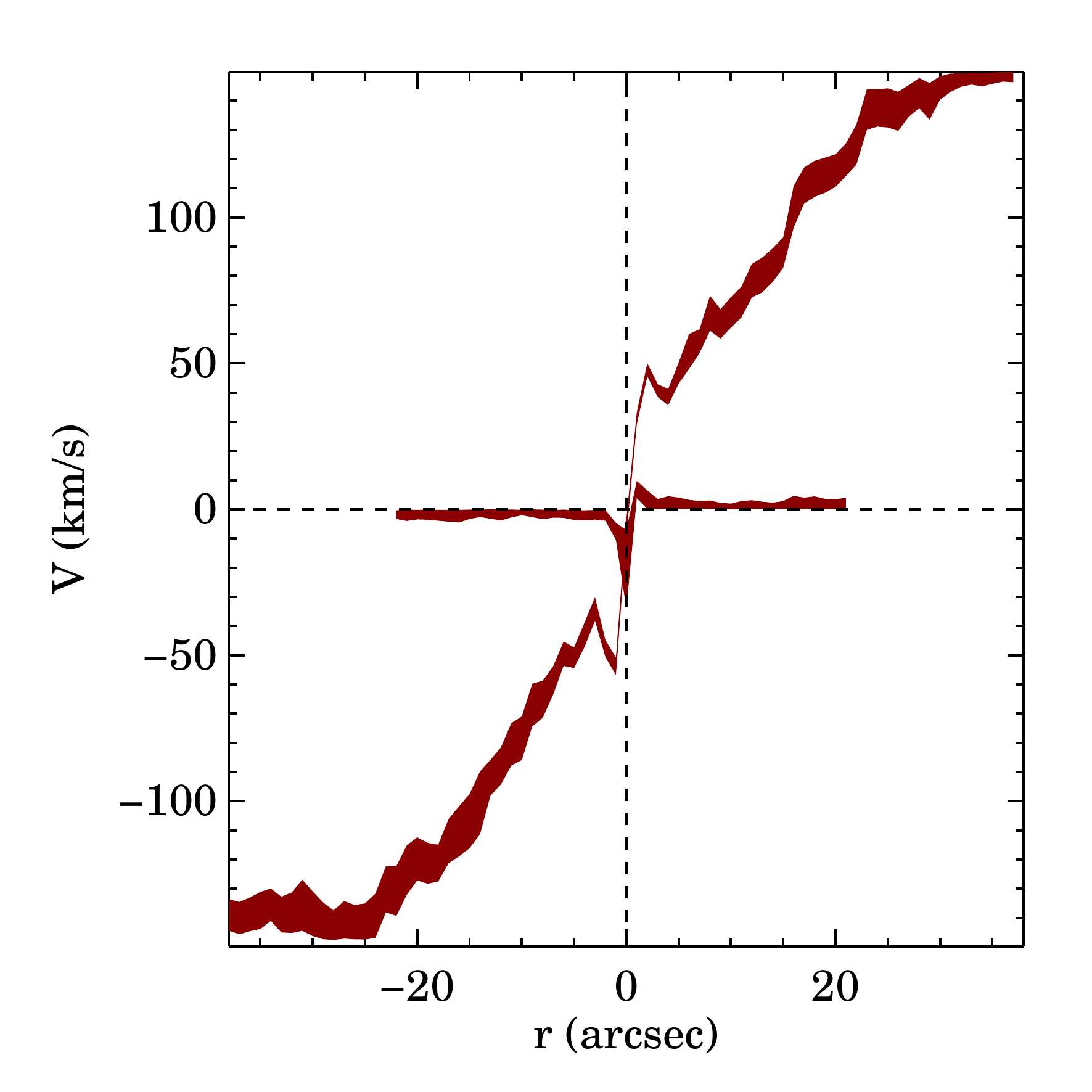}
\includegraphics[width=0.58\linewidth]{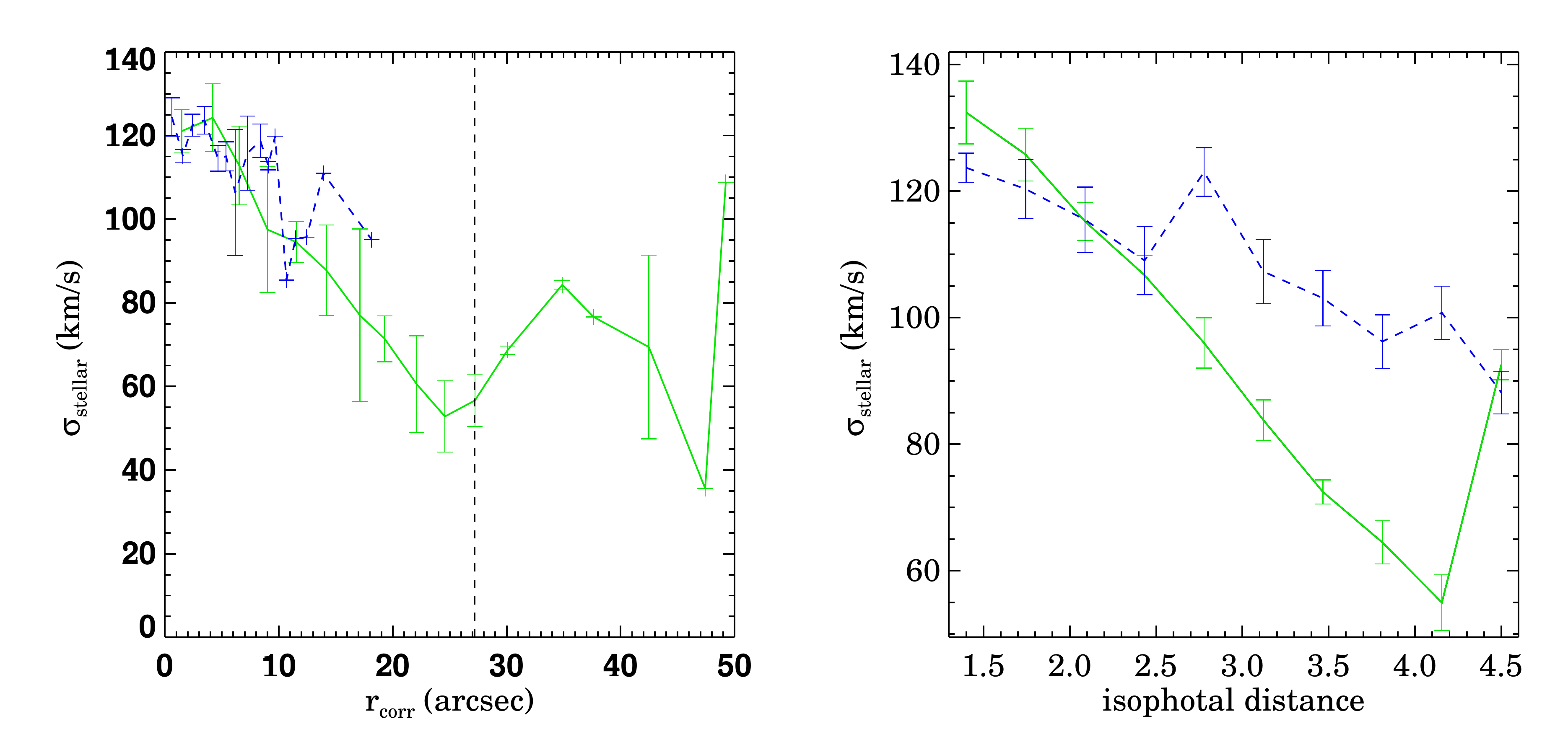}
\caption{Figure \ref{fig:gasvel} continued.}
\label{fig:5375}
\end{figure*}

\begin{figure*}
\includegraphics[width=0.32\linewidth]{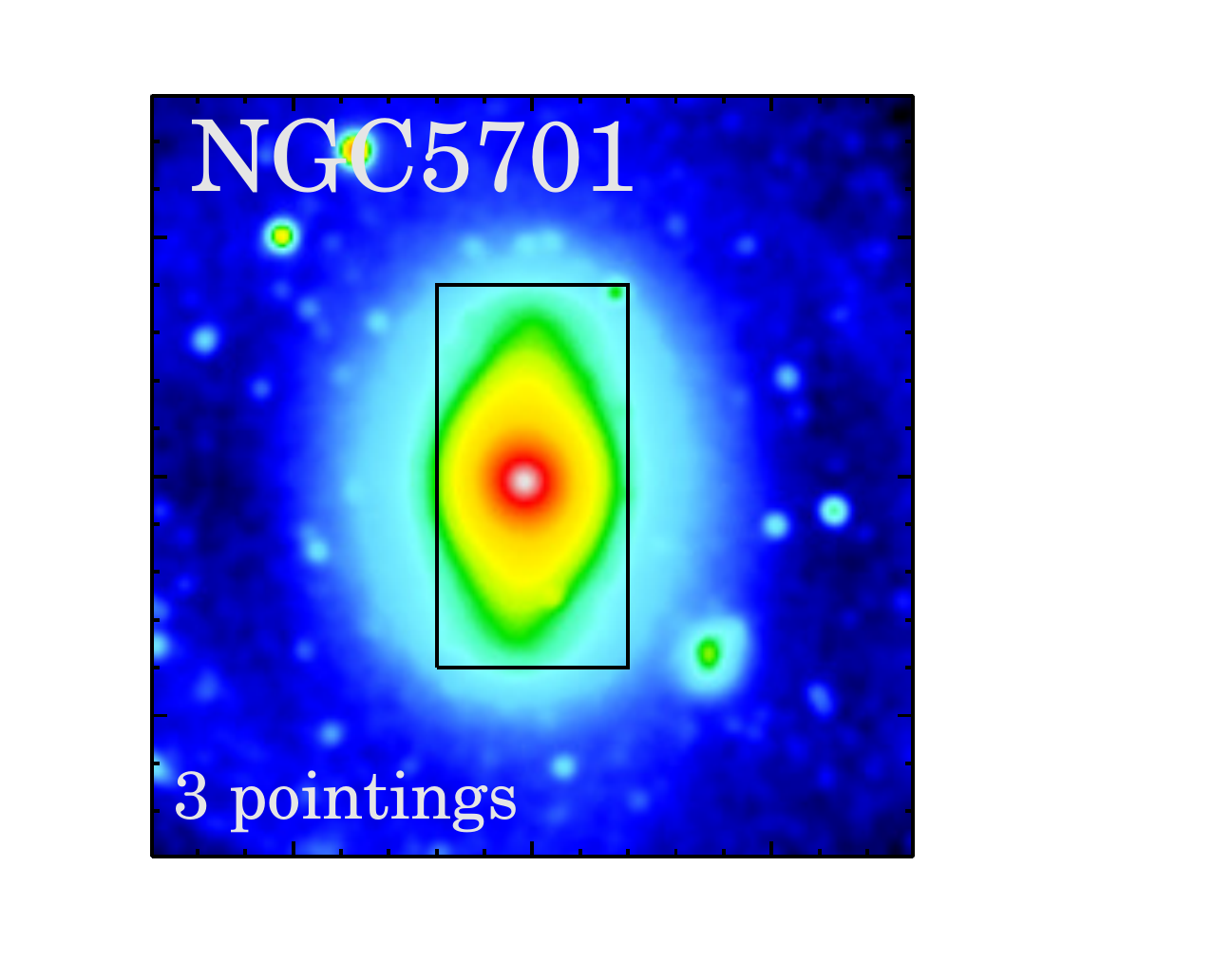}
\includegraphics[width=0.32\linewidth]{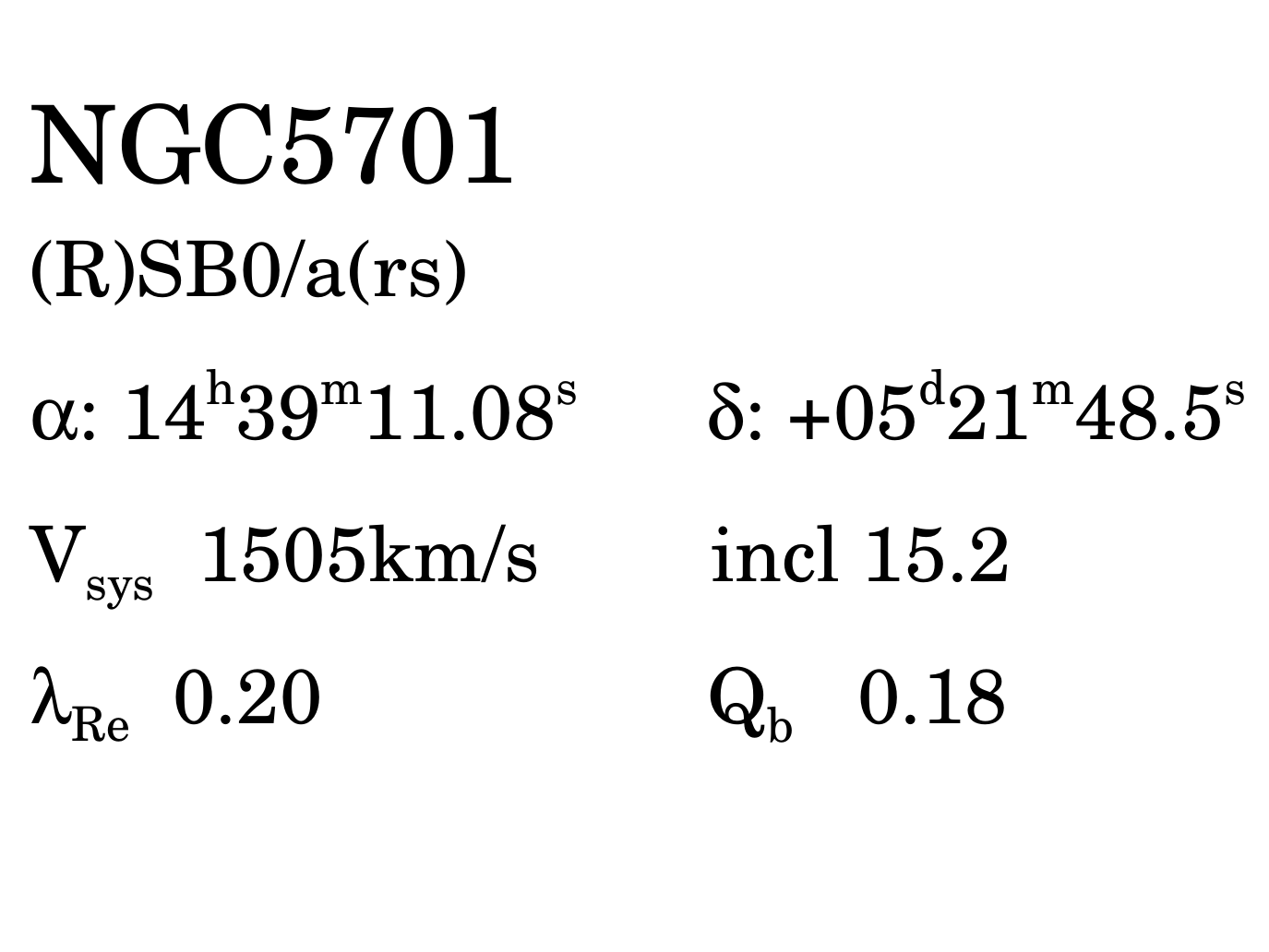}
\includegraphics[width=0.88\linewidth]{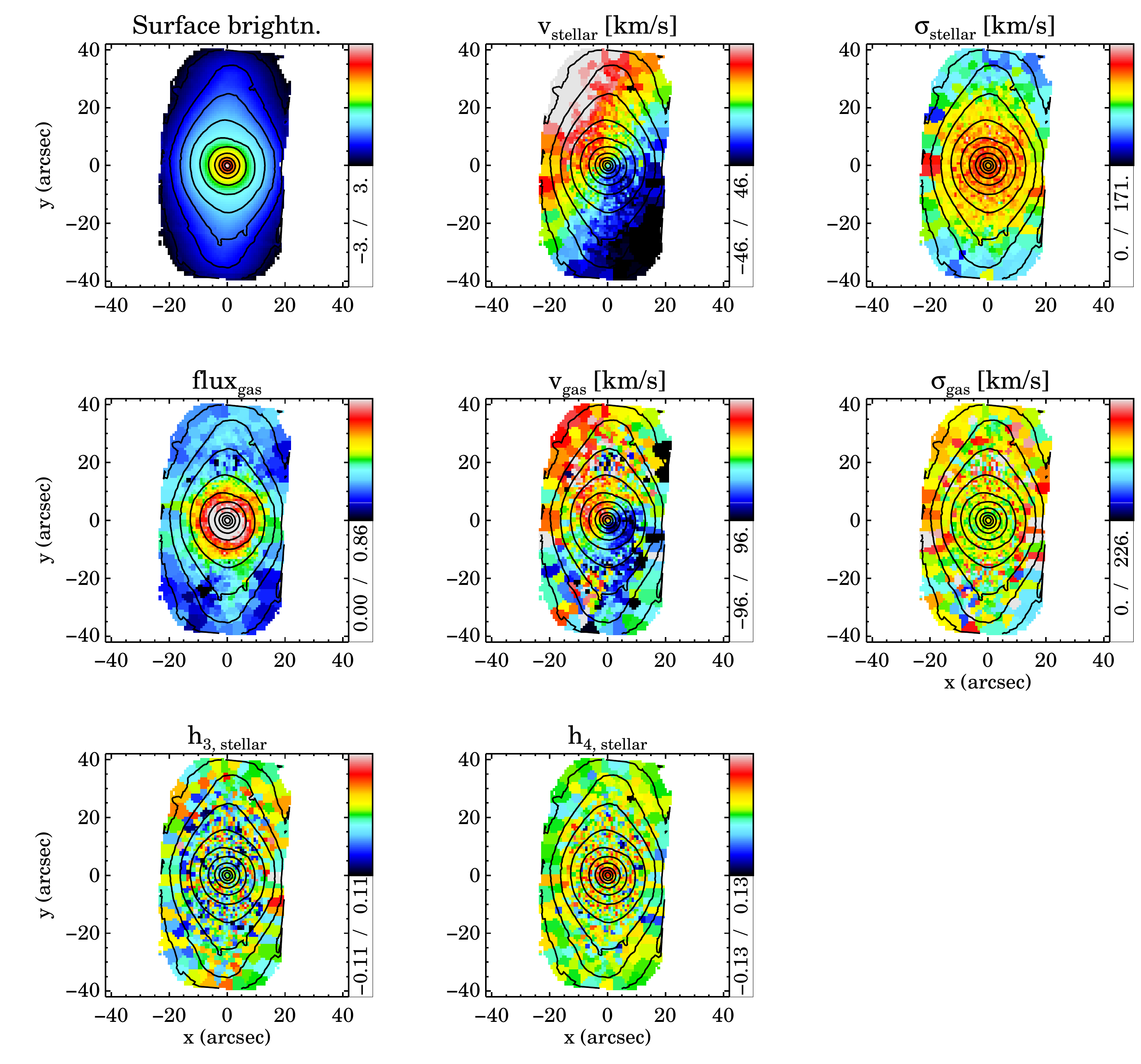}
\includegraphics[width=0.27\linewidth]{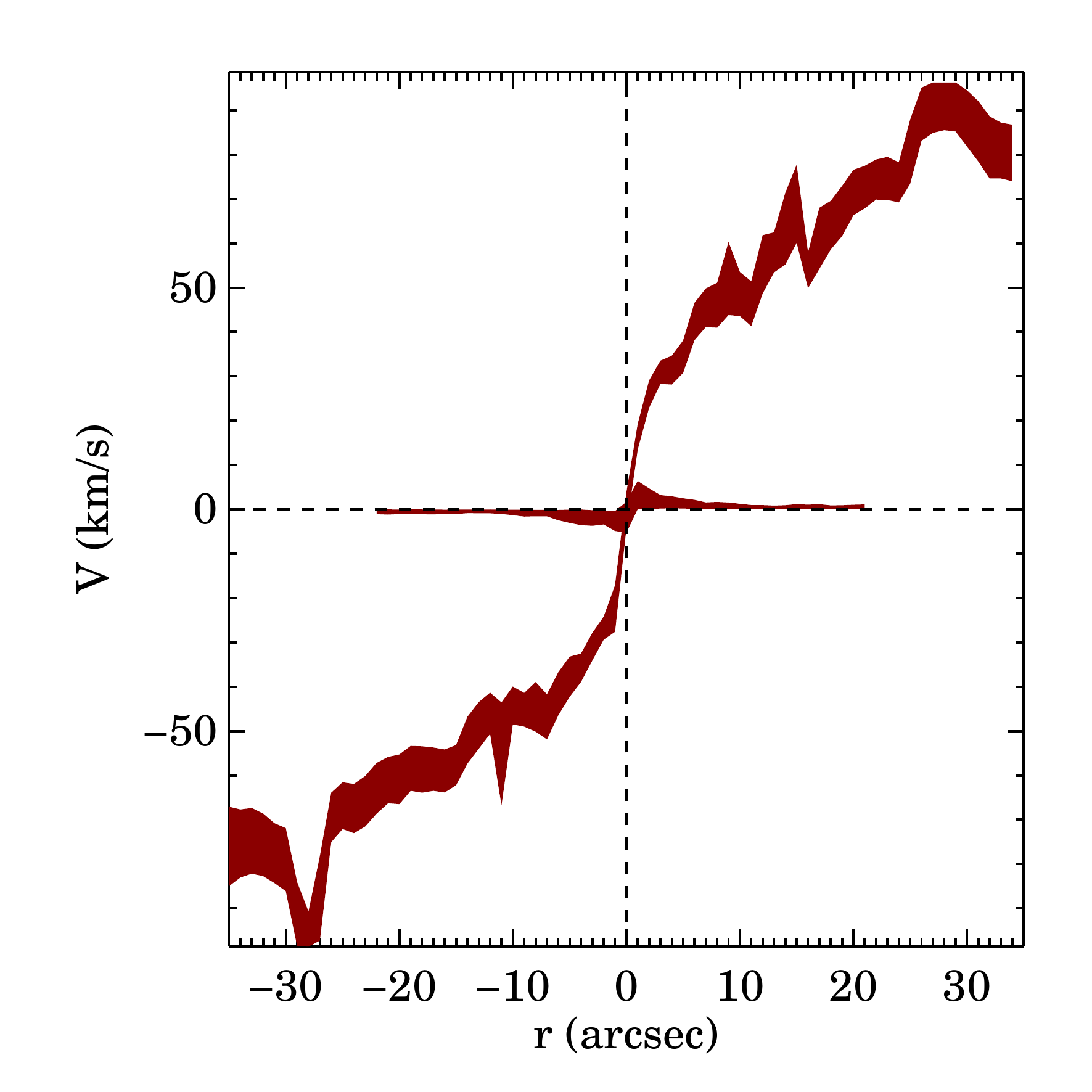}
\includegraphics[width=0.58\linewidth]{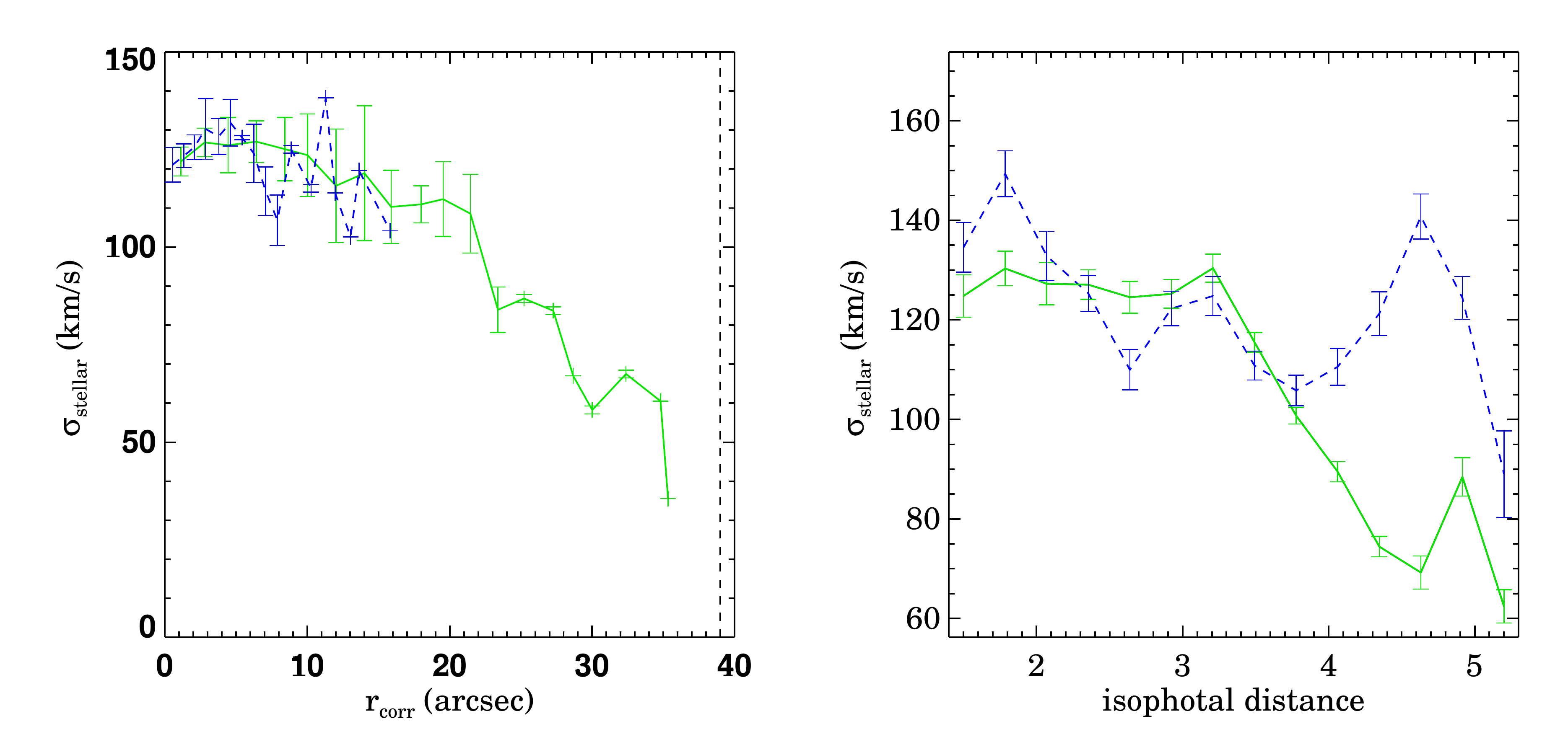}
\caption{Figure \ref{fig:gasvel} continued.}
\label{fig:gassig16}
\end{figure*}

We show maps of the stellar and ionised-gas kinematics for the entire BaLROG 
sample of galaxies in figures~\ref{fig:gasvel} to \ref{fig:gassig16}. In each figure
 we show different maps of each galaxy, top to bottom and left to right:  first row: (i) S$^4$G image of the 
 galaxy with an estimate of the final SAURON mosaic and the number
 of pointings indicated in the left lower corner, (ii) fundamental parameters of
 the galaxy along with the systemic velocity, inclination, stellar angular momentum within one effective radius and the bar strengths measured; second row: (i) surface brightness derived from the SAURON cube (collapsed in wavelength, shown in logarithmic scale), (ii) stellar mean velocity V (in km 
s$^{-1}$), (iii) stellar velocity 
dispersion $\sigma$ (in km s$^{-1}$); third row: (i) flux of the ionised gas, based on [O{\sc iii}] (shown as square-root-scaled), (ii) mean radial ionised gas velocity, (iii) ionised gas velocity dispersion (in  
km s$^{-1}$); fourth row: (i) Gauss-Hermite moments h3 and (ii) h4; fifth row: (i) major and minor axis rotation curves of the stellar velocity, (ii) radial profile (inclination corrected) of the stellar velocity dispersion of the major and minor axis of the bar, (iii) isphotal profile of the stellar velocity dispersion along bar major and minor axis. The cut levels are 
indicated in a box on the right-hand side of each map.


\label{lastpage}

\end{document}